\documentclass[a4paper,12pt]{article}
\usepackage[utf8]{inputenc}
\usepackage[english]{babel}
\usepackage{amsmath, amssymb,graphicx}
\usepackage{braket}
\usepackage{caption}
\usepackage{subcaption}
\usepackage{ulem}
\usepackage{multicol}
\usepackage{textcomp}
\usepackage{multirow}

\pdfoutput=1
\usepackage{jheppub}

\newcommand{\be}{\begin{equation}}
\newcommand{\ee}{\end{equation}}
\newcommand{\bea}{\begin{eqnarray}}
\newcommand{\eea}{\end{eqnarray}}

\newcommand{\fb}{\mathfrak{b}}

\newcommand{\fg}{\mathfrak{g}}

\newcommand{\cA}{\cal A}
\newcommand{\cL}{\cal L}

\numberwithin{equation}{section}

\title{Phase Transition of Anisotropic Hot Dense QGP in Magnetic
  Field: \\ $z^5$-term Holography for Heavy Quarks}

\author{
Kristina Rannu$^a$
}

\affiliation{
  $^a$Peoples Friendship University of Russia, Miklukho-Maklaya
  str. 6, 117198, \\ Moscow, Russia
}

\emailAdd{rannu-ka@rudn.ru}

\abstract{We present a five-dimensional twice anisotropic holographic
  model for heavy quarks supported by Einstein-dilaton-three-Maxwell
  action. A special feature of the model is the presence of $z^5$-term
  in the metric strain coefficient (warp factor). It's influence on
  the model properties, mainly on the confinement/deconfinement phase
  transition, is considered. Conditions for the direct magnetic
  catalysis are found and the corresponding phase diagrams are
  discussed.
}


\begin{document}

\maketitle



\section{Introduction}\label{introduction}

Theoretical modelling of processes in quark-gluon plasma (QGP) is one
of the most vital and urgent problems of quantum chromodynamics (QCD)
and high-energy physics in general. Comprehension of QGP phase
structure is a step necessary for developing our physical world
picture.

To describe the QGP behavior, first of all its
confinement/deconfinement phase transition, we use a holographic
approach to potential reconstruction method within the AdS$_5$/CFT$_4$
duality \cite{Casalderrey-Solana:2011dxg, Arefeva:2014kyw,
  Arefeva:2021kku}. Wherein all the important QGP properties should be 
encoded in the 5D Lagrangian and the metric serving as an ansatz for
the corresponding Einstein equations of motion (EOM) solution. This
method called a bottom-up approach is broadly applied in high-energy
physics for QGP studies \cite{Erlich:2005qh, Karch:2006pv,
  Gursoy:2008bu, Gursoy:2007cb, Gursoy:2007er, Gursoy:2009jd,
  Andreev:2006nw, Gursoy:2008za, Gursoy:2010fj, Colangelo:2011sr,
  Arefeva:2019yzy, Hajilou:2021wmz, Li:2013oda, Li:2012ay, Mia:2010zu,
  Dudal:2018ztm, Dudal:2017max, Li:2014dsa, Yang:2015aia,
  Chelabi:2015cwn, Fang:2015ytf, Arefeva:2018jyu, Fang:2018axm,
  Chen:2019rez, He:2013qq, He:2010ye, Chen:2018msc} and was used in
the previous works of our current wide research on holographic QGP
(HQGP) \cite{AR-2018, ARS-2019, ARS-2019qfthep, Arefeva:2019vov,
  Arefeva:2019dvl, Arefeva:2019qen, APS, Arefeva:2020aan,
  ARS-Light-2020, Arefeva:2020vhf, ARS-Heavy-2020, Arefeva:2020bjk,
  Arefeva:2021mag, Arefeva:2021jpa, Arefeva:2021btm, Arefeva:2021kku,
  Arefeva:2022avn, Arefeva:2022bhx, Rannu:2022fxw, ARS-Light-2022,
  ARS-Heavy-2023, Arefeva:2023ter, Arefeva:2023fky,
  Arefeva:2023uji}. This consideration is its further development.
 
It is approved that QGP produced in heavy-ion collisions (HIC) is
anisotropic for $1$--$5$ fm/$c \sim 10^{-24}$ s
\cite{Strickland:2013uga} and exists in strong magnetic field, $eB
\sim 0.3$ GeV$^2$ \cite{Skokov:2009qp, Voronyuk:2011jd,Bzdak:2011yy,
  Deng:2012pc}. Therefore we include sources of two different
anisotropies into the model ansatz: primary anisotropy, representing
spacial anisotropy of QGP, and anisotropy associated with magnetic
field. We also focus on respectively large values of chemical
potential in our consideration, as our research is mainly oriented to
future experiments on heavy-ion collisions at high values of baryon
density such as NICA project \cite{Arefeva:2016rob}.

The picture of confinement/deconfinement (phase diagram) actually
consists of a combination of two different phase transition types. The
1-st order phase transition originates from the thermodynamical
properties of the solution obtained, namely the behavior of free
energy as a temperature function, and temporal Wilson loops
determining the so called crossover. Both phase transition types are
sensible to the quarks mass \cite{Brown:1990ev, Philipsen:2016hkv},
with heavy \cite{AR-2018, ARS-2019, ARS-Heavy-2020, ARS-Heavy-2023}
and light quarks \cite{ARS-Light-2020, ARS-Light-2022} reacting to
magnetic field in different way. Magnetic field also affects heavy and
light quarks differently. According to Lattice calculations
\cite{DElia:2010abb, DElia:2018xwo, Bali:2012zg} the magnetic
catalysis (MC), i.e. the rise of the phase transition temperature with
the magnetic field increasing, should take place for heavy quarks,
while for light quarks, on the contrary, the phase transition
temperature drop with magnetic field increasing, i.e. inverse magnetic
catalysis (IMC), is expected.

The quark mass is included into the model effectively via the warp
factor exponent. In this work we try to add the $z^5$-term and
investigate it's effect on the holographic solution properties and the
phase diagram features. Such an expansion was proposed in
\cite{Bohra:2020qom} as a promising way to preserve such aspects of
QCD in a strong magnetic field as the string tension, the entanglement
entropy etc. Our important goal is to improve the magnetic behavior of
our previous heavy quarks model \cite{ARS-Heavy-2020}.

This paper is organised as follows. In Sect.\ref{model} the 5-dim
holographic model of hot dense anisotropic QCD in magnetic field is
presented and the corresponding 5-dim BH solution reconstructing heavy
quarks model is obtained. In Sect.\ref{thermodynamics} thermodynamic
properties of the model in magnetic field are considered and the 1-st
order phase transition features are revealed. Sect.\ref{phase}
temporal Wilson loops are calculated and the resulting phase diagram
is composed. The main conclusions of this investigation and subjects
for further research are given in Sect.\ref{conclusions}. Plots
showing the solution behavior are stored in
Appendix~\ref{appendixA}. Temperature and free energy dependence on
chemical potential and the magnetic field strength are presented on
plots in Appendix~\ref{appendixB}.

\section{Model}\label{model}

We consider the metric ansatz and the Lagrangian in Einstein frame
used in \cite{ARS-Heavy-2020}: 
\begin{gather}
  ds^2 = \cfrac{L^2}{z^2} \, \fb(z) \left[
    - \, g(z) \, dt^2 + dx_1^2 
    + \left( \cfrac{z}{L} \right)^{2-\frac{2}{\nu}} dx_2^2
    + e^{c_B z^2} \left( \cfrac{z}{L} \right)^{2-\frac{2}{\nu}} dx_3^2
    + \cfrac{dz^2}{g(z)} \right] \! , \label{eq:5.01} \\
  {\cL} = \sqrt{-g} \left[ R 
    - \cfrac{f_0(\phi)}{4} \, F_0^2 
    - \cfrac{f_1(\phi)}{4} \, F_1^2
    - \cfrac{f_3(\phi)}{4} \, F_3^2
    - \cfrac{1}{2} \, \partial_{\mu} \phi \, \partial^{\mu} \phi
    - V(\phi) \right], \label{eq:5.02}
\end{gather}  
\begin{gather}
  \begin{split}
    \phi &= \phi(z), \\
    \mbox{Electric anzats $F_0$:} \quad
    A_0 &= A_t(z), \quad A_{i = 1,2,3,4} = 0, \\
    \mbox{Magnetic anzats $F_k$:} \quad
    F_1 &= q_1 \, dx^2 \wedge dx^3, \quad 
    F_3 = q_3 \, dx^1 \wedge dx^2,
  \end{split}\label{eq:5.03}
\end{gather}
where $\nu$ and $c_B$ parametrize spacial and magnetic field
anisotropies respectively, $\fb(z)$ is the warp factor, $\phi(z)$ is
the scalar field, $f_0(\phi)$, $f_1(\phi)$ and $f_3(\phi)$ are the
coupling functions associated with the Maxwell fields $A_{\mu}$, $F_1$
and $F_3$ correspondingly, $q_1$ and $q_3$ are constant ``charges''
and $V(\phi)$ is the scalar field potential.

Varying Largrangian (\ref{eq:5.02}) over metric (\ref{eq:5.01}), the
scalar field and vector potential we get the following EOM:
\begin{gather}
  \begin{split}
    \phi'' &+ \phi' \left(
      \cfrac{g'}{g} + \cfrac{3 \fb'}{2 \fb} 
      - \cfrac{\nu + 2}{\nu z} + c_B z
    \right)
    + \left( \cfrac{z}{L} \right)^2 \cfrac{(A_t')^2}{2 \fb g} \ 
    \cfrac{\partial f_0}{\partial \phi} \ - \\
    &- \left( \cfrac{L}{z} \right)^{2-\frac{4}{\nu}} 
    \cfrac{e^{-c_Bz^2} \, q_1^2}{2 \fb g} \ 
    \cfrac{\partial f_1}{\partial \phi} 
    - \left( \cfrac{z}{L} \right)^{\frac{2}{\nu}} 
    \cfrac{q_3^2}{2 \fb g} \ \cfrac{\partial f_3}{\partial \phi}
    - \left( \cfrac{L}{z} \right)^2 \cfrac{\fb}{g} \,
    \cfrac{\partial V}{\partial \phi} = 0,
  \end{split}\label{eq:5.09} \\
  A_t'' + A_t' \left(
    \cfrac{\fb'}{2 \fb} + \cfrac{f_0'}{f_0} 
    + \cfrac{\nu - 2}{\nu z} + c_B z
  \right) = 0, \label{eq:5.10} \\
  g'' + g' \left(
    \cfrac{3 \fb'}{2 \fb} - \cfrac{\nu + 2}{\nu z} - c_B z
  \right) 
  - 2 g \left(
    \cfrac{3 \fb'}{2 \fb} - \cfrac{2}{\nu z} + c_B z
  \right) c_B z
  - \left( \cfrac{z}{L} \right)^2 \cfrac{f_0 (A_t')^2}{\fb} 
  = 0, \label{eq:5.11} \\
  \fb'' - \cfrac{3 (\fb')^2}{2 \fb} + \cfrac{2 \fb'}{z}
  - \cfrac{4 \fb}{3 \nu z^2} \left(
    1 - \cfrac{1}{\nu}
    + \left( 1 - \cfrac{3 \nu}{2} \right) c_B z^2
    - \cfrac{\nu c_B^2 z^4}{2}
  \right)
  + \cfrac{\fb \, (\phi')^2}{3} = 0, \label{eq:5.12} \\
  2 g' \left( 1 - \cfrac{1}{\nu} \right) 
  + 3 g \left( 1 - \cfrac{1}{\nu} \right) \left(
    \cfrac{\fb'}{\fb} - \cfrac{4 \left( \nu + 1 \right)}{3 \nu z}
    + \cfrac{2 c_B z}{3}
  \right)
  + \left( \cfrac{L}{z} \right)^{1-\frac{4}{\nu}} 
  \cfrac{L \, e^{-c_Bz^2} q_1^2 \, f_1}{\fb} = 0, \label{eq:5.13} \\
  \begin{split}
    2 g' \left( 1 - \cfrac{1}{\nu} + c_B z^2 \right) 
    &+ 3 g \left[ \Big( 1 - \cfrac{1}{\nu} + c_B z^2 \Big) 
      \left(
        \cfrac{\fb'}{\fb} - \cfrac{4}{3 \nu z} + \cfrac{2 c_B z}{3}
      \right)
      - \cfrac{4 \left( \nu - 1 \right)}{3 \nu z} \right] + \\
    &+ \left( \cfrac{L}{z} \right)^{1-\frac{4}{\nu}}
    \cfrac{L \, e^{-c_Bz^2} q_1^2 \, f_1}{\fb}
    - \left( \cfrac{z}{L} \right)^{1+\frac{2}{\nu}}
    \cfrac{L \, q_3^2 \, f_3}{\fb} = 0,
  \end{split}\label{eq:5.14} \\
  \begin{split}
    \cfrac{\fb''}{\fb} + \cfrac{(\fb')^2}{2 \fb^2}
    + \cfrac{3 \fb'}{\fb} \left(
      \cfrac{g'}{2 g} - \cfrac{\nu + 1}{\nu z} + \cfrac{2 c_B z}{3}
    \right)
    &- \cfrac{g'}{3 z g} \left( 5 + \cfrac{4}{\nu} - 3 c_B z^2 \right)
    + \cfrac{8}{3 z^2}
    \left( 1 + \cfrac{3}{2 \nu} + \cfrac{1}{2\nu^2} \right) - \\
    &- \cfrac{4 c_B}{3}
    \left( 1 + \cfrac{3}{2 \nu} - \cfrac{c_B z^2}{2} \right)
    + \cfrac{g''}{3 g} 
    + \cfrac{2}{3} \left( \cfrac{L}{z} \right)^2 \cfrac{\fb V}{g} 
    = 0.
  \end{split}\label{eq:5.15}
\end{gather}
Subtracting (\ref{eq:5.14}) from (\ref{eq:5.13}) we get the expression
for the coupling function $f_3$:
\begin{gather}
  f_{3} = 2 \left( \cfrac{L}{z} \right)^{\frac{2}{\nu}} \fb g \
  \cfrac{c_B z}{q_3^2} \left(
    \cfrac{g'}{g} + \cfrac{3 \fb'}{2 \fb} - \cfrac{2}{\nu z} + c_B z
  \right). \label{eq:5.16}
\end{gather}

To set a correspondence to Bohra et al. solution \cite{Bohra:2020qom} 
we take the warp factor and the coupling function for the electric
Maxwell field $A_t$ as
\begin{gather}
  \fb(z) = e^{2{\cA}(z)} =  e^{ - \, 2 a z^2
    \, - \, 2 d q_3^2 z^5}, \label{eq:5.18} \\
  f_0 = e^{-(c+q_3^2)z^2} \,
  \cfrac{z^{-2+\frac{2}{\nu}}}{\sqrt{\fb}} \label{eq:5.22}
\end{gather}
with $a = 0.15$ and $c = 1.16$. The $z^5$-term coefficient $d$
serves as the model parameter whose effect on the solution properties
is investigated.

\begin{figure}[t!]
  \centering
  \includegraphics[scale=0.5]{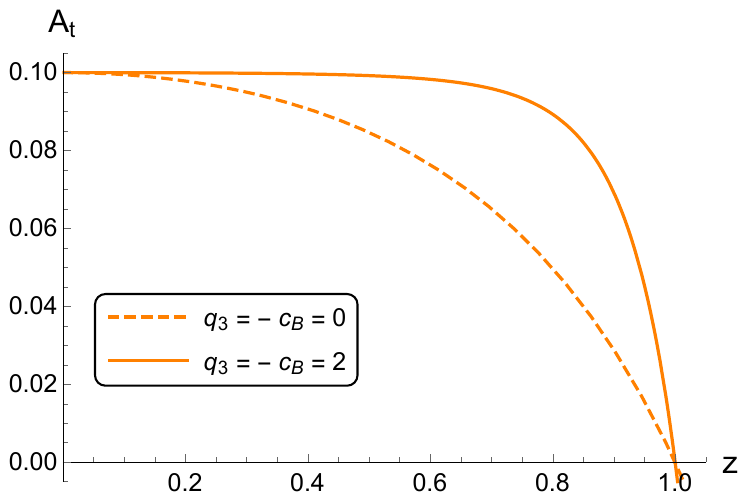} \quad
  \includegraphics[scale=0.5]{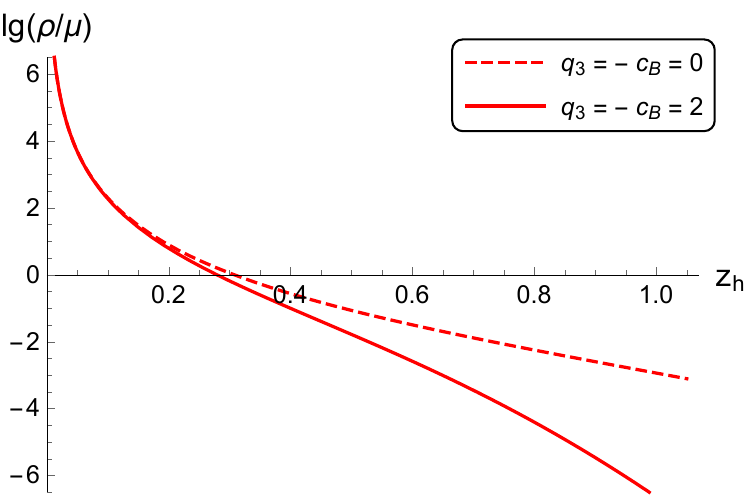} \\
  A \hspace{200pt} B
  \caption{Electric potential $A_t(z)$ (A) and density $\rho(z_h)/\mu$
    in logarithmic scale (B) for the ``heavy quarks'' with (solid
    lines) and without (dashed lines) magnetic field, $q_3 = - \, c_B
    = 0, \, 2$ (B); $c = 1.16$, $z_h = 1$.}
  \label{Fig:Atrho-z5}
\end{figure}

We solve system (\ref{eq:5.09})--(\ref{eq:5.16}) with usual boundary
conditions
\begin{gather}
  A_t(0) = \mu, \quad A_t(z_h) = 0, \label{eq:5.19} \\
  g(0) = 1, \quad g(z_h) = 0, \label{eq:5.20} \\
  \phi(z_0) = 0, \label{eq:5.21}
\end{gather}
where $z_0$ serves to fit the string tension behavior
\cite{ARS-Light-2020}, and get
\begin{gather}
  A_t (z) = \mu \, \cfrac{e^{(c-\frac{c_B}{2}+q_3^2)z^2} 
    - e^{(c-\frac{c_B}{2}+q_3^2)z_h^2}}{1 -
    e^{(c-\frac{c_B}{2}+q_3^2)z_h^2}}, \label{eq:5.23} \\
  g(z) = e^{c_B z^2} \left[ 1 - \cfrac{I_1(z)}{I_1(z_h)}
    + \cfrac{\mu^2 \bigl( 2 c - c_B + 2 q_3^2 \bigr) I_2(z)}{L^2
      \left(1 - e^{(c-\frac{c_B}{2}+q_3^2) z_h^2} \right)^2}
    \left( 1 - \cfrac{I_1(z)}{I_1(z_h)} \,
      \cfrac{I_2(z_h)}{I_2(z)} \right) \right], \label{eq:5.26} \\
  I_1(z) = \int_0^z e^{3\left(a-\frac{c_B}{2}+dq_3^2\xi^3\right)\xi^2}
  \xi^{1+\frac{2}{\nu}} \, d \xi, \qquad 
  I_2(z) = \int_0^z
  e^{3\left(a+\frac{c-2c_B+q_3^2}{2}+dq_3^2\xi^3\right)\xi^2}
  \xi^{1+\frac{2}{\nu}} \, d \xi, \label{eq:5.27}
\end{gather}
\begin{gather}
  \begin{split}
    \phi(z) = \int_{z_0}^z &\Biggl(
    \cfrac{4 (\nu - 1)}{\nu^2 \xi^2} 
    + 36a - 2 c_B \left( 3 - \cfrac{2}{\nu} \right)
    + 2 \left( 12 a^2 - c_B^2 \right) \xi^2 + \\
    &\qquad \qquad + 180 d q_3^2 \xi^3
    + 120 a d q_3^2 \xi^5
    + 150 d^2 q_3^4 \xi^8 \Biggr)^{1/2} d \xi,
  \end{split} \label{eq:5.32} \\
  \begin{split}
    f_1(\phi(z)) &= 4 \left( \cfrac{z}{L} \right)^{2-\frac{4}{\nu}}
    e^{-(2a-c_B)z^2-2d q_3^2 z^5} \, \cfrac{\nu - 1}{q_1^2 \nu z} \,
    \left\{
      \left( \cfrac{\nu + 1}{\nu z}
        + \cfrac{3}{2} \ (2 a - c_B) z
        + \cfrac{15}{2} \ d q_3^2 z^4 \right) \times \right. \\
    &\times \left[ 1 - \cfrac{I_1(z)}{I_1(z_h)}
      - \cfrac{\mu^2 (2 c - c_B + 2 q_3^2)}{L^2 \left(
          1 - e^{\left(c-\frac{c_B}{2}+q_3^2\right)z_h^2} \right)^2}
      \ \cfrac{I_1(z) I_2(z_h) - I_1(z_h) I_2(z)}{I_1(z_h)}
    \right] + \\
    &+ \left. \cfrac{e^{3\left(a-\frac{c_B}{2}\right)z^2+3d q_3^2 z^5}
        z^{1+\frac{2}{\nu}}}{2 I_1(z_h)}
      \left[ 1 - \cfrac{\mu^2 (2 c - c_B + 2 q_3^2)}{L^2 \left(
            1 - e^{\left(c-\frac{c_B}{2}+q_3^2\right)z_h^2} \right)^2}
        \left( e^{\left(c-\frac{c_B}{2}+q_3^2\right)z^2} I_1(z_h) - I_2
          (z_h) \right) \right] \right\},
  \end{split} \label{eq:5.33} \\
  \begin{split}
    f_3(\phi(z)) &= - \, 2 \left( \cfrac{z}{L} \right)^{-\frac{2}{\nu}}
    e^{-(2a-c_B)z^2-2d q_3^2 z^5} \, \cfrac{c_B}{q_3^2} \,
    \left\{ \left( \cfrac{2}{\nu}
        + 3 \, (2 a - c_B) z^2
        + 15 \, d q_3^2 z^5 \right) \times \right. \\
    &\times \left[ 1 - \cfrac{I_1(z)}{I_1(z_h)}
      - \cfrac{\mu^2 (2 c - c_B + 2 q_3^2)}{L^2 \left(
          1 - e^{\left(c-\frac{c_B}{2}+q_3^2\right)z_h^2} \right)^2}
        \ \cfrac{I_1(z) I_2(z_h) - I_1(z_h) I_2(z)}{I_1(z_h)}
      \right] + \\
      &+ \left. \cfrac{e^{3\left(a-\frac{c_B}{2}\right)z^2+3d q_3^2 z^5} 
          z^{2+\frac{2}{\nu}}}{I_1(z_h)}
      \left[ 1 - \cfrac{\mu^2 (2 c - c_B + 2 q_3^2)}{L^2 \left(
            1 - e^{\left(c-\frac{c_B}{2}+q_3^2\right)z_h^2} \right)^2}
        \left( e^{\left(c-\frac{c_B}{2}+q_3^2\right)z^2} I_1(z_h) - I_2
          (z_h) \right) \right] \right\},
  \end{split} \label{eq:5.34} \\
  \begin{split}
    V(\phi(z)) = &-  \cfrac{e^{(2a-c_B)z^2+2d q_3^2 z^5}}{2
      L^2 \nu^2} 
    \left\{ 2 \left( 2 + 6 \nu + 4 \nu^2 + \bigl( 
        6 a (3 + 2 \nu) - (7 + 6 \nu) c_B \bigr) \nu z^2 -
      \right. \right. \\
    &- \left. \cfrac{c_B^2 \nu^2 z^4}{4}
      + 15 (3 - \nu) d q_3^2 \nu z^5 
      + \left( 6 a - \cfrac{5}{2} \, c_B + 15 d q_3^2 z^3
      \right)^2 \nu^2 z^4 \right) \times \\
    &\times \left[ 1 - \cfrac{I_1(z)}{I_1(z_h)}
      - \cfrac{\mu^2 (2 c - c_B + 2 q_3^2)}{L^2 \left(
          1 - e^{\left(c-\frac{c_B}{2}+q_3^2\right)z_h^2} \right)^2}
      \ \cfrac{I_1(z) I_2(z_h) - I_1(z_h) I_2(z)}{I_1(z_h)}
    \right] + \\
    &+ 2 \left( 1 + 2 \nu + 2 (3 a - c_B) \nu z^2 
      + 15 d q_3^2 \nu z^5 \right)
    \cfrac{e^{3\left(a-\frac{c_B}{2}\right)z^2+3d q_3^2 z^5} 
      \nu z^{2+\frac{2}{\nu}}}{I_1(z_h)} \ \times \\
    &\times \left[ 1 - \cfrac{\mu^2 (2 c - c_B +
        2 q_3^2)}{L^2 \left( 1 -
          e^{\left(c-\frac{c_B}{2}+q_3^2\right)z_h^2} \right)^2}
      \left( e^{\left(c-\frac{c_B}{2}+q_3^2\right)z^2} I_1(z_h) - I_2
        (z_h) \right) \right] + \\
    &+ \left. \cfrac{\mu^2 (2 c - c_B + 2 q_3^2)^2
        e^{\left(3a+c-2c_B+q_3^2\right)z^2+3d q_3^2 z^5} \nu^2
        z^{4+\frac{2}{\nu}}}{L^2 \left(
          1 - e^{\left(c-\frac{c_B}{2}+q_3^2\right)z_h^2} \right)^2}
    \right\}.
  \end{split} \label{eq:5.35}
\end{gather}

Density is the coefficient in $A_t$ expansion:
\begin{gather}
  A_t(z) = \mu - \rho \,z^2 + \dots \ \Longrightarrow \
  \rho = - \, \cfrac{\mu \bigl( 2 c - c_B + 2 q_3^2 \bigr)}{2
    \left(1 - e^{(c-\frac{c_B}{2}+q_3^2) z_h^2}
    \right)}. \label{eq:5.25}
\end{gather}
They both are rather stable relative to magneic field
(Fig.\ref{Fig:Atrho-z5}).

We have two constants describing magnetic field in this model: amount 
of the magnetic field source $q_3$ and the metric parameter $c_B$ for 
the magnetic field back-reaction to the 5D curvature. Generally these 
constants are independent from each other, so different forms of their 
connection can be possible within the model constructed. Let us now 
consider the simplest and most obvious set of this connection special 
cases and check which of them are physical.

Blackening function $g(z)$
(Fig.\ref{Fig:gz-q3cB-nu1-mu0-z5}--\ref{Fig:gz-q3cB-nu45-mu0-z5}),
coupling function $f_3\bigl(\phi(z)\bigr)$
(Fig.\ref{Fig:f3z-q3cB-nu1-mu0-z5}--\ref{Fig:f3z-q3cB-nu45-mu0-z5}),
scalar field $\phi(z)$
(Fig.\ref{Fig:phiz-q3cB-nu1-mu0-z5}--\ref{Fig:phiz-q3cB-nu45-mu0-z5})
and scalar potenial $V\bigl(\phi(z)\bigr)$
(Fig.\ref{Fig:Vphi-q3cB-nu1-mu0-z5}--\ref{Fig:Vphi-q3cB-nu45-mu0-z5})
are plotted in Appendix\ref{appendixA}. Their analysis is summarised
in Table.\ref{Tab:solution}, and the main conclusion is that $c_B < 0$
is required.

\begin{table}[h!]
  \begin{center}
    \begin{tabular}{|c|c|c|c|c|}
      \hline
      $d$ & $c_B$, $q_3$ & $g(z)$ & $f_3(z)$ & $V(z)$ \\
      \hline
      \multirow{4}{*}{$d = 0.06$} & $c_B = - \, 0.5$ \qquad $c_B =
      0.5$ & stable \ stable & NEC noNEC & stable \quad stable \\
      \ & $c_B < 0$ $c_B > 0$, $q_3 = 0.5$ & stable unstable & NEC
      noNEC & stable unstable \\
      \ & $q_3 = - \, c_B$ \qquad $q_3 = c_B$ & stable unstable &
      noNEC NEC & stable unstable \\
      \ & $q_3^2 = - \, c_B$ \qquad $q_3^2 = c_B$ & stable unstable &
      NEC noNEC & stable unstable \\
      \hline
      \multirow{4}{*}{$d = 0.01$} & $c_B = - \, 0.5$ \qquad $c_B =
      0.5$  & stable \ stable & NEC noNEC & stable \quad stable \\
      \ & $c_B < 0$ $c_B > 0$, $q_3 = 0.5$ & stable unstable & NEC
      noNEC & stable unstable \\
      \ & $q_3 = - \, c_B$ \qquad $q_3 = c_B$ & stable unstable & NEC
      noNEC & stable unstable \\
      \ & $q_3^2 = - \, c_B$ \qquad $q_3^2 = c_B$ & stable unstable &
      NEC noNEC & stable unstable \\
      \hline
    \end{tabular}
    \caption{Solution properties depending on magnetic field
      parameters $c_B$, $q_3$.}
    \label{Tab:solution}
  \end{center}
\end{table}


\section{Thermodynamics}\label{thermodynamics}

For metric (\ref{eq:5.01}) and the warp factor (\ref{eq:5.18})
temperature and entropy are:
\begin{gather}
  \begin{split}
    T = \cfrac{|g'|}{4 \pi} \, \Bigl|_{z=z_h} &= \left|
      - \, \cfrac{e^{(3a-\frac{c_B}{2})z_h^2+3dq_3^2 z_h^5} \,
        z_h^{1+\frac{2}{\nu}}}{4 \pi \, I_1(z_h)} \right. \times \\
    &\times \left. \left[ 
        1 - \cfrac{\mu^2 \bigl(2 c - c_B + 2 q_3^2 \bigr) 
          \left(e^{(c-\frac{c_B}{2}+q_3^2) z_h^2} \, I_1(z_h) - I_2(z_h)
          \right)}{L^2 \left(1 - e^{(c-\frac{c_B}{2}+q_3^2) z_h^2}
          \right)^2} \right] \right|, 
  \end{split} 
  \label{eq:5.30} \\
  s = \cfrac{1}{4} \left( \cfrac{L}{z_h} \right)^{1+\frac{2}{\nu}}
  e^{-(3a-\frac{c_B}{2})z_h^2-3dq_3^2 z_h^5}. \label{eq:5.31}
\end{gather}

\begin{figure}[t!]
  \centering
  \includegraphics[scale=0.23]{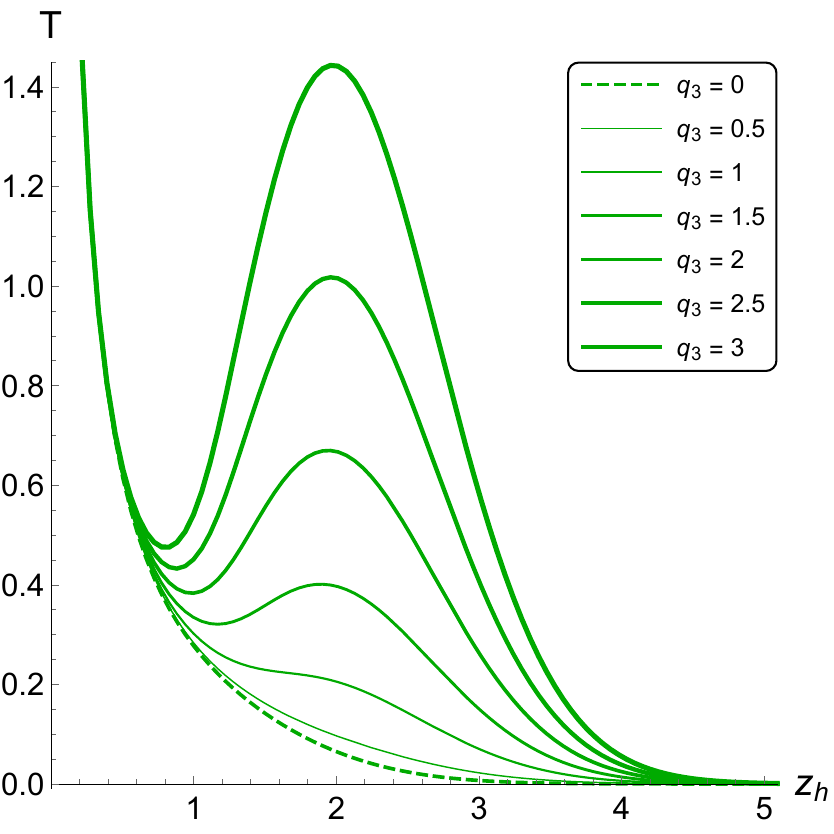}
  \quad
  \includegraphics[scale=0.23]{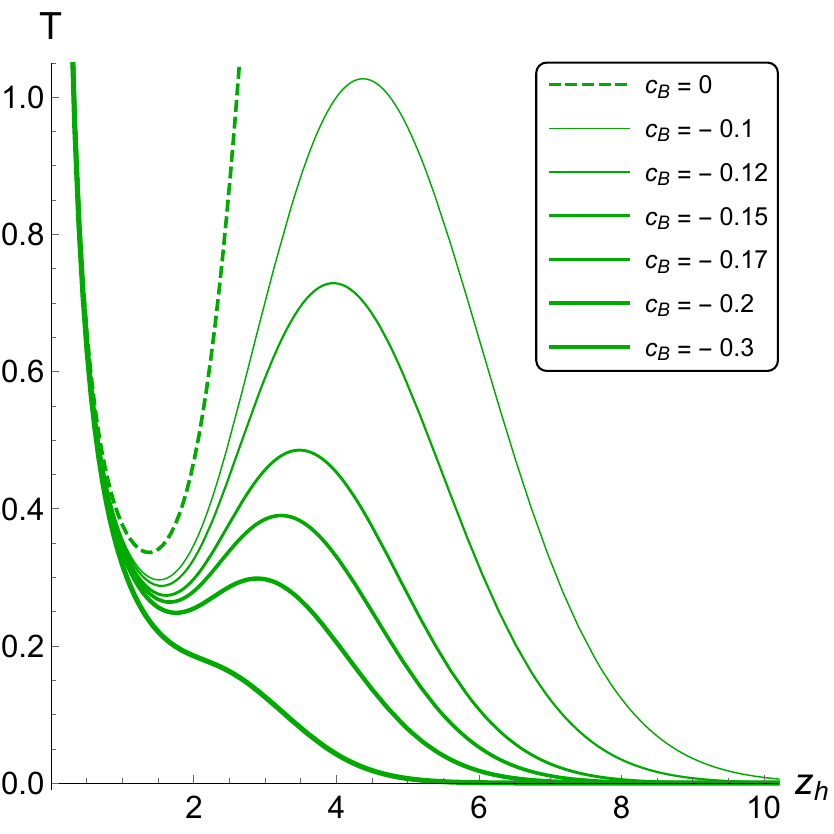}
  \quad
  \includegraphics[scale=0.23]{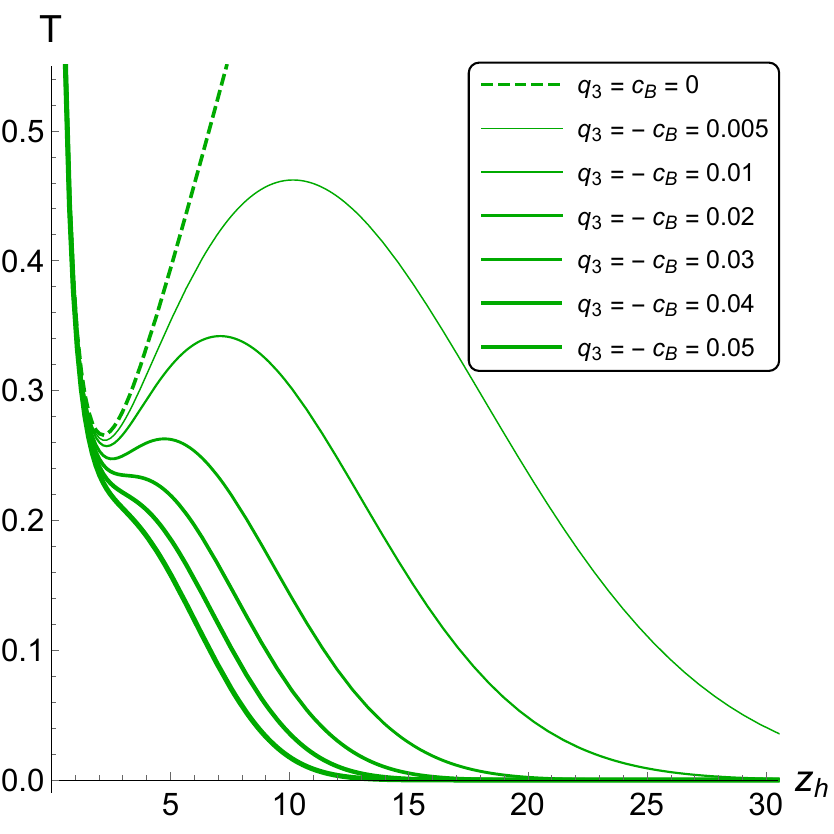}
  \quad
  \includegraphics[scale=0.23]{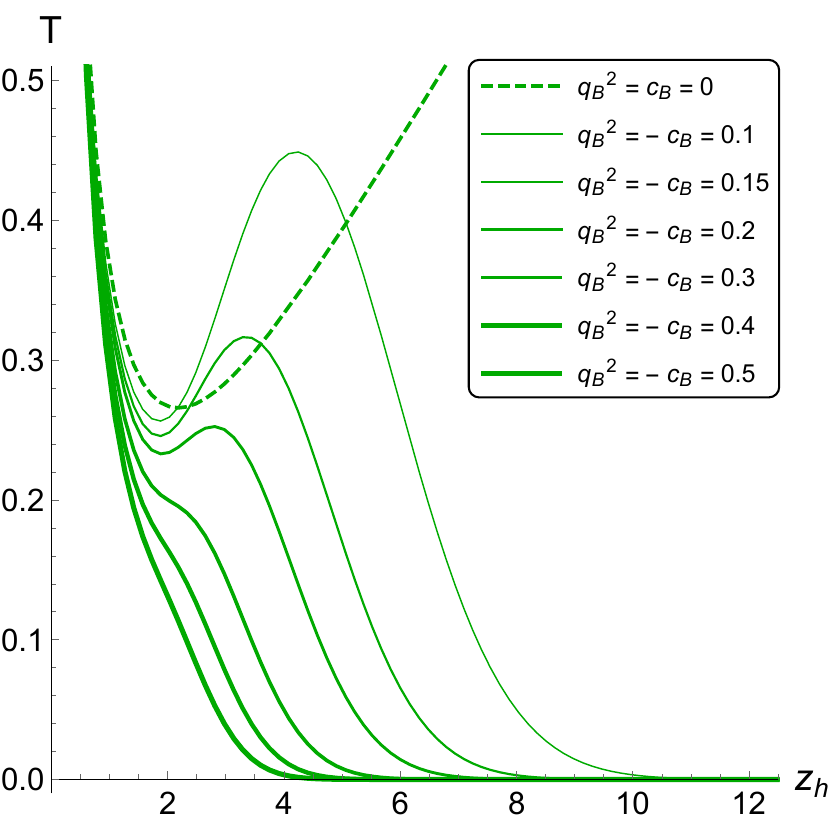} \\
  \includegraphics[scale=0.23]{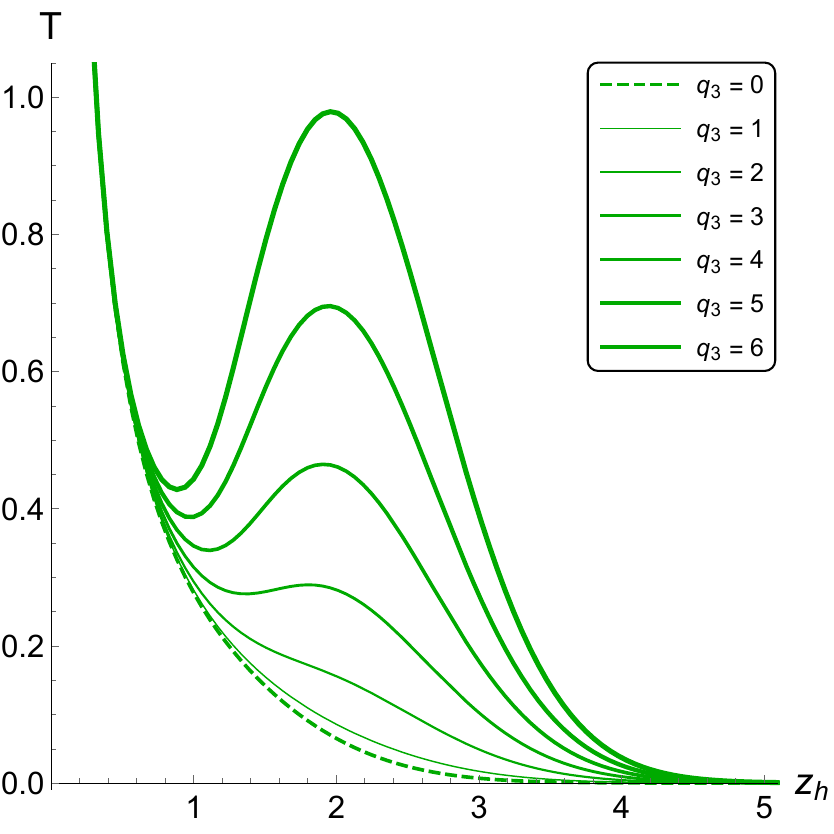}
  \quad
  \includegraphics[scale=0.23]{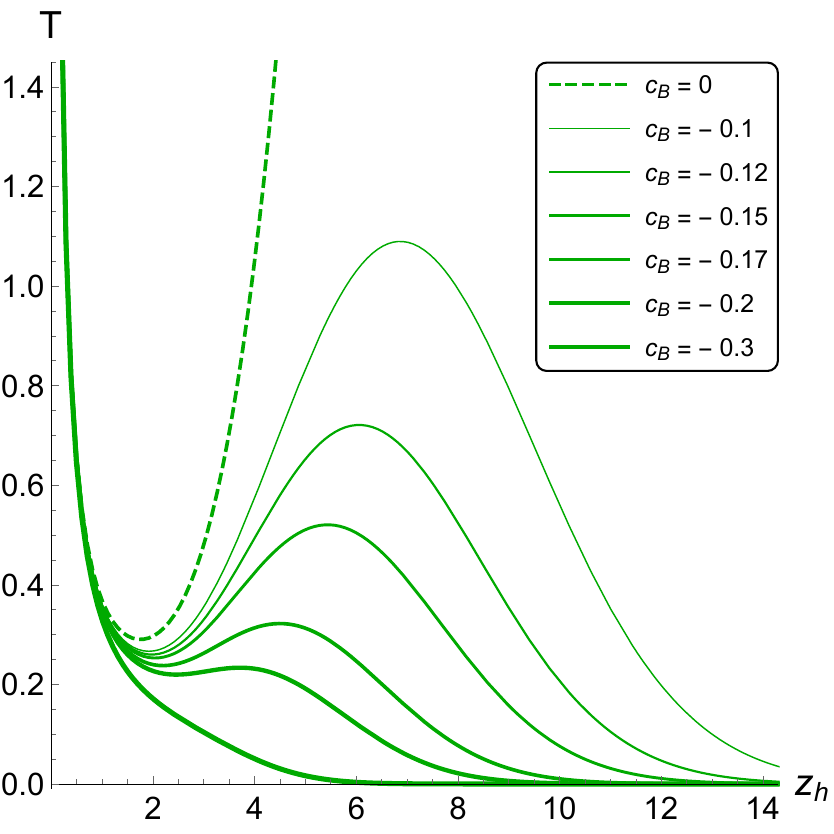}
  \quad
  \includegraphics[scale=0.23]{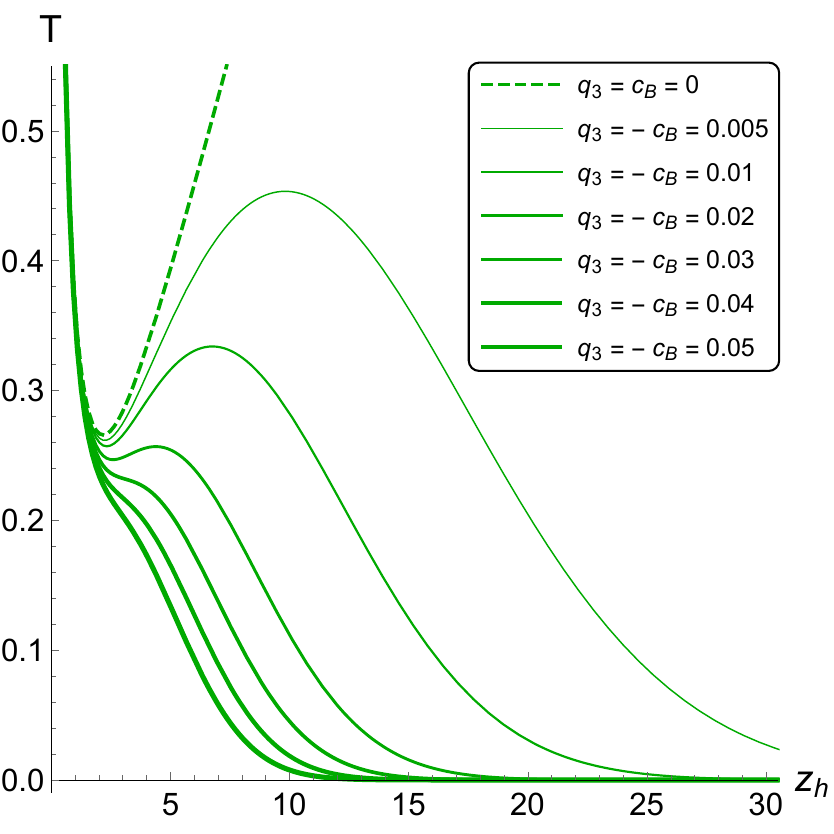}
  \quad
  \includegraphics[scale=0.23]{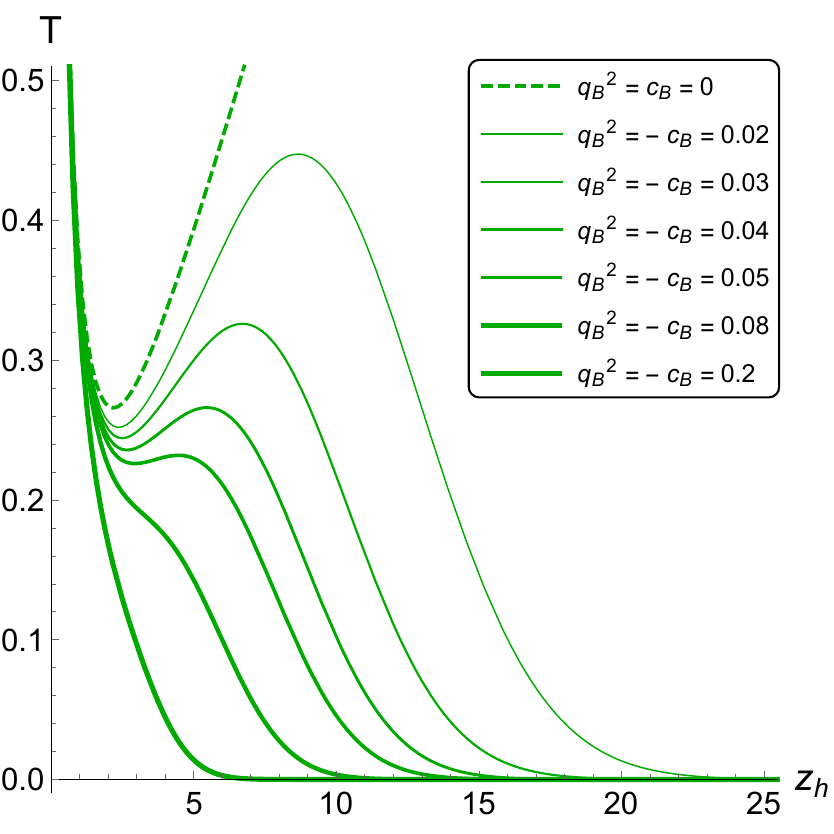} \\
  A \hspace{95pt} B \hspace{95pt} C \hspace{95pt} D
  \caption{Temperature $T(z_h)$ in magnetic field with different $q_3$
    (A) for $c_B = - \, 0.5$, with different $c_B$ for $q_3 = 0.5$ (B), 
    for different $q_3 = - \, c_B$ (C), for different $q_3^2 = - \,
    c_B$ (D) for $d = 0.06 > 0.05$ (1-st line) and $d = 0.01 < 0.05$
    (2-nd line) in primary isotropic case $\nu = 1$, $a = 0.15$, $c =
    1.16$, $\mu = 0$.}
  \label{Fig:Tzh-q3cB-nu1-mu0-z5}
\end{figure}
\begin{figure}[t!]
  \centering
  \includegraphics[scale=0.23]{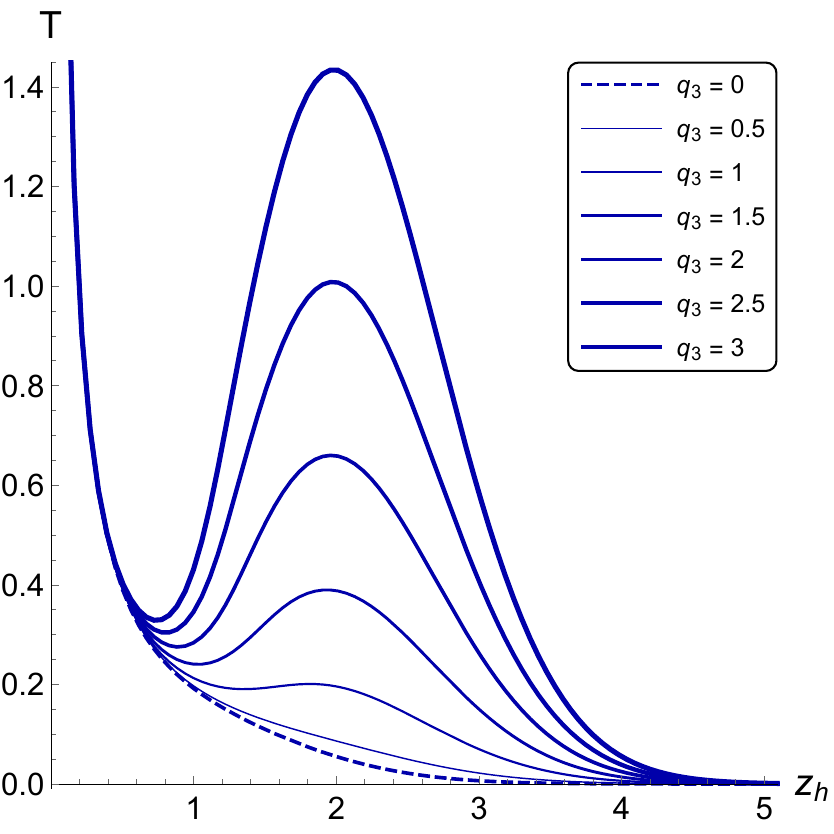}
  \quad
  \includegraphics[scale=0.23]{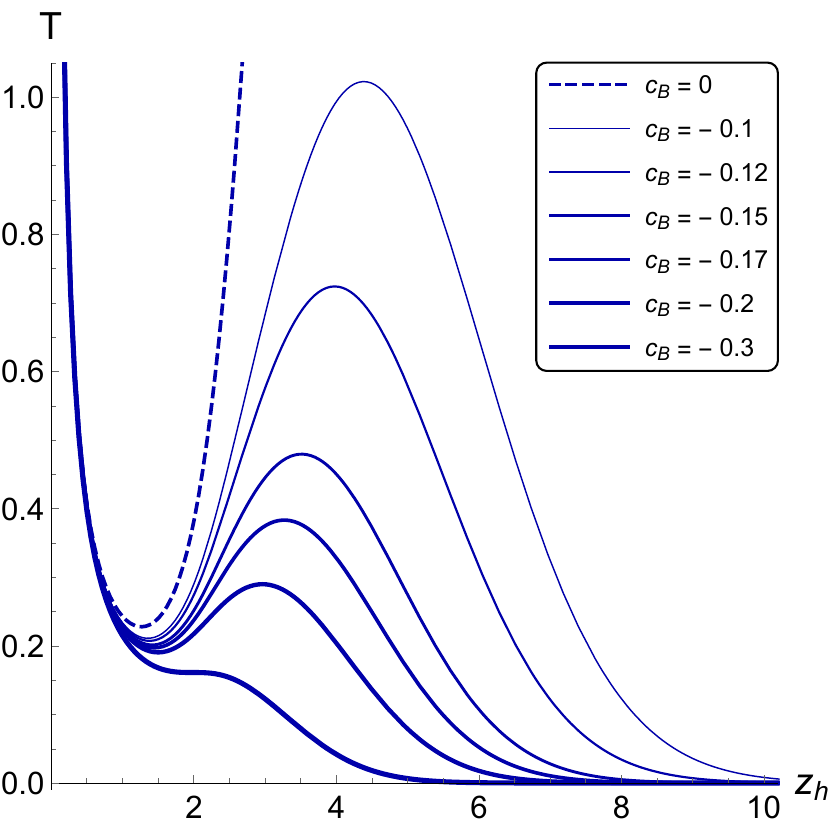}
  \quad
  \includegraphics[scale=0.23]{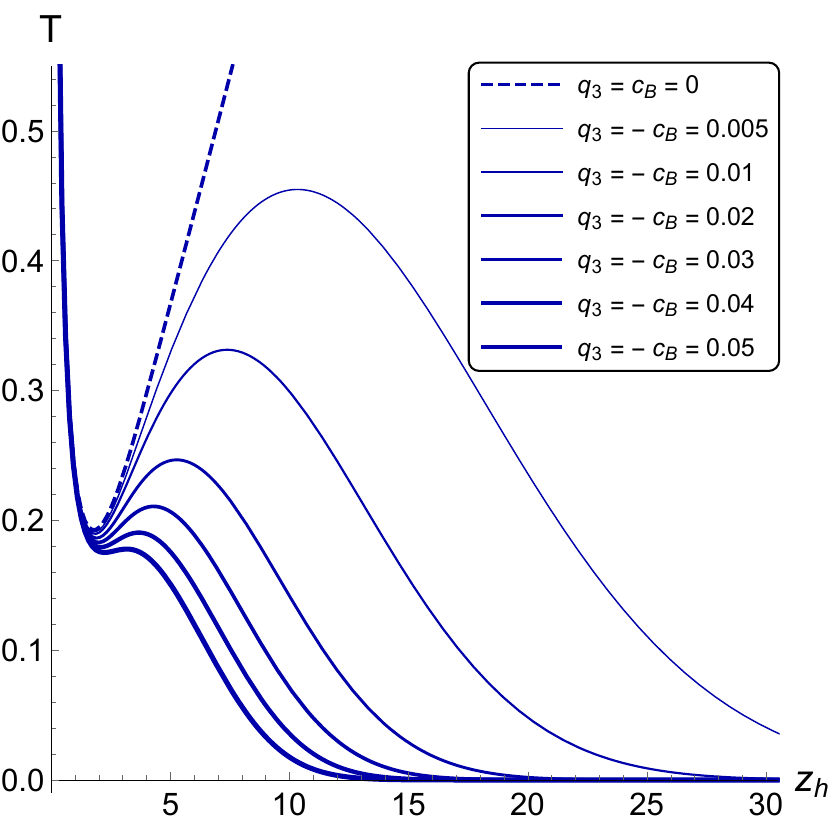}
  \quad
  \includegraphics[scale=0.23]{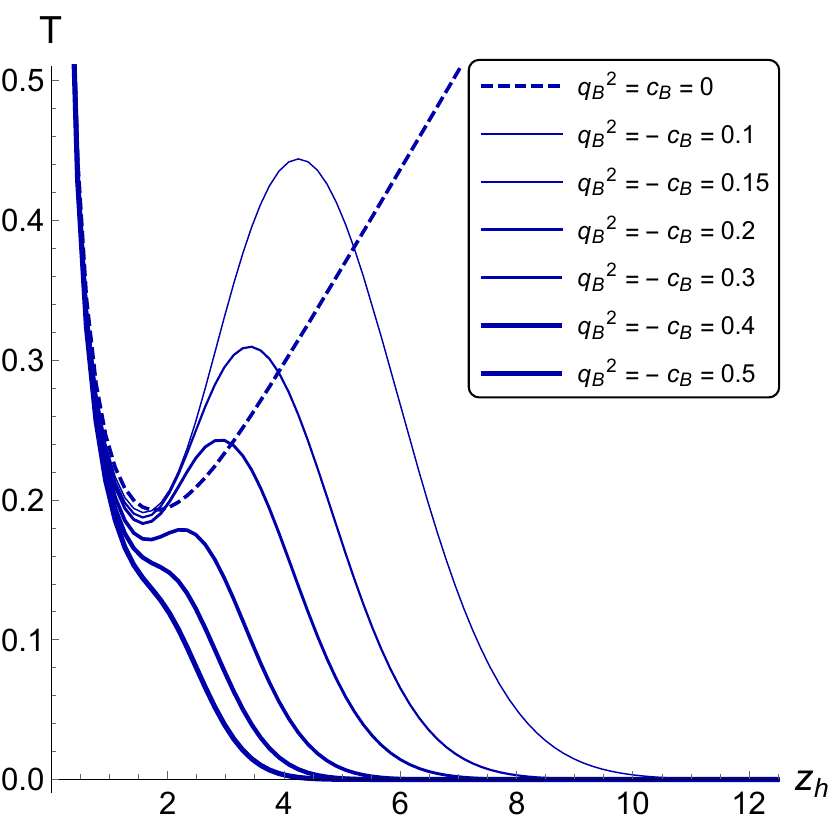} \\
  \includegraphics[scale=0.23]{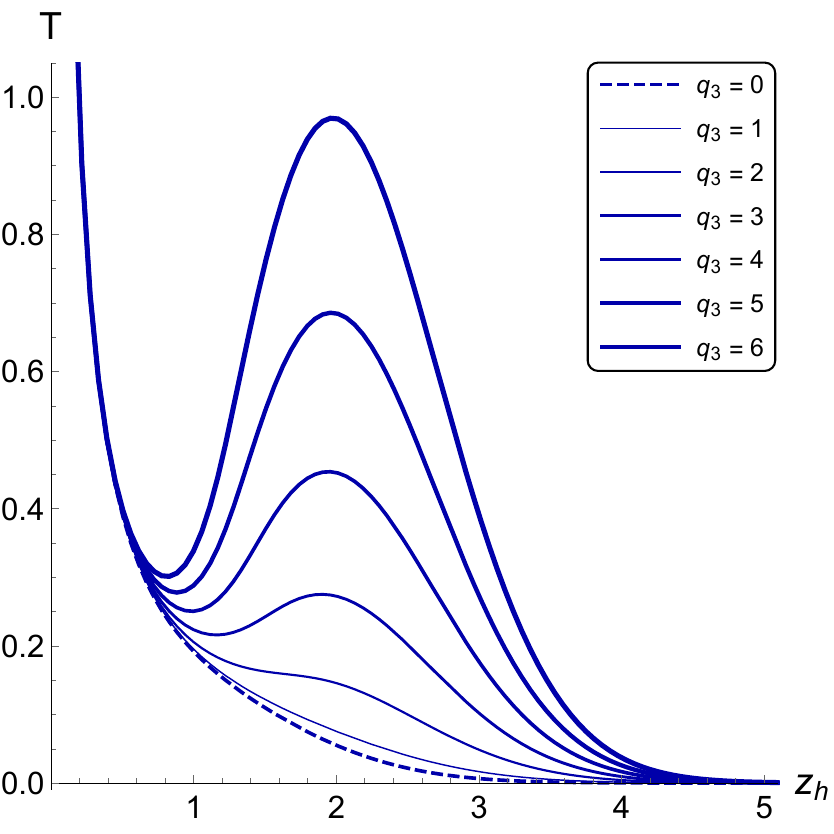}
  \quad
  \includegraphics[scale=0.23]{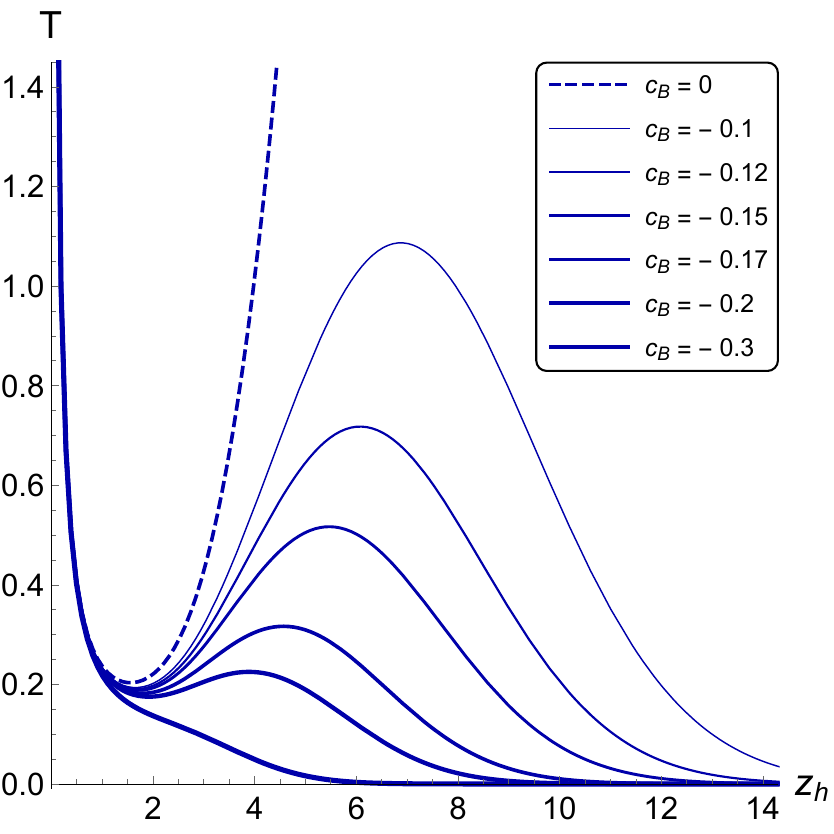}
  \quad
  \includegraphics[scale=0.23]{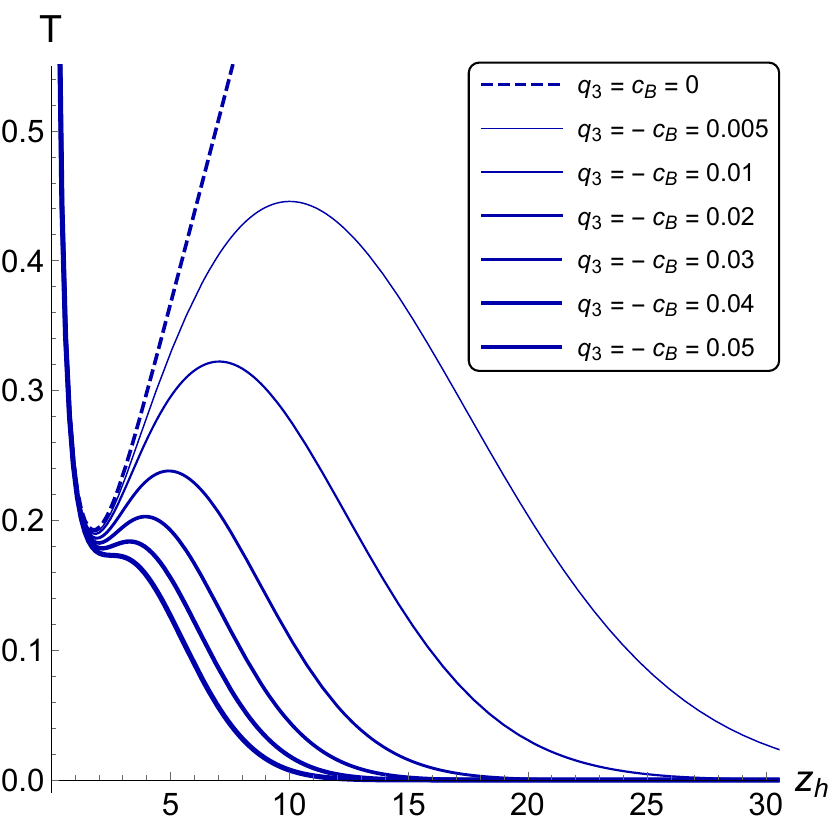}
  \quad
  \includegraphics[scale=0.23]{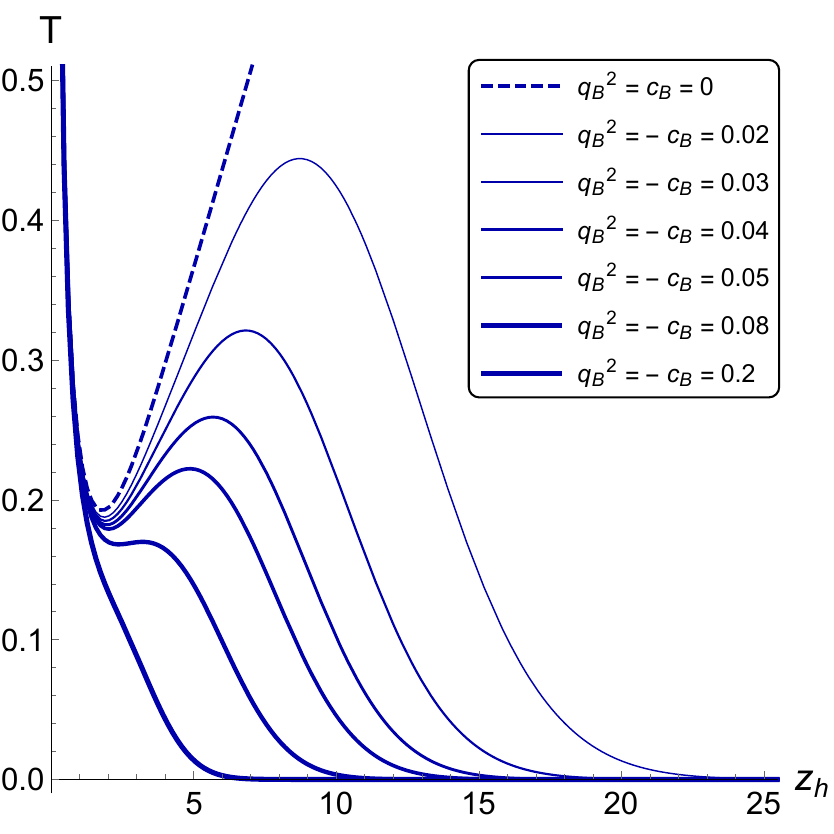} \\
  A \hspace{95pt} B \hspace{95pt} C \hspace{95pt} D
  \caption{Temperature $T(z_h)$ in magnetic field with different $q_3$
    (A) for $c_B = - \, 0.5$, with different $c_B$ for $q_3 = 0.5$ (B), 
    for different $q_3 = - \, c_B$ (C), for different $q_3^2 = - \,
    c_B$ (D) for $d = 0.06 > 0.05$ (1-st line) and $d = 0.01 < 0.05$
    (2-nd line) in primary anisotropic case $\nu = 4.5$, $a = 0.15$,
    $c = 1.16$, $\mu = 0$.}
  \label{Fig:Tzh-q3cB-nu45-mu0-z5}
\end{figure}

In model \cite{Bohra:2020qom} type of magnetic catalysis (direct or
inverse) depends on $z^5$ coefficient with borderline value $d \approx
0.05$. As we look for the magnetic catalysis effect both regions $d >
0.05$ and $d < 0.05$ should be checked. We also try different regimes
of undependent and connected $c_B$ and $q_3$
constants. Fig.\ref{Fig:Tzh-q3cB-nu1-mu0-z5}--\ref{Fig:Tzh-q3cB-nu45-mu0-z5}
show, that direct magnetic catalysis within the current model can be
expected for $T(z_h,q_3)$ with fixed $c_B < 0$ for both $d < 0.05$ and
$d > 0.05$.

\begin{figure}[b!]
  \centering 
  \includegraphics[scale=0.27]{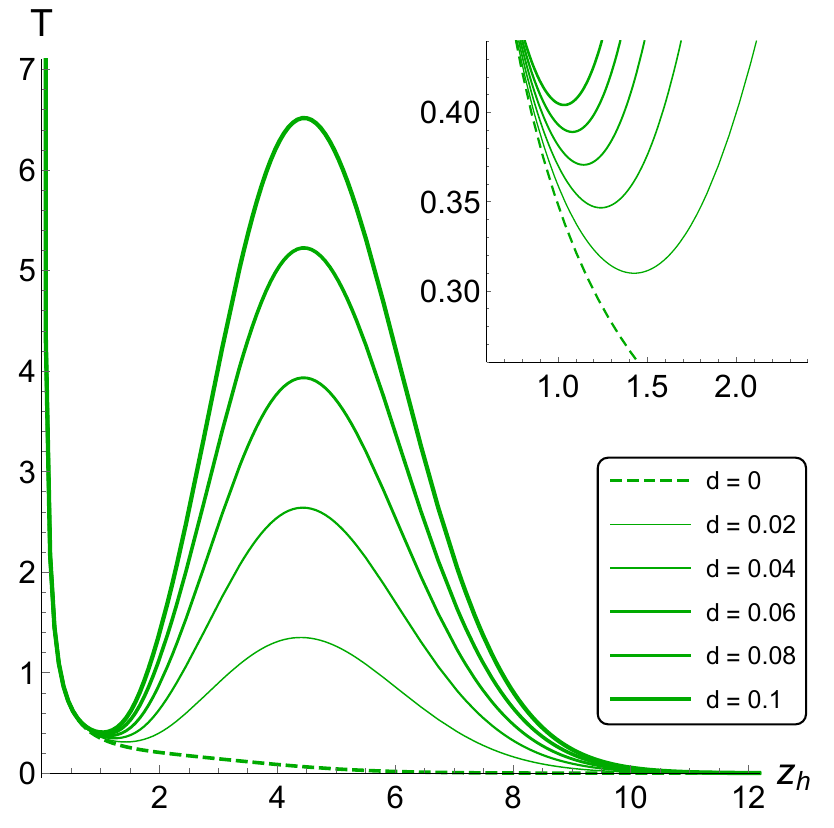} \quad
  \includegraphics[scale=0.27]{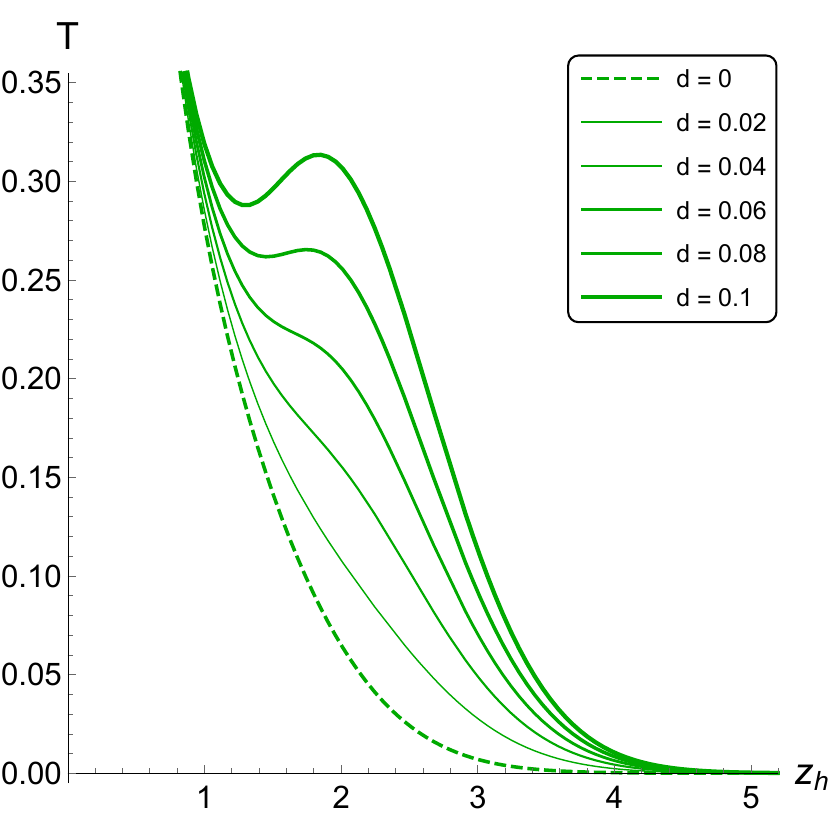} \quad
  \includegraphics[scale=0.27]{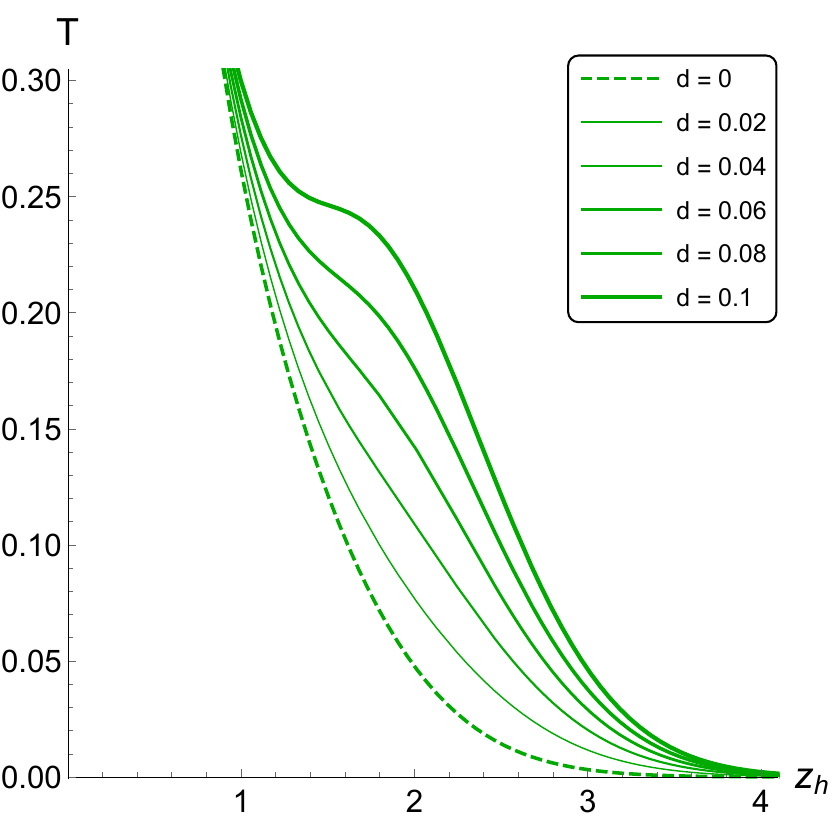} \\
  \includegraphics[scale=0.27]{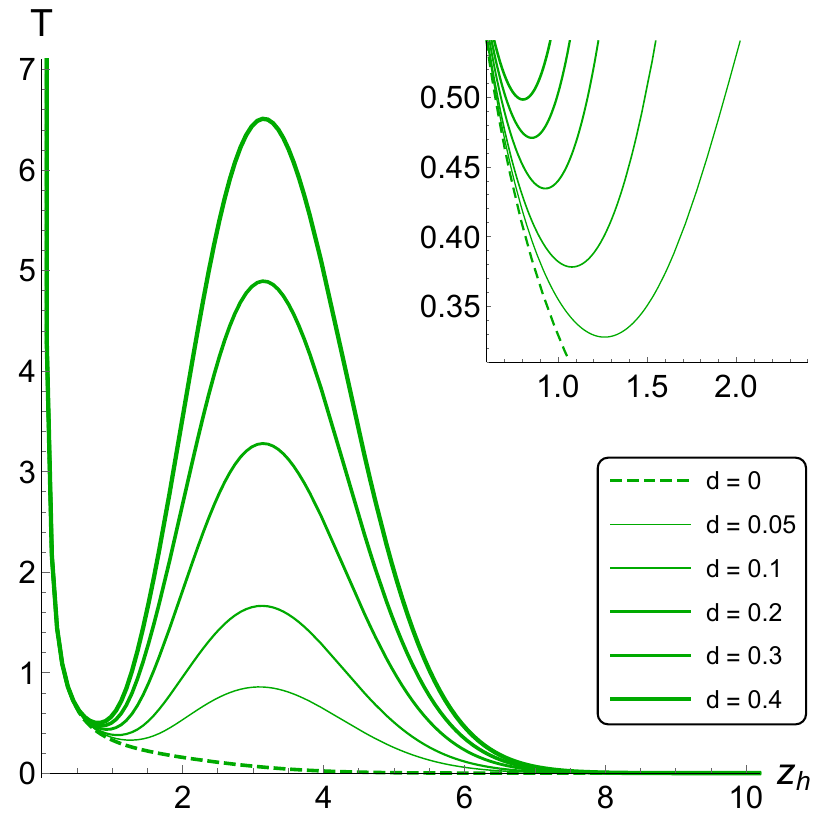} \quad
  \includegraphics[scale=0.27]{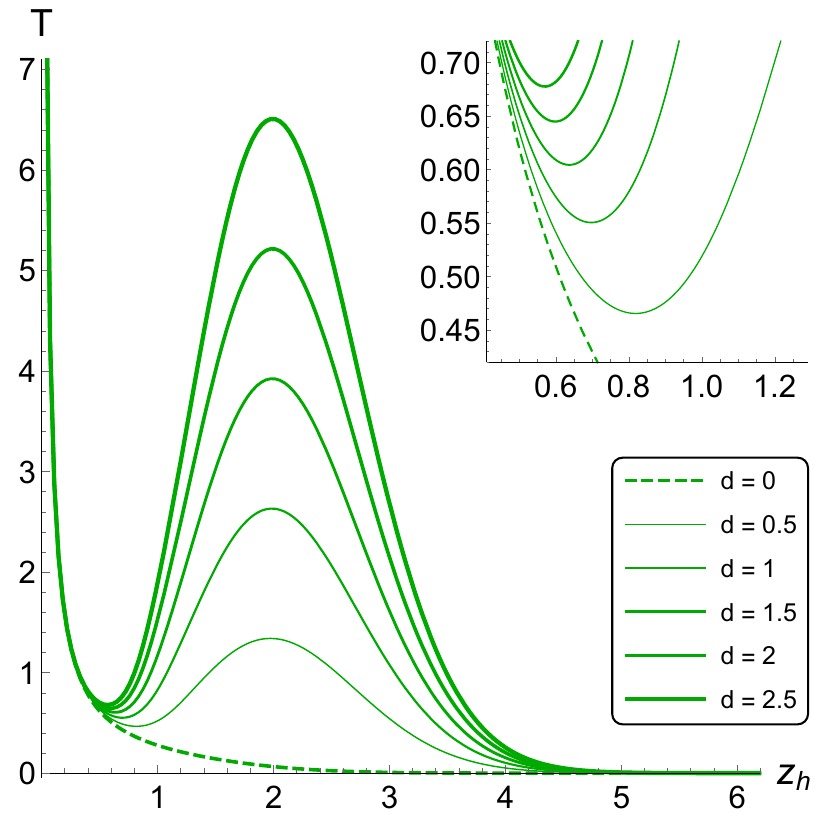} \quad
  \includegraphics[scale=0.27]{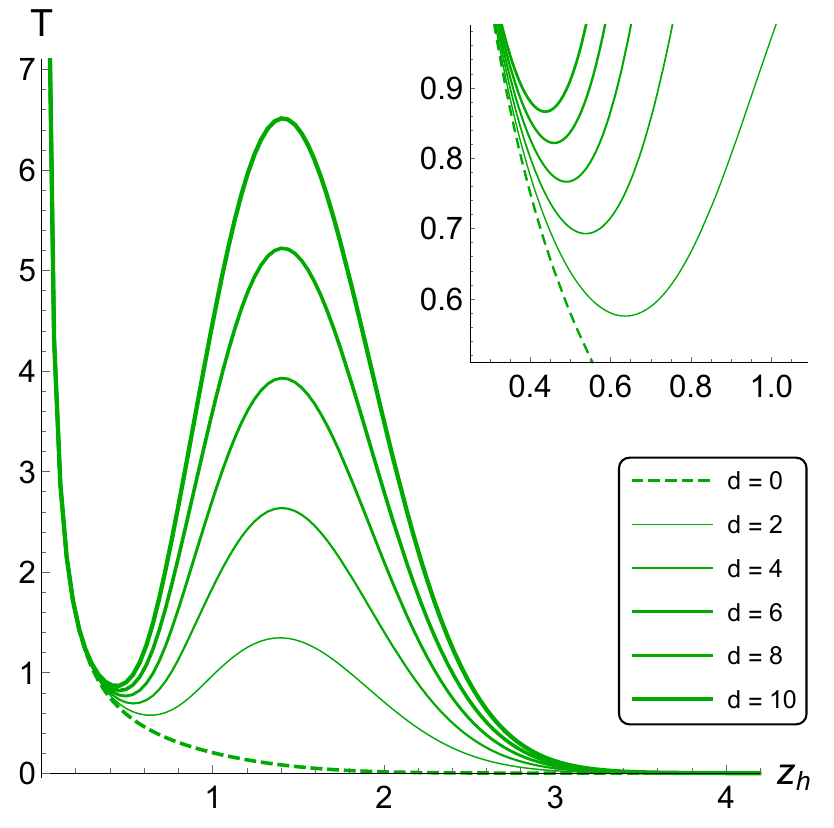} \\
  A \hspace{110pt} B \hspace{110pt} C
  \caption{Temperature $T(z_h)$ in magnetic field with a fixed set of
    coefficient $d = 0, \, 0.02, \, 0.04, \, 0.06, \, 0.1$ (1-st line) 
    for $c_B = - \, 0.1$ (A), $c_B = - \, 0.5$ (B) and $c_B = - \,
    0.6$ (C); with different $d$ and close $T$-values (2-nd line) for
    $c_B = - \, 0.2$ (A), $c_B = - \, 0.5$ (B) and $c_B = - \, 1$ (C);
    $\nu = 1$, $a = 0.15$, $c = 1.16$, $q_3 = 1$, $\mu = 0$.}
  \label{Fig:Tzhd-cB-nu1-q31-mu0-z5}
\end{figure}

For further investigation we need to choose $c_B$ and $d$ fixed
values, so let us first consider their influence on thermodynamics
properties of the system in primary isotropic case $\nu = 1$ with
normalization to the magnetic field $q_3 = 1$. We do not seek for MC
effect now (Fig.\ref{Fig:Tzh-q3cB-nu1-mu0-z5}.), we are interested in
the back-reaction on metric $c_B$ and correction coefficient $d$
influence.

\begin{figure}[h!]
  \centering 
  \includegraphics[scale=0.27]{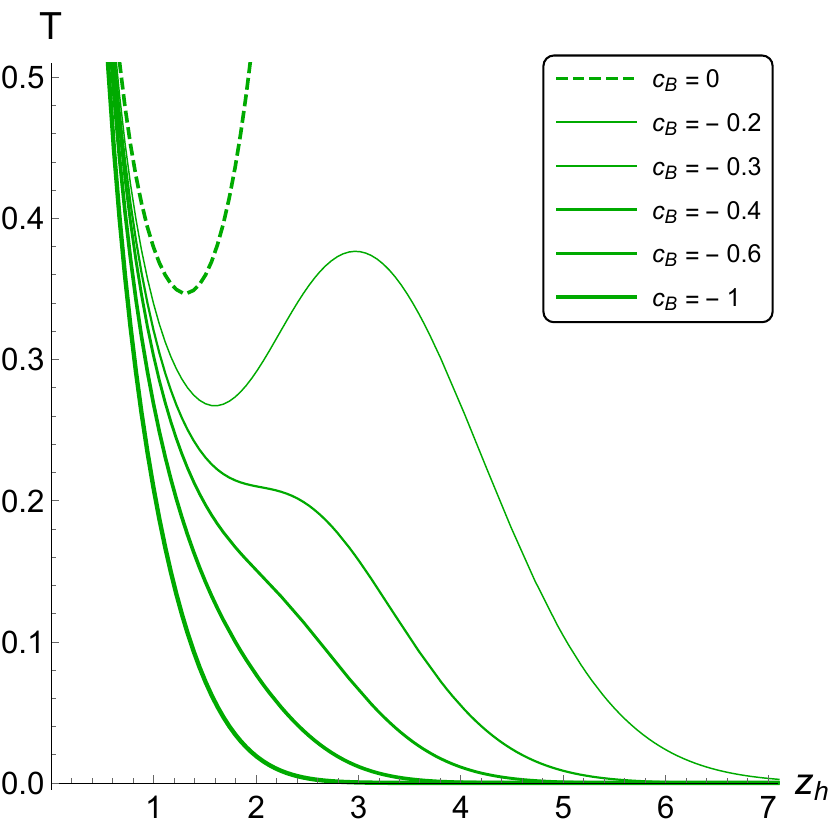} \quad
  \includegraphics[scale=0.27]{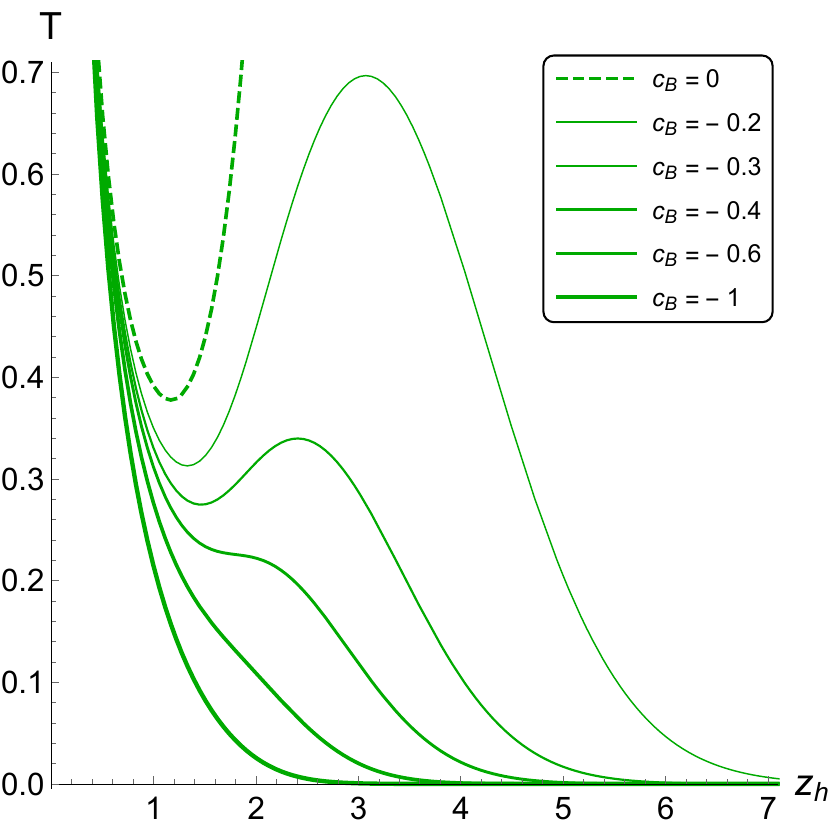} \quad
  \includegraphics[scale=0.27]{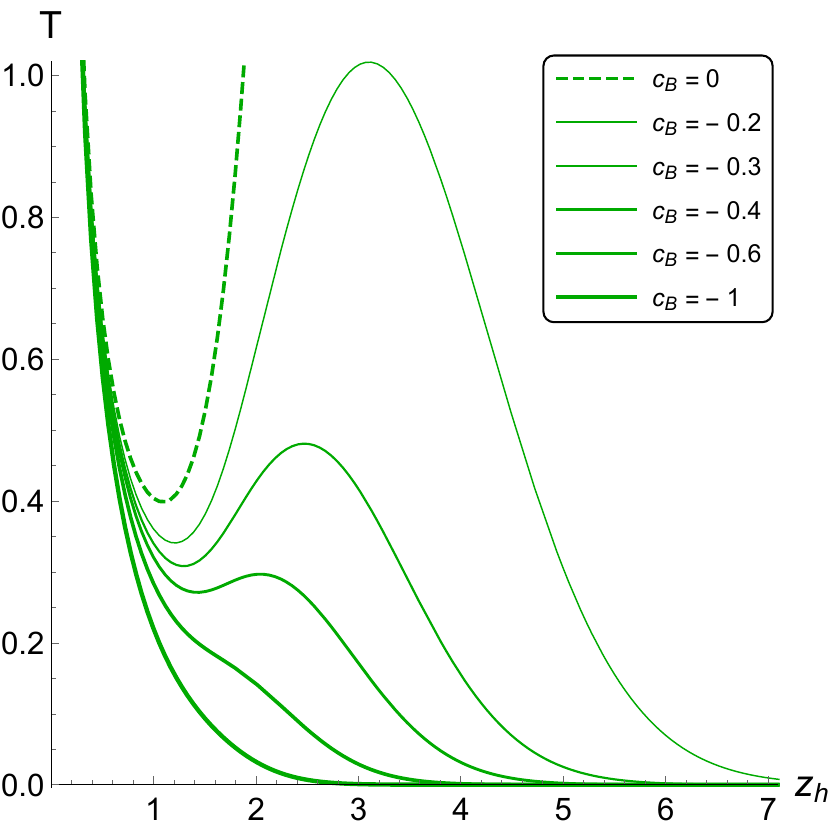} \\
  A \hspace{110pt} B \hspace{110pt} C \\
  \includegraphics[scale=0.27]{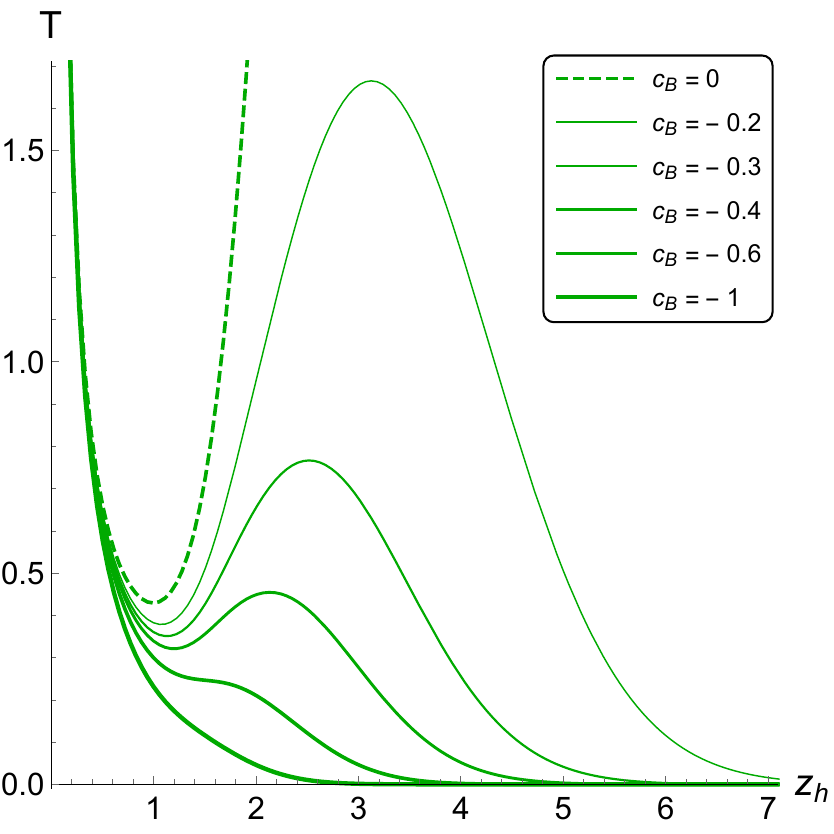} \quad
  \includegraphics[scale=0.27]{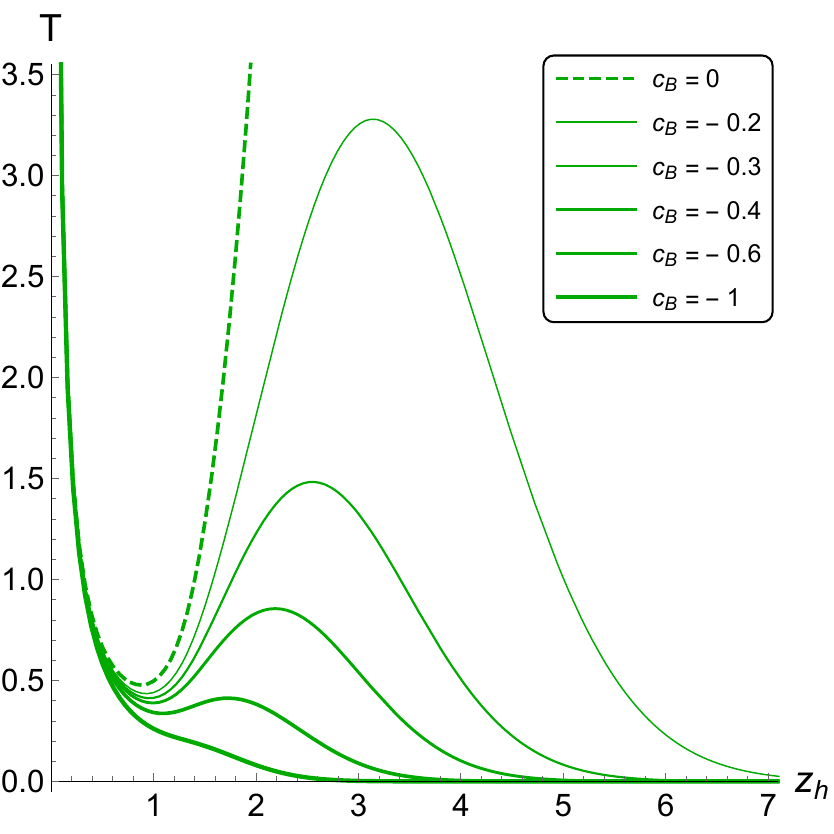} \quad
  \includegraphics[scale=0.27]{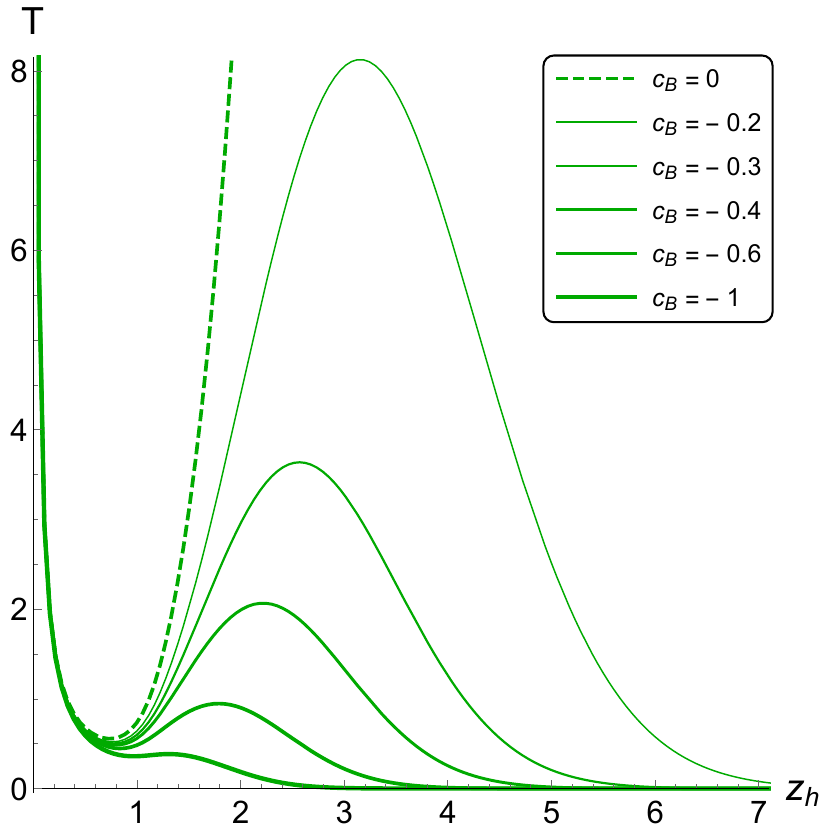} \\
  D \hspace{110pt} E \hspace{110pt} F
  \caption{Temperature $T(z_h)$ in magnetic field with different
    backreaction $c_B$ for coefficient $d = 0.02$ (A), $d = 0.04$ (B),
    $d = 0.06$ (C), $d = 0.1$ (D), $d = 0.2$ (E), $d = 0.5$ (F); $\nu
    = 1$, $a = 0.15$, $c = 1.16$, $q_3 = 1$, $\mu = 0$.}
  \label{Fig:TzhcB-d-nu1-q31-mu0-z5}
\end{figure}
\begin{figure}[h!]
  \centering 
  \includegraphics[scale=0.27]{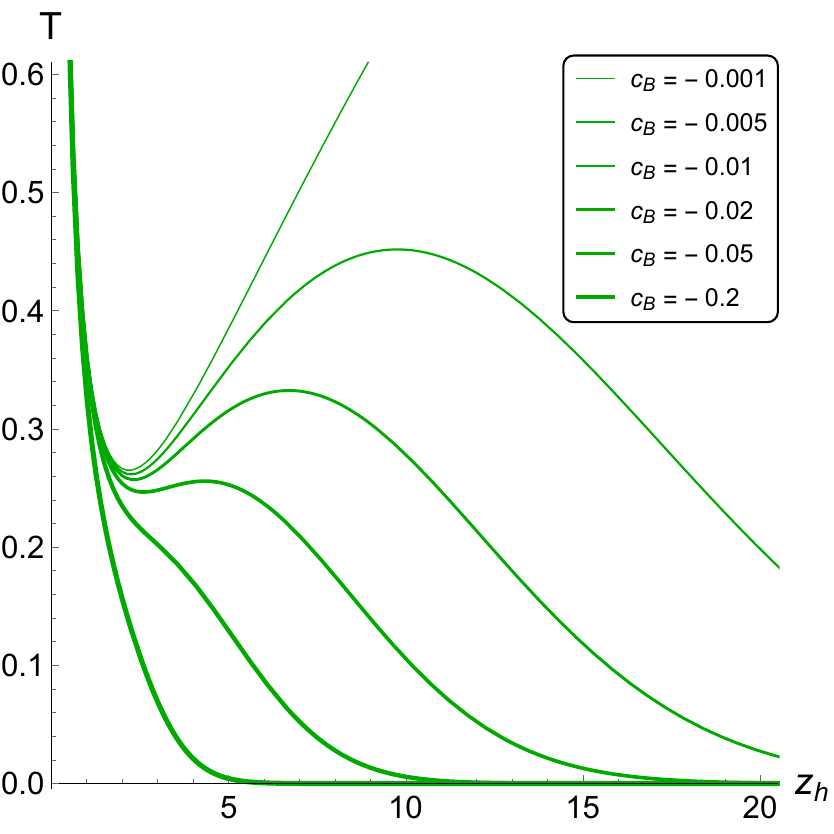} \quad
  \includegraphics[scale=0.27]{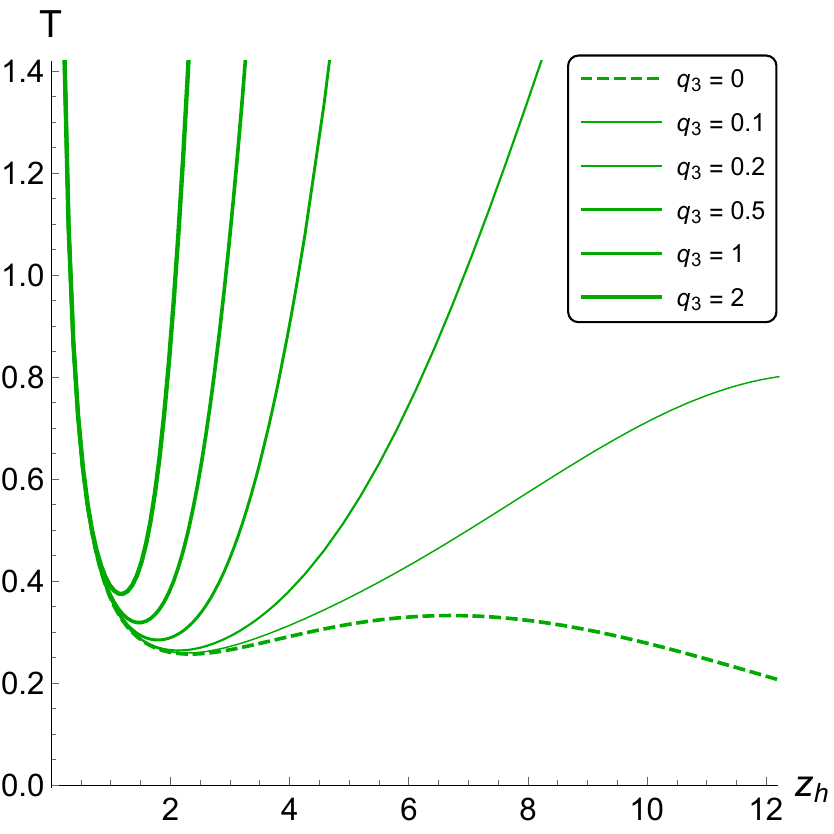} \quad
  \includegraphics[scale=0.27]{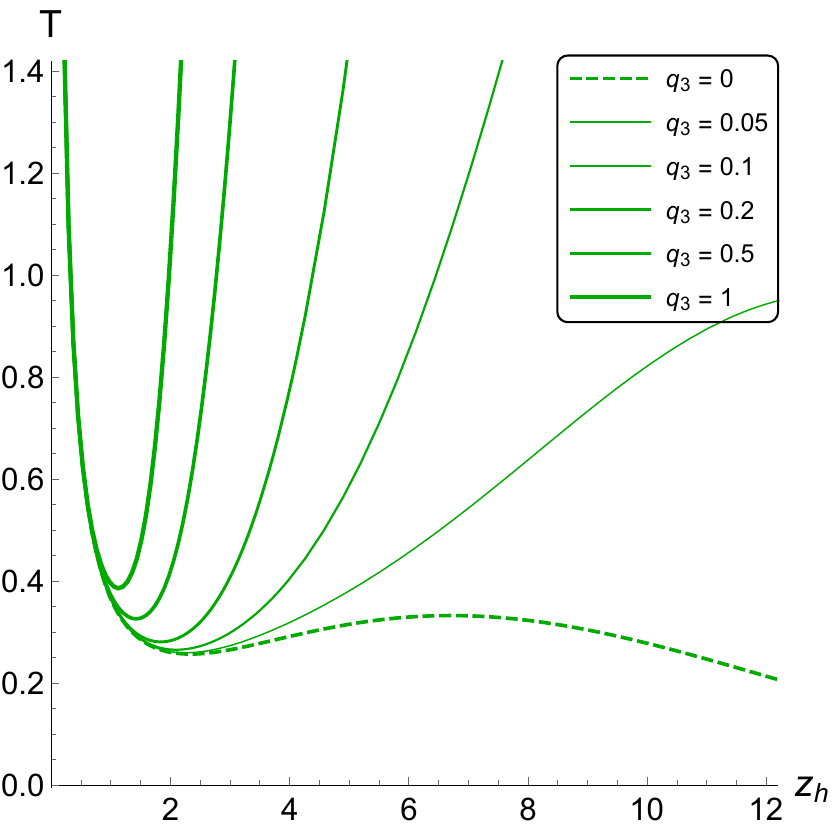} \\
  A \hspace{110pt} B \hspace{110pt} C \\
  \includegraphics[scale=0.27]{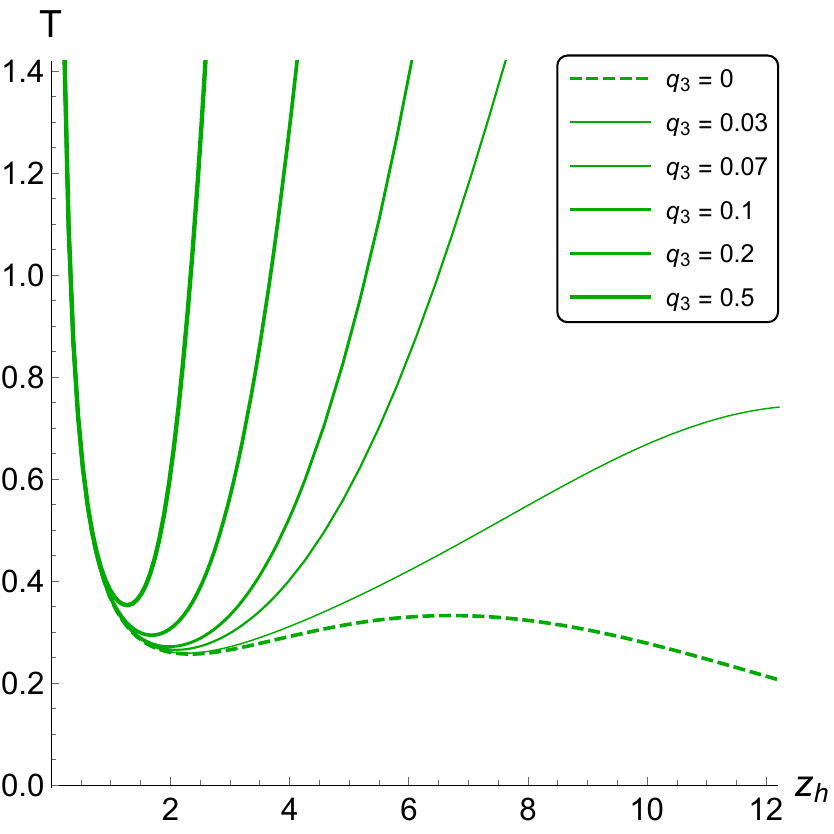} \quad
  \includegraphics[scale=0.27]{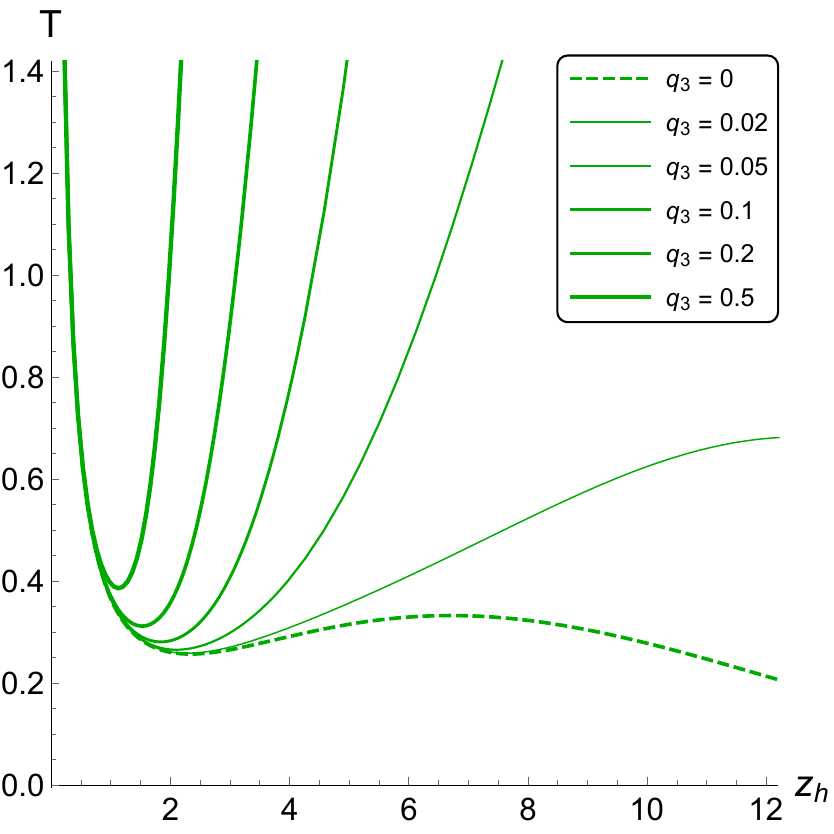} \quad
  \includegraphics[scale=0.27]{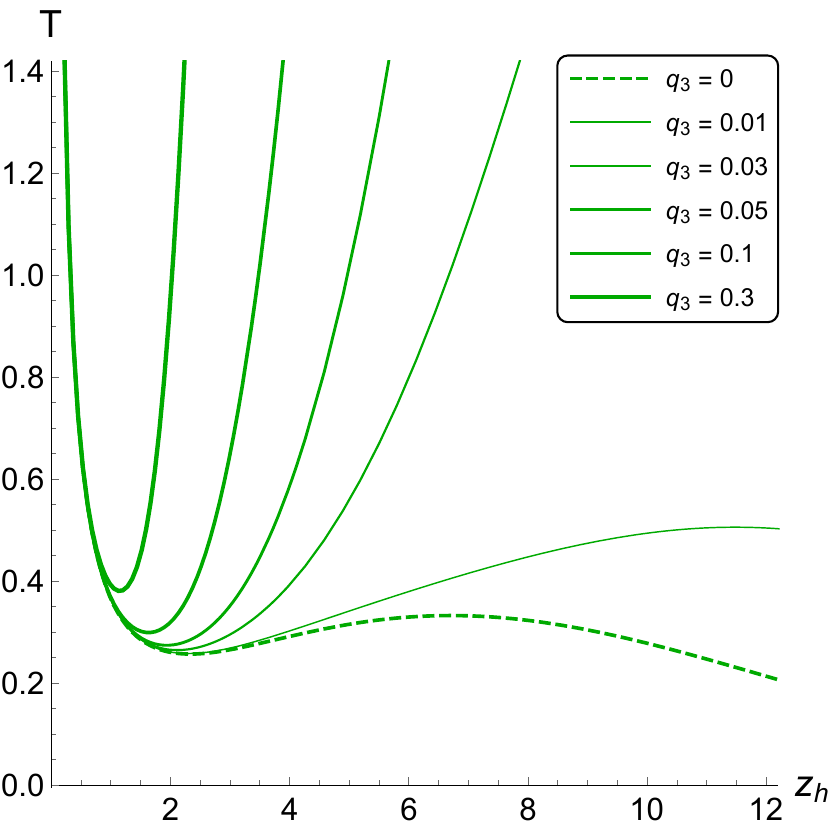} \\
  D \hspace{110pt} E \hspace{110pt} F
  \caption{Temperature $T(z_h)$ in zero magnetic field limit $q_3 = 0$
    for different back-reaction $c_B$, $d = 0.1$ (A) and in different
    magnetic field $q_3$ for $d = 0.01$ (B), $d = 0.05$ (C), $d = 0.1$
    (D), $d = 0.2$ (E) and $d = 0.5$ (F), $c_B = - \, 0.01$; $\nu =
    1$, $a = 0.15$, $c = 1.16$, $\mu = 0$.}
  \label{Fig:Tzhq30-nu1-d001-mu0-z5}
\end{figure}

Fig.\ref{Fig:Tzhd-cB-nu1-q31-mu0-z5} shows the temperature dependence
$T(z_h,d)$ for different back-reaction on metric $c_B < 0$. It's larger
absolute value, meaning stronger magnetic field influence, makes
multivalued temperature behavior degenerate sooner, thus leading to
monotony (Fig.\ref{Fig:Tzhd-cB-nu1-q31-mu0-z5}, 1-st line). As to the
coefficient $d$, it stabilises the multivalued temperature behavior
and consequently the opportunity of the background phase transition. To
keep the multivalued behavior within the same interval under stronger
magnetic field deformation the larger $d$-value is required
(Fig.\ref{Fig:Tzhd-cB-nu1-q31-mu0-z5}, 2-nd line). Besides the local
minimum and maximum of temperature are shifted to the left, to
the region of lower $z_h$ values by stronger magnetic field
deformation, while coefficicient $d$ has no notable effect on
$T_{min}$ and $T_{max}$ positions.

Larger $d$ slows down the suppression of the phase transition by
magnetic field (Fig.\ref{Fig:TzhcB-d-nu1-q31-mu0-z5}). Temperature at
it's local maximum and minimum grows and retains multivalued behavior
with increasing absolute values of the metric back-reaction $c_B$.

To explore the properties of the current model lets us choose
intermediate values of the parameters, so that the magnetic field
influence on metric isn't unlikely weak. On the other hand, we need to
preserve zero magnetic field limit, i.e. the 1-st order phase
transition should exist at $\mu = 0$ for $q_3 = 0$. Therefore $c_B
\gtrsim - \, 0.02$ is required
(Fig.\ref{Fig:Tzhq30-nu1-d001-mu0-z5}.A). Larger $d$ makes
temperature more sensitive to the magnetic field
(Fig.\ref{Fig:Tzhq30-nu1-d001-mu0-z5}.B-F), that is especially
obvious on small $q_3$ values. Here and futher $d = 0$, $d = 0.01$ 
and $d = 0.1$ for $c_B = - \, 0.01$ are considered.

Phase diagram depicts temperature $T(z_h)$ dependence on chemical
potential $\mu$, that, in turn, depends on magnetic field $q_3$ and
coefficient $d$. All the plots discussed below are given in
Appendix~\ref{appendixB}. In $z^2$ limit $d = 0$ temperature $T(z_h)$
has local minimum, whose value decreases with increasing $\mu$ till
$\mu_{crit}$: $T_{min}(\mu_{crit}) = 0$
(Fig.\ref{Fig:Tzhmu-q3-nu1-d0-z5}.A). For $\mu > \mu_{crit}$
temperature curve $T(z_h) > 0$ splits into two parts, of which only
the left one, with smaller $z_h$, is significant. It decreases
monotonously, so the 1-st order phase transition shouldn't occur any
more. As the magnetic field increases, the local minimum for $\mu = 0$
gradually becomes less pronounced and eventually disappears. For
larger $\mu$ temperature function is three-valued as the unstable
branch corresponding to small black holes (larger $z_h$) changes from
increasing to decreasing
(Fig.\ref{Fig:Tzhmu-q3-nu1-d0-z5}.E.F). Therefore in strong magnetic
field the 1-st order phase transition doesn't exist for near zero
$\mu$ and when appears, it has more complex character than usual.

Turning $z^5$ warp factor term on maintains a local temperature minimum 
even in strong magnetic field
(Fig.\ref{Fig:Tzhmu-q3-nu1-d001-z5}). Two-valued function  $T(z_h)$
becomes not single-valued, but four-valued instead
(Fig.\ref{Fig:Tzhmu-q3-nu1-d001-z5}.F).

In primary isotropic media $\nu = 1$ this particular effect disappears
as $d$ increases further thus saving 1-st order phase transition for
strong magnetic field values (Fig.\ref{Fig:Tzhmu-q3-nu1-d01-z5}.E,F),
whereas for intermediate ones it disappears before the transition
temperature reaches zero
(Fig.\ref{Fig:Tzhmu-q3-nu1-d01-z5}.C,D). Anisotropy $\nu = 4.5$
doesn't meet such difficulties
(Fig.\ref{Fig:Tzhmu-q3-nu45-d01-z5}.C,D). It also shifts the 1-st
order phase transition $T(z_h)$ values down, but doesn't seem to have
a significant effect on other temperature aspects
(Fig.\ref{Fig:Tzhmu-q3-nu45-d0-z5}--\ref{Fig:Tzhmu-q3-nu45-d01-z5}).

\begin{figure}[t!]
  \centering
  \includegraphics[scale=0.3]{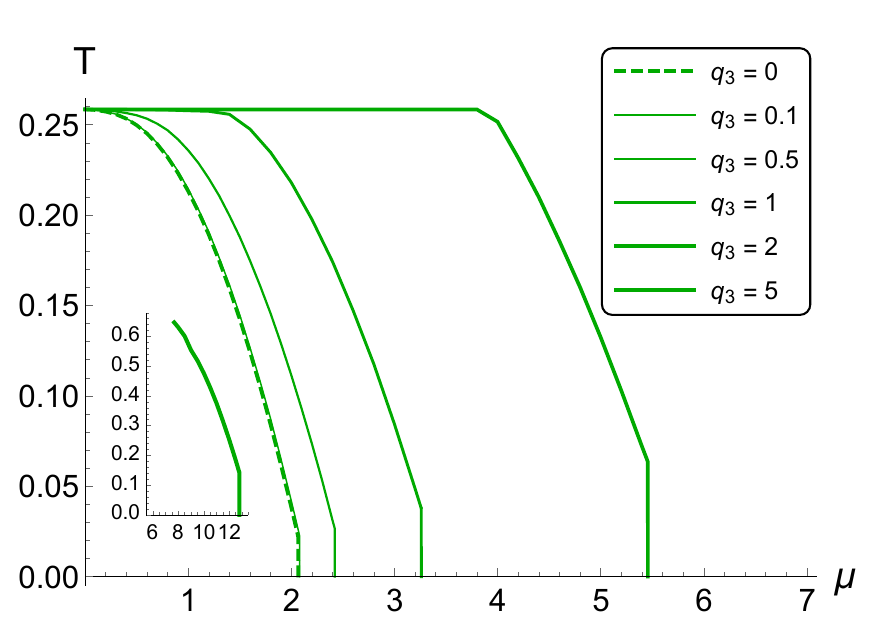} \quad
  \includegraphics[scale=0.3]{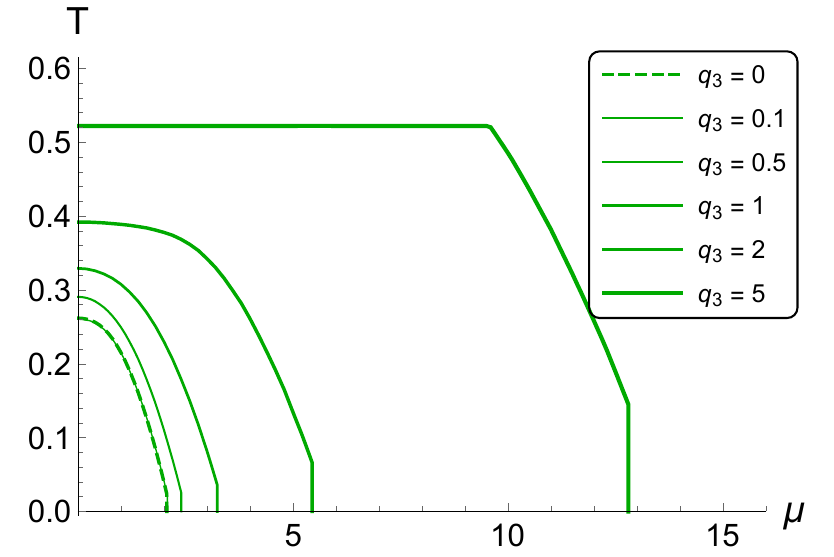} \quad
  \includegraphics[scale=0.3]{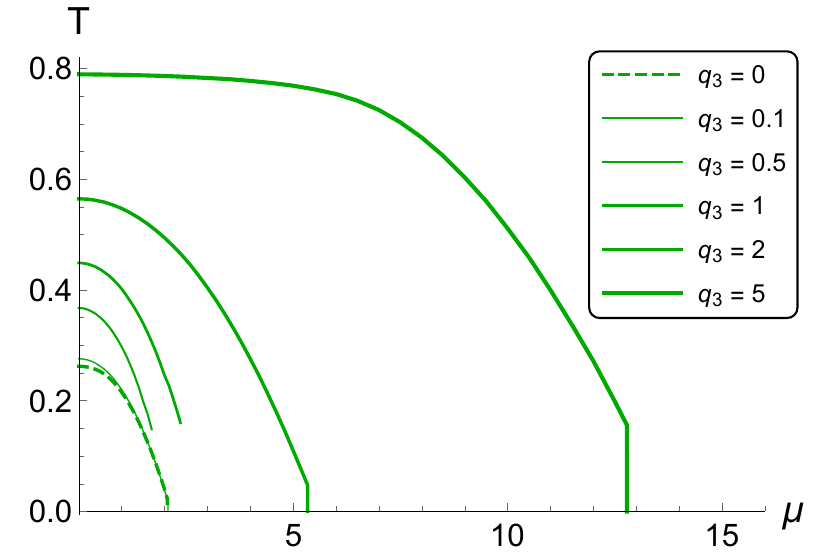} \\
  \includegraphics[scale=0.3]{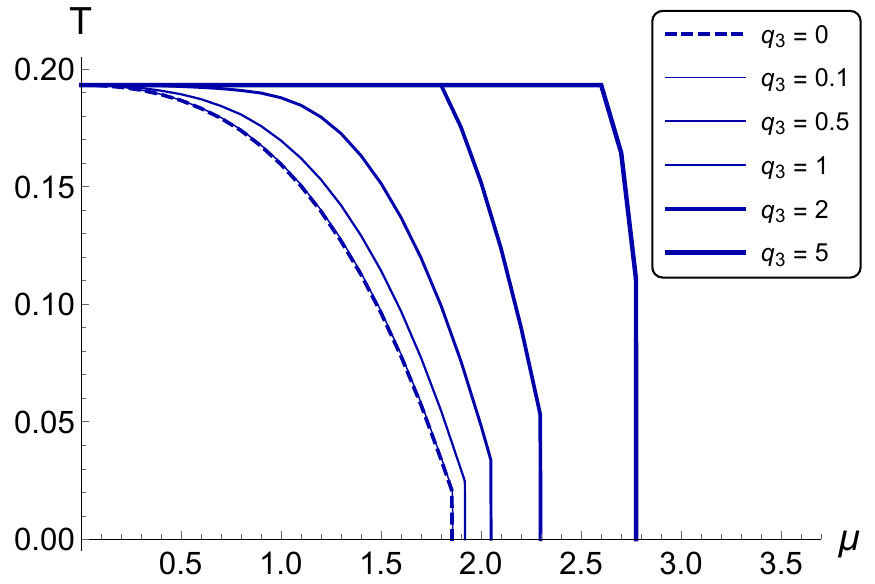} \quad
  \includegraphics[scale=0.3]{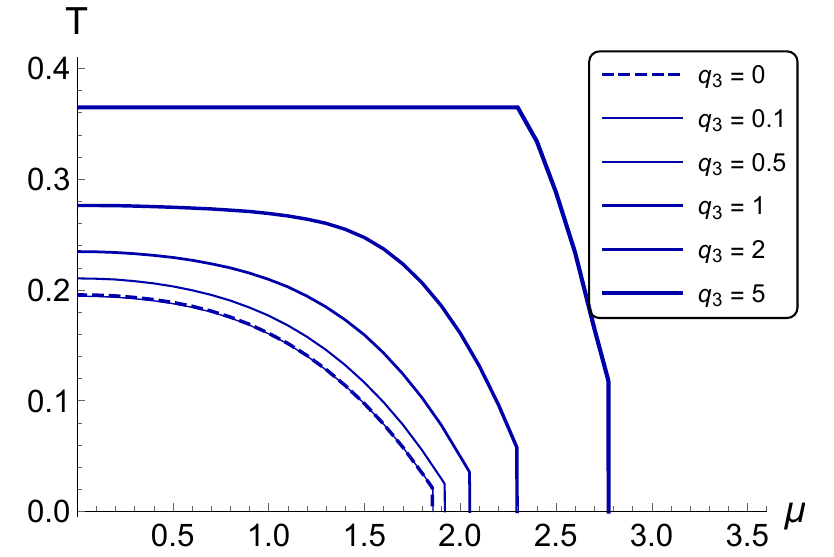} \quad
  \includegraphics[scale=0.3]{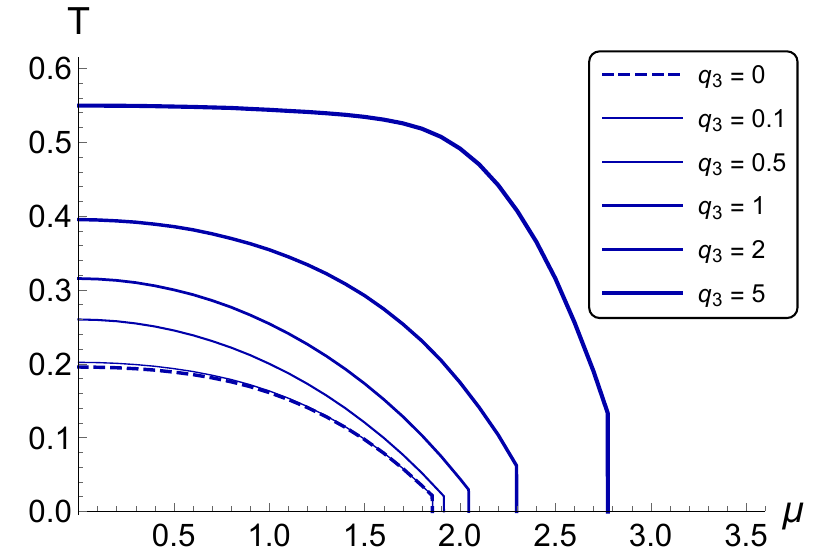} \\
  A \hspace{125pt} B \hspace{125pt} C
  \caption{1-st order phase transition $T(\mu)$ in magnetic field with
    different $q_3$ for $d = 0$ (A), $d = 0.01$ (B) and $d = 0.1$ (c)
    in primary isotropic case $\nu = 1$ (1-st line) and primary
    anisotropic case $\nu = 4.5$ (2-nd line); $a = 0.15$, $c = 1.16$,
    $c_B = - \, 0.01$.}
  \label{Fig:Tmuq3-z5}
\end{figure}

To study phase transition picture a free energy investigation
\begin{gather}
  F = - \int s \, d T = \int_{z_h}^{\infty} s \, T' dz
  \label{eq:5.36}
\end{gather}
is needed. It confirms the conclusions above regarding the 1-st order 
phase transition
(Fig.\ref{Fig:FTmu-q3-nu1-d0-z5}--\ref{Fig:FTmu-q3-nu45-d01-z5}). It
also shows that the transition mainly occurs from a small black hole
to thermal gas, i.e. Hawking-Page phase transition, $F(T_{HP}) =
0$. The expected exceptions are cases of three- and four-valued
temperature for small $d$ in strong magnetic field
(Fig.\ref{Fig:FTmu-q3-nu1-d0-z5}.E,F,
\ref{Fig:FTmu-q3-nu1-d001-z5}.E,F, \ref{Fig:FTmu-q3-nu45-d001-z5}.F)
and of non-monotonic temperature behavior smoothing for larger $d$ in
intermediate magnetic field for $\nu = 1$
(Fig.\ref{Fig:FTmu-q3-nu1-d01-z5}.C,D).

Direct magnetic catalysis is observed for the 1-st order phase
transition both in primary isotropic and anisotropic cases
(Fig.\ref{Fig:Tmuq3-z5}). Maximum chemical potential value also
increases with magnetic field, but is hardly sensitive to $z^5$
coefficient $d$. For $\nu = 1$ the difference for $\mu_{crit}$ between
zero and non-zero $d$ cases is more noticeable for a larger magnetic
field (Fig.\ref{Fig:Tmuq3-z5}, 1-st line), but for $\nu = 4.5$ it is
hardly visible (Fig.\ref{Fig:Tmuq3-z5}, 2-nd line). Coefficient $d$
increases the phase transition temperature and makes the 1-st order
phase transition curve decrease not so sharp near the maximum
accessible $\mu$ values. This last effect is more significant in
anisotropic media $\nu = 4.5$.


\section{Temporal Wilson loops and phase diagram}\label{phase}

To calculate the expectation value of the temporal Wilson loop
\begin{gather}
  W[C_\vartheta] = e^{-S_{\vartheta,t}}, \label{eq:5.37}
\end{gather}
oriented along vector $\vec n$
\begin{gather}
  n_{x1} = \cos \vartheta \sin \alpha, \quad
  n_{x2} = \sin \vartheta \sin \alpha, \quad 
  n_{x3} = \cos \alpha, \label{eq:5.38}
\end{gather}
we use our metric (\ref{eq:5.01}) as a background:
\begin{gather}
  ds^2 = G_{\mu\nu}dx^{\mu}dx^{\nu}
  = \cfrac{L^2}{z^2} \ \fb_s(z) \left[
    - \, g(z) dt^2 + \fg_1 dx^2 + \fg_2 dx_2^2 + \fg_3 dx_3^2
    + \cfrac{dz^2}{g(z)} \right], \label{eq:5.39}
\end{gather}
where $\fb_s(z) = \fb(z) \exp\bigl(\sqrt{2/3} \
  \phi(z) \bigr)$ is the 
model warp factor in string frame and $\fg_i$ are the corresponding
$G_{\mu\nu}$-metric components.

\begin{figure}[b!]
  \centering 
  \includegraphics[scale=0.3]{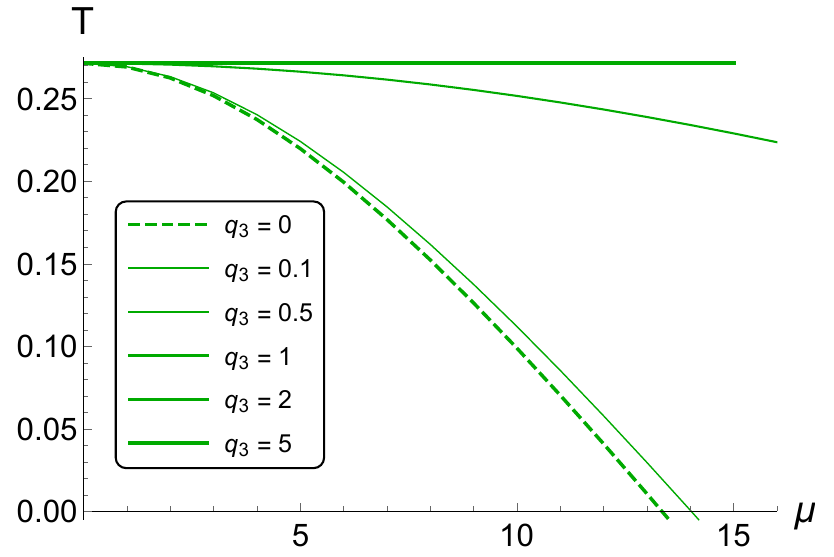}
  \quad
  \includegraphics[scale=0.3]{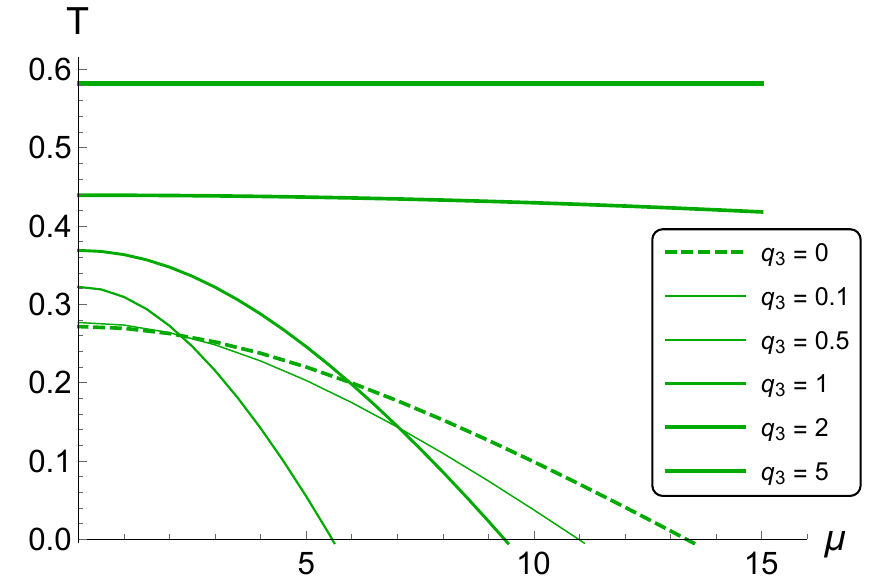}
  \quad
  \includegraphics[scale=0.3]{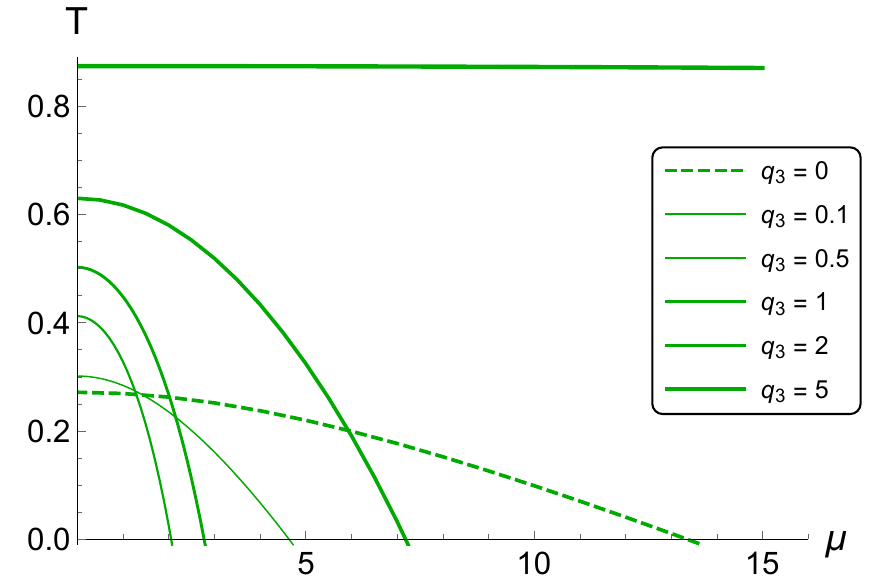} \\
  \includegraphics[scale=0.3]{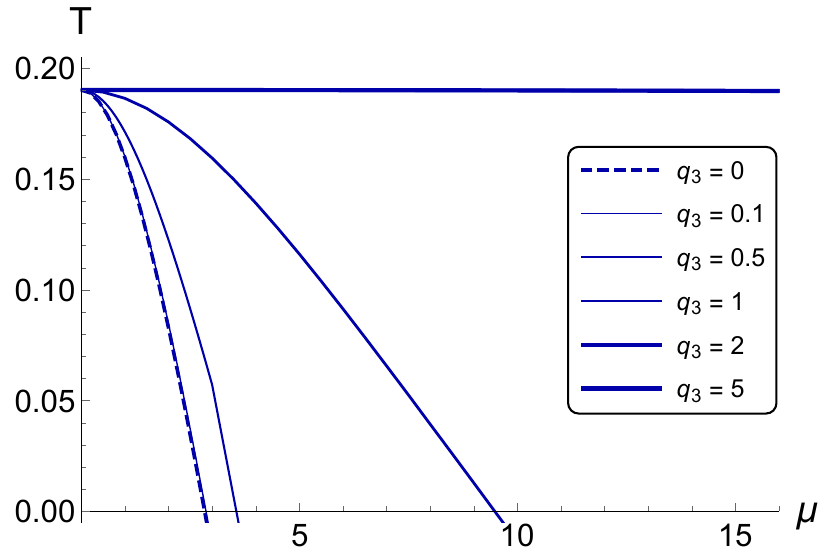}
  \quad
  \includegraphics[scale=0.3]{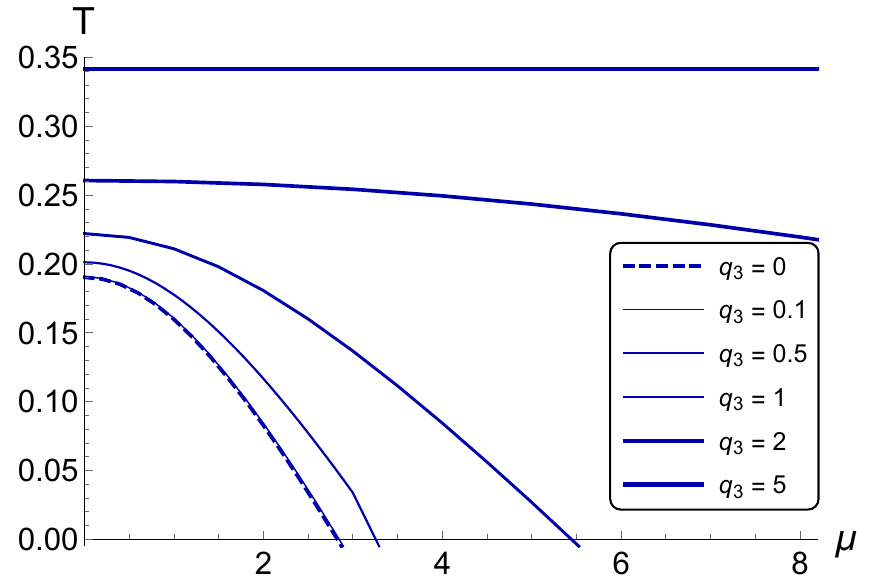}
  \quad
  \includegraphics[scale=0.3]{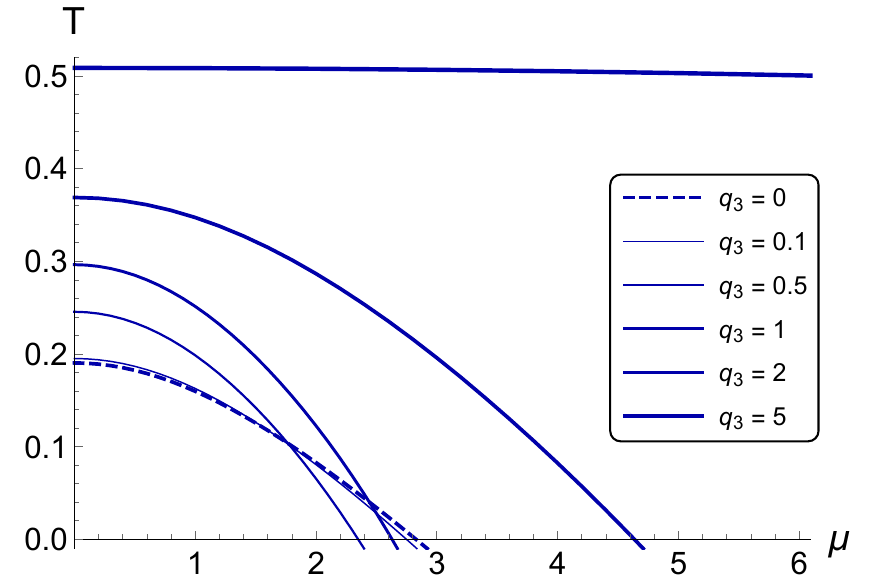} \\
  A \hspace{120pt} B \hspace{120pt} C \\
  \caption{Wilson loop phase transition curves $T(\mu)$ for different
    magnetic field $q_3$ for coefficient $d = 0$ (A), $d = 0.01$ (B)
    and $d = 0.1$ (C) in primary isotropic case $\nu = 1$ (1-st line)
    and primary anisotropic case $\nu = 4.5$ (2-nd line), $a = 0.15$,
    $c = 1.16$, $c_B = - \, 0.01$.}
  \label{Fig:WLmuq3-cB-001-d-z5}
\end{figure}

We are interested in the turning points for Wilson loops WL$x_1$,
WL$x_2$ and WL$x_3$, oriented along $x_1$, $x_2$ and $x_3$ axes,
respectively. They are defined by equations:
\begin{gather}
  \hspace{-75pt}
  \mbox{WL}x_1: \quad - \, 4 a z - 10 d q_3^2 z^4 + \sqrt{\cfrac23} \
  \phi'(z) + \cfrac{g'}{2 g} - \cfrac{2}{z} \ \Big|_{z = z_{DWx_1}}
  \hspace{-15pt} = 0, \label{eq:5.53} \\
  \hspace{-55pt}
  \mbox{WL}x_2: \quad - \, 4 a z - 10 d q_3^2 z^4 + \sqrt{\cfrac23} \
  \phi'(z) + \cfrac{g'}{2 g} - \cfrac{\nu + 1}{\nu z} \ \Big|_{z =
    z_{DWx_2}} \hspace{-15pt} = 0, \label{eq:5.54} \\
  \mbox{WL}x_3: \quad - \, 4 a z - 10 d q_3^2 z^4 + \sqrt{\cfrac23} \
  \phi'(z) + \cfrac{g'}{2 g} - \cfrac{\nu + 1}{\nu z} + c_B z \
  \Big|_{z = z_{DWx_3}} \hspace{-15pt} = 0. 
  \label{eq:5.55}
\end{gather}
These expressions basically are the same that were used in previous
considerations \cite{ARS-Heavy-2020, ARS-Light-2022}, written for the
current model's warp-factor. For the detailed derivation the reader is
referred to \cite{AR-2018, ARS-2019}. In this work we consider WL$x_3$
(\ref{eq:5.55}) only, as WL$x_1$ (\ref{eq:5.53}) and WL$x_2$
(\ref{eq:5.54}) are its particular cases for $\nu = 1$ and $c_B = 0$.


For the phase transition curves provided by the temporal Wilson loops
magnetic catalysis effect also takes place
(Fig.\ref{Fig:WLmuq3-cB-001-d-z5}). In the $z^2$ limit $T(\mu = 0)$ is
the same for any magnetic field, and larger $d$ leads to larger
temperature along the entire phase transition curve. Its slope
decreases with increasing magnetic field and eventually becomes
horizontal. As the magnetic field increases, chemical potential range
$[0, \mu_{crit}]$ shrinks and then starts to increase again. Together
with the curve slope changing this leads to the fact that at large
$\mu$ values the phase transition temperature drops for stronger
magnetic field, i.e. inverse magnetic catalysis is locally
observed. This effect is especially noticeable for primary isotropy
(Fig.\ref{Fig:WLmuq3-cB-001-d-z5}, 1-st line) and is quite reduced in
anisotropic media (Fig.\ref{Fig:WLmuq3-cB-001-d-z5}, 2-nd line).

\begin{figure}[h!]
  \centering 
  \includegraphics[scale=0.3]{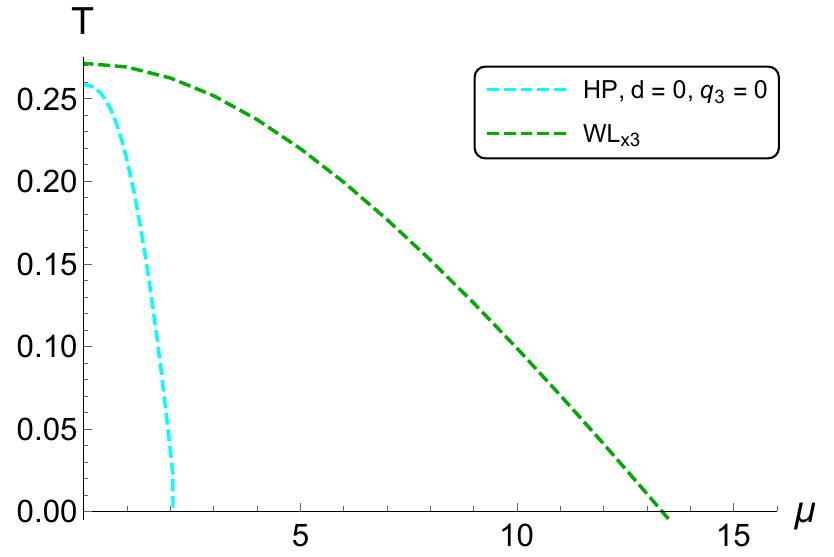} \quad
  \includegraphics[scale=0.3]{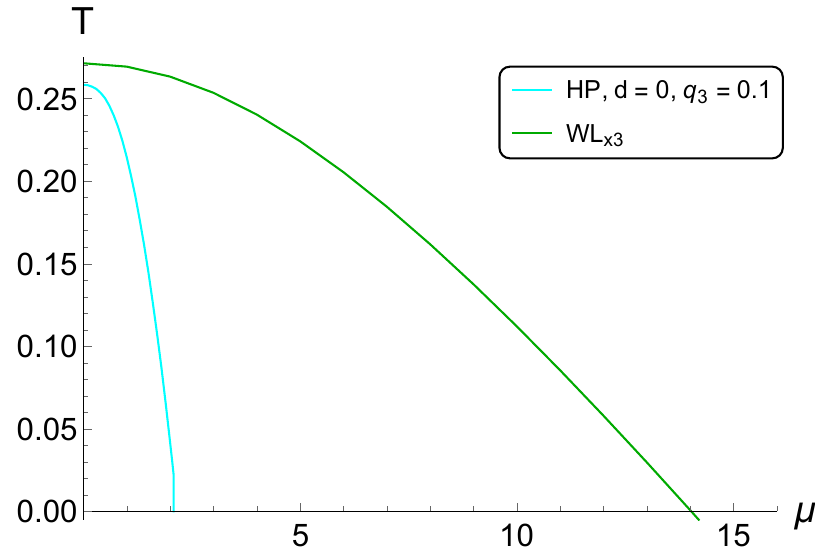} \quad
  \includegraphics[scale=0.3]{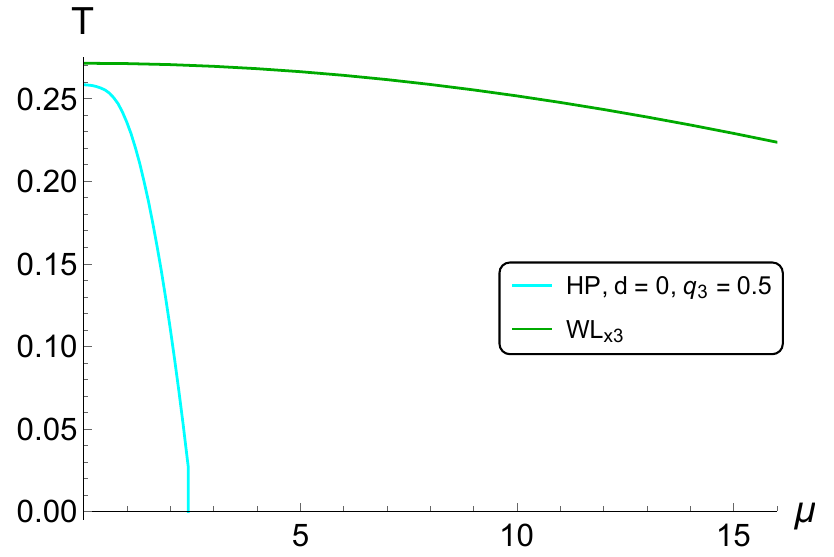} \\
  A \hspace{120pt} B \hspace{120pt} C \\
  \includegraphics[scale=0.3]{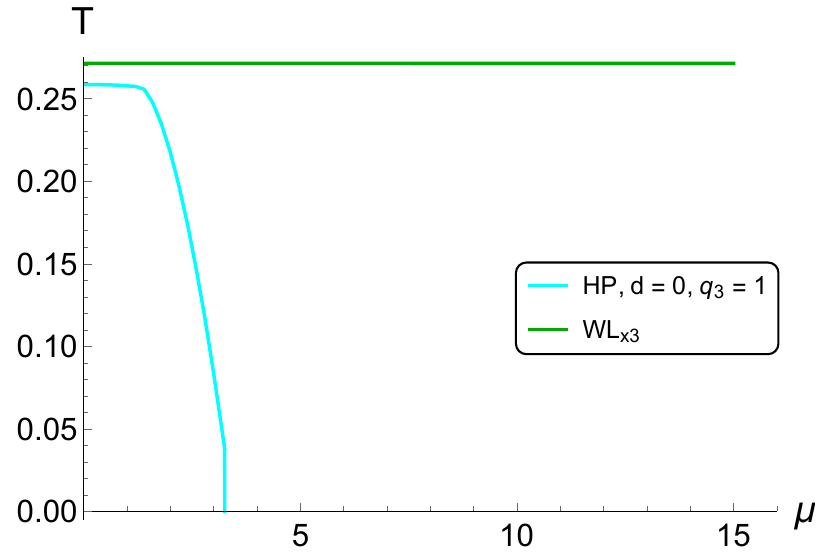} \quad
  \includegraphics[scale=0.3]{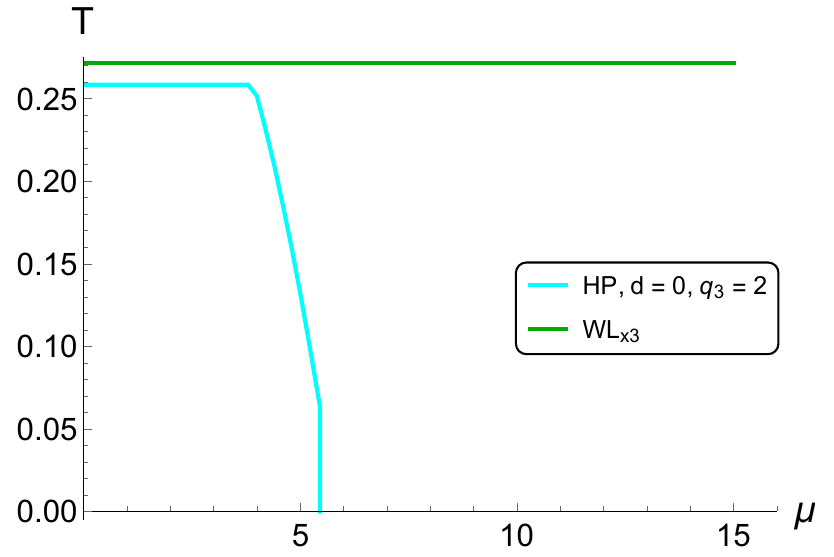} \quad
  \includegraphics[scale=0.3]{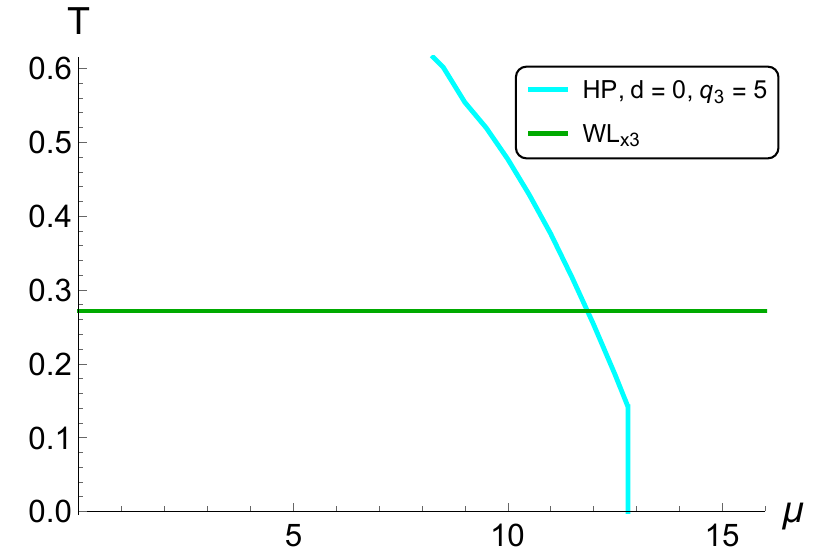} \\
  D \hspace{120pt} E \hspace{120pt} F
  \caption{Phase diagram $T(\mu)$ as a combination of the 1-st order
    phase transition (HP) and a crossover (WL) in magnetic field for
    $q_3 = 0$ (A), $q_3 = 0.01$ (B), $q_3 = 0.05$ (C), $q_3 = 1$ (D),
    $q_3 = 2$ (E), $q_3 = 5$ (F); $\nu = 1$, $a = 0.15$, $c = 1.16$,
    $c_B = - \, 0.01$, $d = 0$.}
  \label{Fig:PDq3-cB-001-nu1-d0-z5}
\end{figure}

\begin{figure}[h!]
  \centering 
  \includegraphics[scale=0.3]{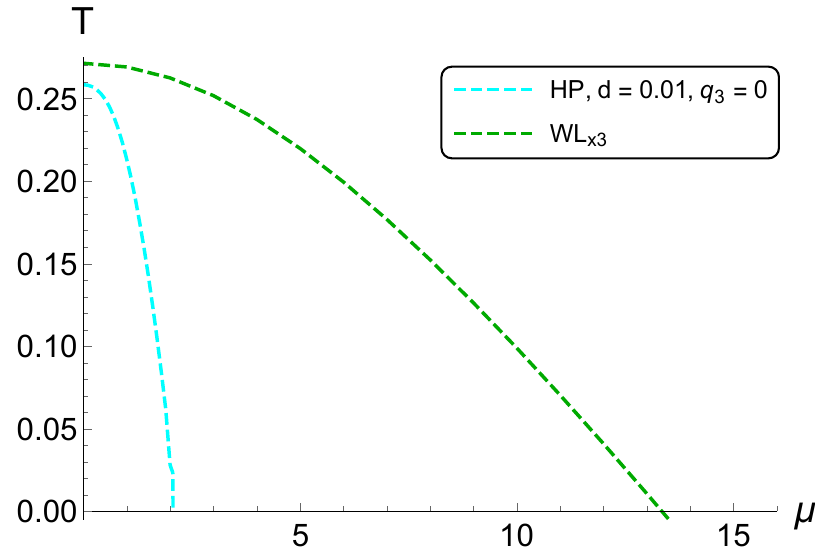} \quad
  \includegraphics[scale=0.3]{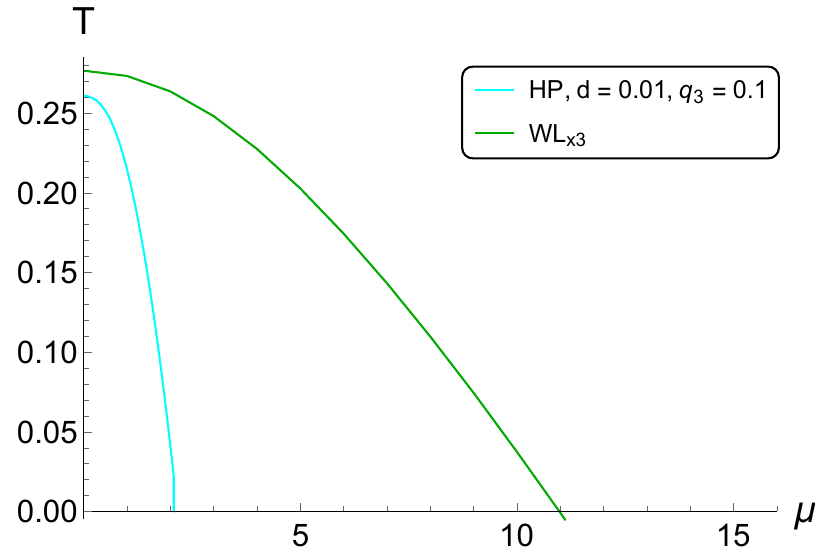} \quad
  \includegraphics[scale=0.3]{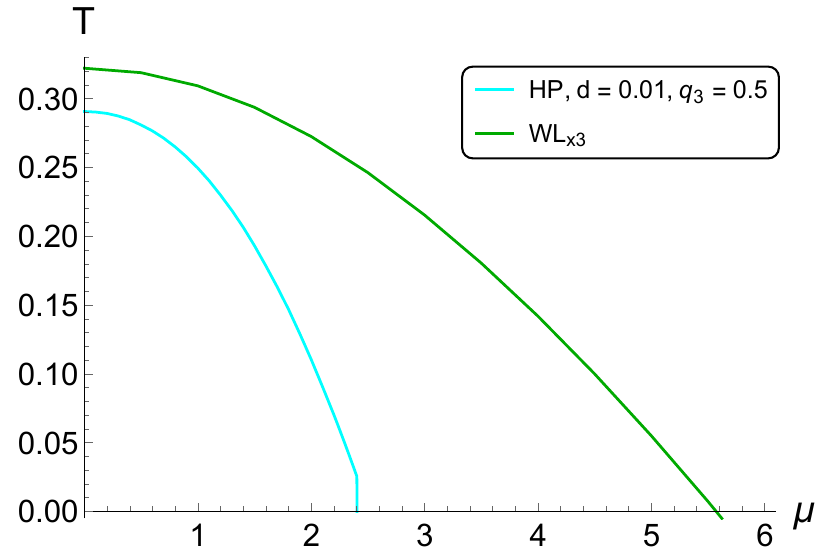} \\
  A \hspace{120pt} B \hspace{120pt} C \\
  \includegraphics[scale=0.3]{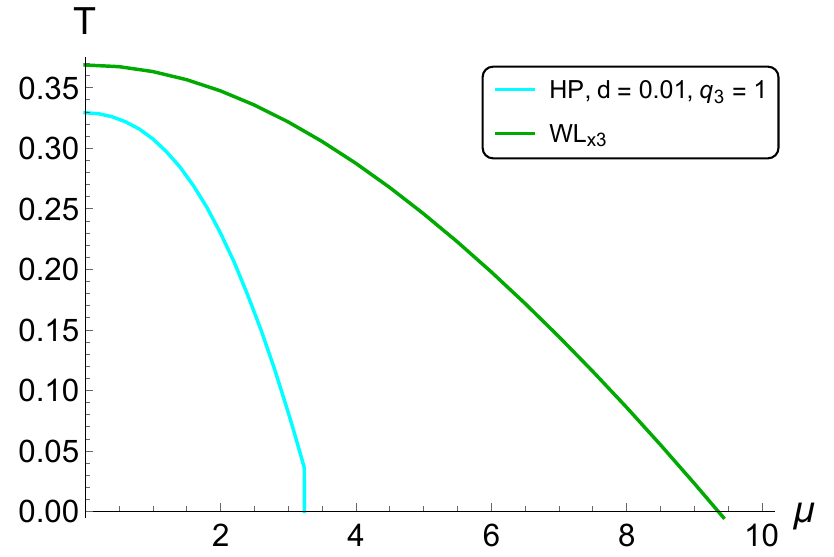} \quad
  \includegraphics[scale=0.3]{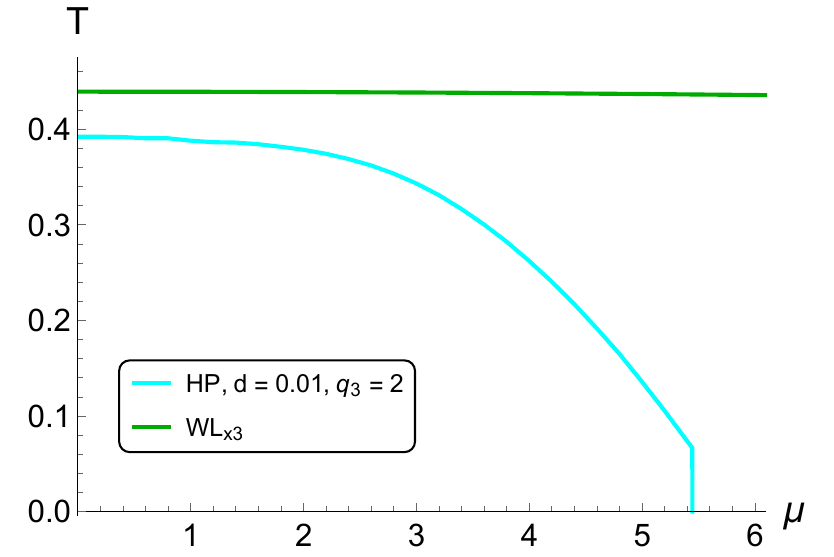} \quad
  \includegraphics[scale=0.3]{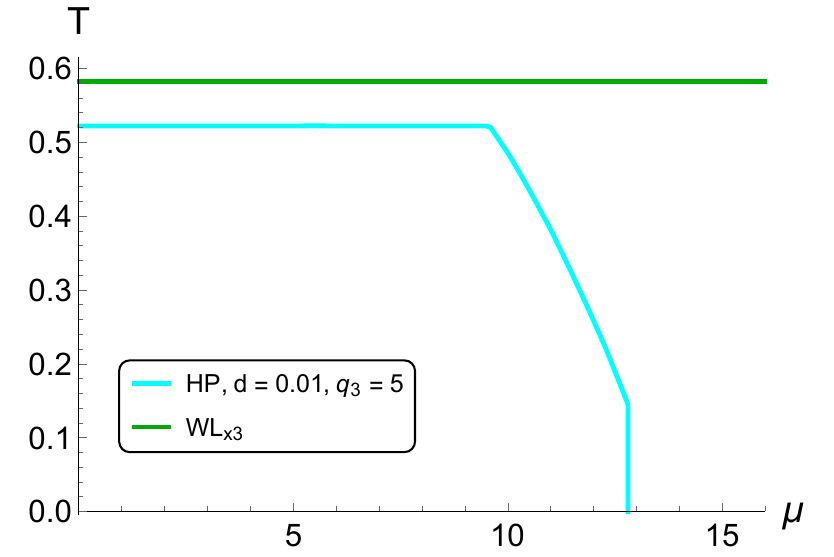} \\
  D \hspace{120pt} E \hspace{120pt} F
  \caption{Phase diagram $T(\mu)$ as a combination of the 1-st order
    phase transition (HP) and a crossover (WL) in magnetic field for
    $q_3 = 0$ (A), $q_3 = 0.01$ (B), $q_3 = 0.05$ (C), $q_3 = 1$ (D),
    $q_3 = 2$ (E), $q_3 = 5$ (F); $\nu = 1$, $a = 0.15$, $c = 1.16$,
    $c_B = - \, 0.01$, $d = 0.01$.}
  \label{Fig:PDq3-cB-001-nu1-d001-z5}
\end{figure}

\begin{figure}[h!]
  \centering 
  \includegraphics[scale=0.3]{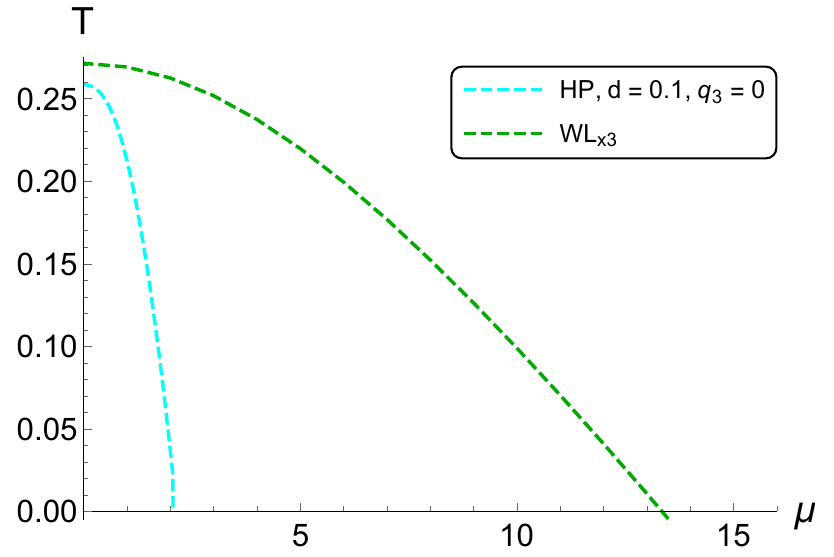} \quad
  \includegraphics[scale=0.3]{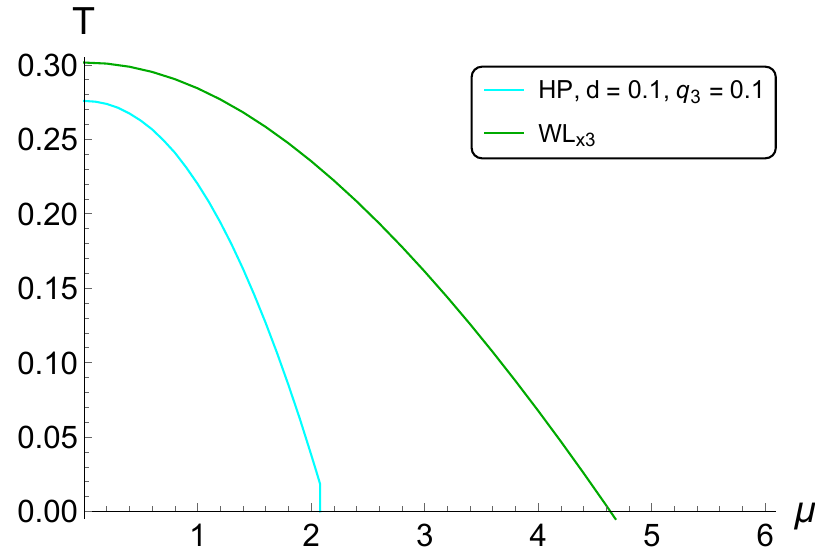} \quad
  \includegraphics[scale=0.3]{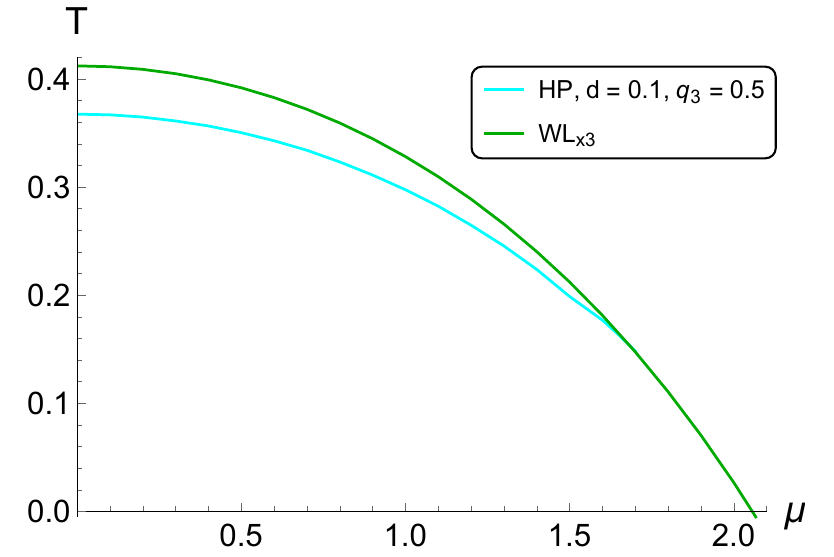} \\
  A \hspace{120pt} B \hspace{120pt} C \\
  \includegraphics[scale=0.3]{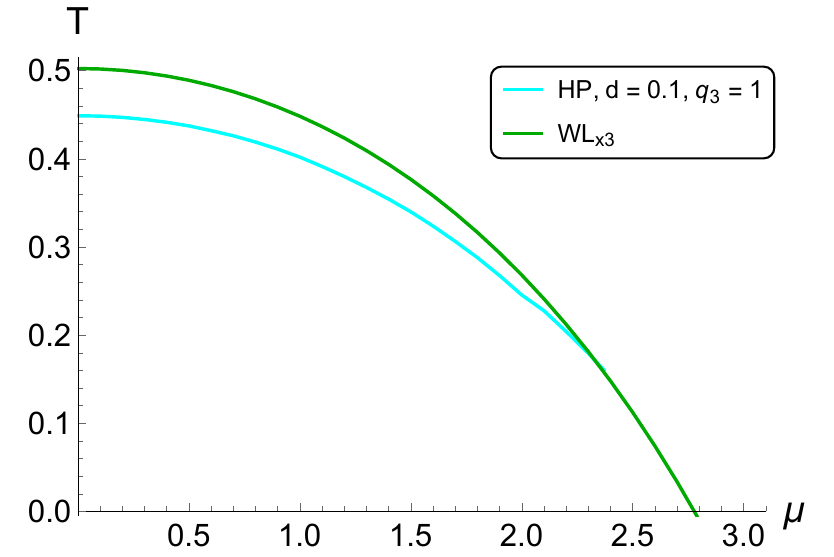} \quad
  \includegraphics[scale=0.3]{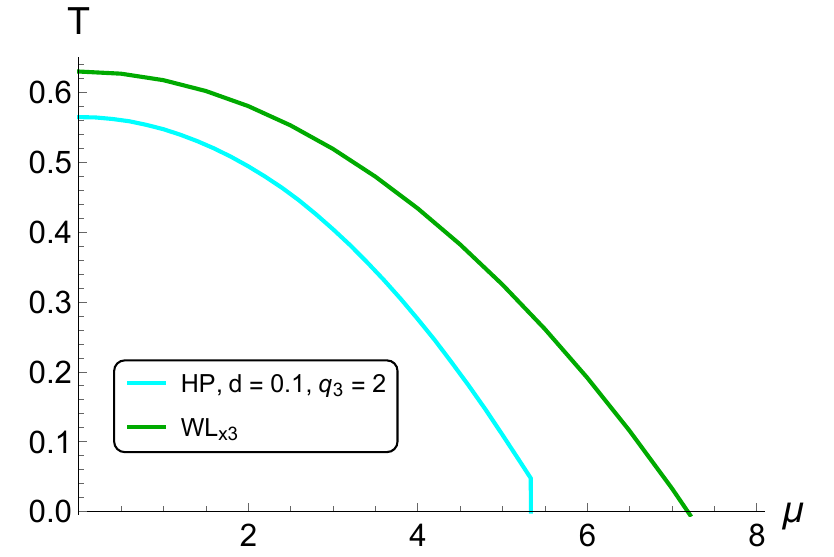} \quad
  \includegraphics[scale=0.3]{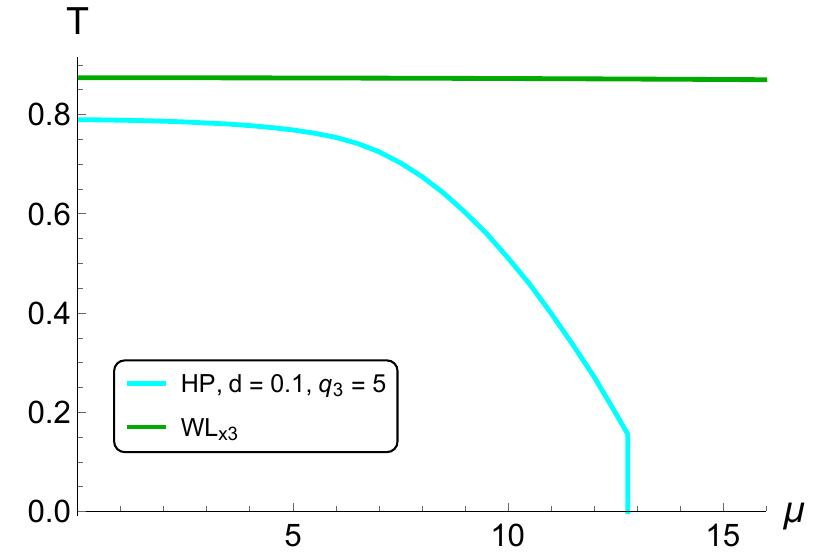} \\
  D \hspace{120pt} E \hspace{120pt} F
  \caption{Phase diagram $T(\mu)$ as a combination of the 1-st order
    phase transition (HP) and a crossover (WL) in magnetic field for
    $q_3 = 0$ (A), $q_3 = 0.01$ (B), $q_3 = 0.05$ (C), $q_3 = 1$ (D),
    $q_3 = 2$ (E), $q_3 = 5$ (F); $\nu = 1$, $a = 0.15$, $c = 1.16$,
    $c_B = - \, 0.01$, $d = 0.1$.}
  \label{Fig:PDq3-cB-001-nu1-d01-z5}
\end{figure}

\begin{figure}[h!]
  \centering 
  \includegraphics[scale=0.3]{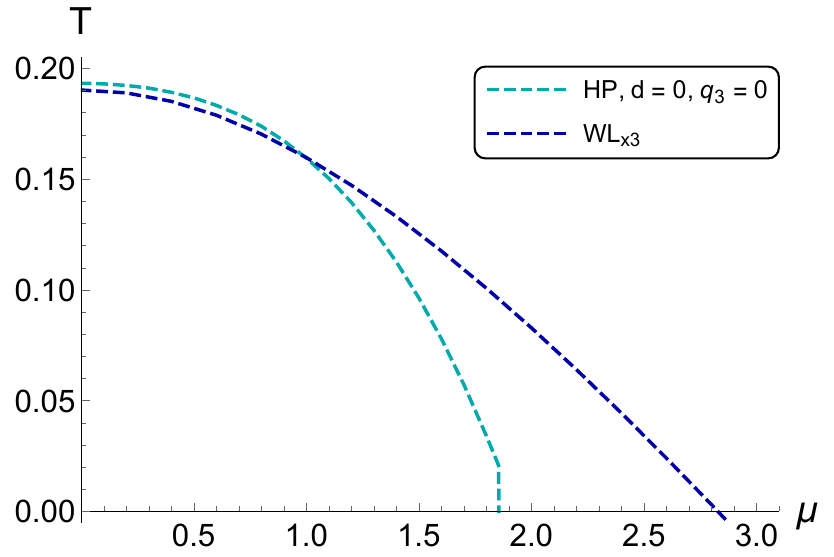} \quad
  \includegraphics[scale=0.3]{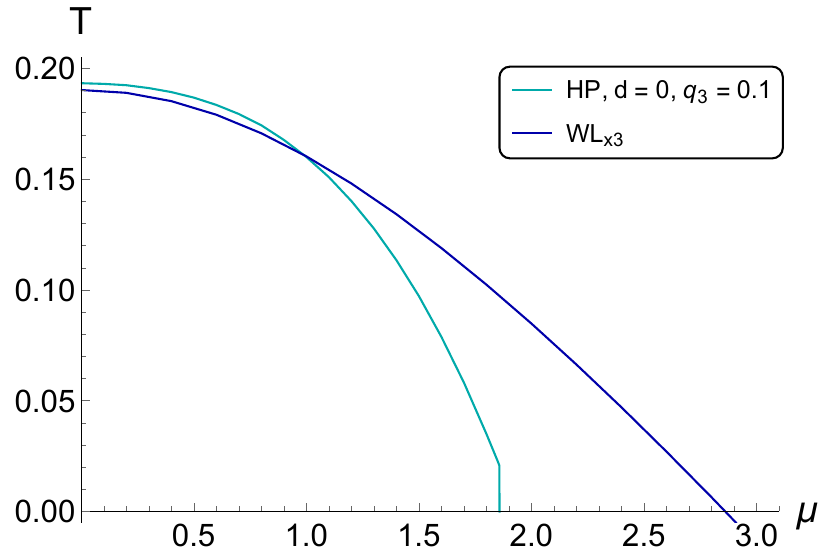} \quad
  \includegraphics[scale=0.3]{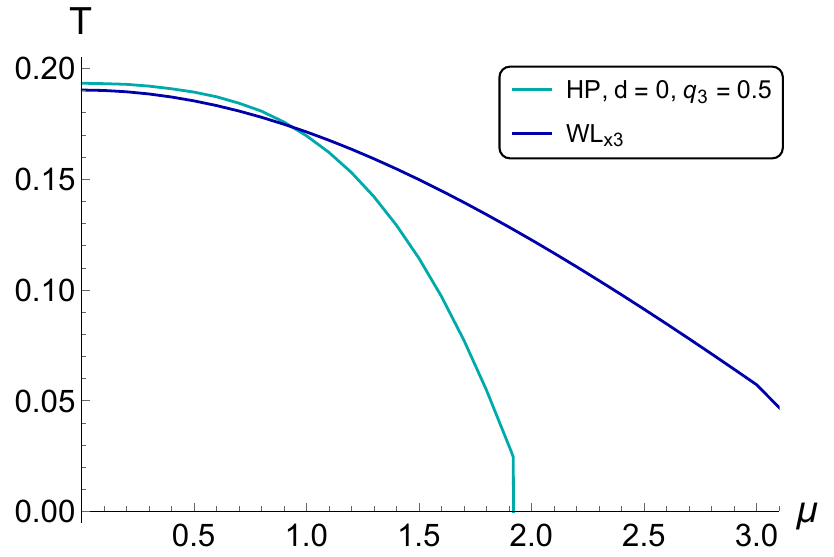} \\
  A \hspace{120pt} B \hspace{120pt} C \\
  \includegraphics[scale=0.3]{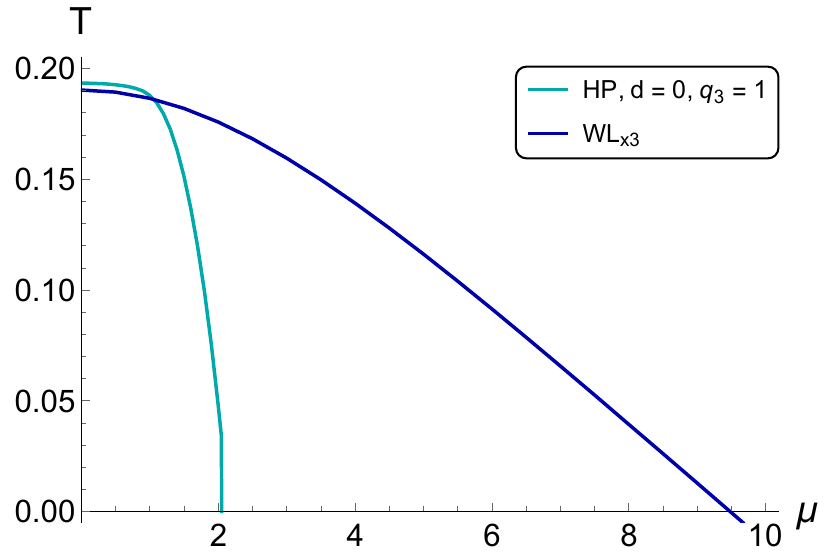} \quad
  \includegraphics[scale=0.3]{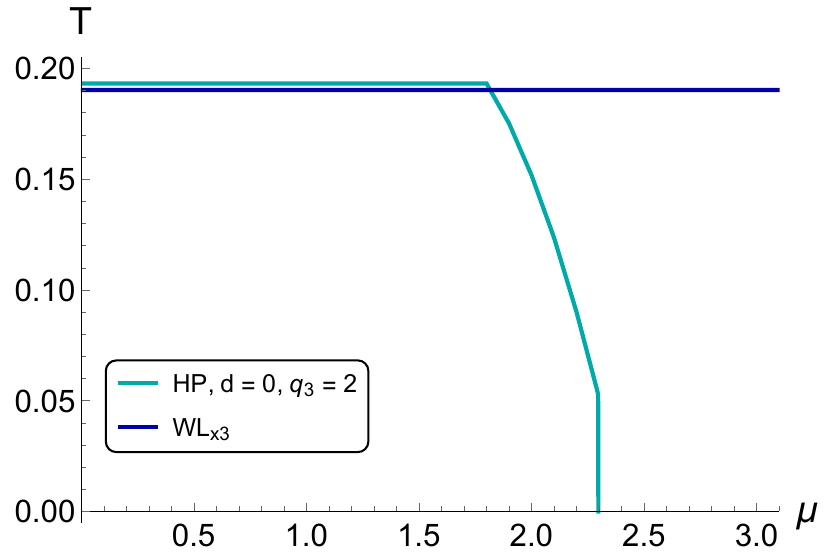} \quad
  \includegraphics[scale=0.3]{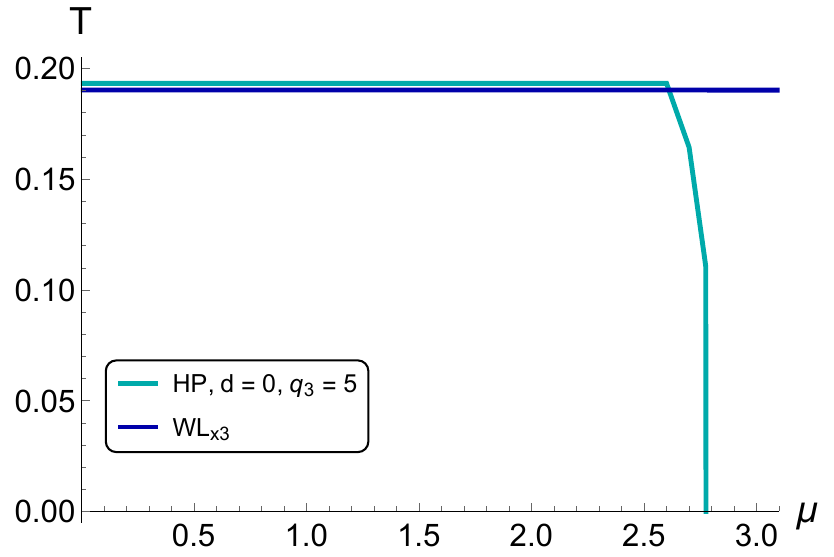} \\
  D \hspace{120pt} E \hspace{120pt} F
  \caption{Phase diagram $T(\mu)$ as a combination of the 1-st order
    phase transition (HP) and a crossover (WL) in magnetic field for
    $q_3 = 0$ (A), $q_3 = 0.01$ (B), $q_3 = 0.05$ (C), $q_3 = 1$ (D),
    $q_3 = 2$ (E), $q_3 = 5$ (F); $\nu = 4.5$, $a = 0.15$, $c = 1.16$,
    $c_B = - \, 0.01$, $d = 0$.}
  \label{Fig:PDq3-cB-001-nu45-d0-z5}
\end{figure}

\begin{figure}[h!]
  \centering 
  \includegraphics[scale=0.3]{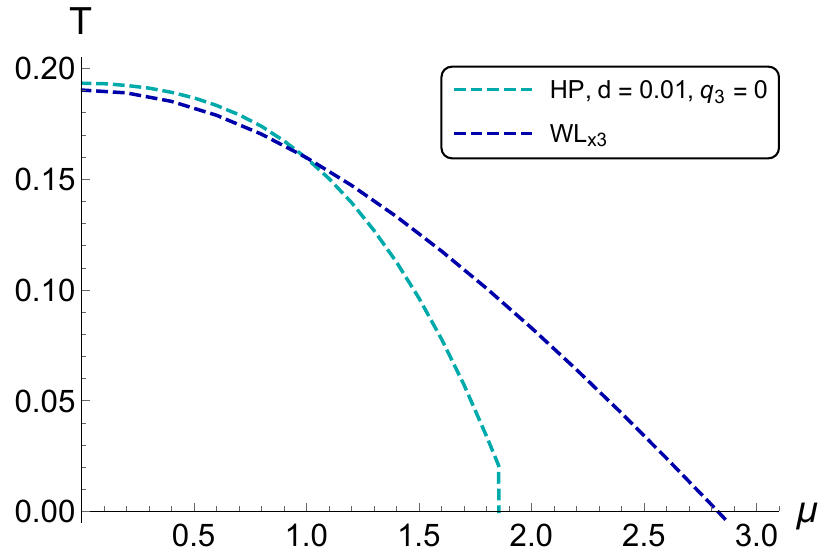} \quad
  \includegraphics[scale=0.3]{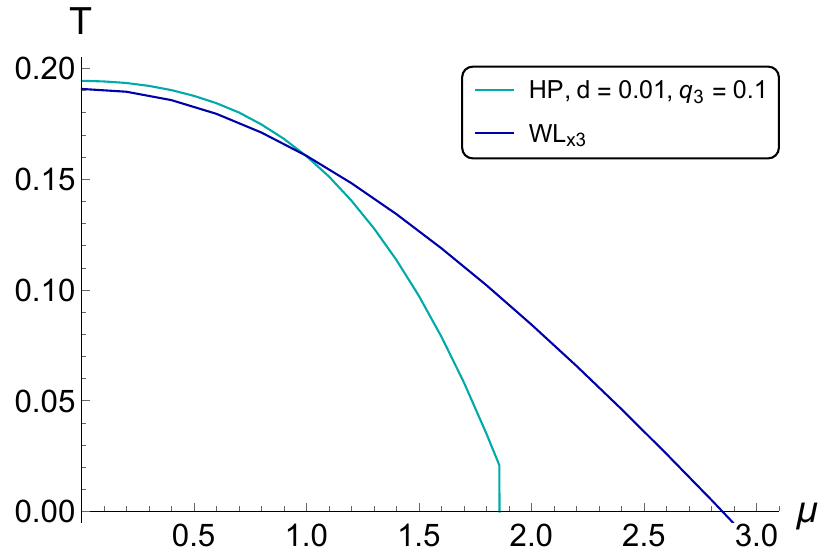} \quad
  \includegraphics[scale=0.3]{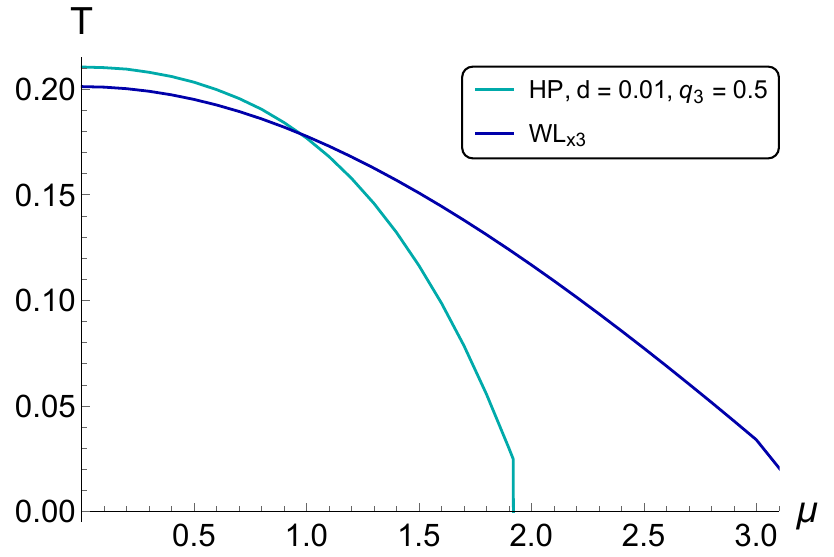} \\
  A \hspace{120pt} B \hspace{120pt} C \\
  \includegraphics[scale=0.3]{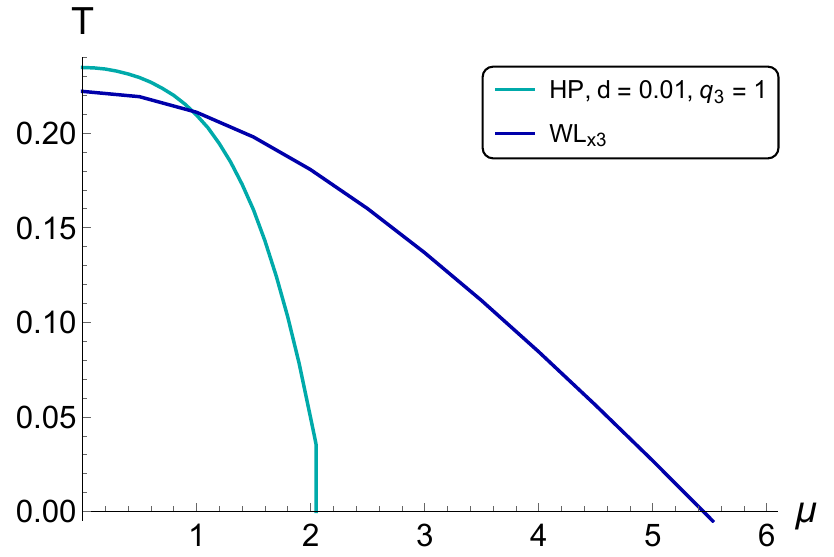} \quad
  \includegraphics[scale=0.3]{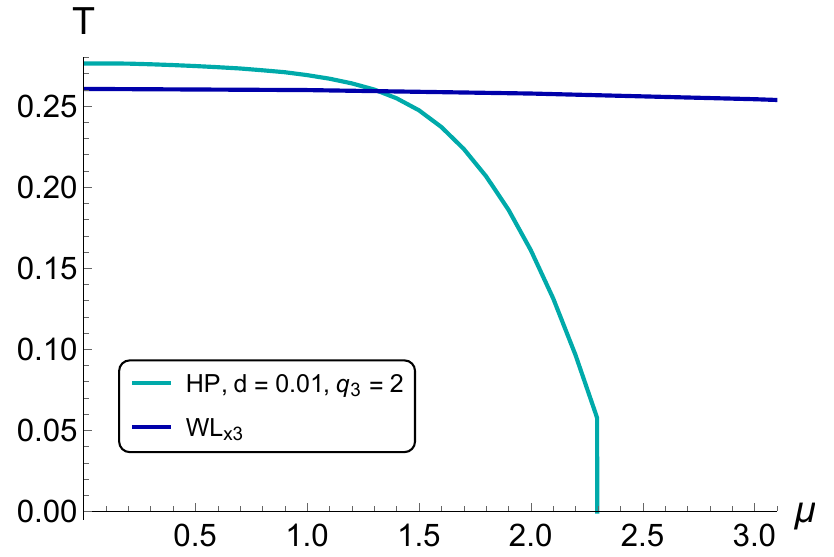} \quad
  \includegraphics[scale=0.3]{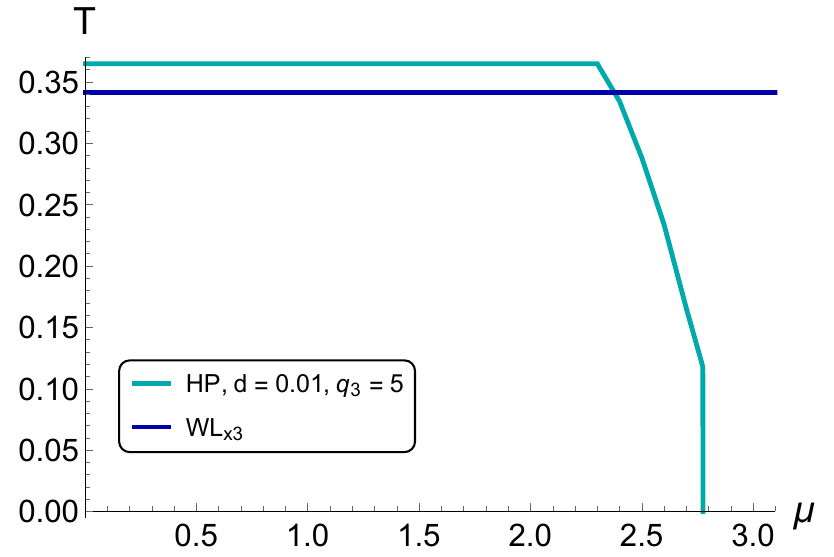} \\
  D \hspace{120pt} E \hspace{120pt} F
  \caption{Phase diagram $T(\mu)$ as a combination of the 1-st order
    phase transition (HP) and a crossover (WL) in magnetic field for
    $q_3 = 0$ (A), $q_3 = 0.01$ (B), $q_3 = 0.05$ (C), $q_3 = 1$ (D),
    $q_3 = 2$ (E), $q_3 = 5$ (F); $\nu = 4.5$, $a = 0.15$, $c = 1.16$,
    $c_B = - \, 0.01$, $d = 0.01$.}
  \label{Fig:PDq3-cB-001-nu45-d001-z5}
\end{figure}

\begin{figure}[h!]
  \centering 
  \includegraphics[scale=0.3]{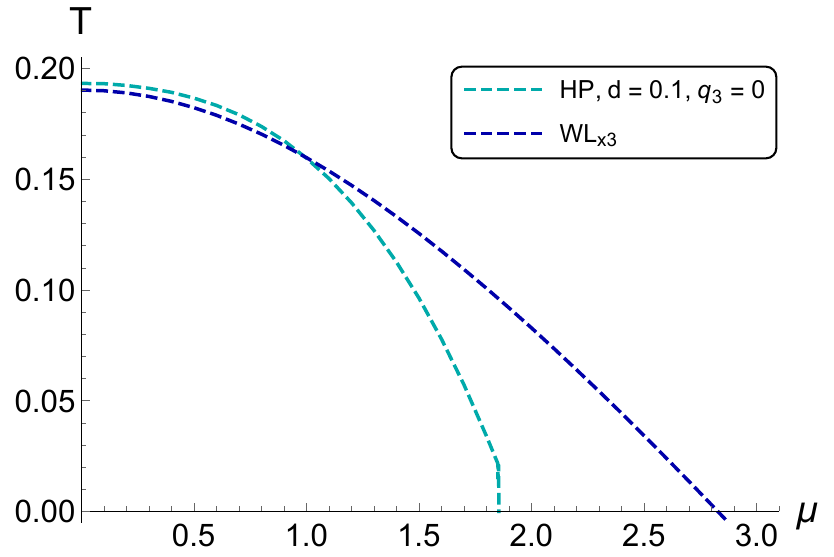} \quad
  \includegraphics[scale=0.3]{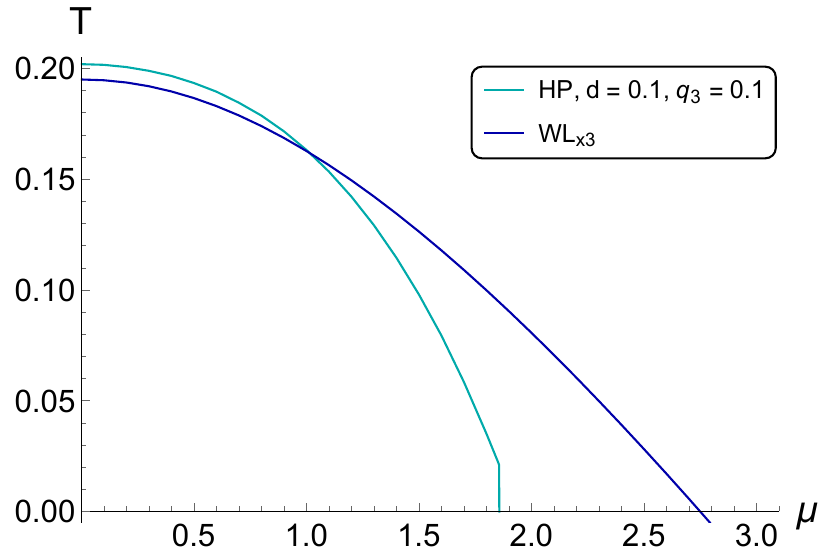} \quad
  \includegraphics[scale=0.3]{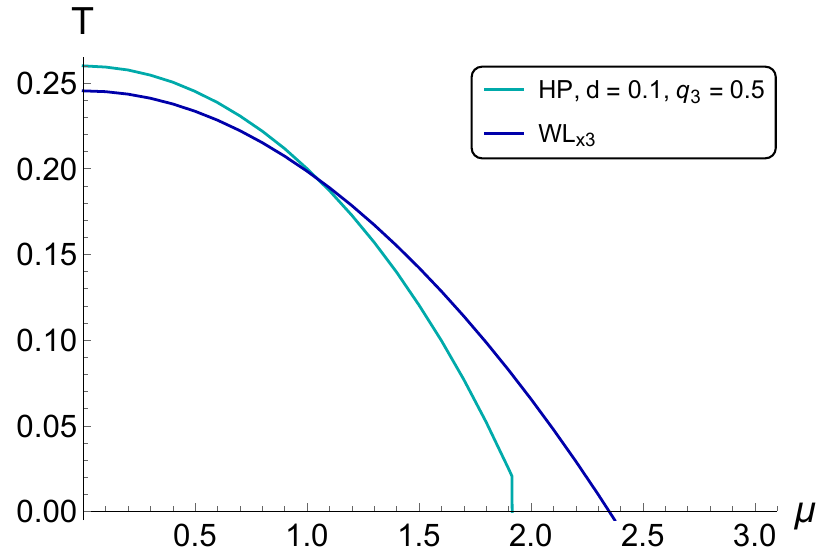} \\
  A \hspace{120pt} B \hspace{120pt} C \\
  \includegraphics[scale=0.3]{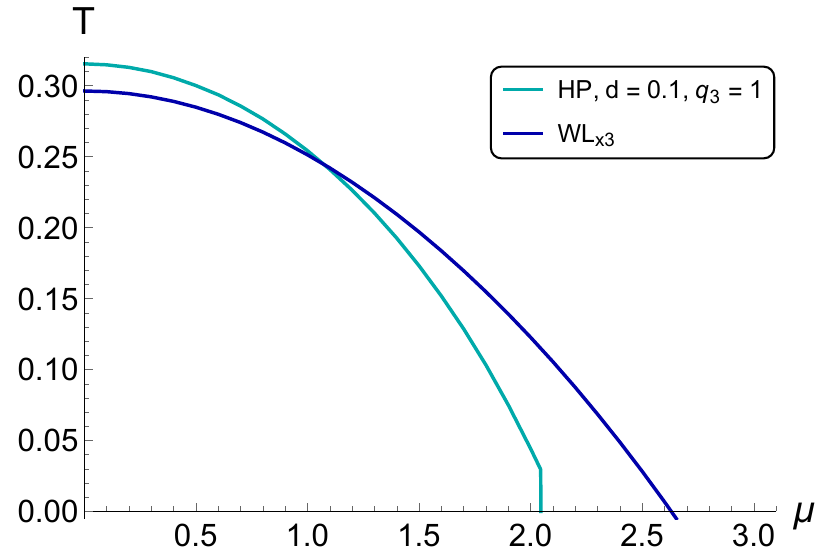} \quad
  \includegraphics[scale=0.3]{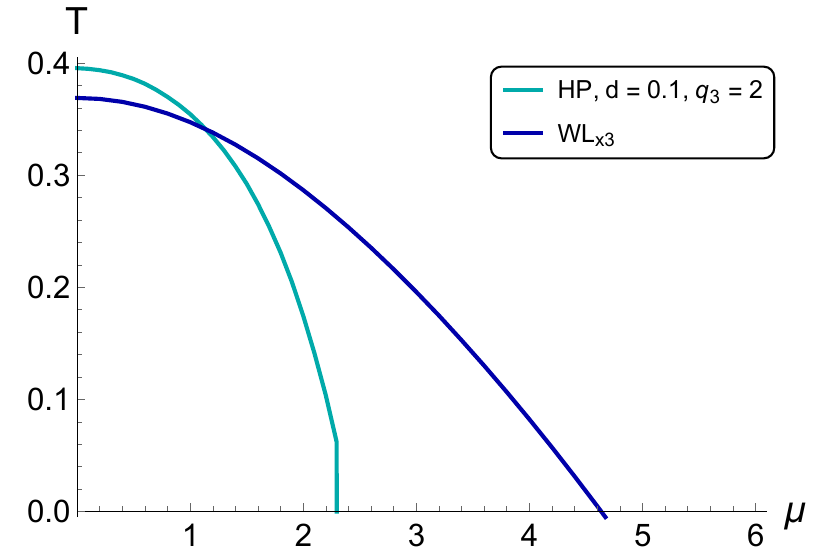} \quad
  \includegraphics[scale=0.3]{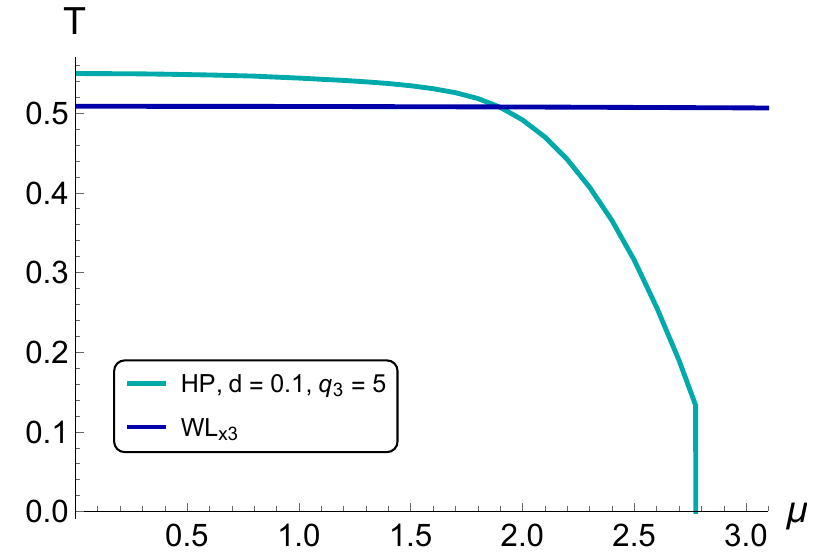} \\
  D \hspace{120pt} E \hspace{120pt} F
  \caption{Phase diagram $T(\mu)$ as a combination of the 1-st order
    phase transition (HP) and a crossover (WL) in magnetic field for
    $q_3 = 0$ (A), $q_3 = 0.01$ (B), $q_3 = 0.05$ (C), $q_3 = 1$ (D),
    $q_3 = 2$ (E), $q_3 = 5$ (F); $\nu = 4.5$, $a = 0.15$, $c = 1.16$,
    $c_B = - \, 0.01$, $d = 0.1$.}
  \label{Fig:PDq3-cB-001-nu45-d01-z5}
\end{figure}

Confronting curves corresponding to the 1-st order phase transition
and temporal Wilson loops shows that in primary isotropic media $\nu =
1$ confinement/deconfinement phase transition is completely defined by
the 1-st order phase transition
(Fig.\ref{Fig:PDq3-cB-001-nu1-d0-z5}--\ref{Fig:PDq3-cB-001-nu1-d01-z5}). The
only exception was found for strong magnetic field in $z^2$ limit
(Fig.\ref{Fig:PDq3-cB-001-nu1-d0-z5}.F). For $\nu = 4.5$ a crossover
for zero and small $\mu$ is observed, and the crossover region is
larger for stronger magnetic field
(Fig.\ref{Fig:PDq3-cB-001-nu45-d0-z5}--\ref{Fig:PDq3-cB-001-nu45-d01-z5}).





\newpage

\

\newpage

\

\newpage

\section{Conslusions}\label{conclusions}

This work investigates the $z^5$-correction in the holographic warp
factor exponent. For this purpose a model describing hot dense
anisotropic quarks-gluon plasma in magnetic field is
constructed. Spatial anisotropy of the QGP produced in HIC is
parametrised by $\nu$, like it was in previous works. Metric
deformation due to magnetic field is described by $c_B$
coefficient. EOM solution requires it to be negative, thus narrowing
the possibilities of magnetic catalysis to the only case of fixed
$c_B$ and the magnetic field variable $q_3$.

Via AdS$_5$/CFT$_4$ correspondence the confinement/deconfinement phase
transition of QGP in 4D theory is mirrored to the collapse of unstable
black hole and presented as the 1-st order phase transition. Free
energy consideration shows that it is mainly realised as the
Hawking-Page transition of the 5D black hole to thermal gas. Stronger
magnetic field leads to the phase transition temperature rise and the
chemical potential range expansion. If $c_B$ absolute value is small
enough then large magnetic field values can be reached. Isotropisation
of the originally anisotropic QGP also leads to larger temperatures
and significantly larger available chemical potentials. The effect of
direct magnetic catalysis is successfully obtained.

Temporal Wilson loops that are also considered. The corresponding
phase transition is presented on the phase diagram as a
crossover. It's temperature and chemical potential range are sensible
to magnetic field. But it turns out to take part in the
confinement/deconfinement phase transition process for large
anisotropy ($\nu = 4.5$) and loose it's influence during
isotropization. The additional flexibility of this picture is provided
by the $z^5$-term coefficient $d$ in the warp factor.

The model obtained in the current work satisfies the stated
requirements and it's fitting capabilities are rather good. At the
same time, it needs further researches such as energy losses, string
tension, running coupling constant, renormgroup flow behavior and
chiral transition.

\section{Acknowledgments}

The author greatly thanks I. Ya. Aref'eva for fruitful
discussions. This work was performed within the scientific project
\textnumero~FSSF-2023-0003 and the ``BASIS'' Science Foundation, grant
\textnumero~22-1-3-18-1.

$$\,$$

\newpage

\appendix 

\section*{Appendix}

\section{Solution. Plots}\label{appendixA}

\begin{figure}[h!]
  \centering
  \includegraphics[scale=0.24]{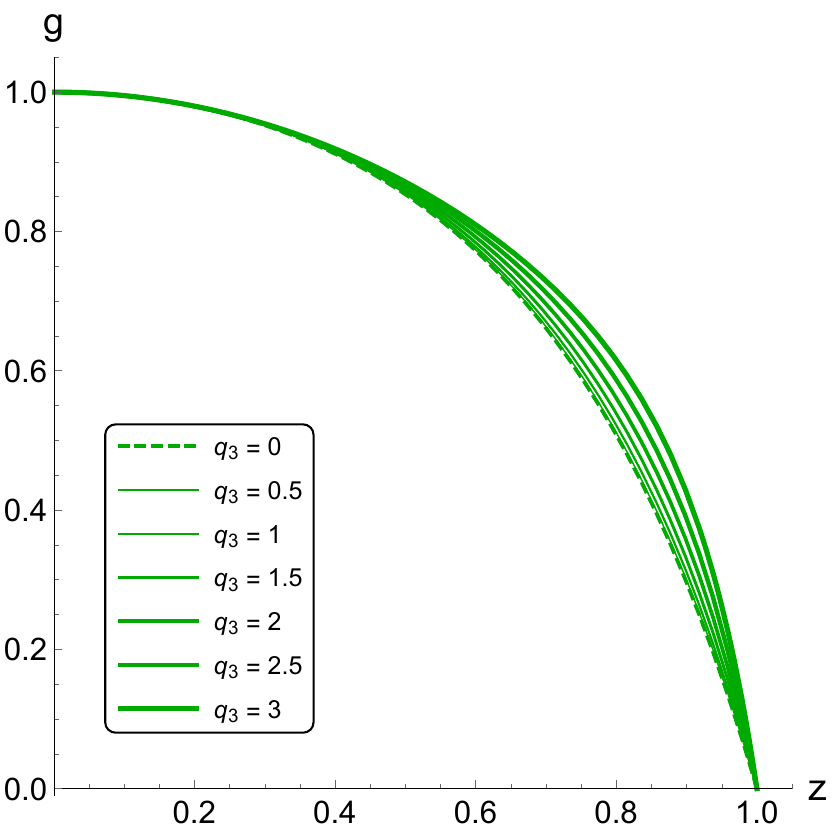} \
  \includegraphics[scale=0.24]{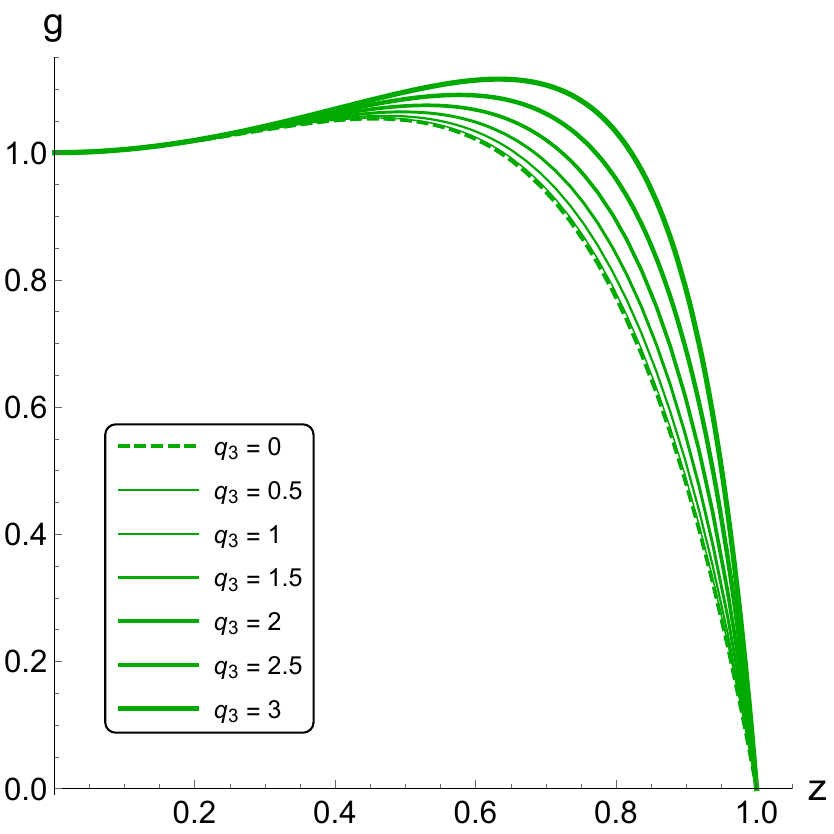} \quad
  \includegraphics[scale=0.24]{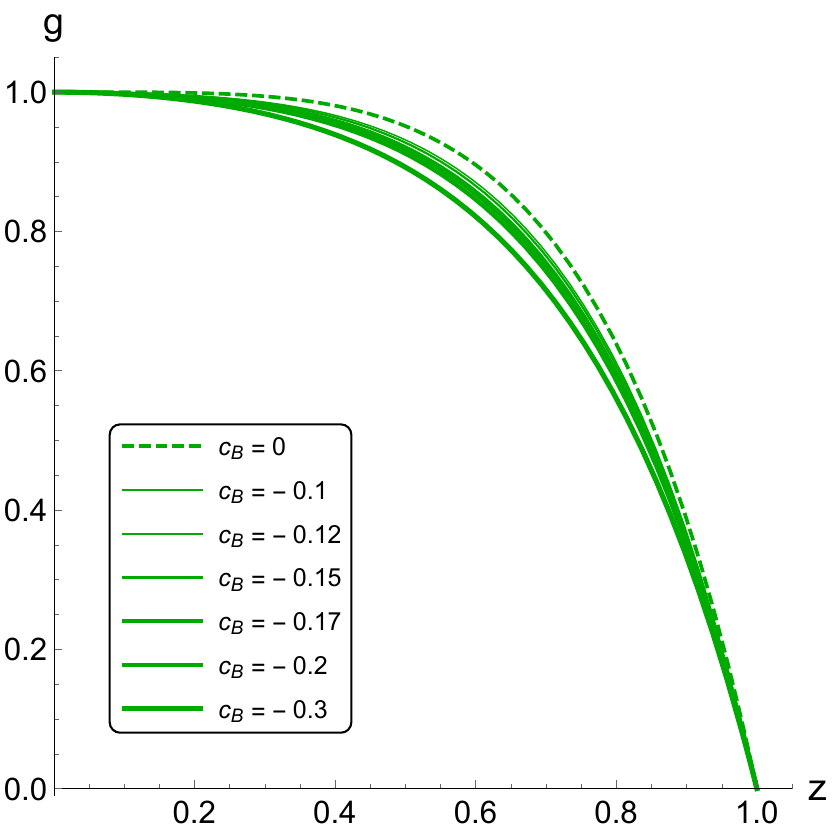} \
  \includegraphics[scale=0.24]{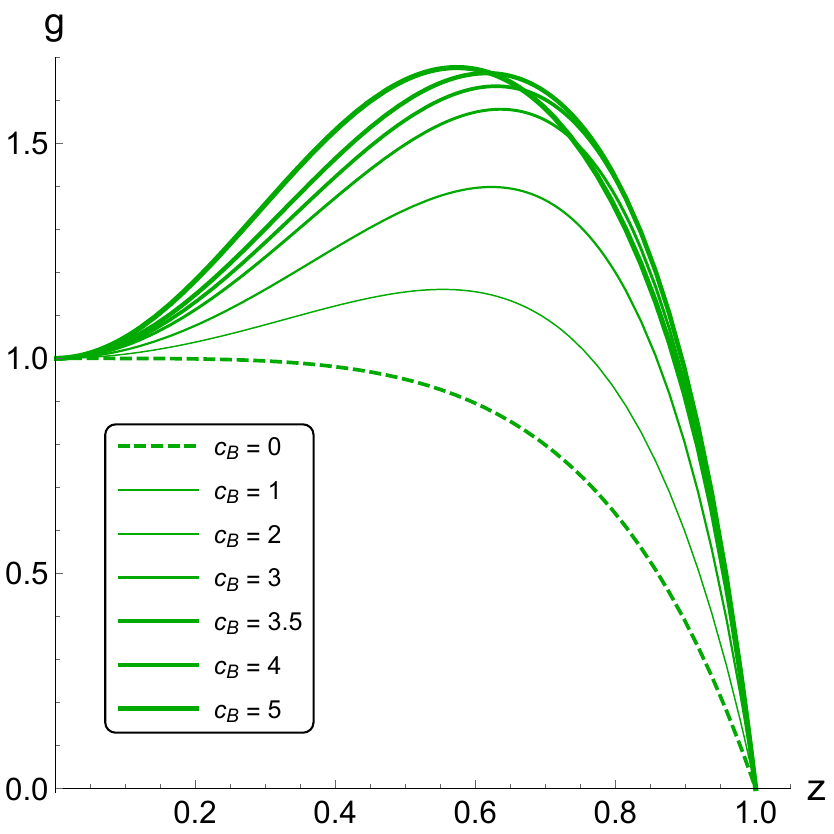} \\
  A \hspace{200pt} B \ \\
  \includegraphics[scale=0.24]{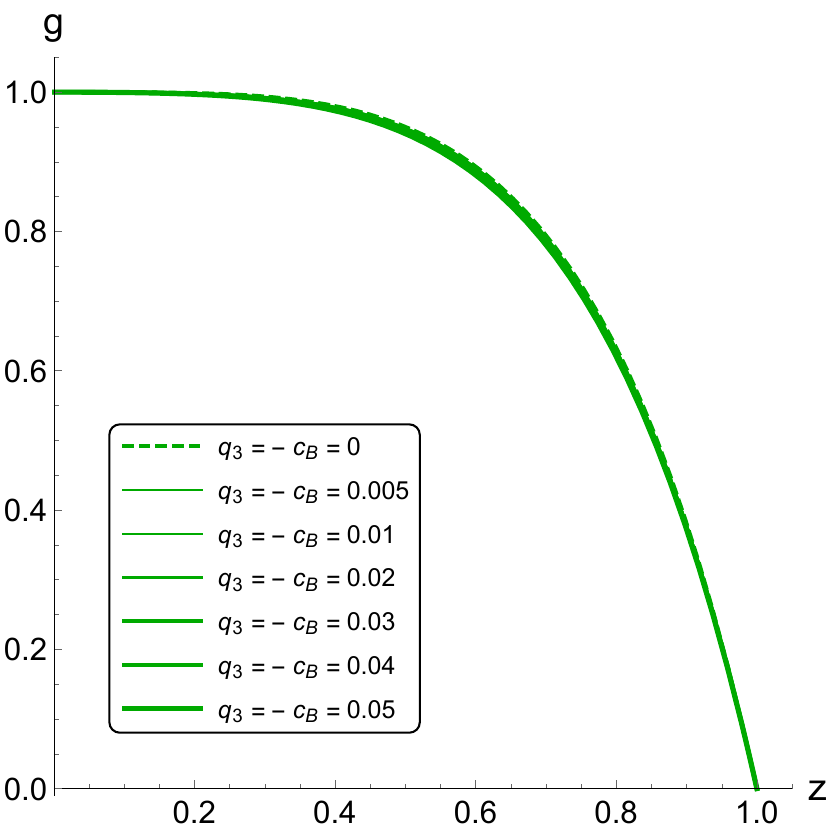} \
  \includegraphics[scale=0.24]{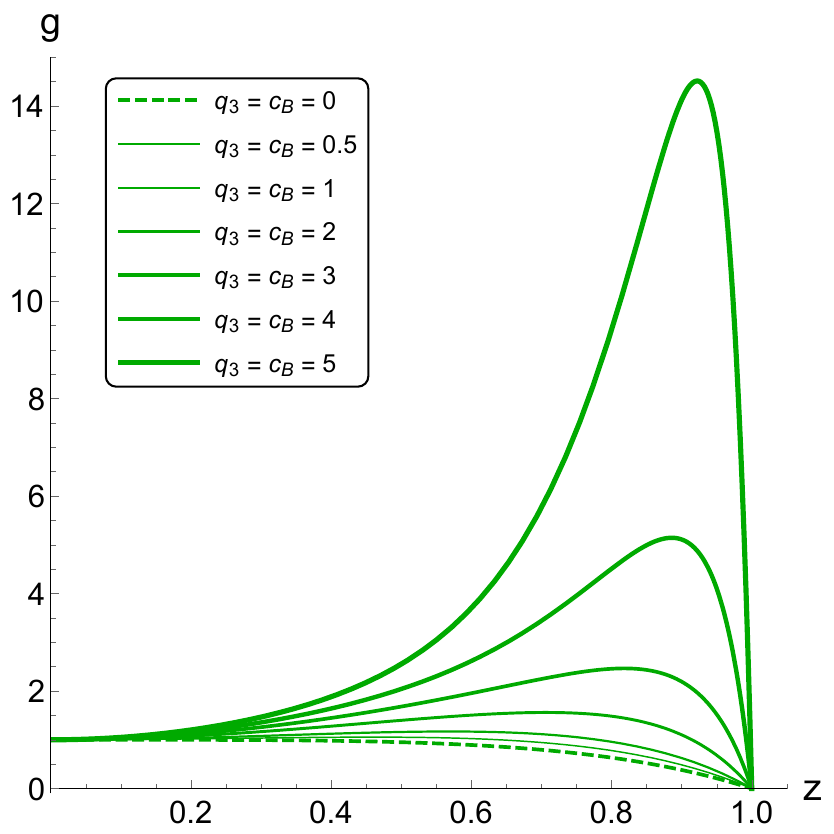} \quad
  \includegraphics[scale=0.24]{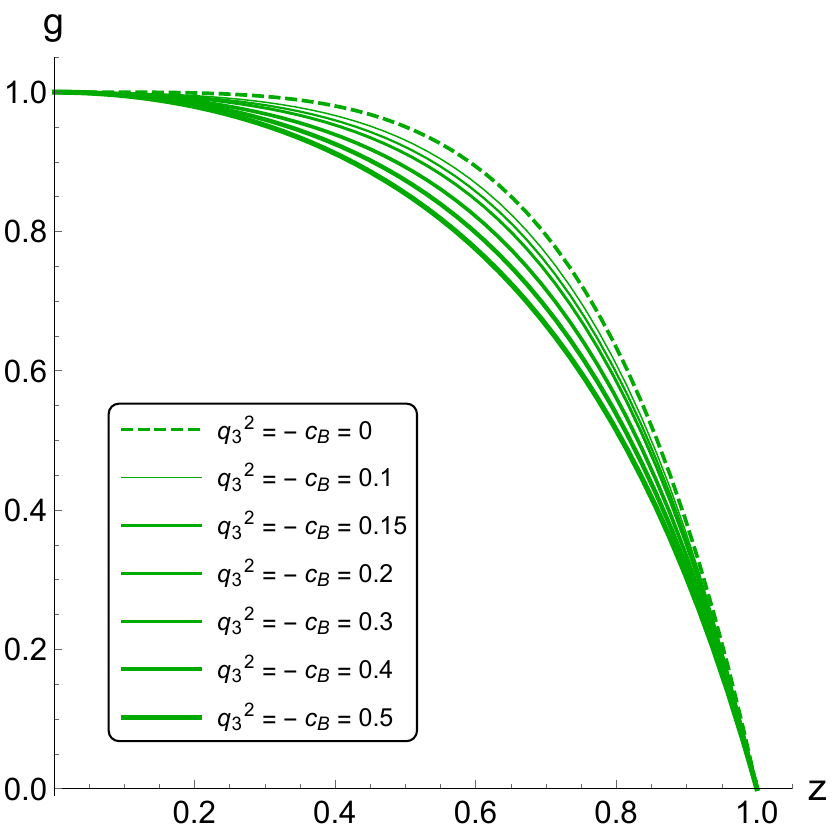} \
  \includegraphics[scale=0.24]{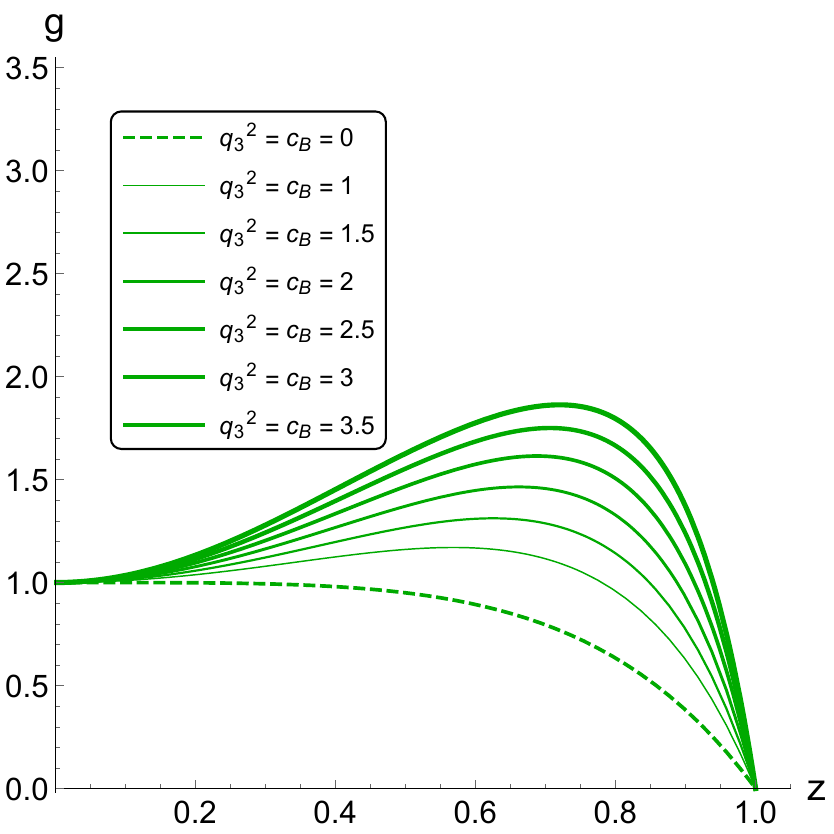} \\
  C \hspace{200pt} D \ \\
  \includegraphics[scale=0.24]{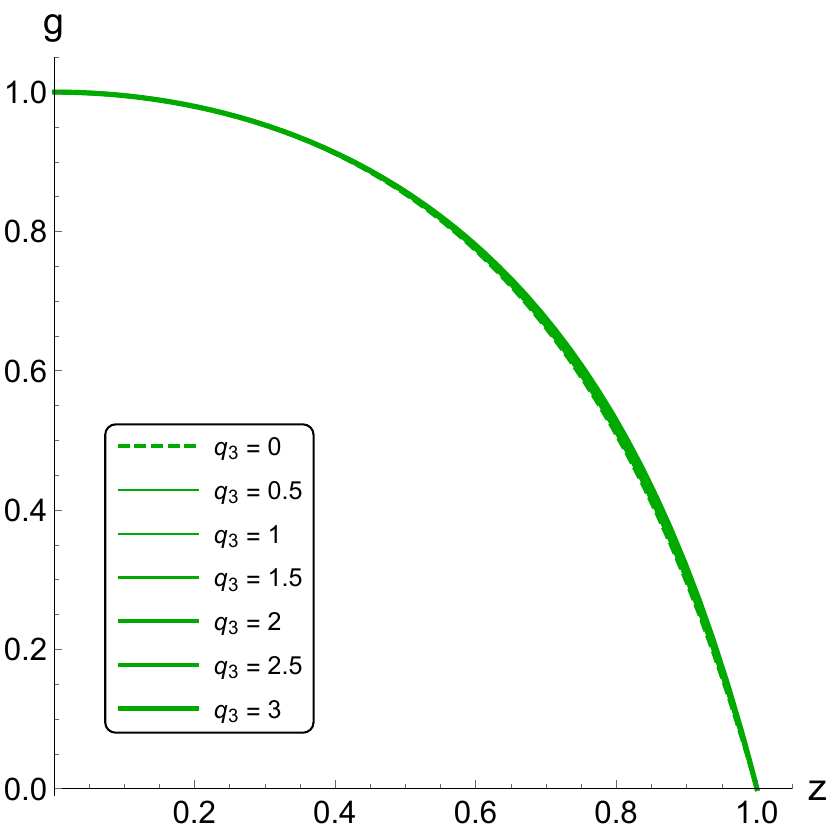} \
  \includegraphics[scale=0.24]{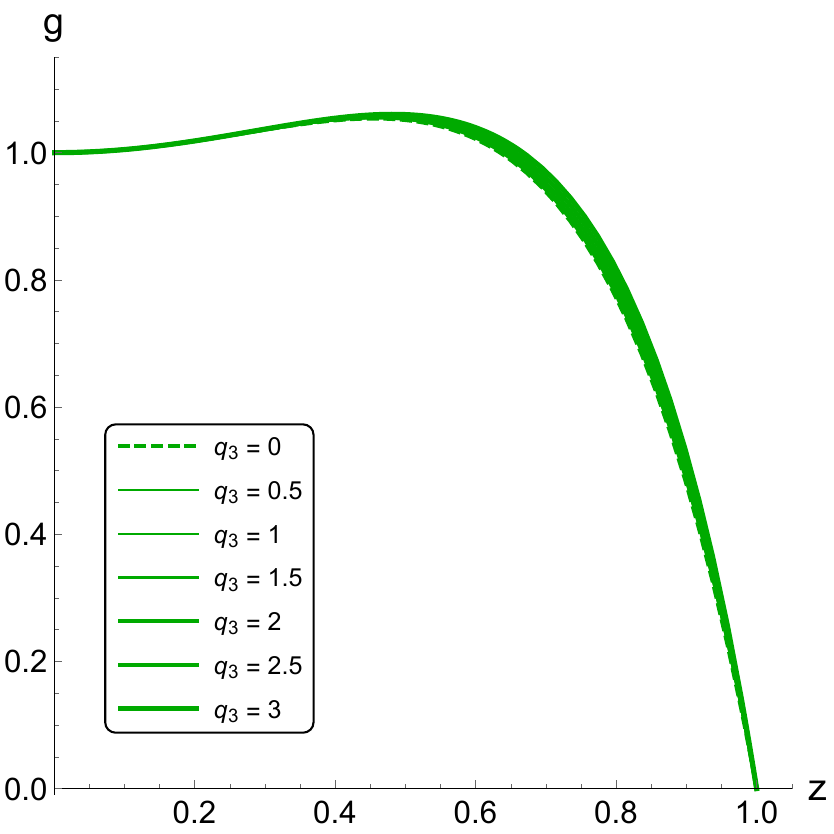} \quad
  \includegraphics[scale=0.24]{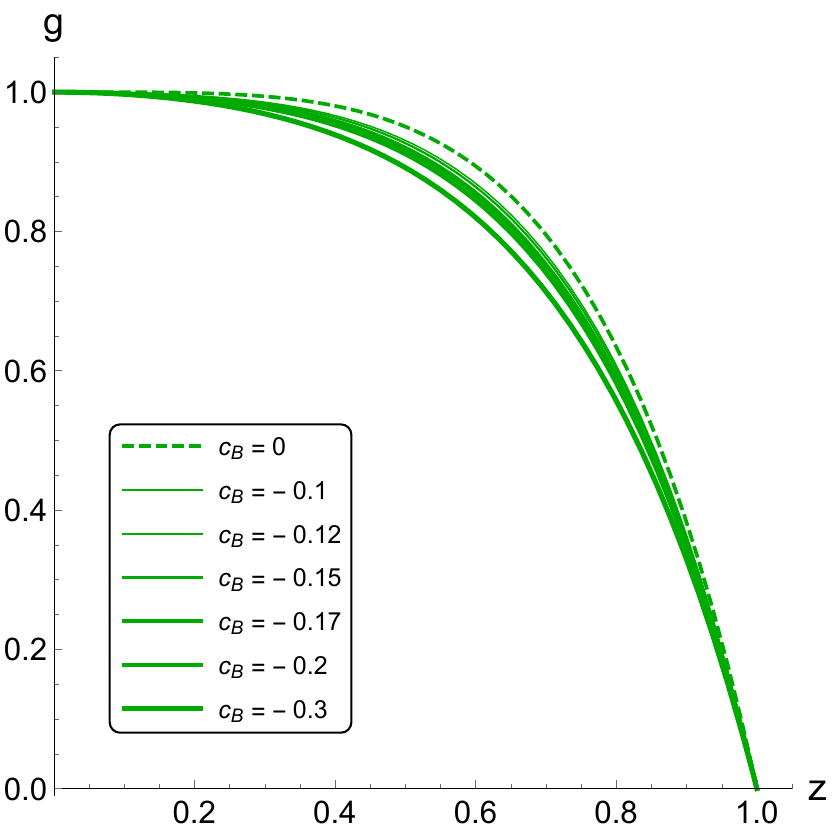} \
  \includegraphics[scale=0.24]{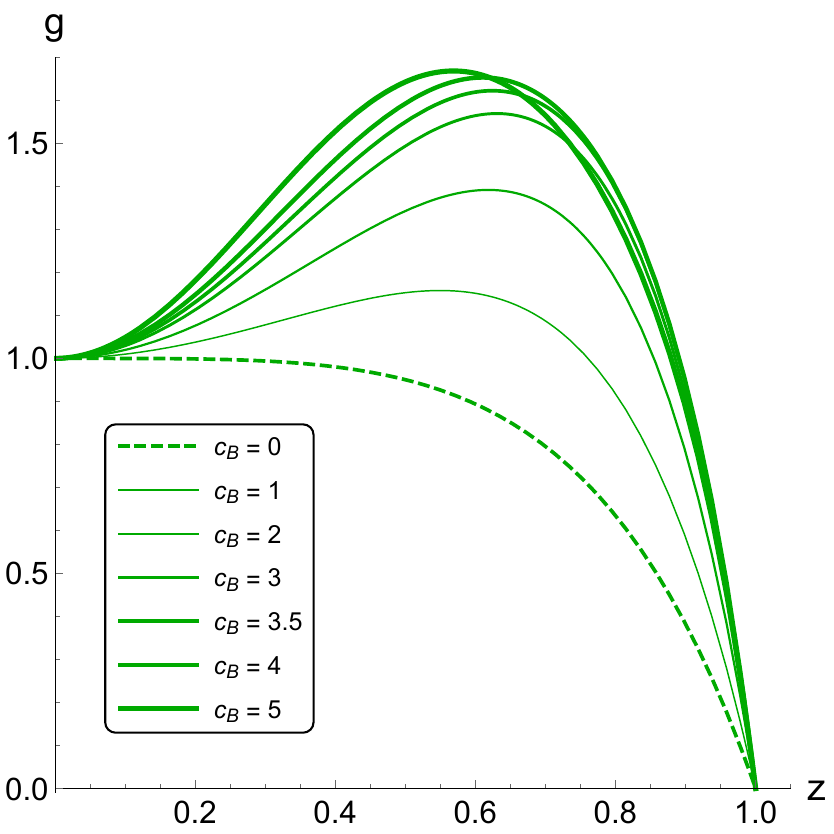} \\
  E \hspace{200pt} F \ \\
  \includegraphics[scale=0.24]{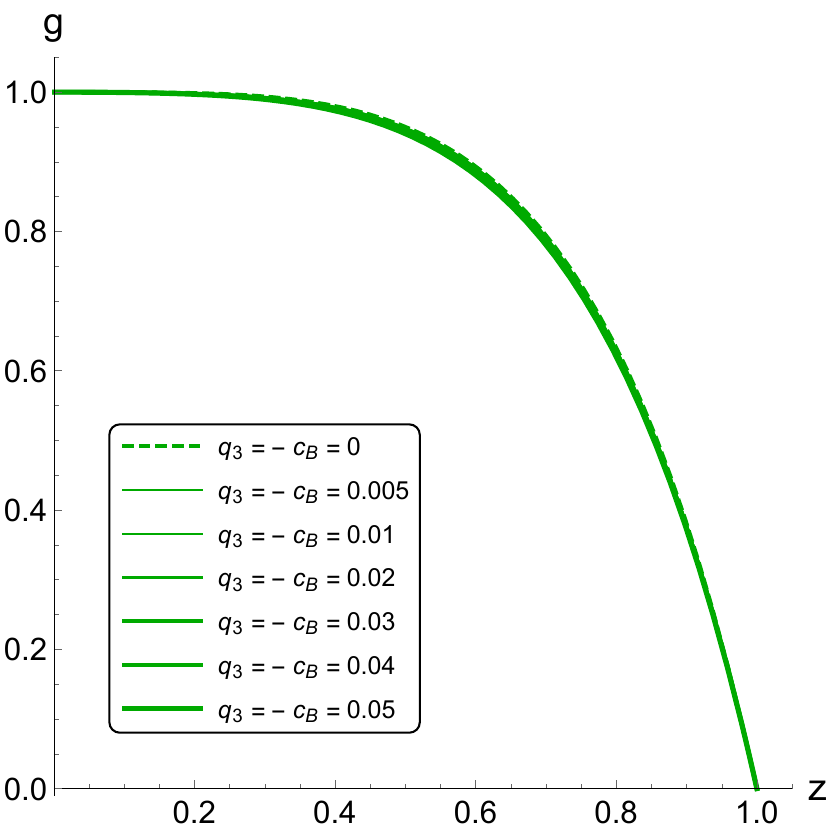} \
  \includegraphics[scale=0.24]{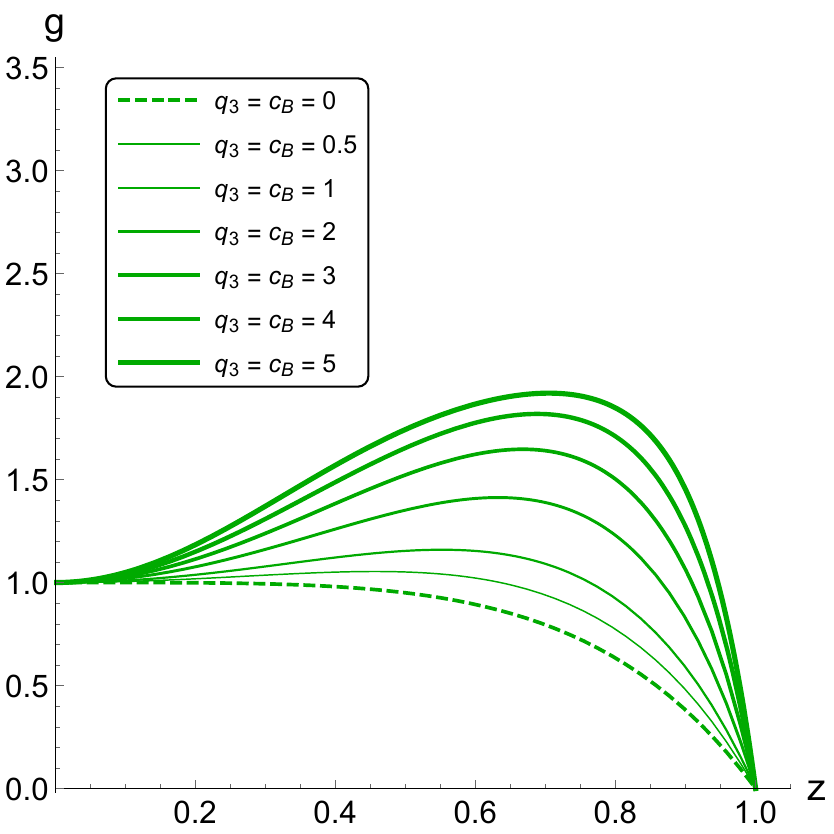} \quad
  \includegraphics[scale=0.24]{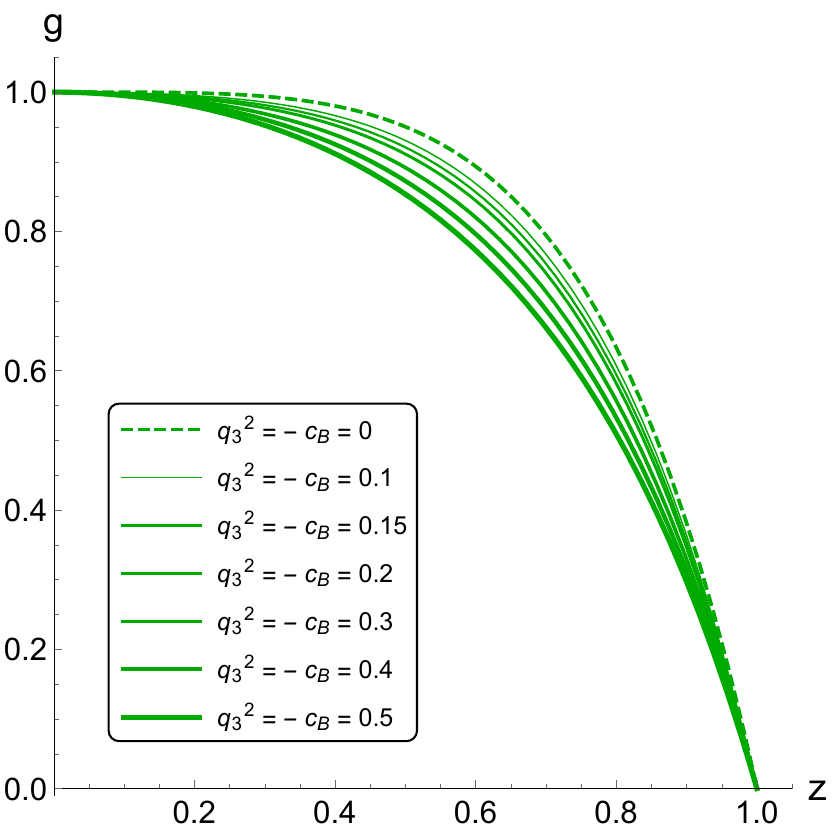} \
  \includegraphics[scale=0.24]{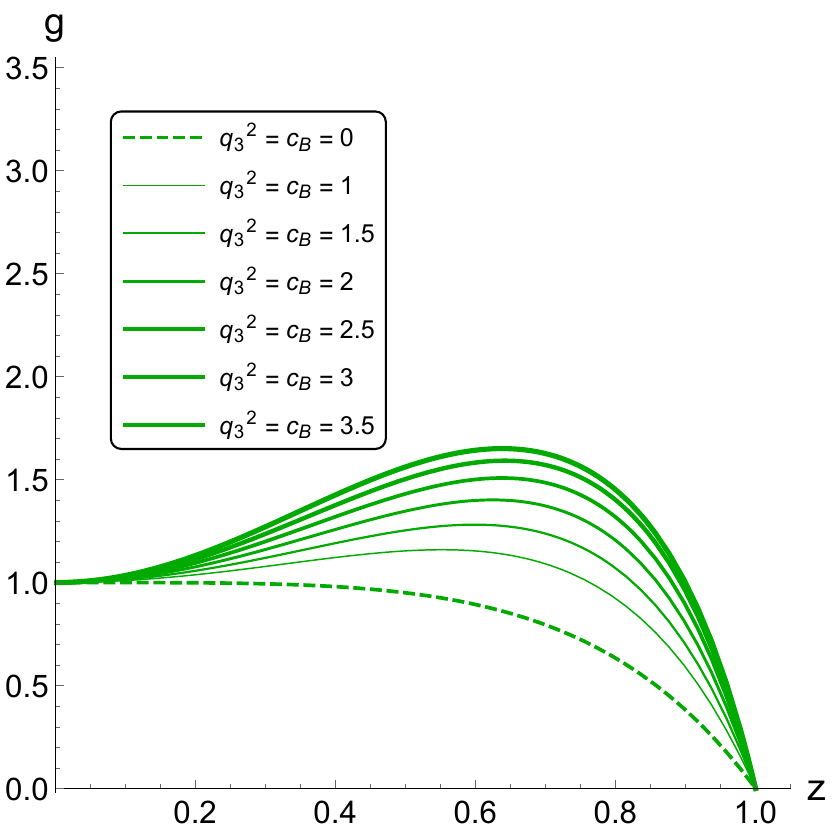} \\
  G \hspace{200pt} H
  \caption{Blackening function $g(z)$ in magnetic field with different
    $q_3$ (A,E) for $c_B = - \, 0.5$ (left) and $c_B = 0.5$ (right);
    with different $c_B$ for $q_3 = 0.5$ (B,F) for $c_B < 0$ (left)
    and $c_B > 0$ (right); for different $q_3 = \pm \, c_B$ (C,G); for
    different $q_3^2 = \pm \, c_B$ (D,H) for $d = 0.06 > 0.05$ (A-D)
    and $d = 0.01 < 0.05$ (E-H) in primary isotropic case $\nu = 1$,
    $a = 0.15$, $c = 1.16$, $\mu = 0$.}
  \label{Fig:gz-q3cB-nu1-mu0-z5}
\end{figure}

\begin{figure}[t!]
  \centering
  \includegraphics[scale=0.24]{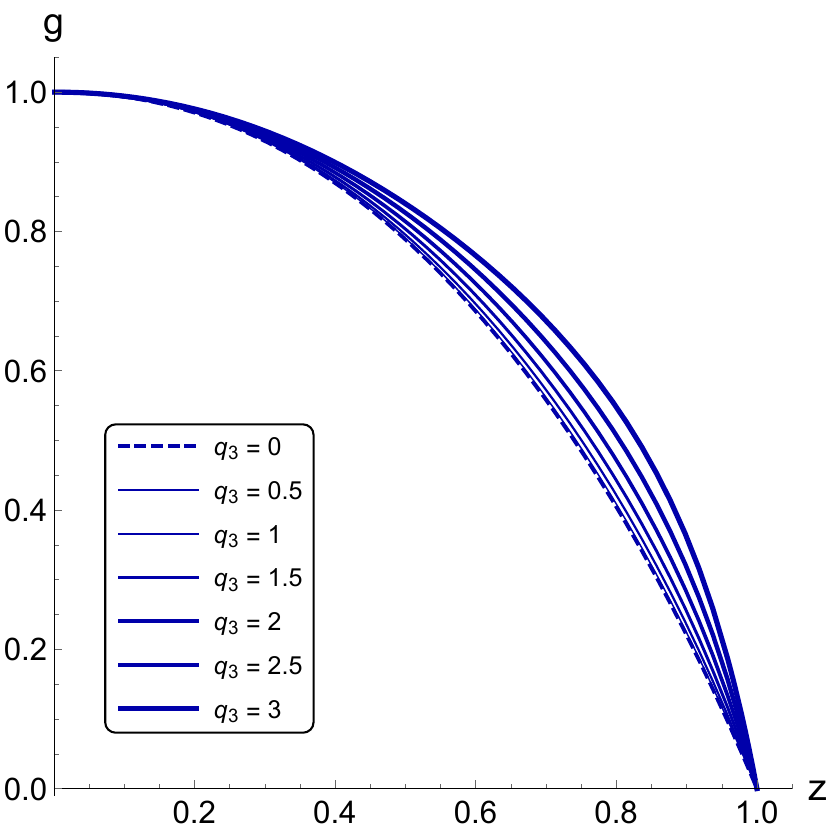} \
  \includegraphics[scale=0.24]{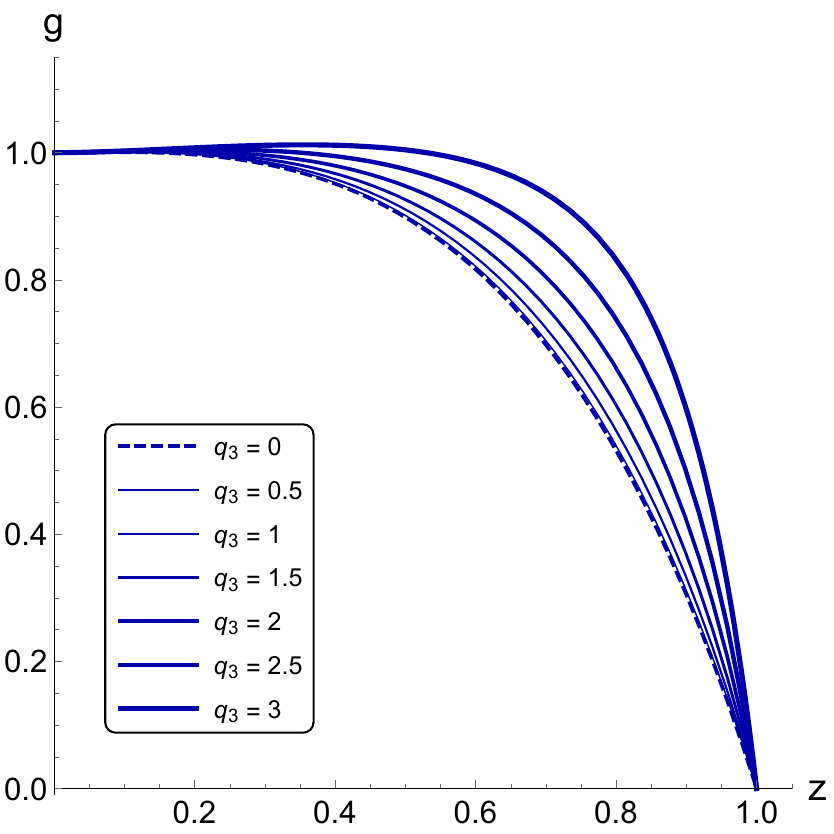} \quad
  \includegraphics[scale=0.24]{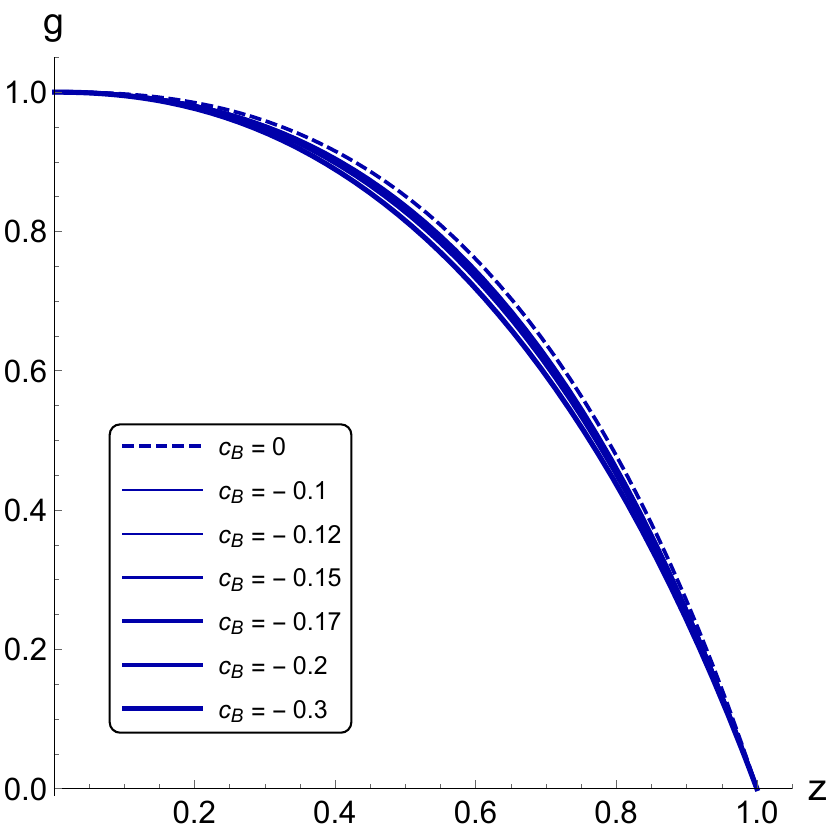} \
  \includegraphics[scale=0.24]{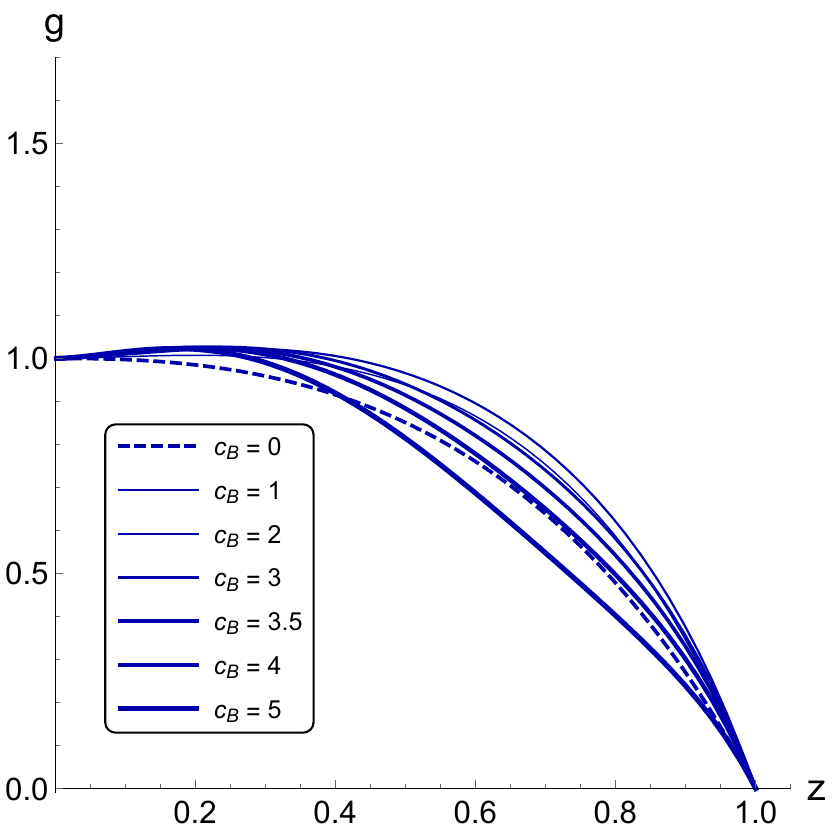} \\
  A \hspace{200pt} B \ \\
  \includegraphics[scale=0.24]{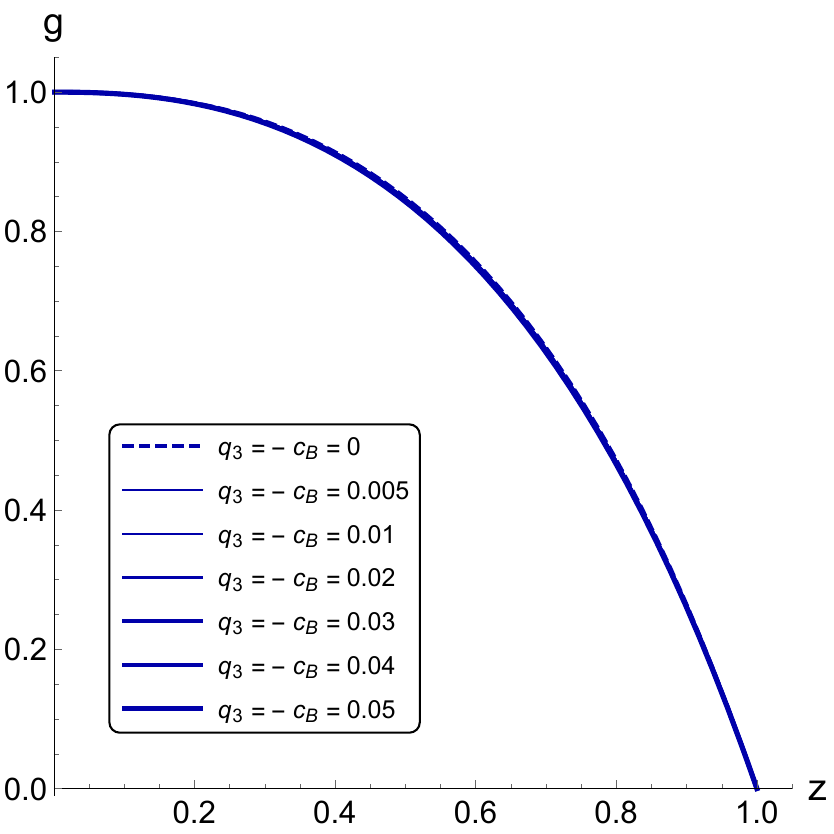} \
  \includegraphics[scale=0.24]{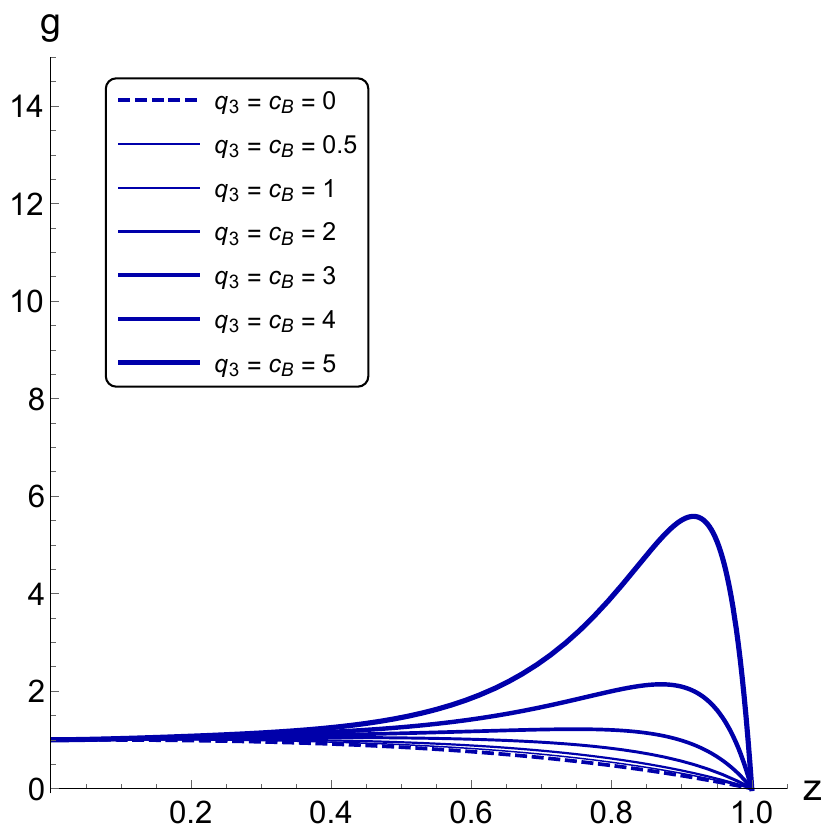} \quad
  \includegraphics[scale=0.24]{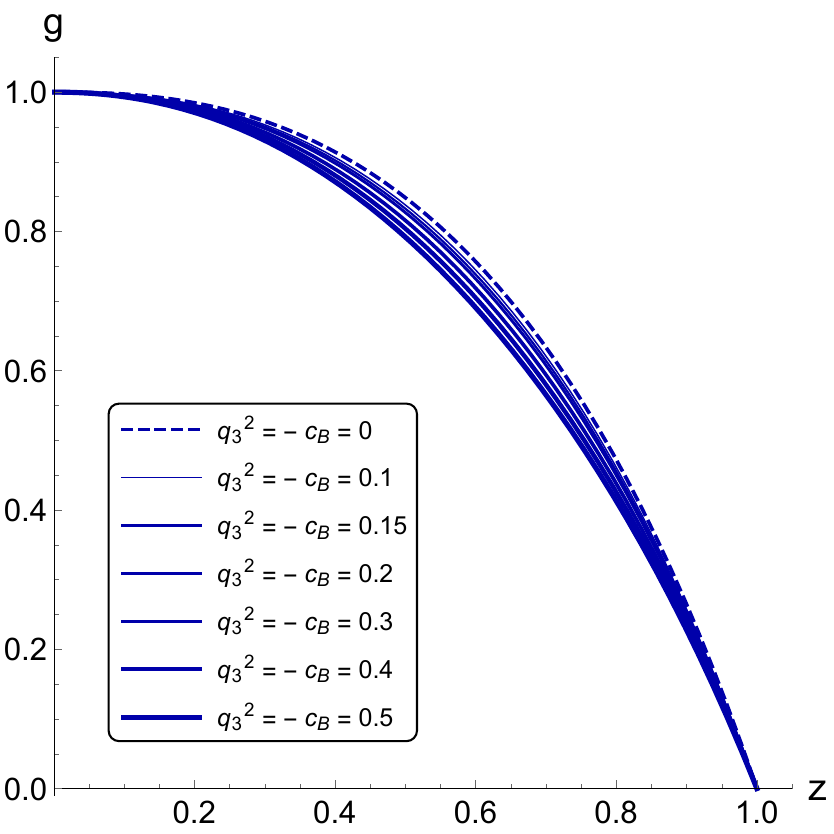} \
  \includegraphics[scale=0.24]{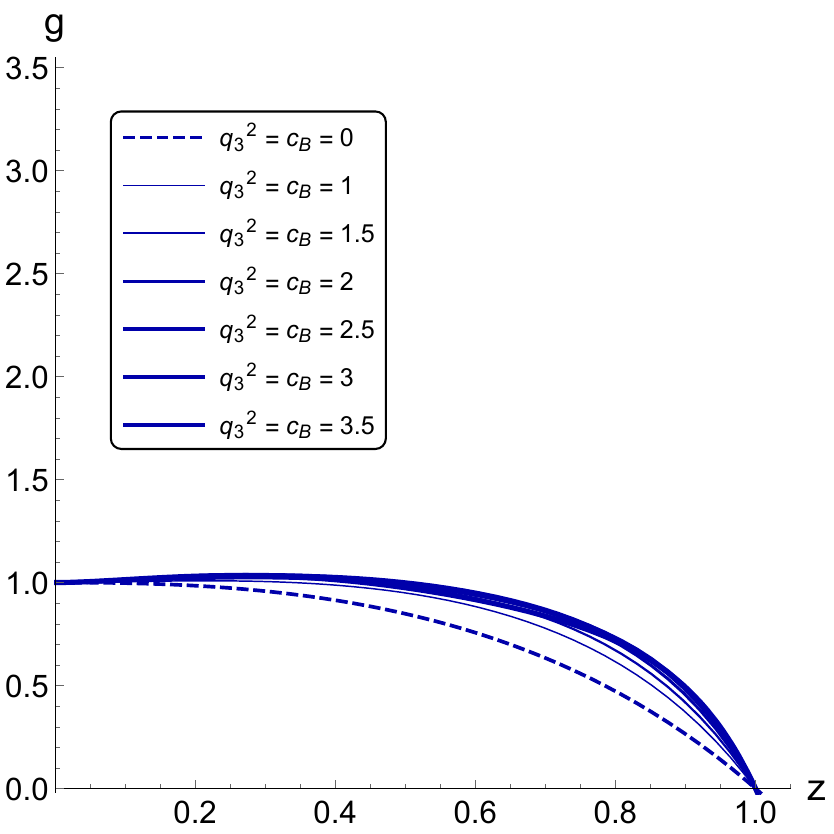} \\
  C \hspace{200pt} D \ \\
  \includegraphics[scale=0.24]{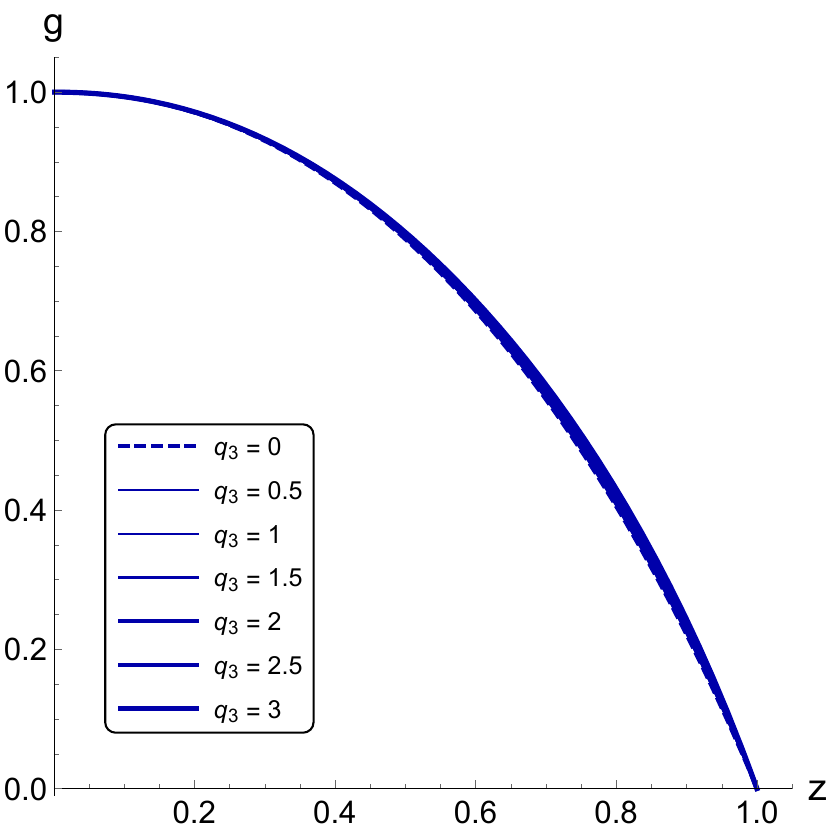} \
  \includegraphics[scale=0.24]{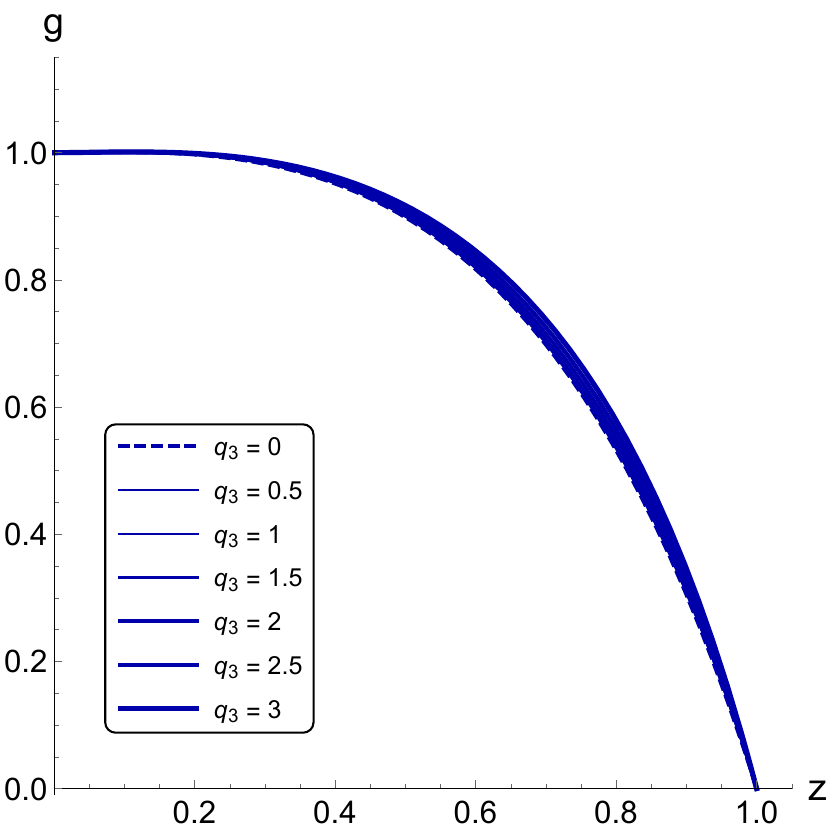} \quad
  \includegraphics[scale=0.24]{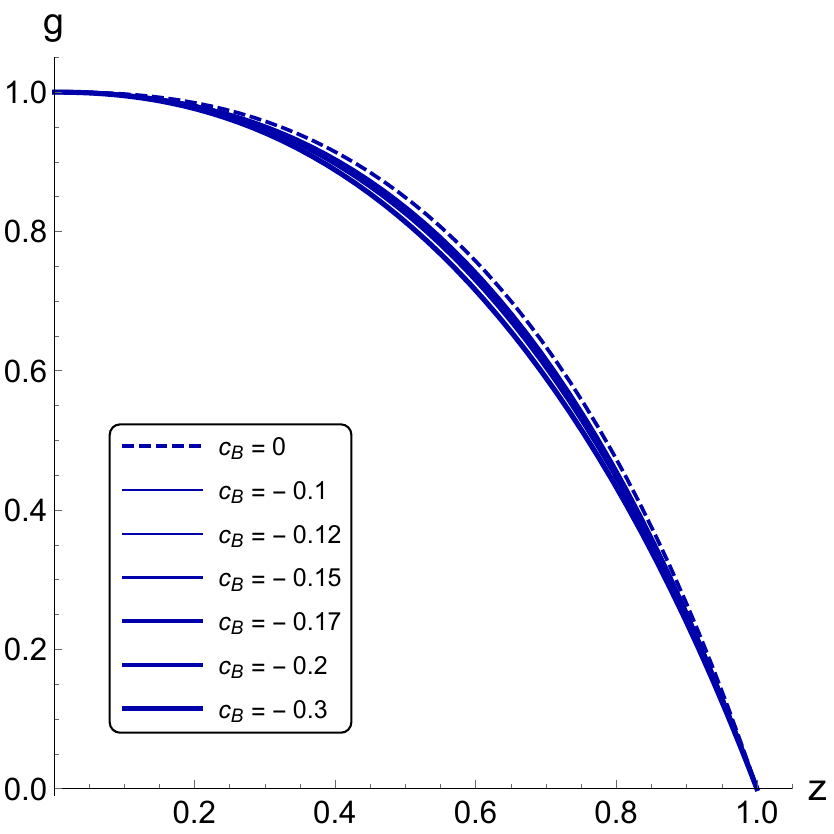} \
  \includegraphics[scale=0.24]{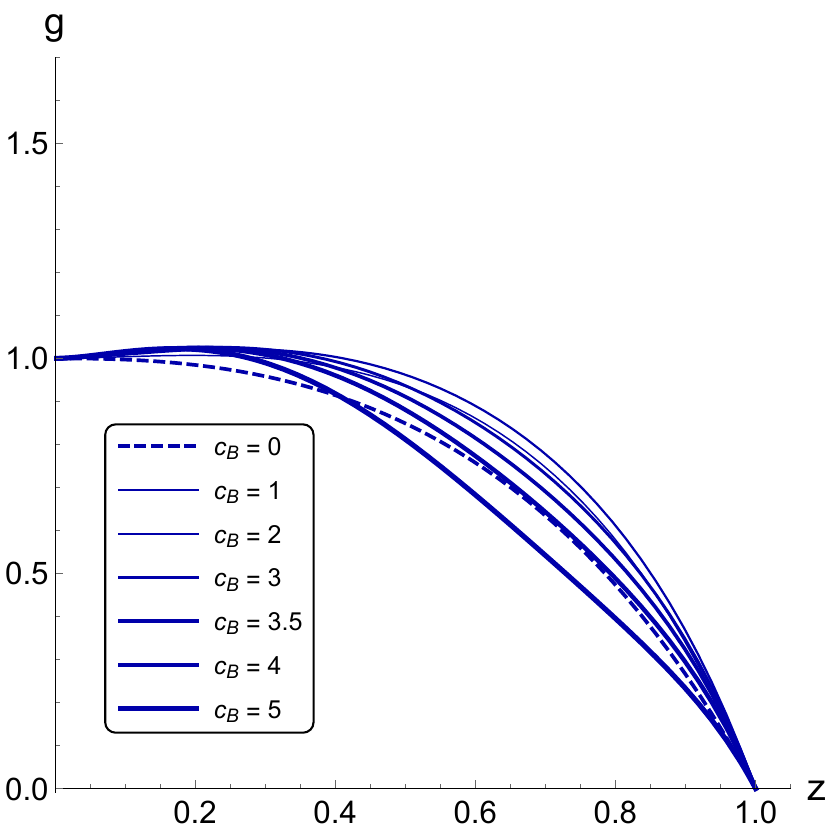} \\
  E \hspace{200pt} F \ \\
  \includegraphics[scale=0.24]{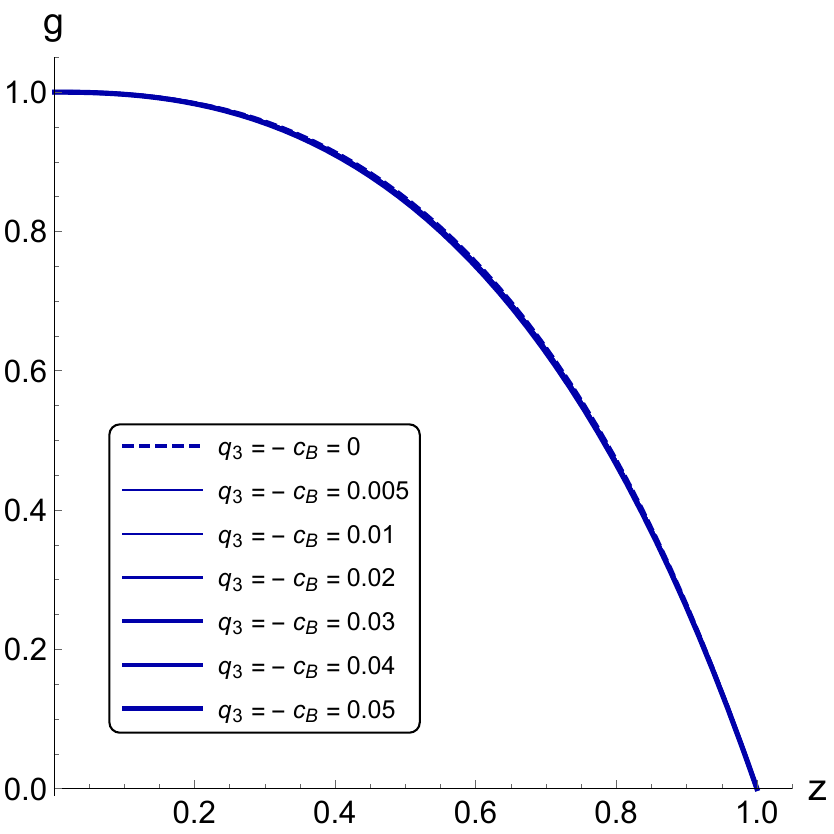} \
  \includegraphics[scale=0.24]{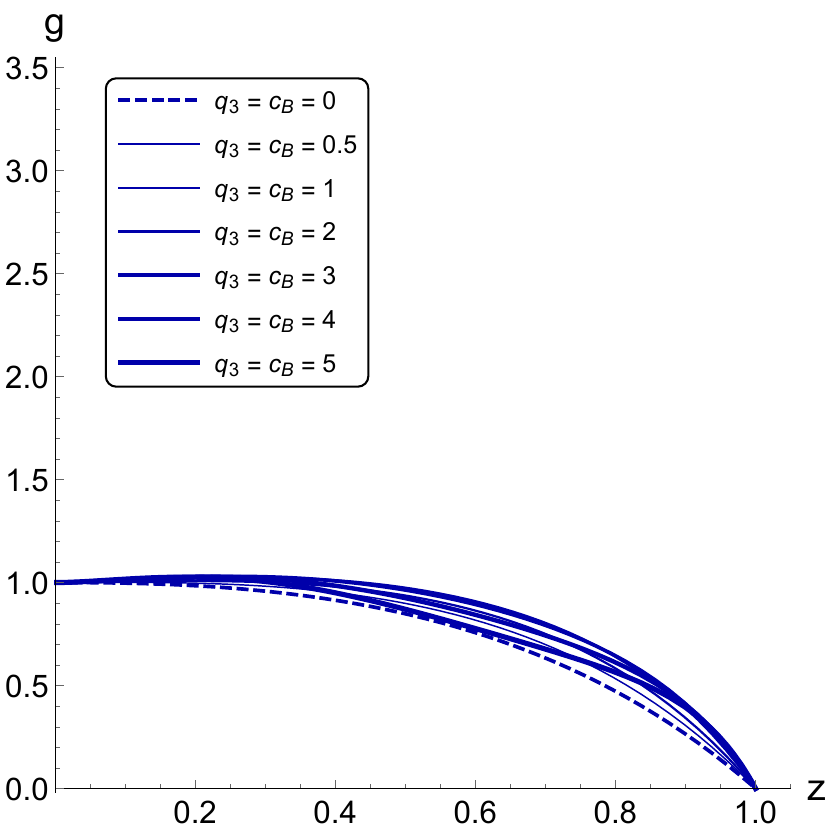} \quad
  \includegraphics[scale=0.24]{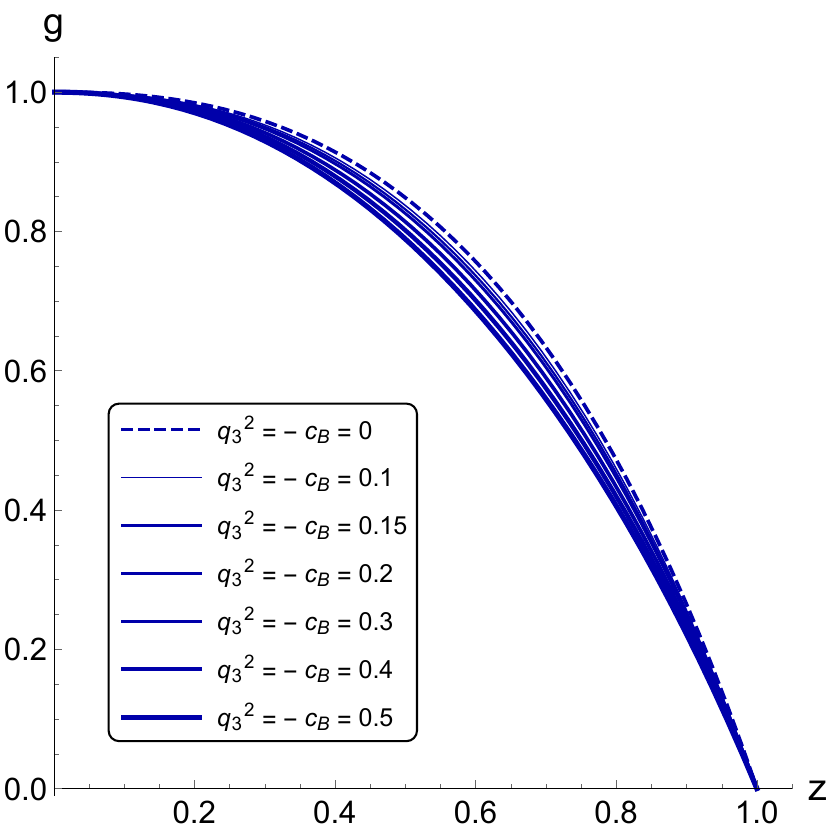} \
  \includegraphics[scale=0.24]{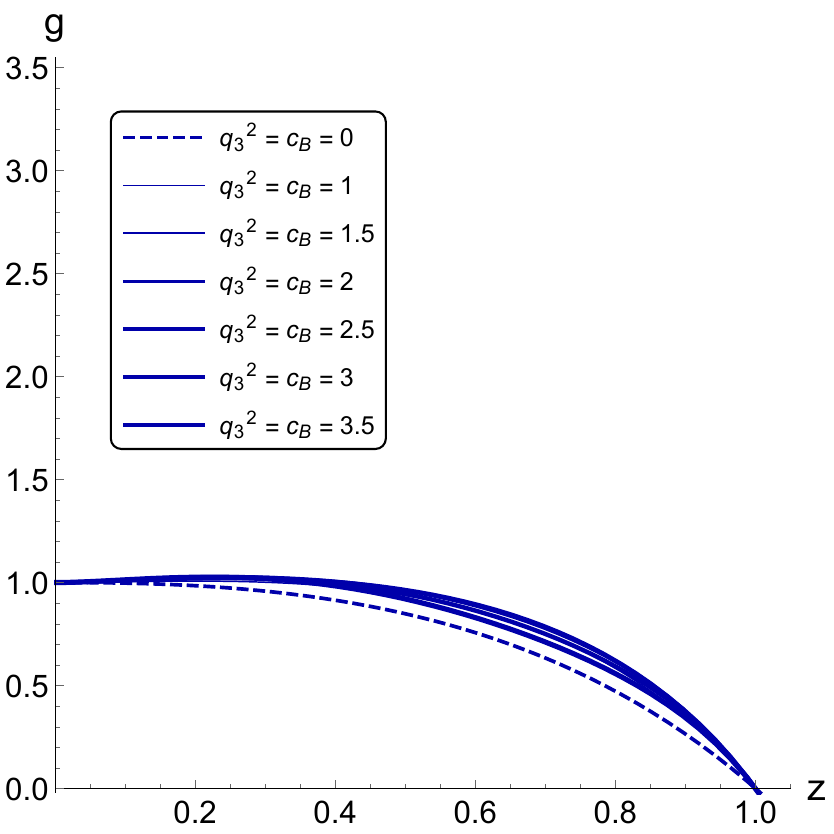} \\
  G \hspace{200pt} H
  \caption{Blackening function $g(z)$ in magnetic field with different
    $q_3$ (A,E) for $c_B = - \, 0.5$ (left) and $c_B = 0.5$ (right);
    with different $c_B$ for $q_3 = 0.5$ (B,F) for $c_B < 0$ (left)
    and $c_B > 0$ (right); for different $q_3 = \pm \, c_B$ (C,G); for
    different $q_3^2 = \pm \, c_B$ (D,H) for $d = 0.06 > 0.05$ (A-D)
    and $d = 0.01 < 0.05$ (E-H) in primary anisotropic case $\nu =
    4.5$, $a = 0.15$, $c = 1.16$, $\mu = 0$.}
  \label{Fig:gz-q3cB-nu45-mu0-z5}
\end{figure}

\begin{figure}[t!]
  \centering
  \includegraphics[scale=0.24]{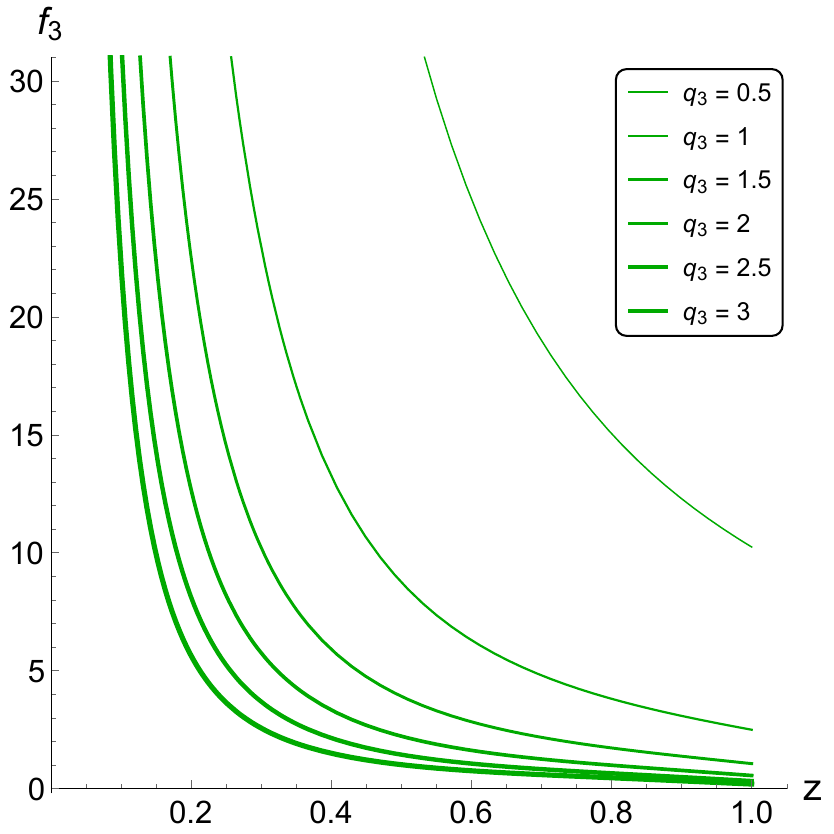} \
  \includegraphics[scale=0.24]{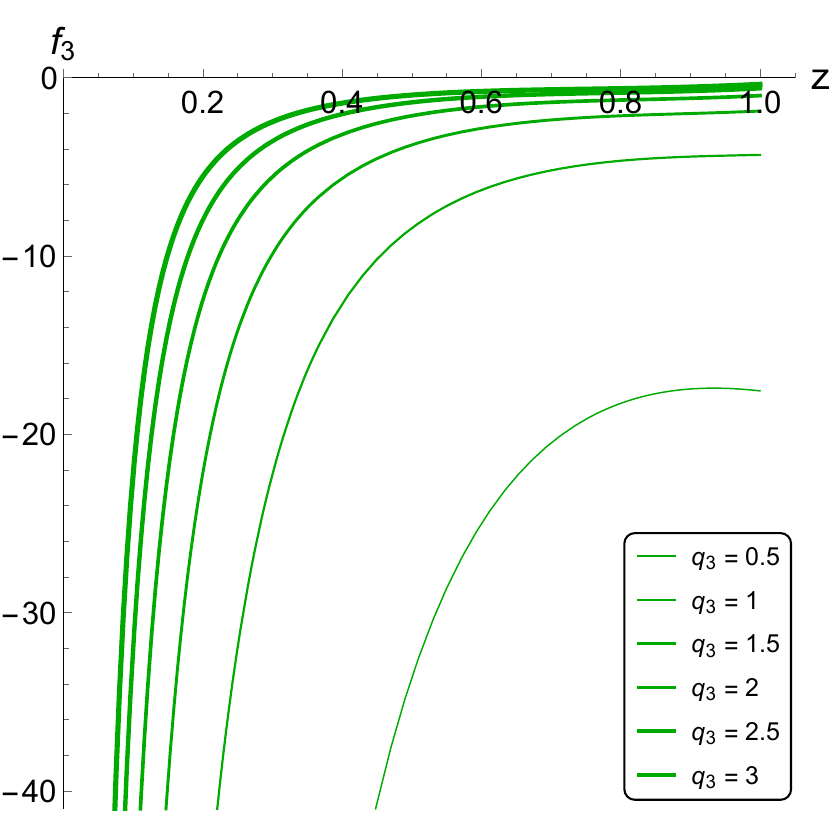} \quad
  \includegraphics[scale=0.24]{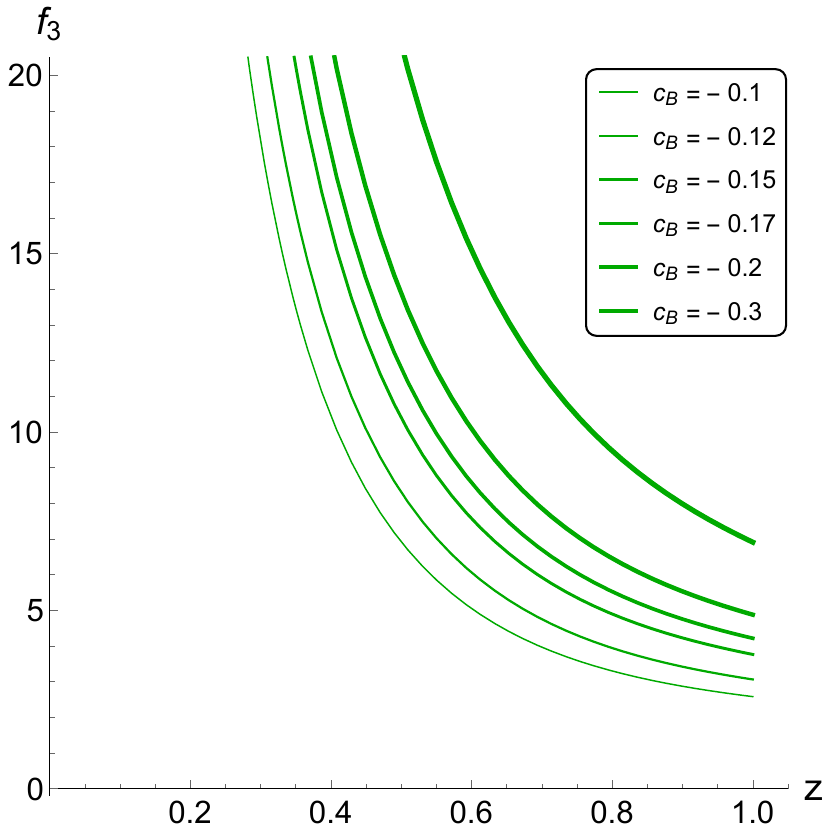} \
  \includegraphics[scale=0.24]{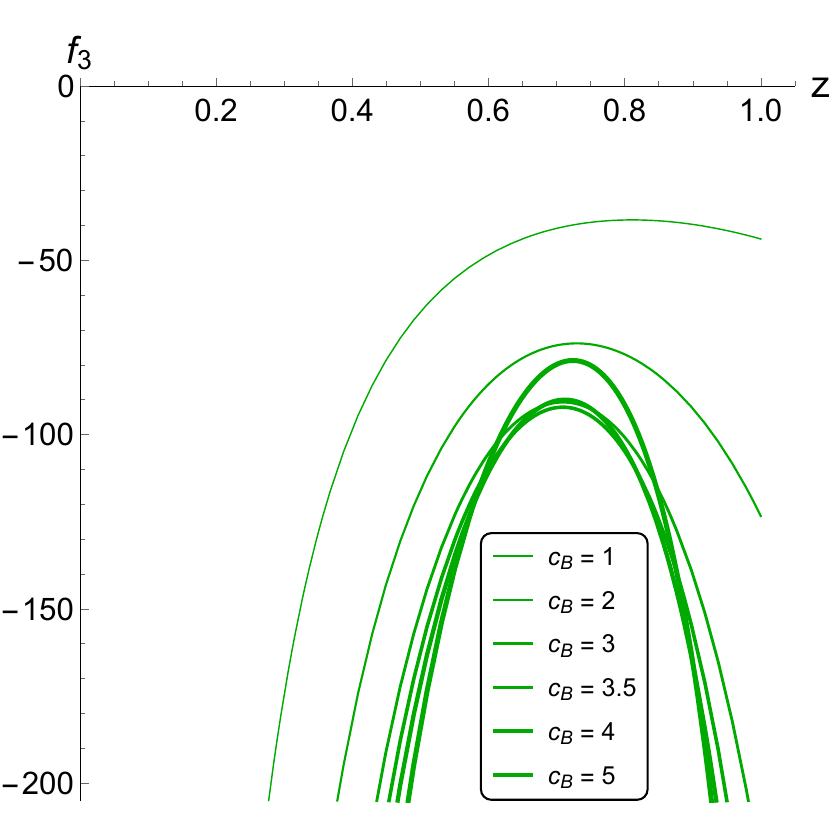} \\
  A \hspace{200pt} B \ \\
  \includegraphics[scale=0.24]{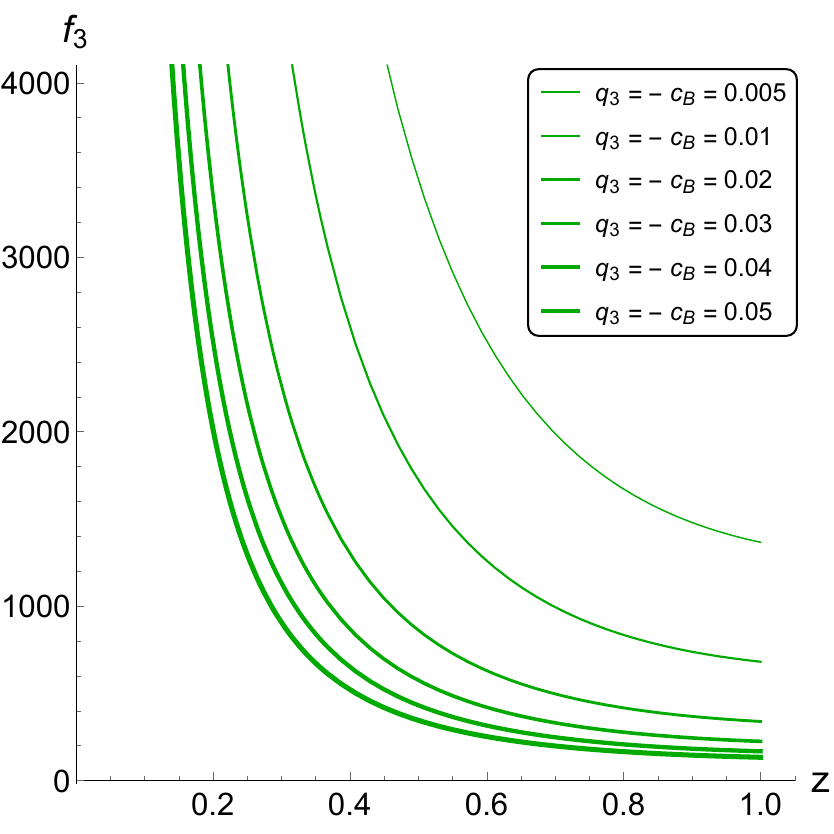} \
  \includegraphics[scale=0.24]{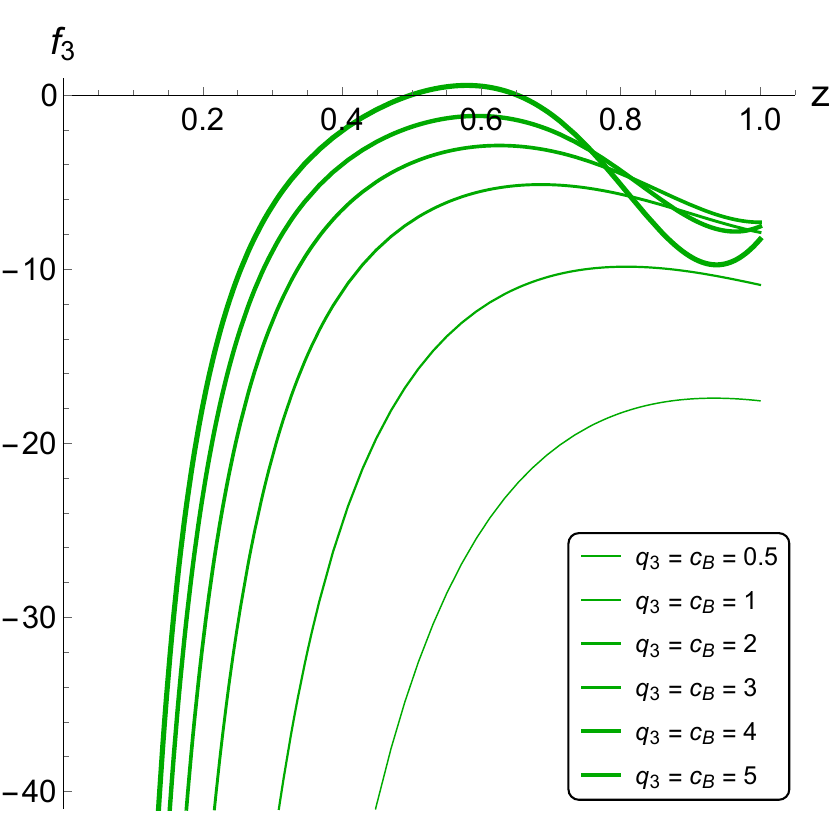} \quad
  \includegraphics[scale=0.24]{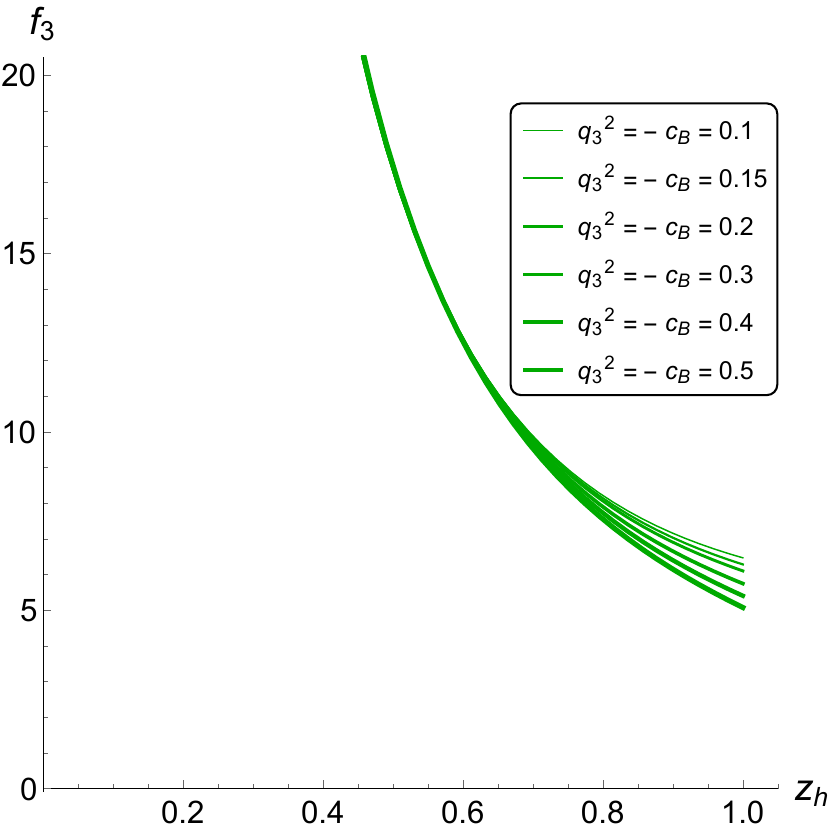} \
  \includegraphics[scale=0.24]{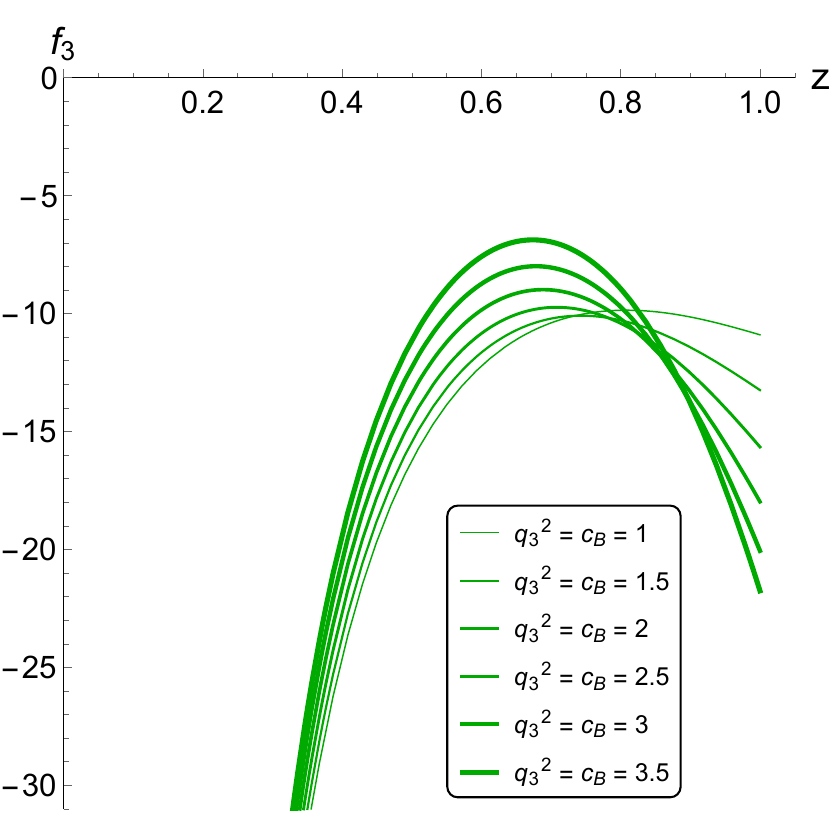} \\
  C \hspace{200pt} D \ \\
  \includegraphics[scale=0.24]{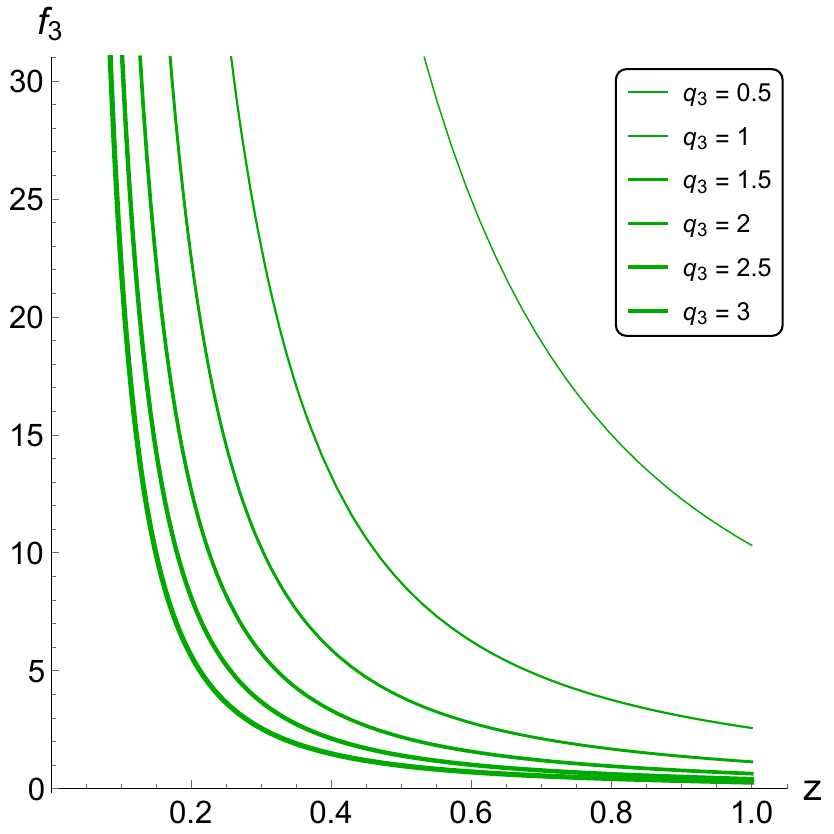} \
  \includegraphics[scale=0.24]{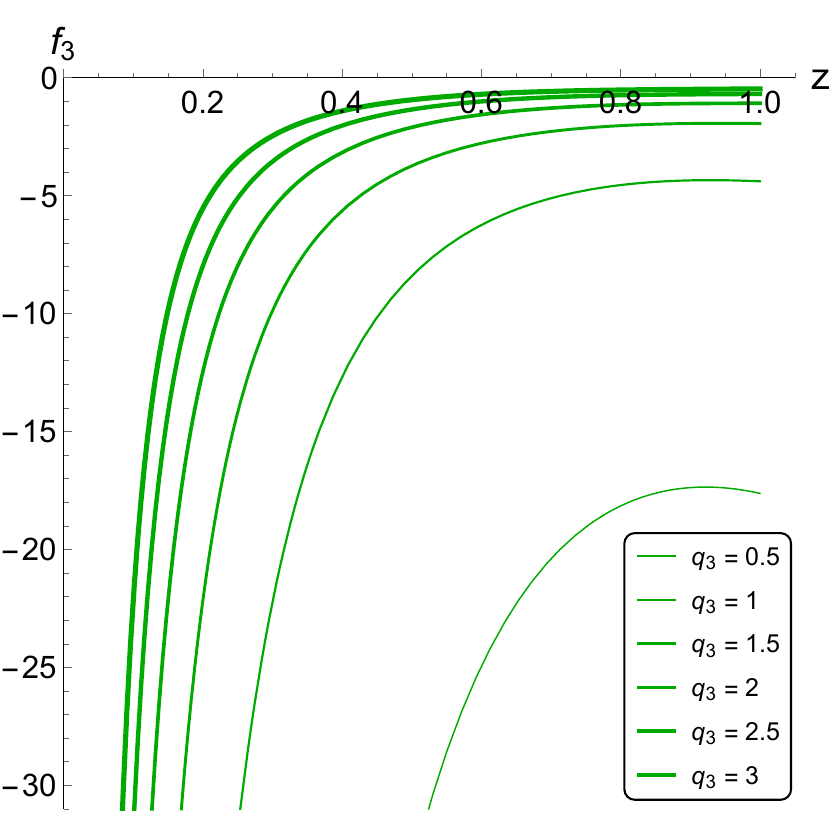} \quad
  \includegraphics[scale=0.24]{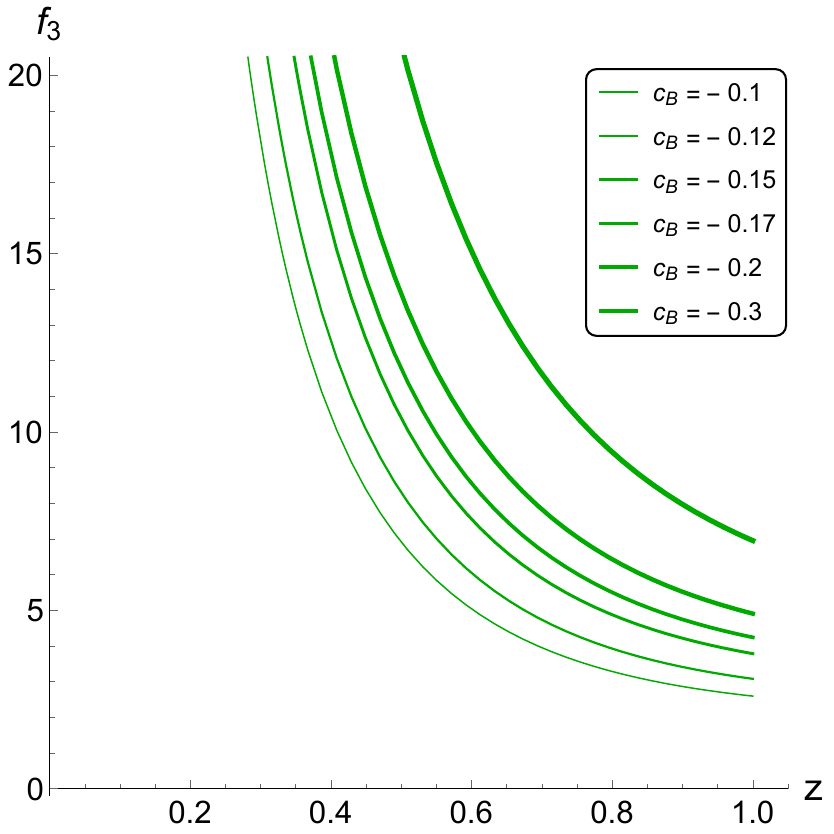} \
  \includegraphics[scale=0.24]{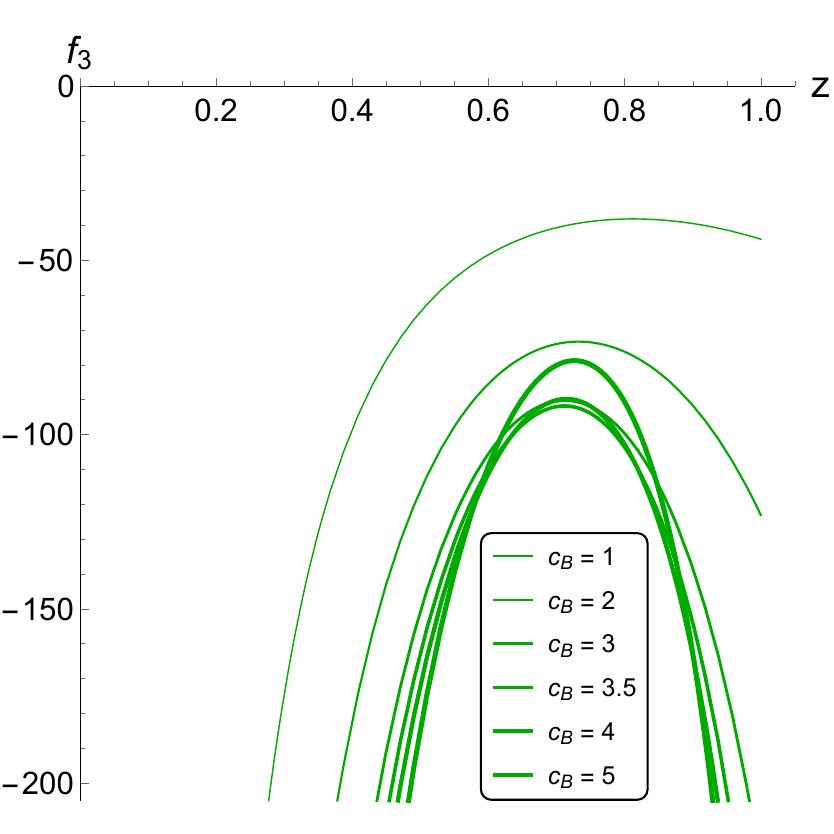} \\
  E \hspace{200pt} F \ \\
  \includegraphics[scale=0.24]{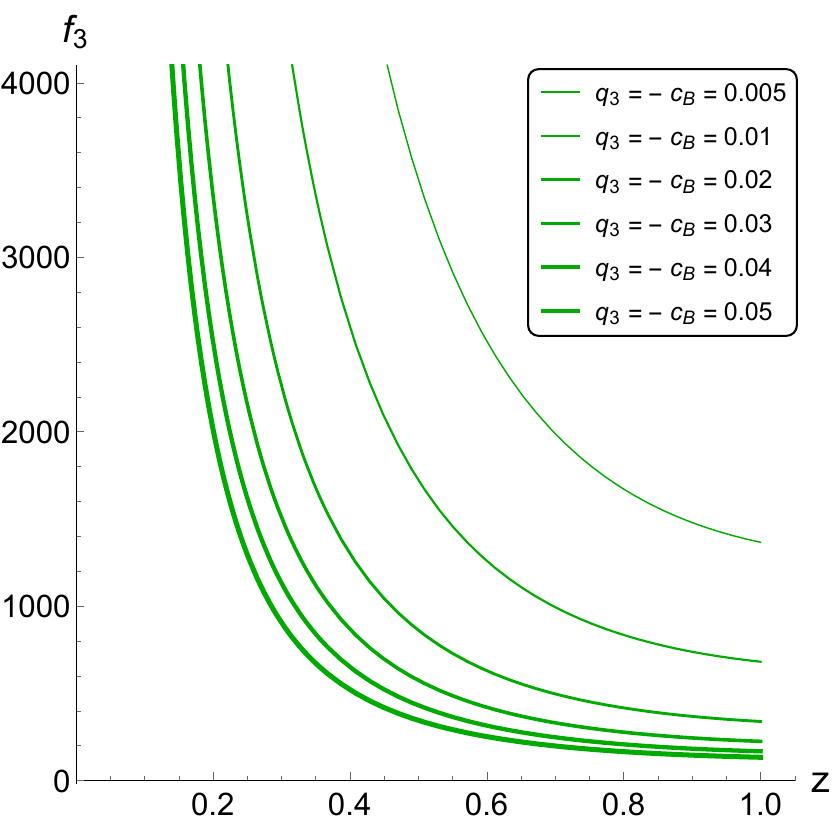} \
  \includegraphics[scale=0.24]{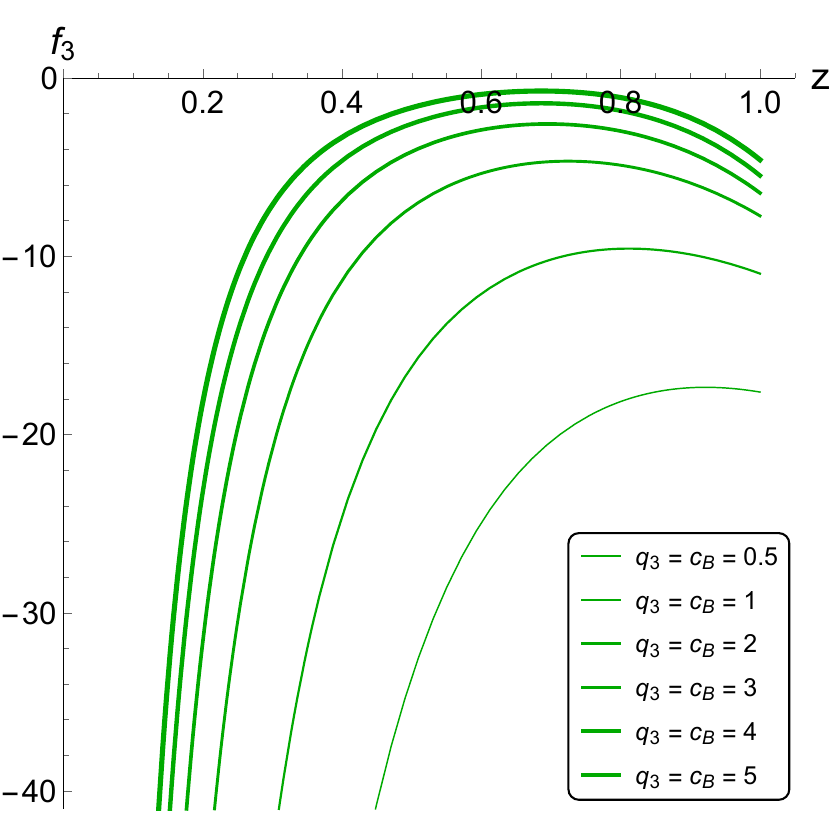} \quad
  \includegraphics[scale=0.24]{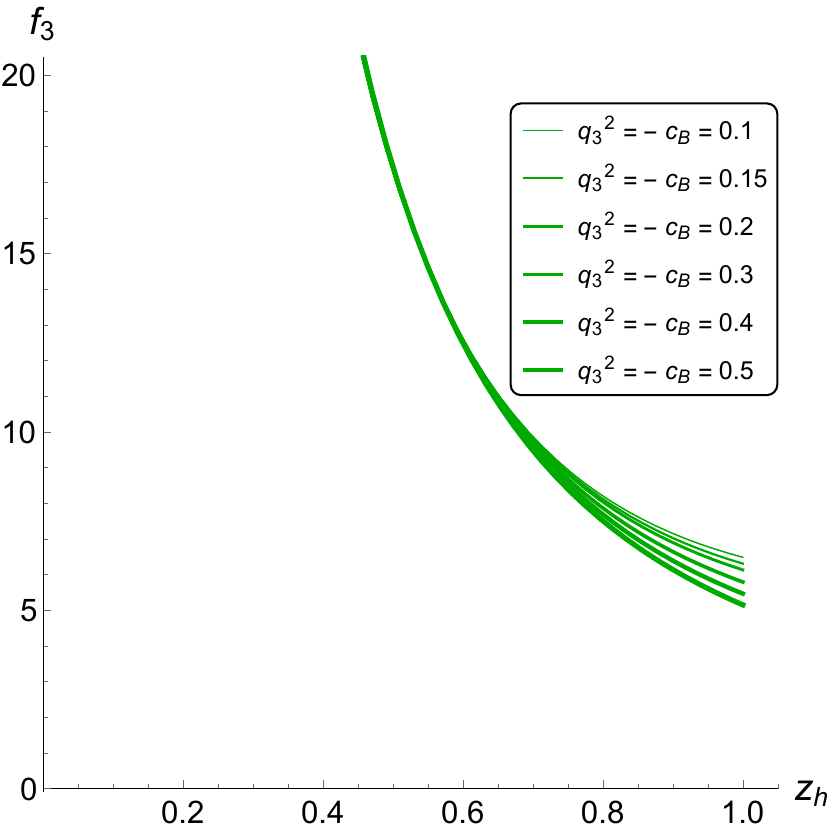} \
  \includegraphics[scale=0.24]{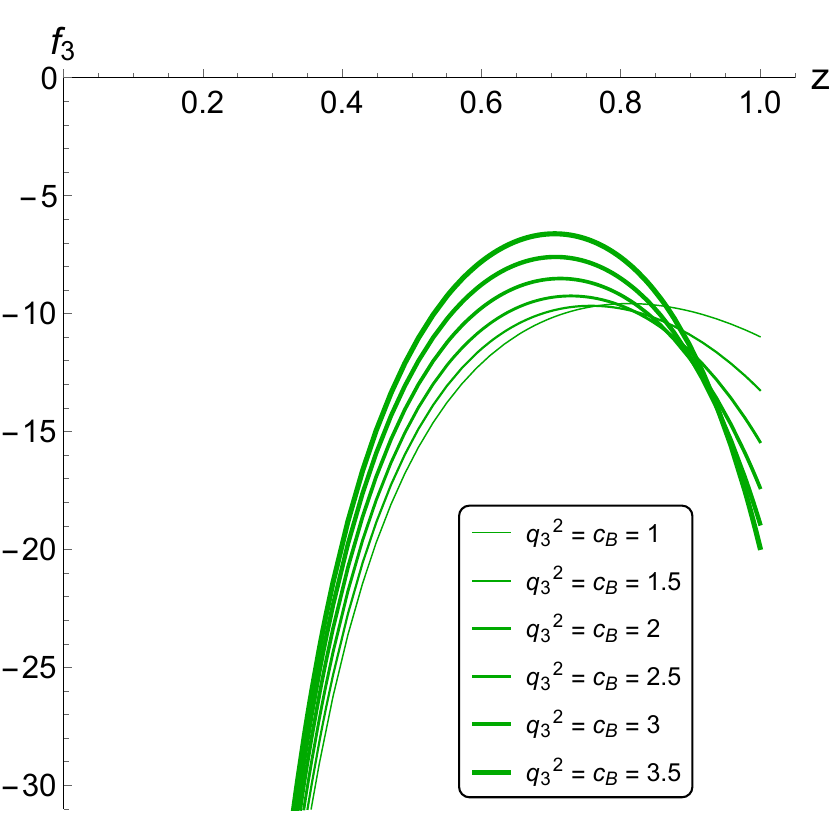} \\
  G \hspace{200pt} H
  \caption{Coupling function $f_3(z)$ in magnetic field with different
    $q_3$ (A,E) for $c_B = - \, 0.5$ (left) and $c_B = 0.5$ (right);
    with different $c_B$ for $q_3 = 0.5$ (B,F) for $c_B < 0$ (left)
    and $c_B > 0$ (right); for different $q_3 = \pm \, c_B$ (C,G); for
    different $q_3^2 = \pm \, c_B$ (D,H) for $d = 0.06 > 0.05$ (A-D)
    and $d = 0.01 < 0.05$ (E-H) in primary isotropic case $\nu = 1$,
    $a = 0.15$, $c = 1.16$, $\mu = 0$.}
  \label{Fig:f3z-q3cB-nu1-mu0-z5}
\end{figure}

\begin{figure}[t!]
  \centering
  \includegraphics[scale=0.24]{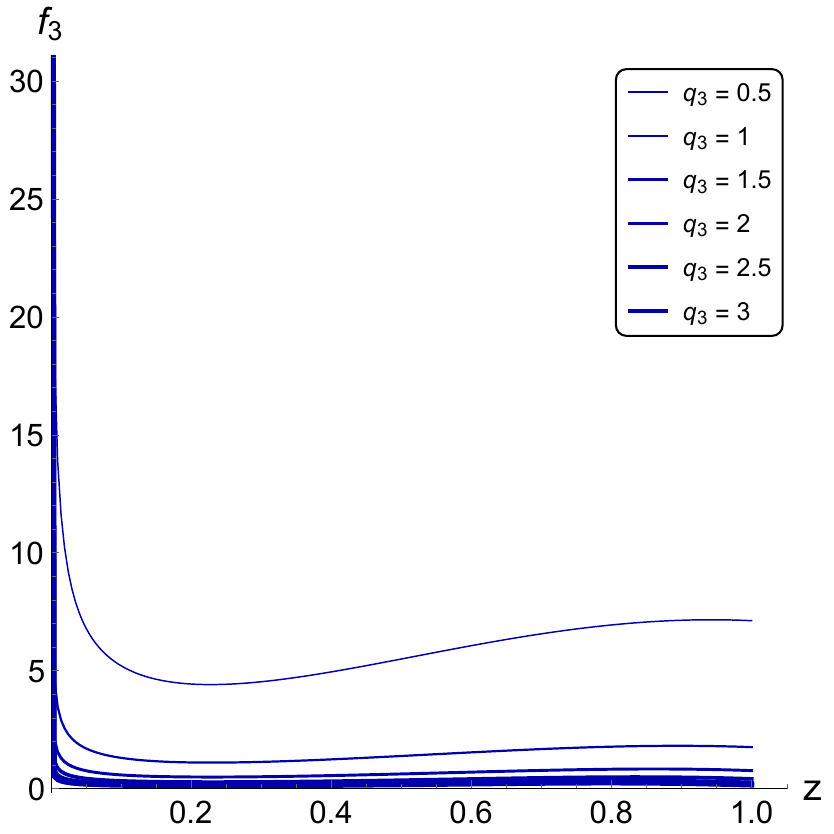} \
  \includegraphics[scale=0.24]{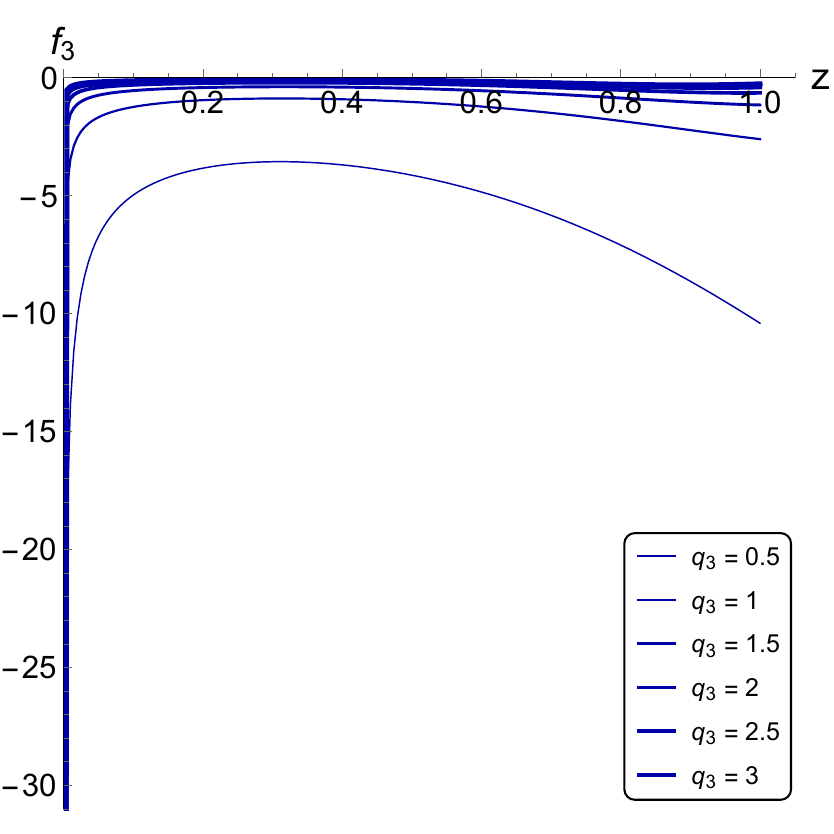} \quad
  \includegraphics[scale=0.24]{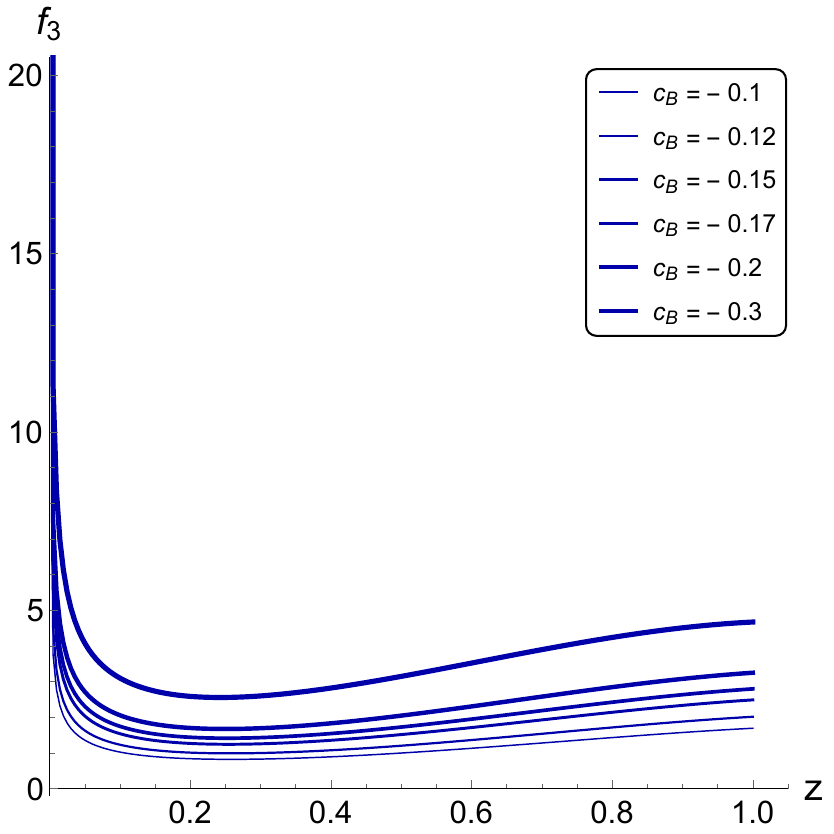} \
  \includegraphics[scale=0.24]{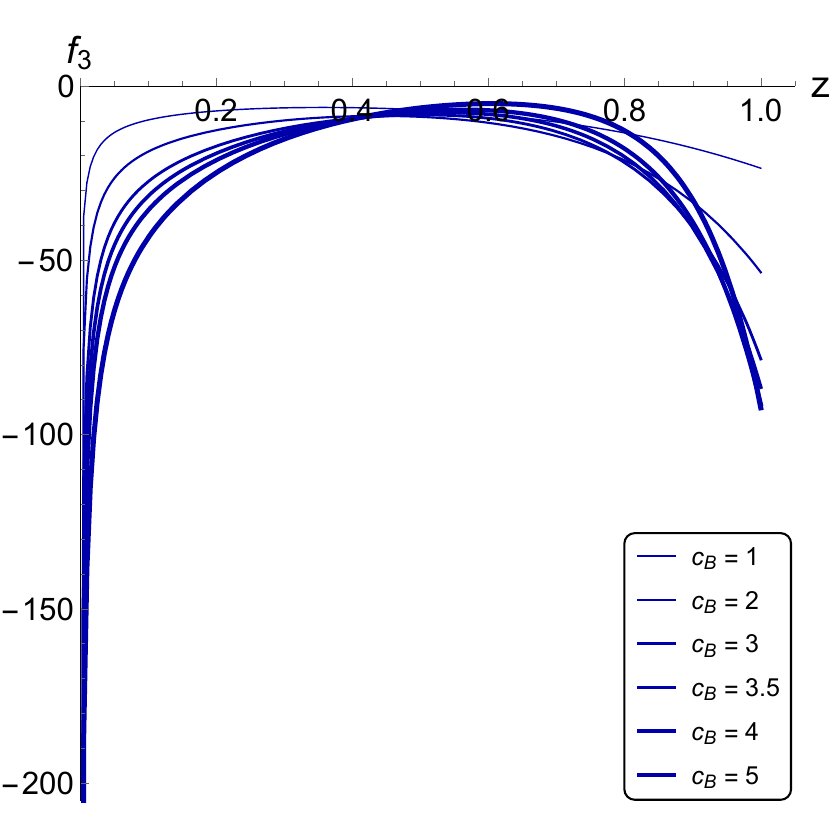} \\
  A \hspace{200pt} B \ \\
  \includegraphics[scale=0.24]{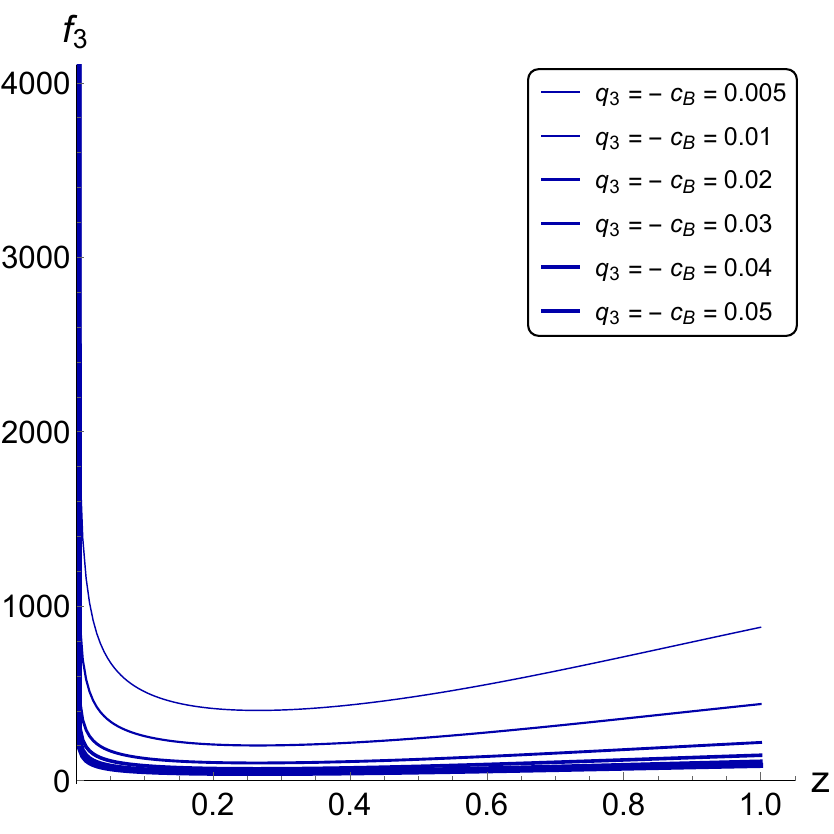} \
  \includegraphics[scale=0.24]{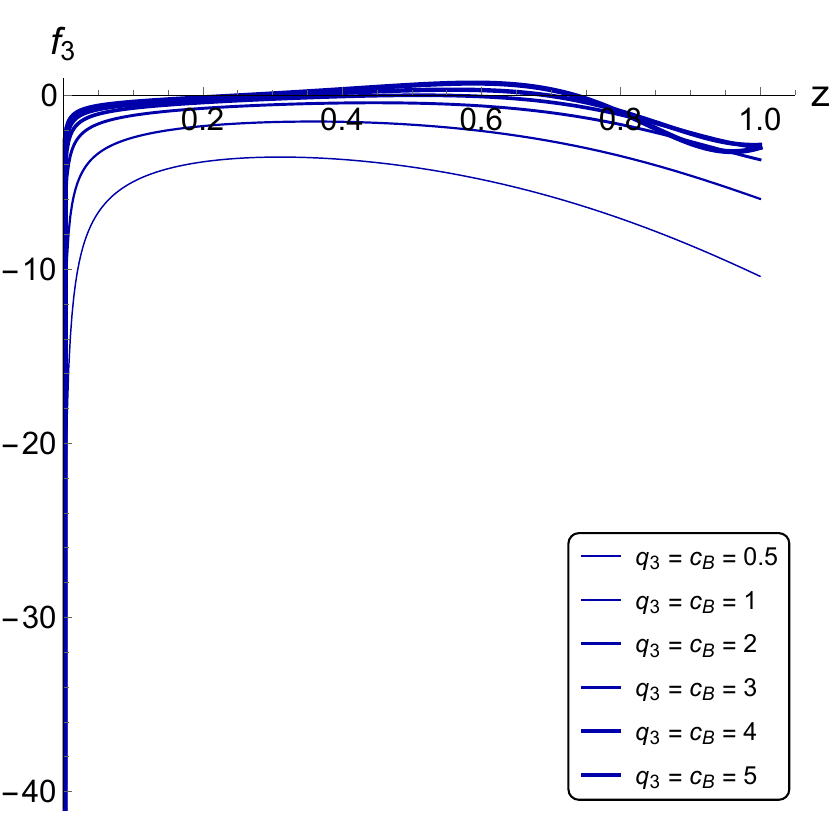} \quad
  \includegraphics[scale=0.24]{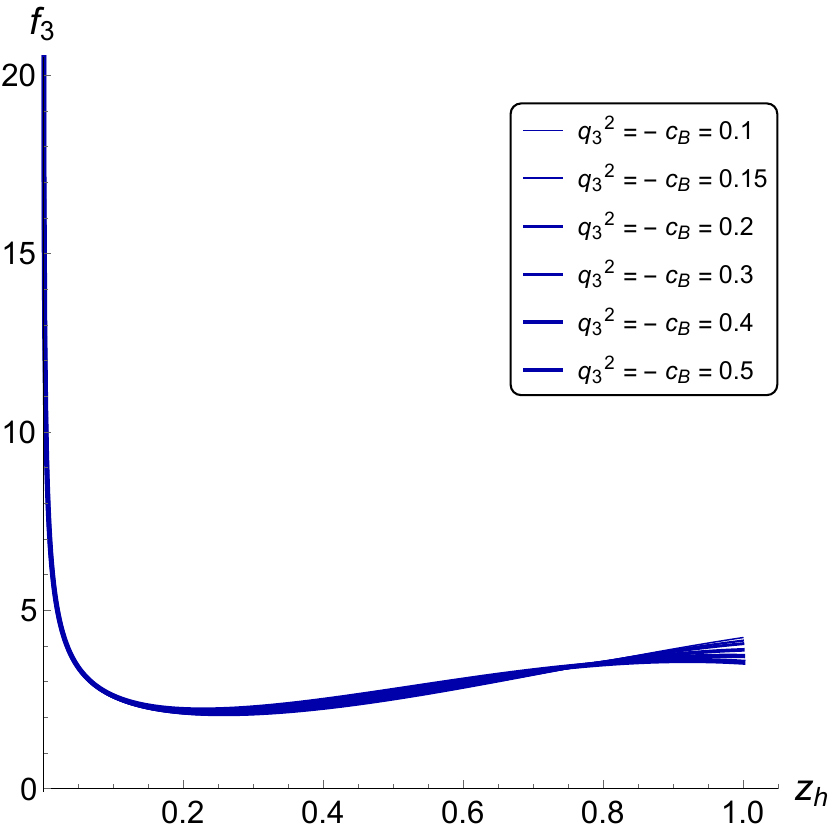} \
  \includegraphics[scale=0.24]{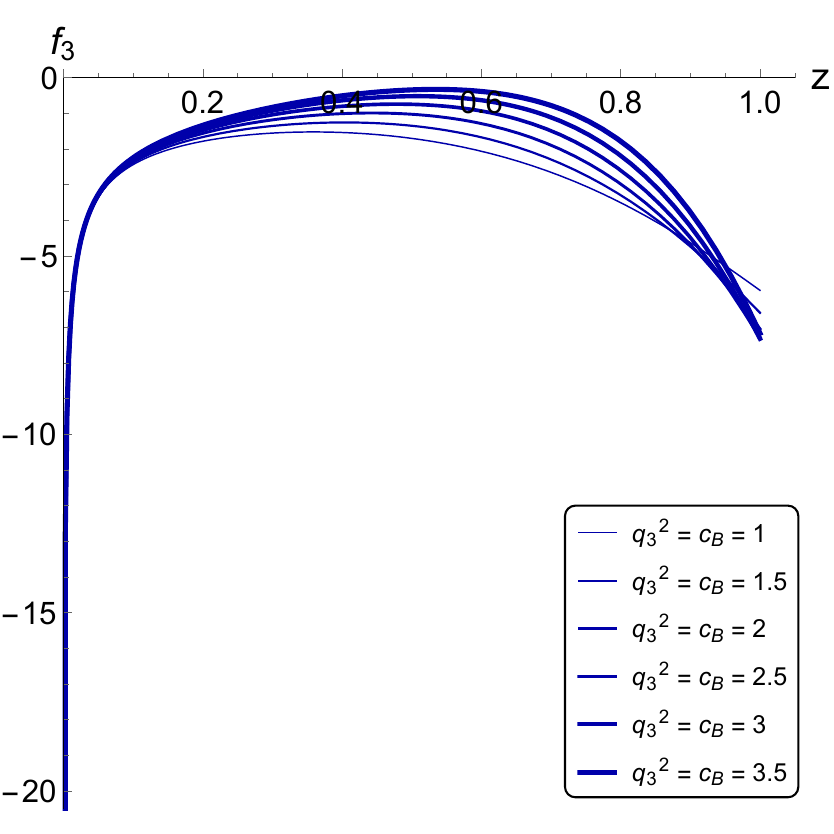} \\
  C \hspace{200pt} D \ \\
  \includegraphics[scale=0.24]{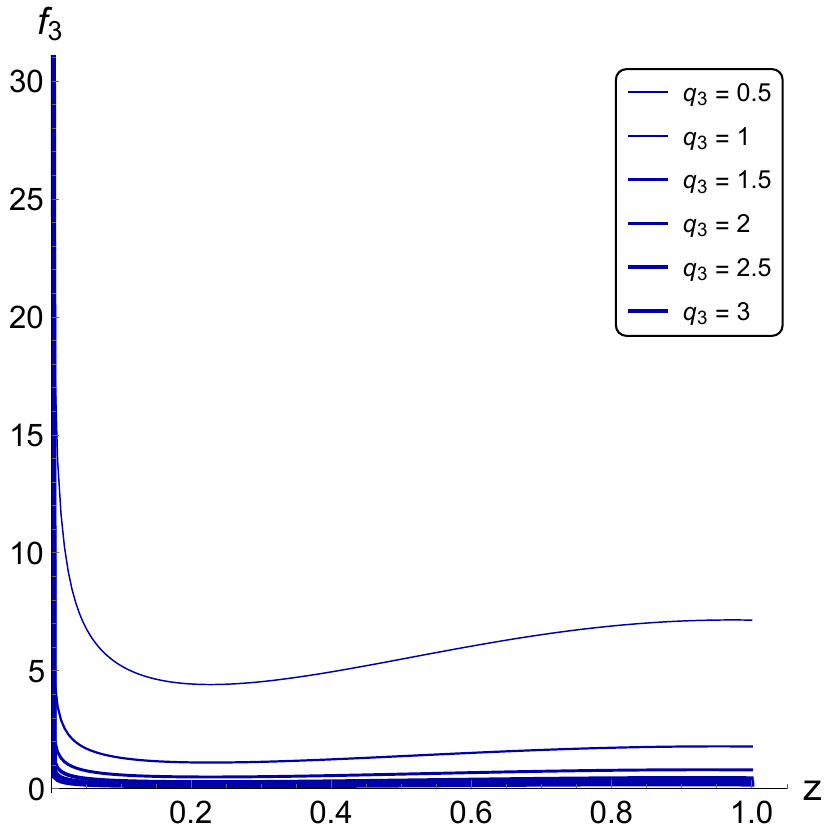} \
  \includegraphics[scale=0.24]{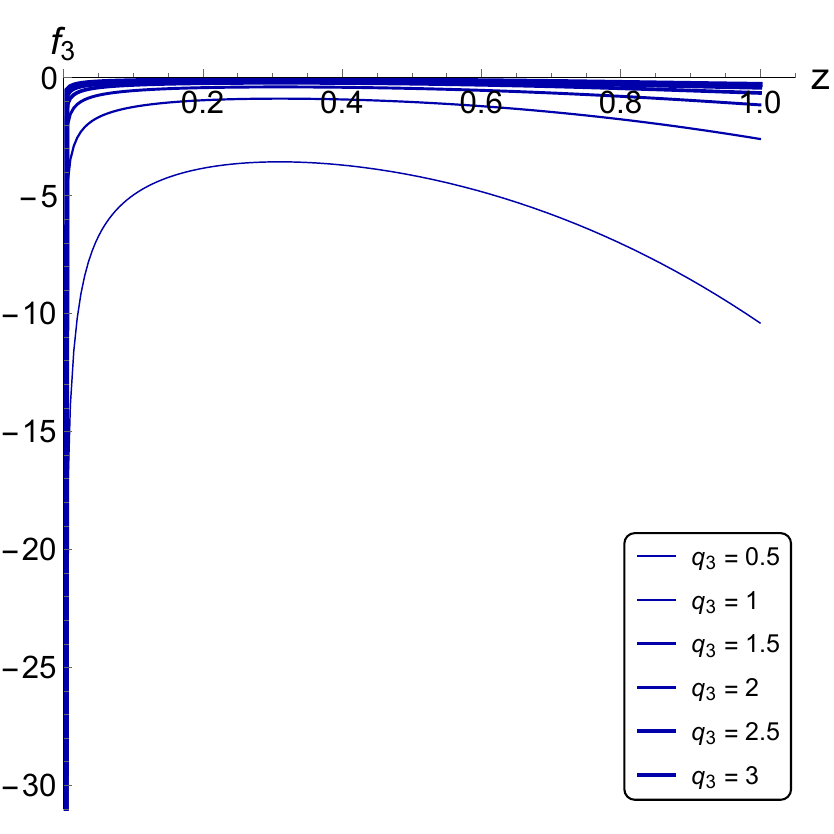} \quad
  \includegraphics[scale=0.24]{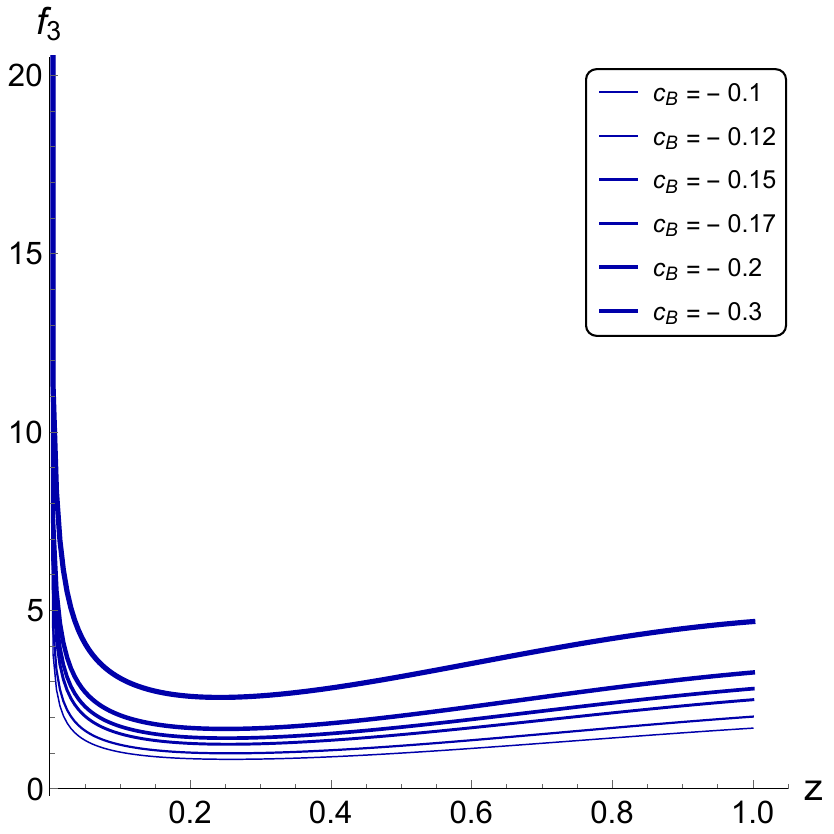} \
  \includegraphics[scale=0.24]{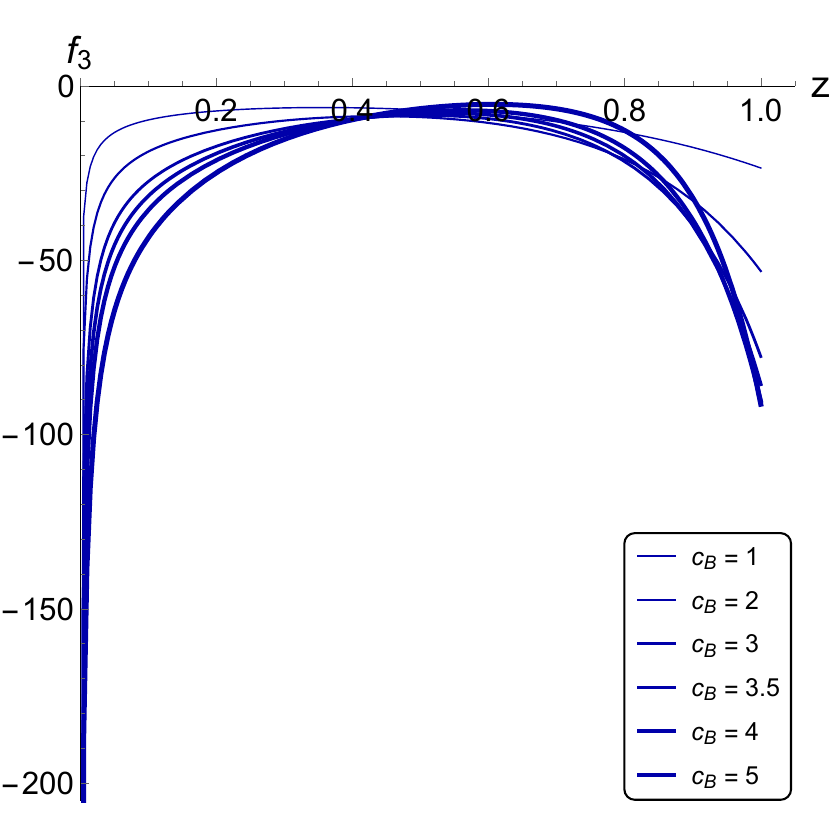} \\
  E \hspace{200pt} F \ \\
  \includegraphics[scale=0.24]{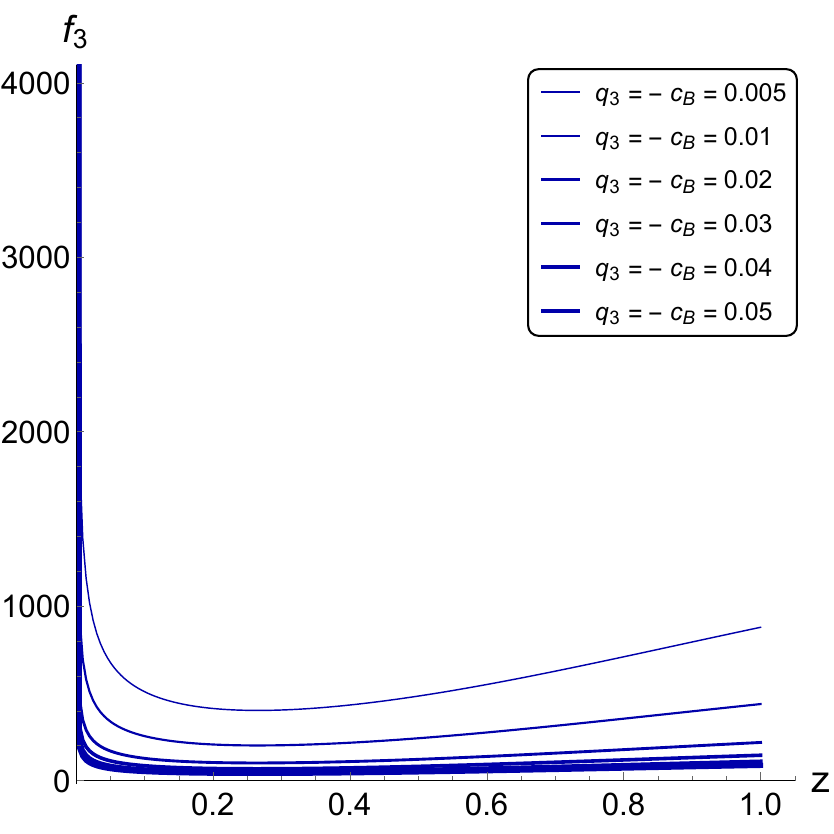} \
  \includegraphics[scale=0.24]{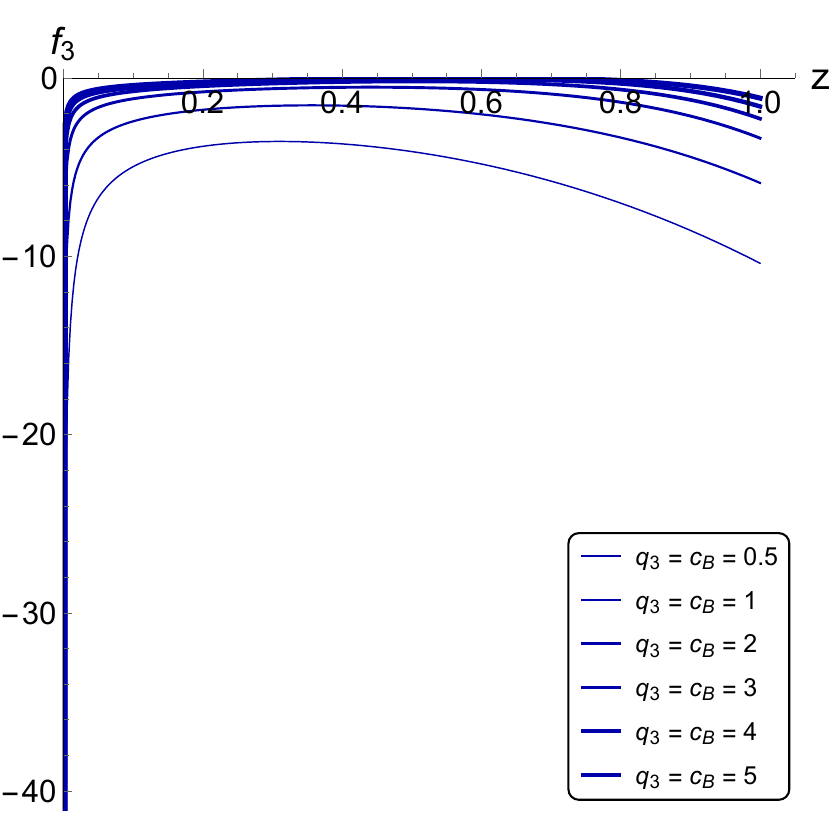} \quad
  \includegraphics[scale=0.24]{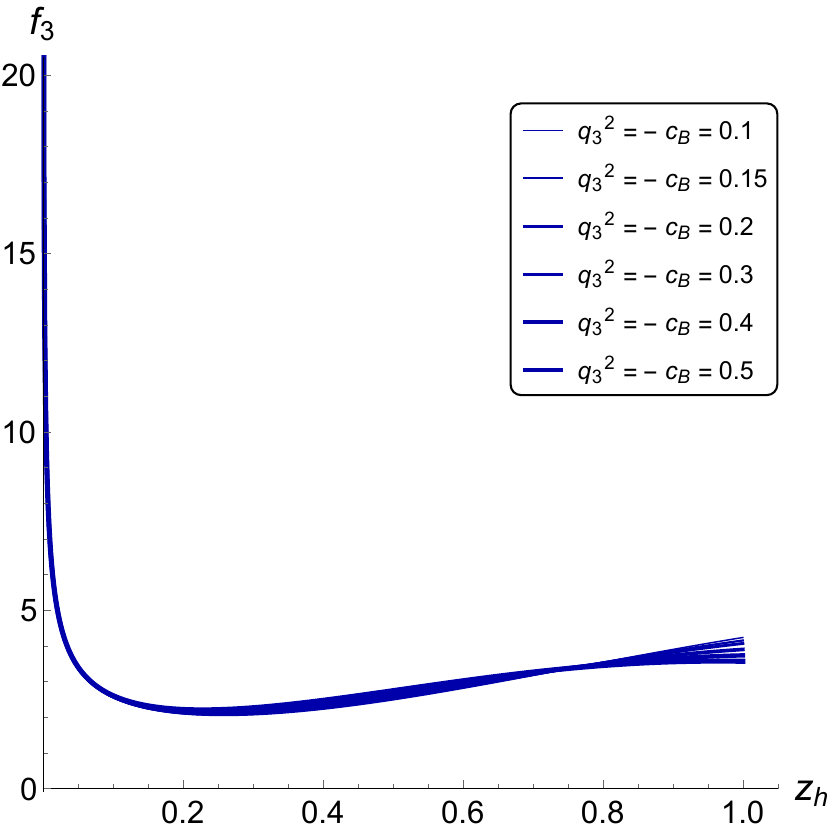} \
  \includegraphics[scale=0.24]{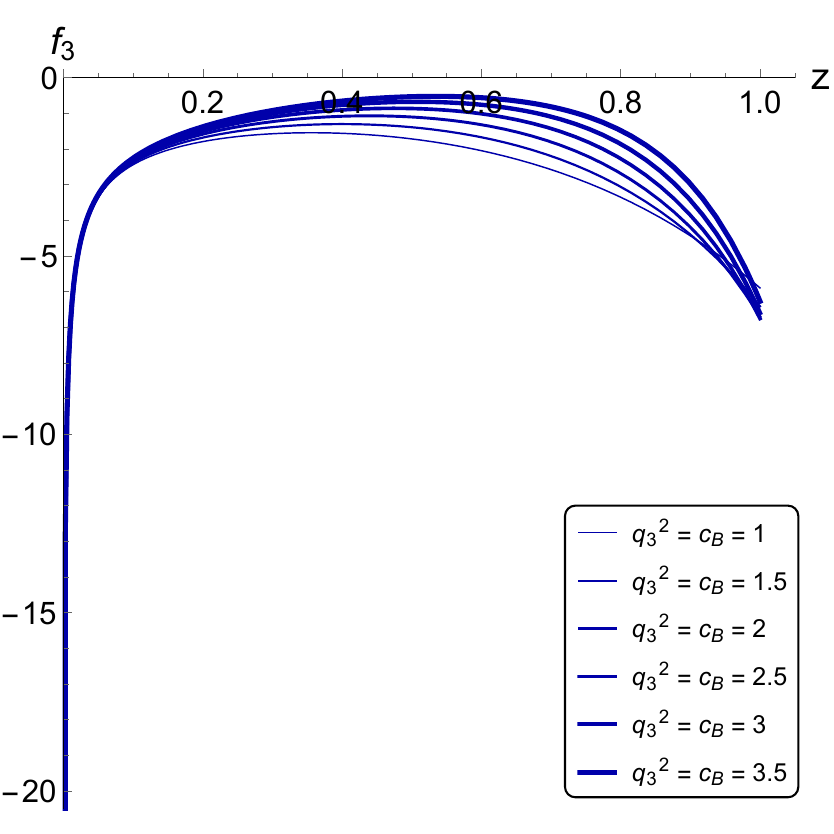} \\
  G \hspace{200pt} H
  \caption{Coupling function $f_3(z)$ in magnetic field with different
    $q_3$ (A,E) for $c_B = - \, 0.5$ (left) and $c_B = 0.5$ (right);
    with different $c_B$ for $q_3 = 0.5$ (B,F) for $c_B < 0$ (left)
    and $c_B > 0$ (right); for different $q_3 = \pm \, c_B$ (C,G); for
    different $q_3^2 = \pm \, c_B$ (D,H) for $d = 0.06 > 0.05$ (A-D)
    and $d = 0.01 < 0.05$ (E-H) in primary anisotropic case $\nu =
    4.5$, $a = 0.15$, $c = 1.16$, $\mu = 0$.}
  \label{Fig:f3z-q3cB-nu45-mu0-z5}
\end{figure}

\begin{figure}[t!]
  \centering
  \includegraphics[scale=0.24]{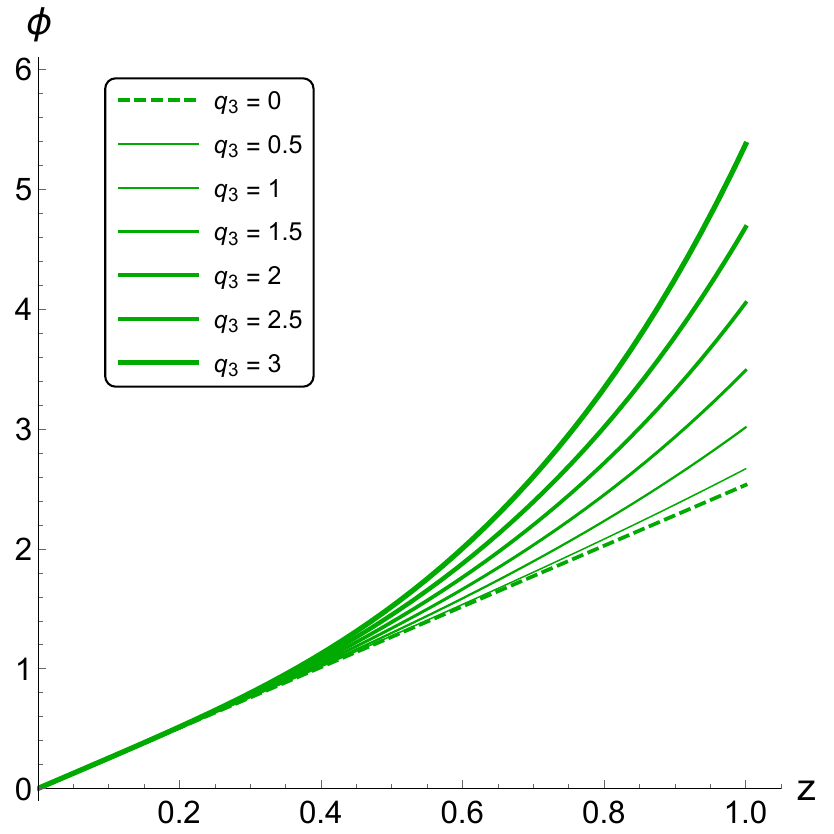} \
  \includegraphics[scale=0.24]{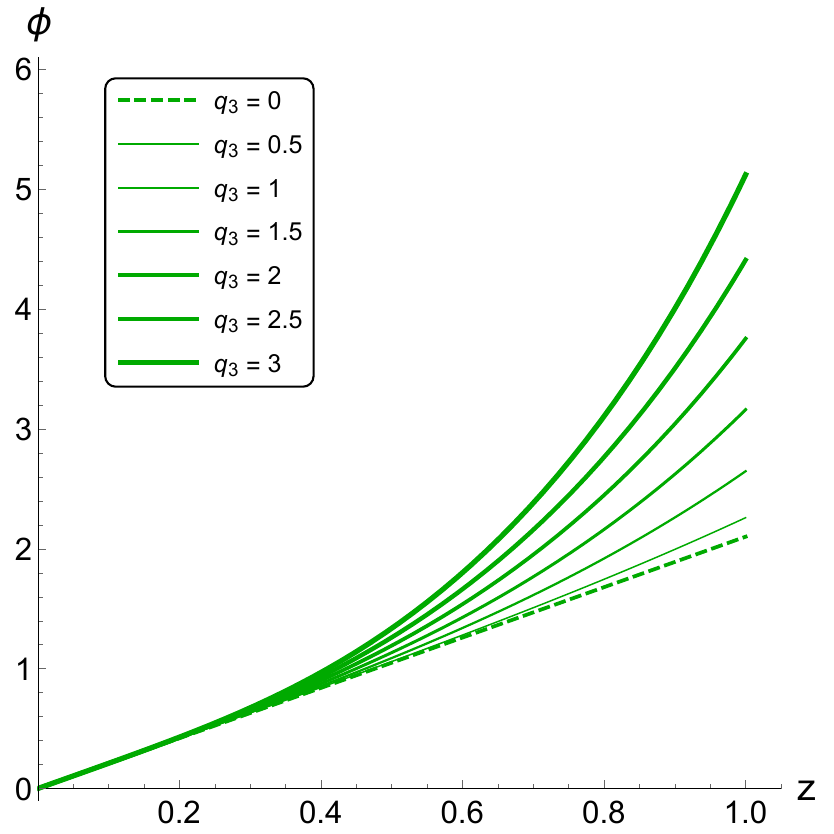} \quad
  \includegraphics[scale=0.24]{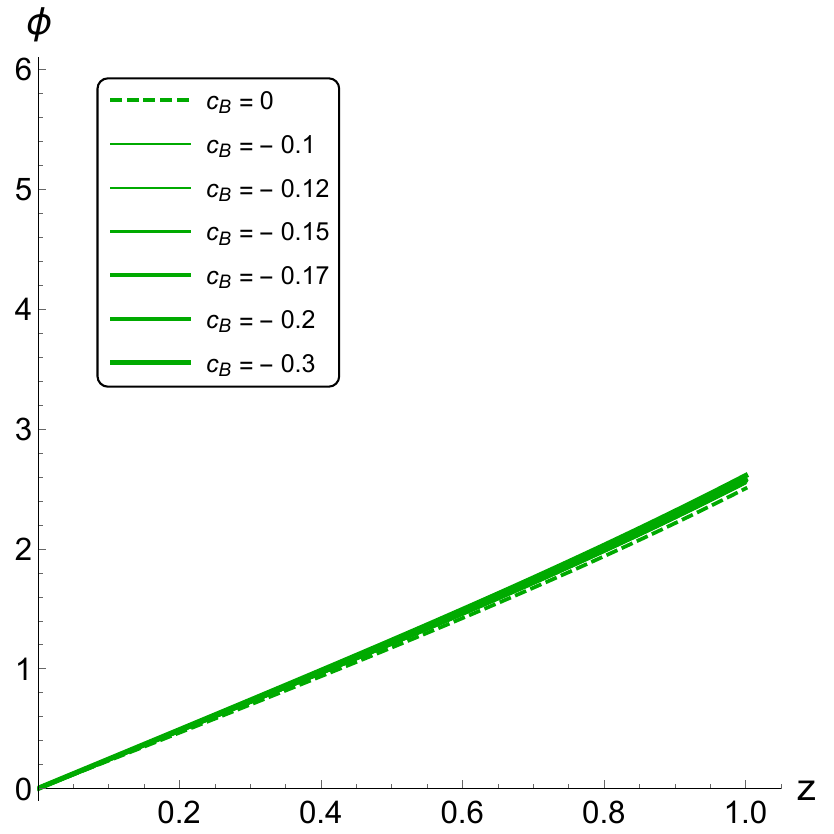} \
  \includegraphics[scale=0.24]{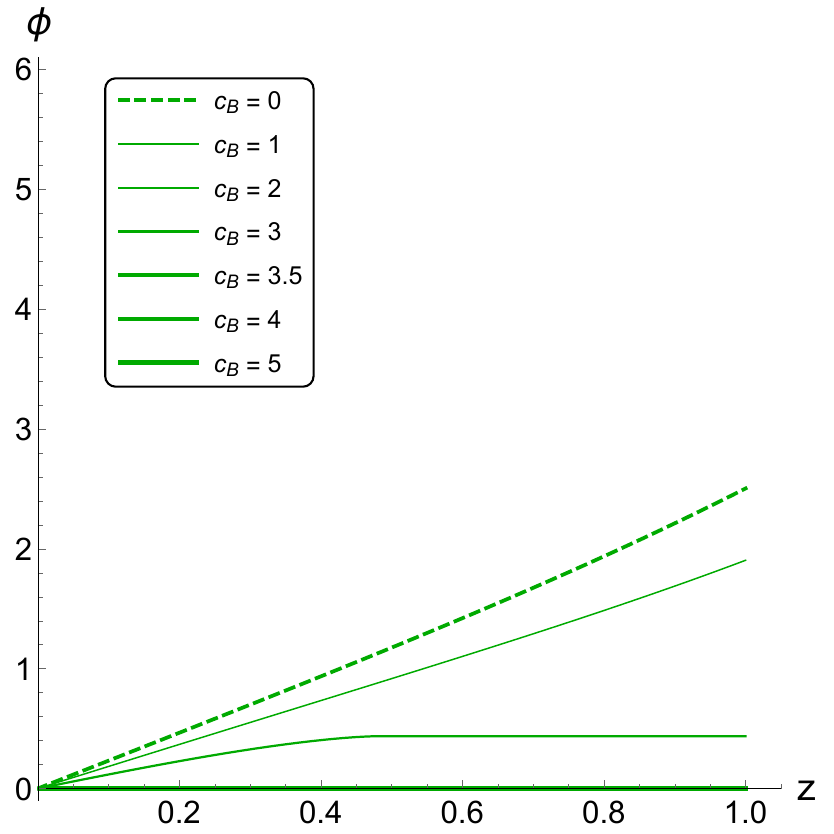} \\
  A \hspace{200pt} B \ \\
  \includegraphics[scale=0.24]{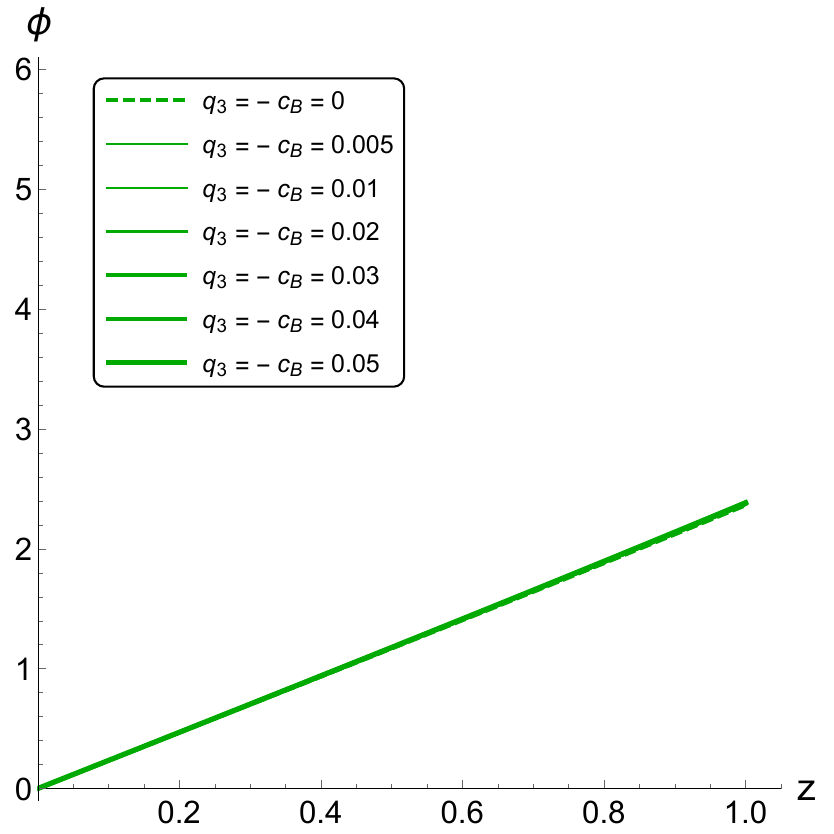} \
  \includegraphics[scale=0.24]{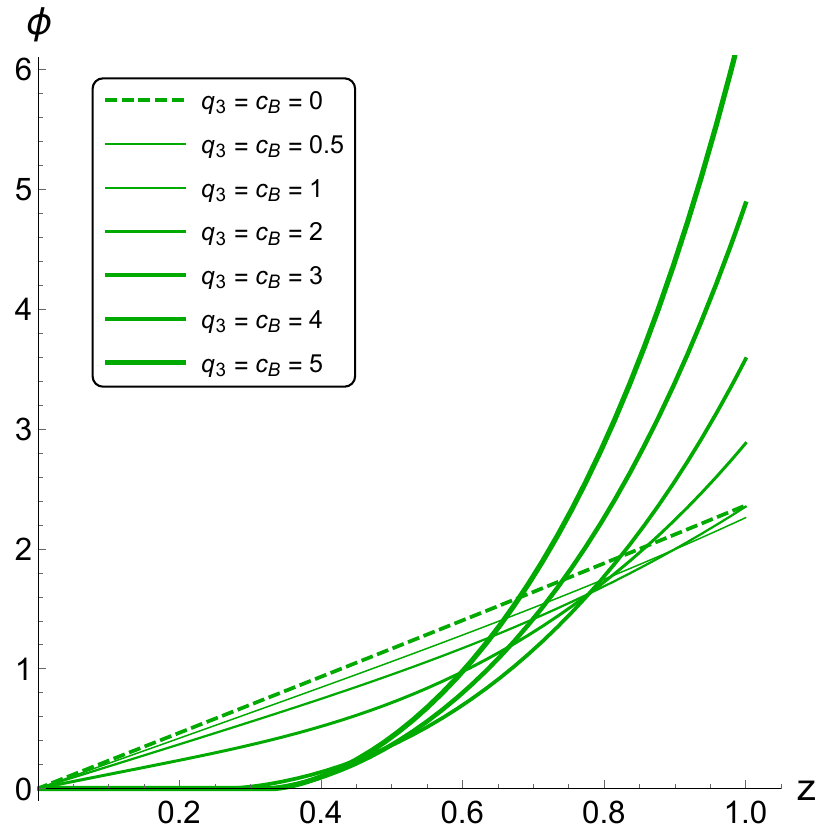} \quad
  \includegraphics[scale=0.24]{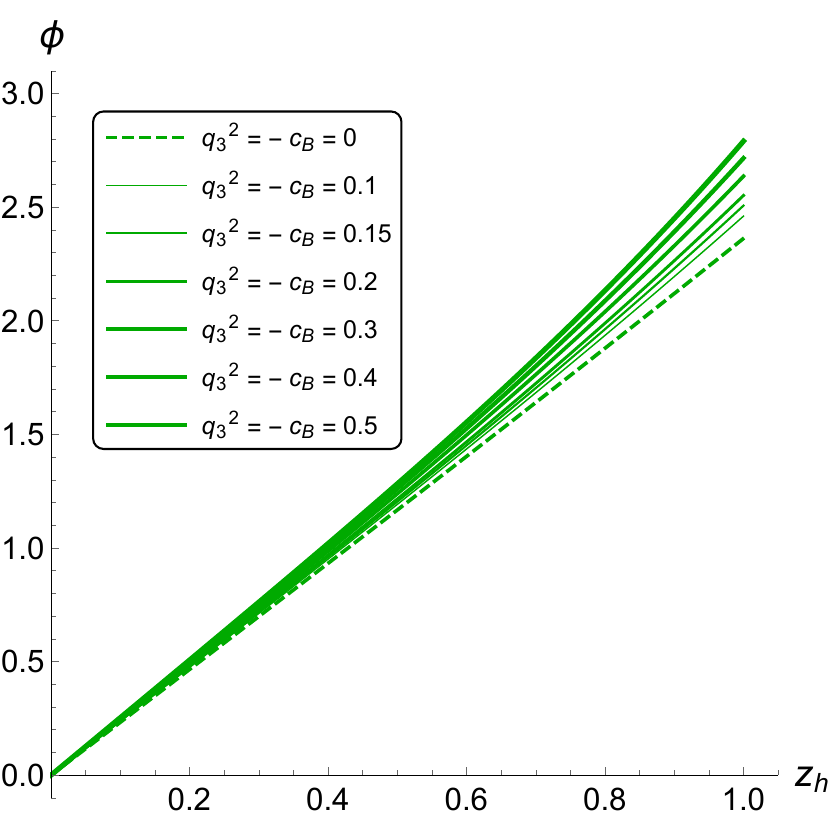} \
  \includegraphics[scale=0.24]{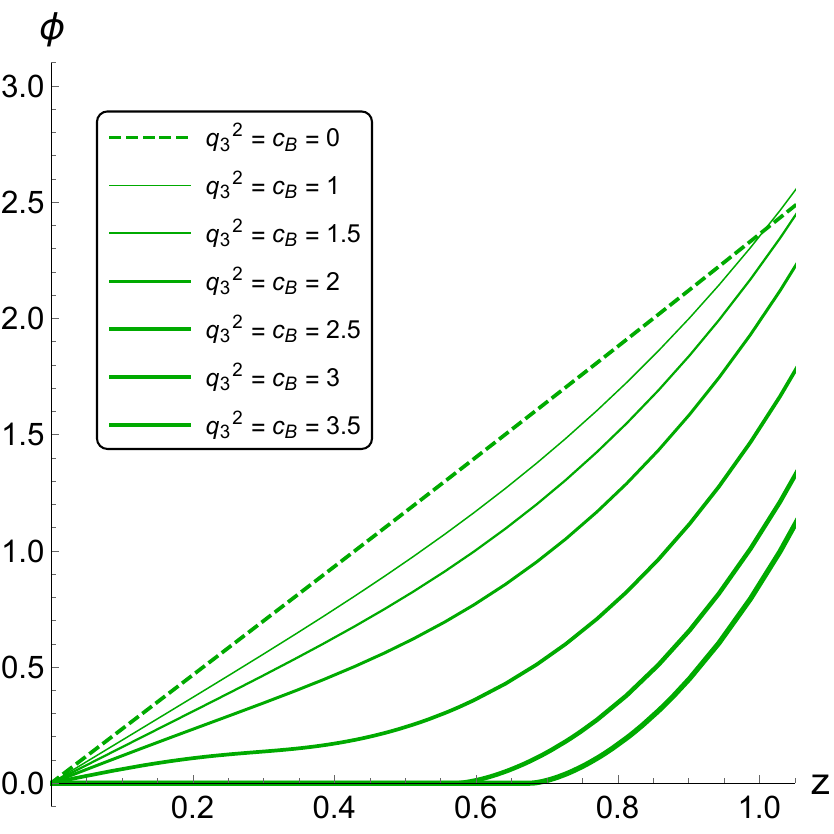} \\
  C \hspace{200pt} D \ \\
  \includegraphics[scale=0.24]{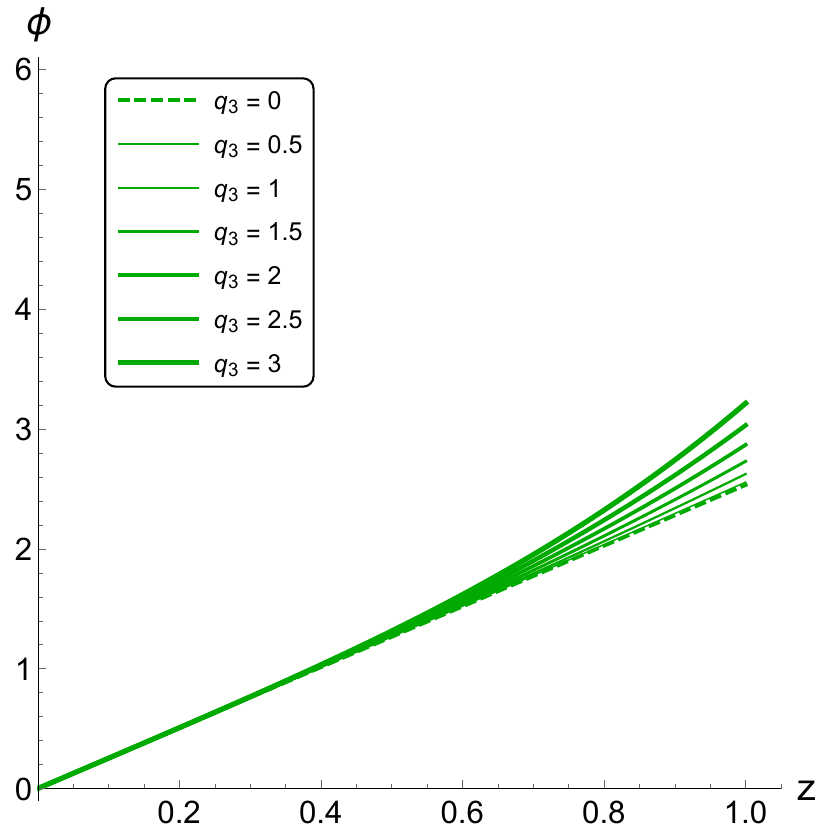} \
  \includegraphics[scale=0.24]{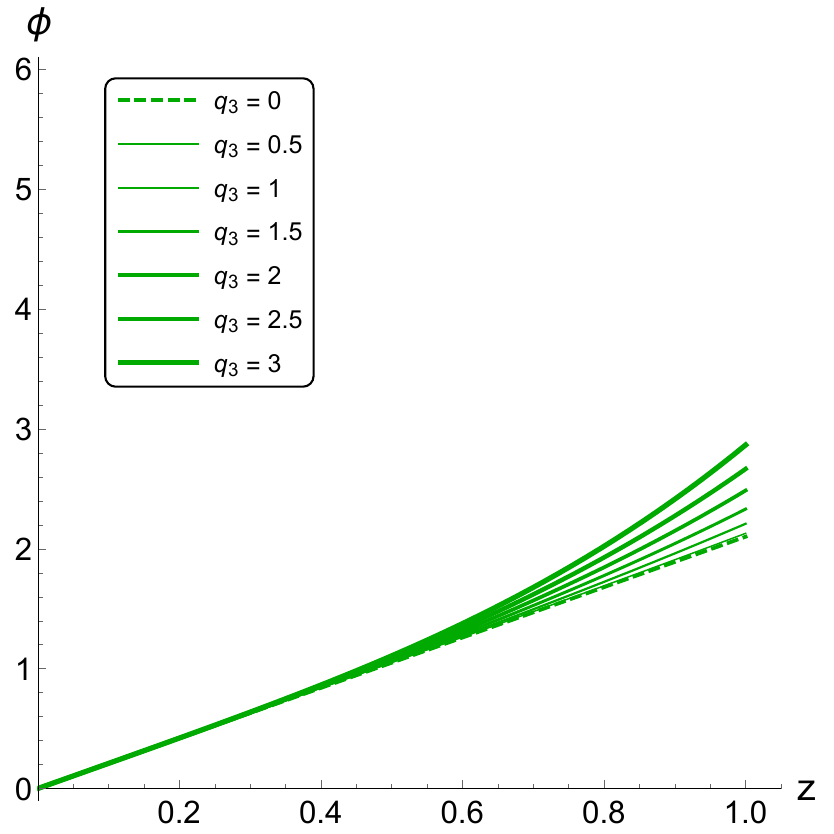} \quad
  \includegraphics[scale=0.24]{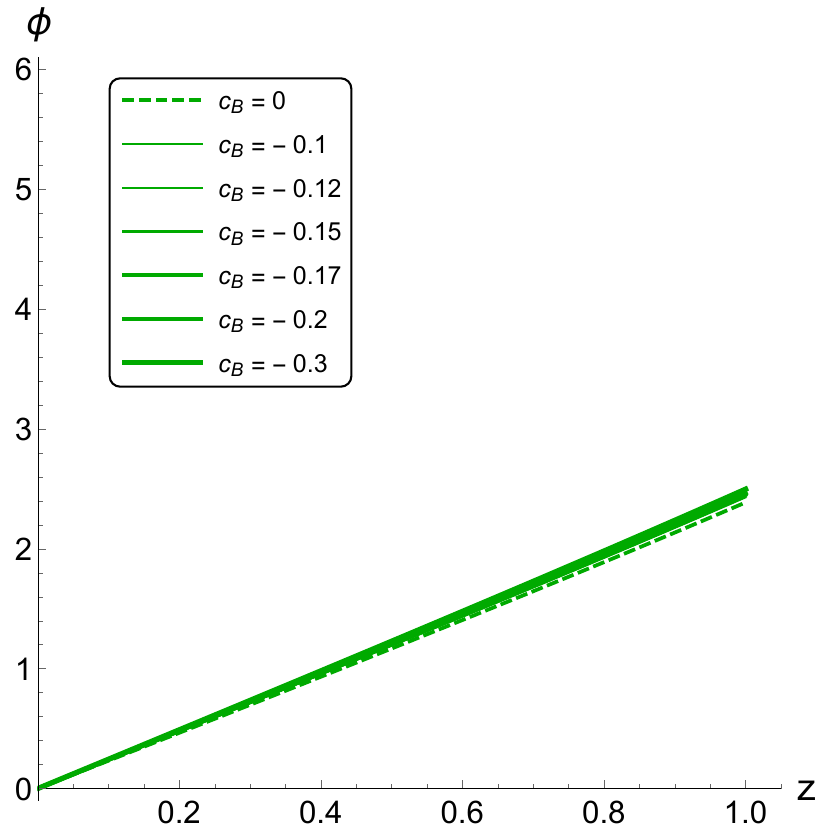} \
  \includegraphics[scale=0.24]{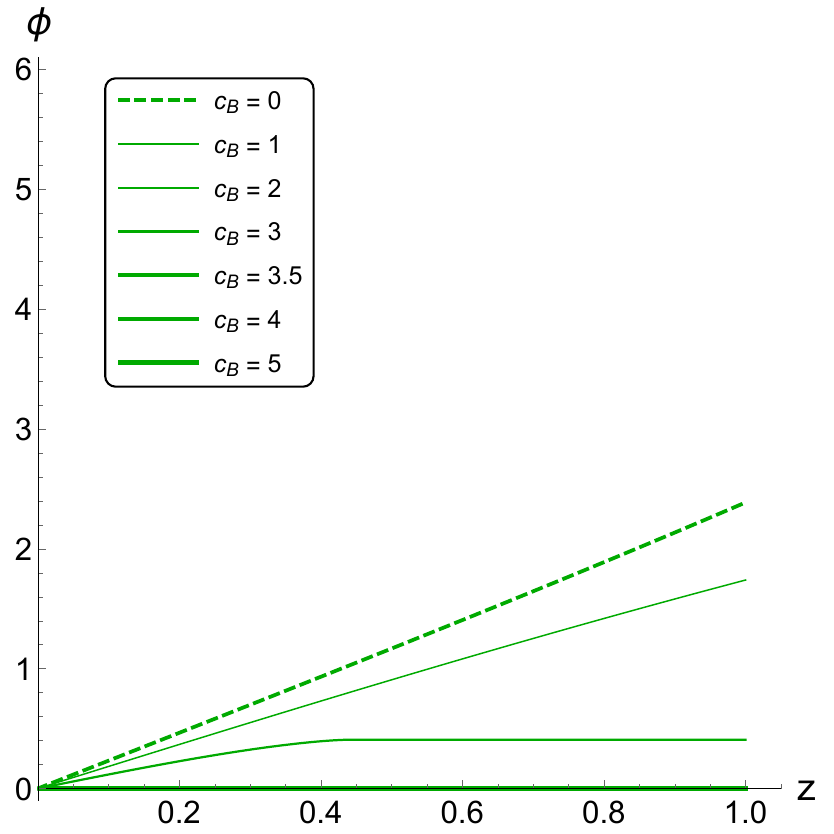} \\
  E \hspace{200pt} F \ \\
  \includegraphics[scale=0.24]{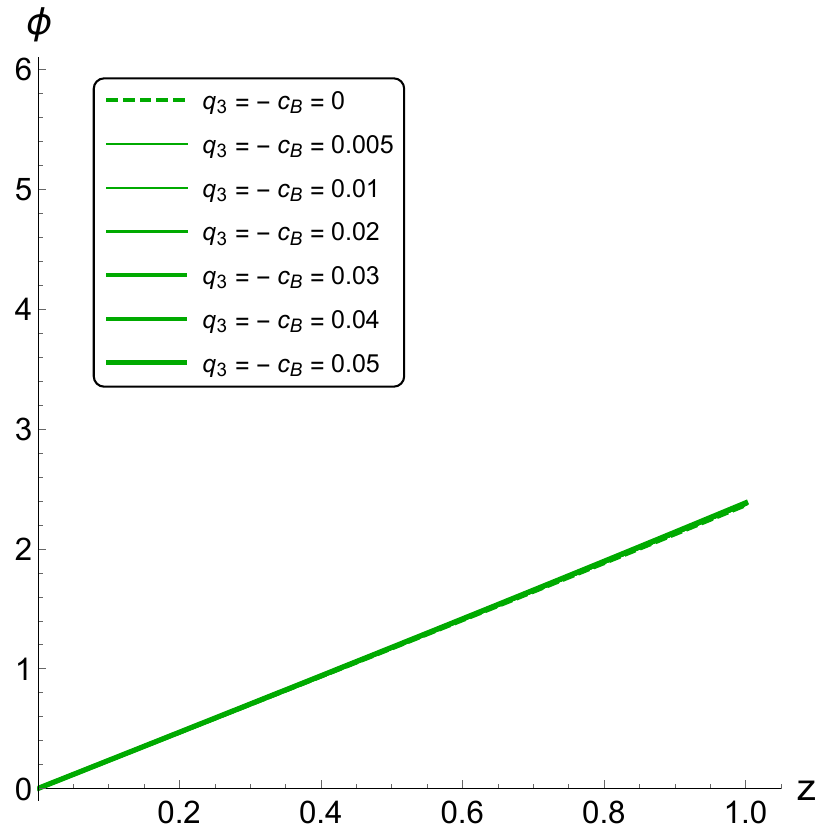} \
  \includegraphics[scale=0.24]{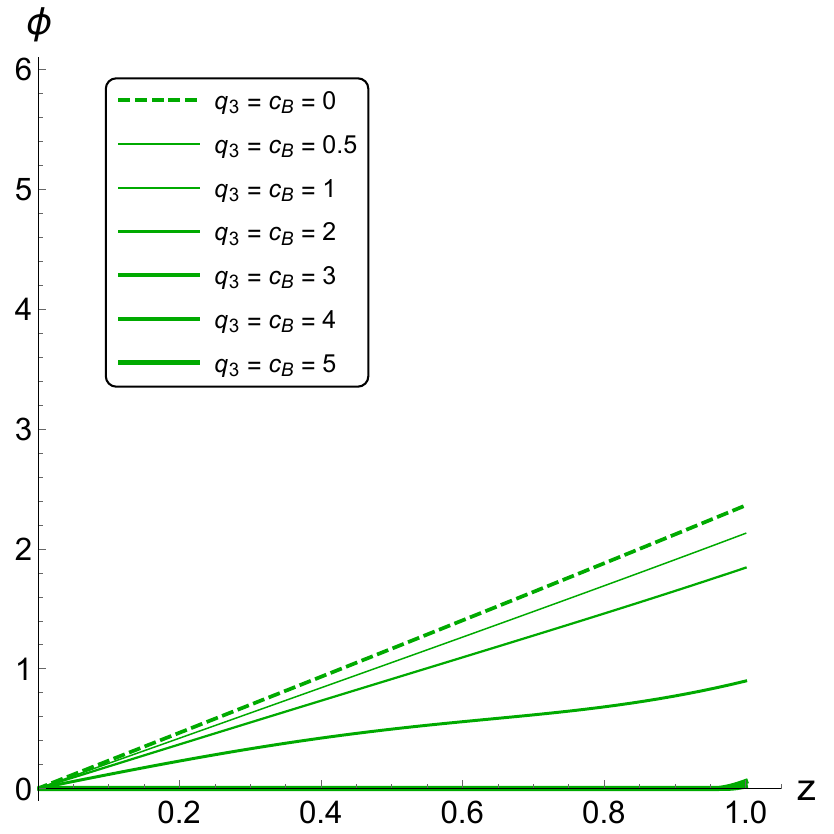} \quad
  \includegraphics[scale=0.24]{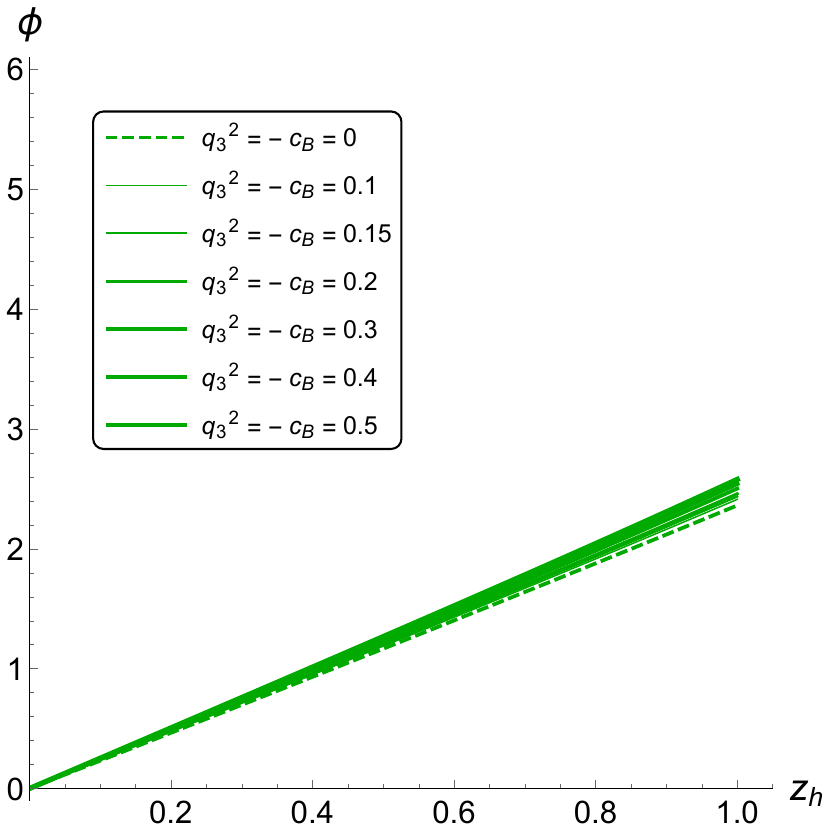} \
  \includegraphics[scale=0.24]{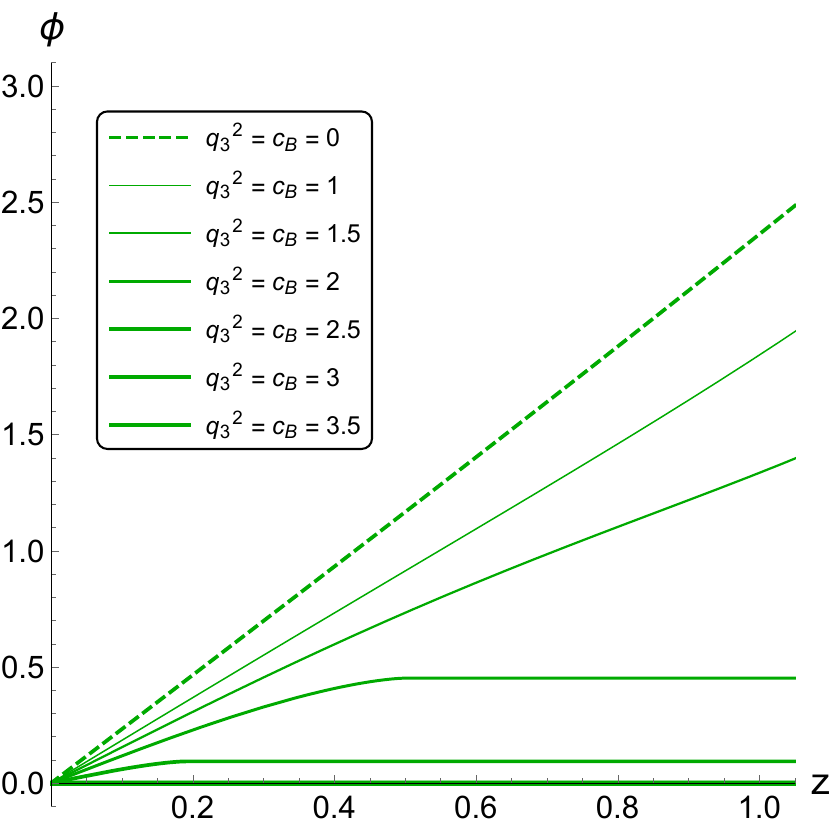} \\
  G \hspace{200pt} H
  \caption{Scalar field $\phi(z)$ in magnetic field with different
    $q_3$ (A,E) for $c_B = - \, 0.5$ (left) and $c_B = 0.5$ (right);
    with different $c_B$ for $q_3 = 0.5$ (B,F) for $c_B < 0$ (left)
    and $c_B > 0$ (right); for different $q_3 = \pm \, c_B$ (C,G); for
    different $q_3^2 = \pm \, c_B$ (D,H) for $d = 0.06 > 0.05$ (A-D)
    and $d = 0.01 < 0.05$ (E-H) in primary isotropic case $\nu = 1$,
    $a = 0.15$, $c = 1.16$.}
  \label{Fig:phiz-q3cB-nu1-mu0-z5}
\end{figure}

\begin{figure}[t!]
  \centering
  \includegraphics[scale=0.24]{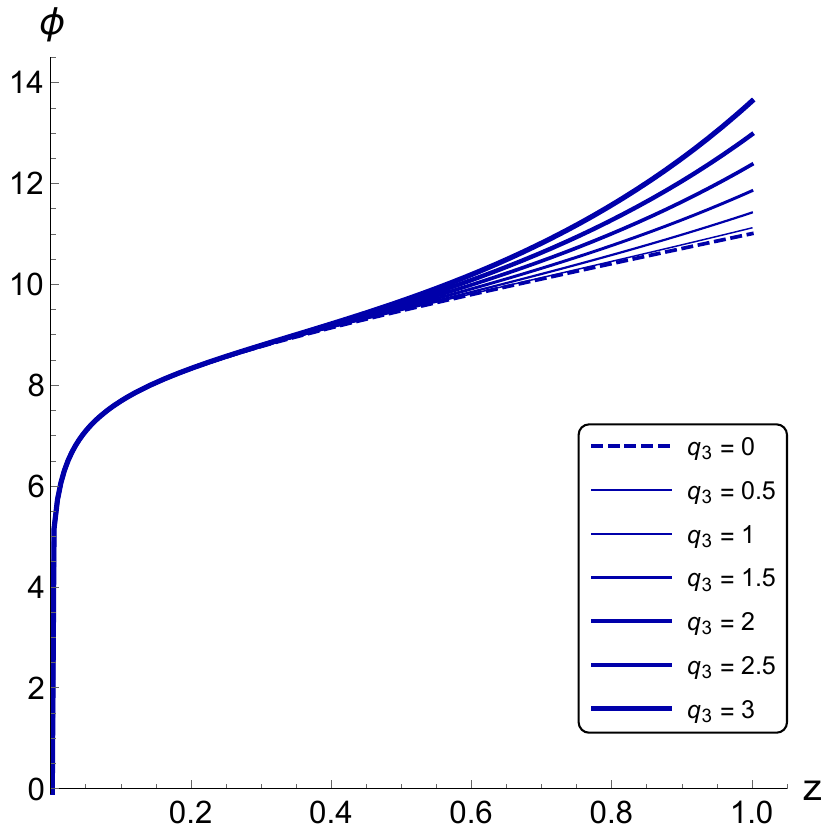} \
  \includegraphics[scale=0.24]{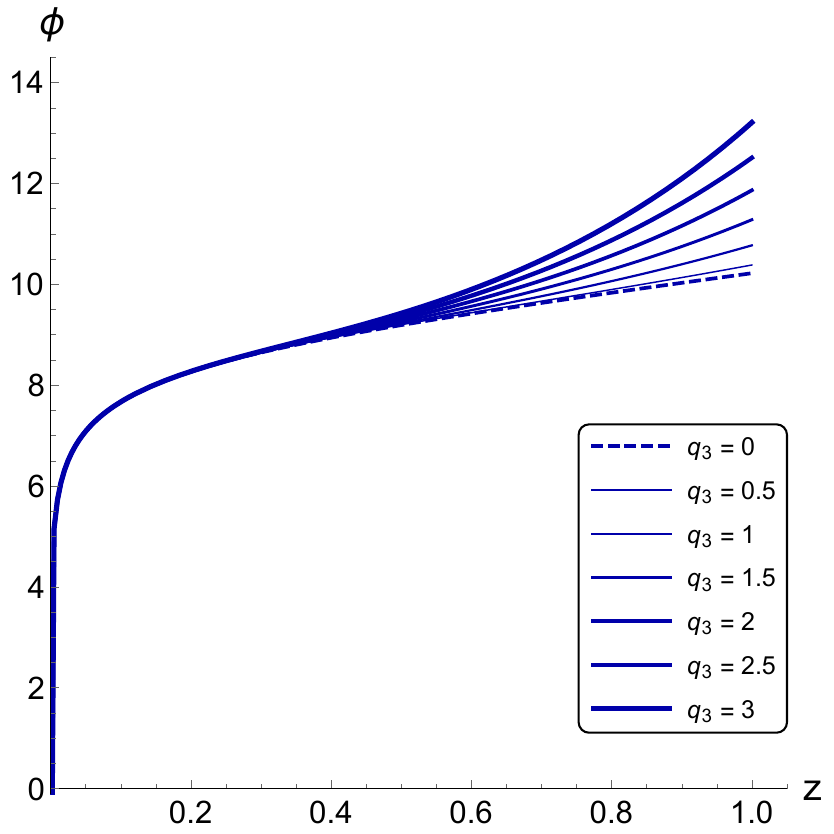} \quad
  \includegraphics[scale=0.24]{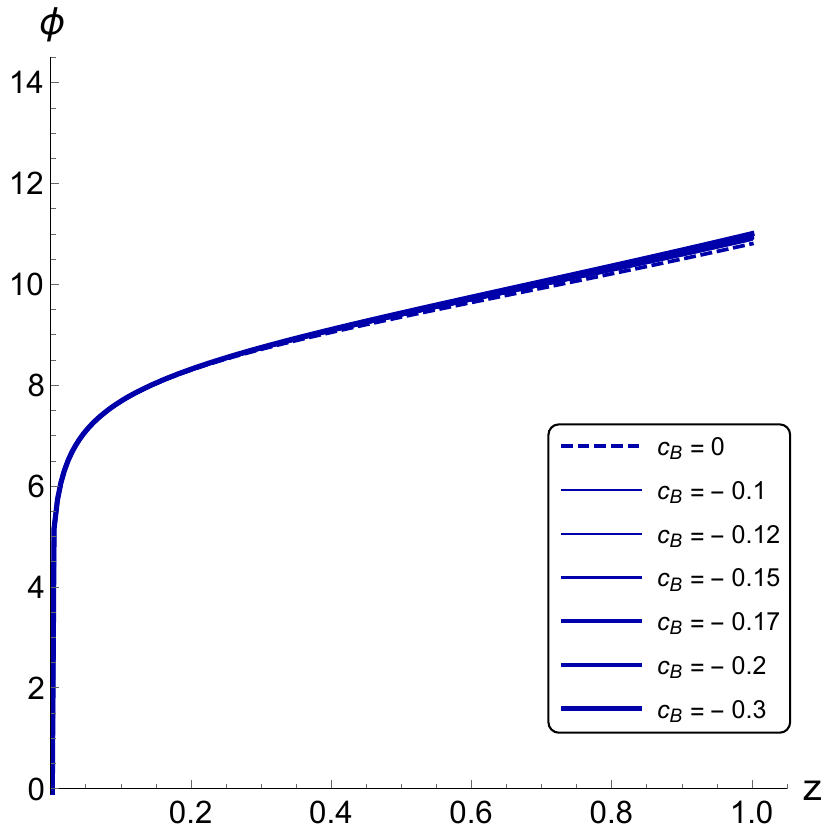} \
  \includegraphics[scale=0.24]{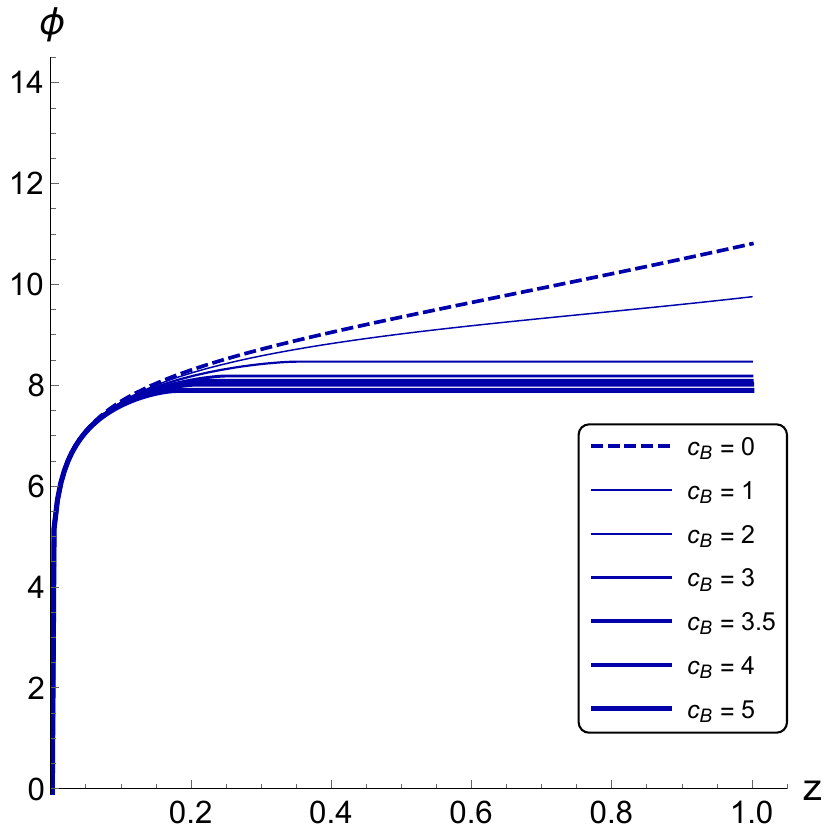} \\
  A \hspace{200pt} B \ \\
  \includegraphics[scale=0.24]{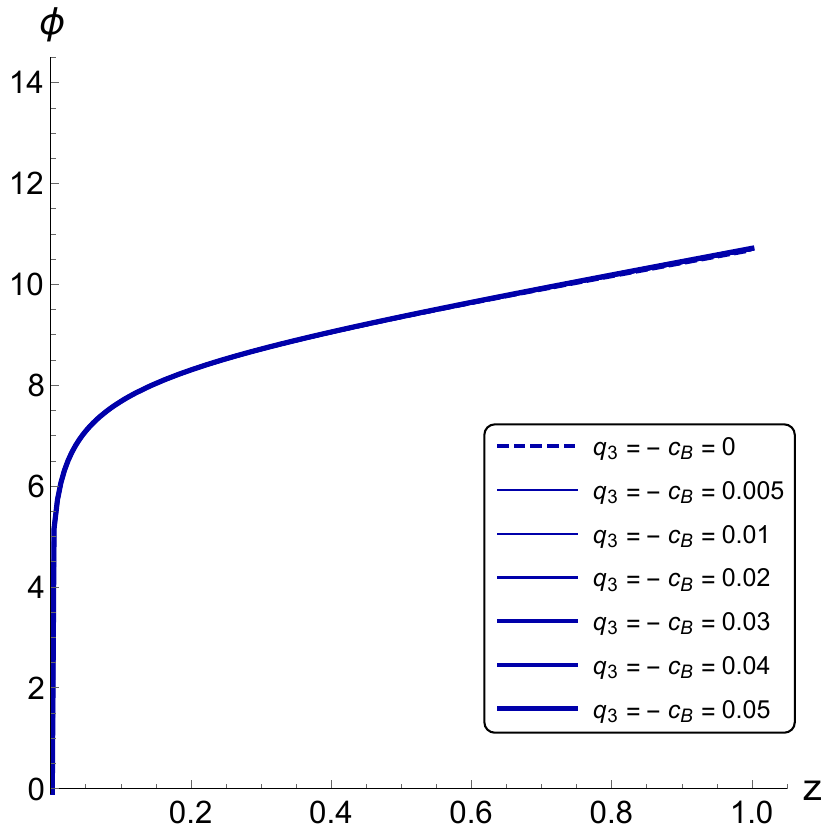} \
  \includegraphics[scale=0.24]{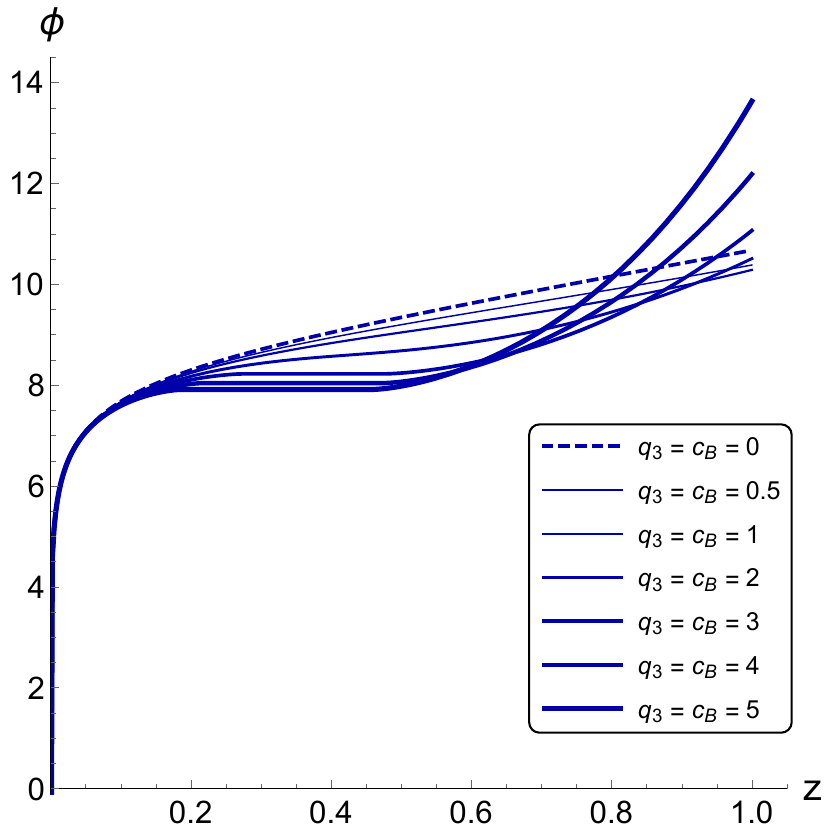} \quad
  \includegraphics[scale=0.24]{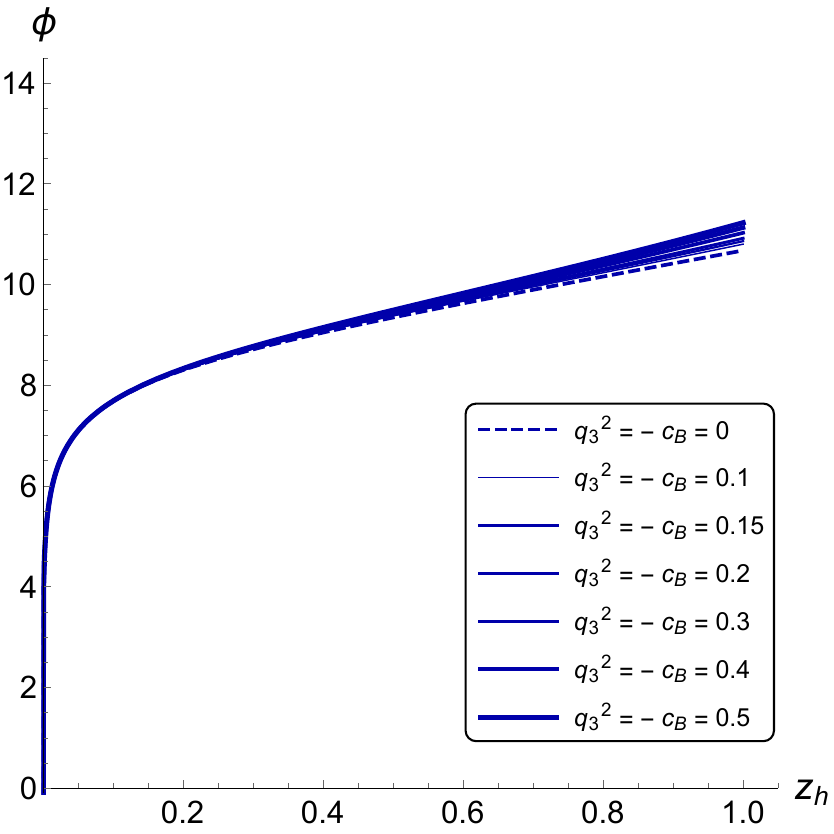} \
  \includegraphics[scale=0.24]{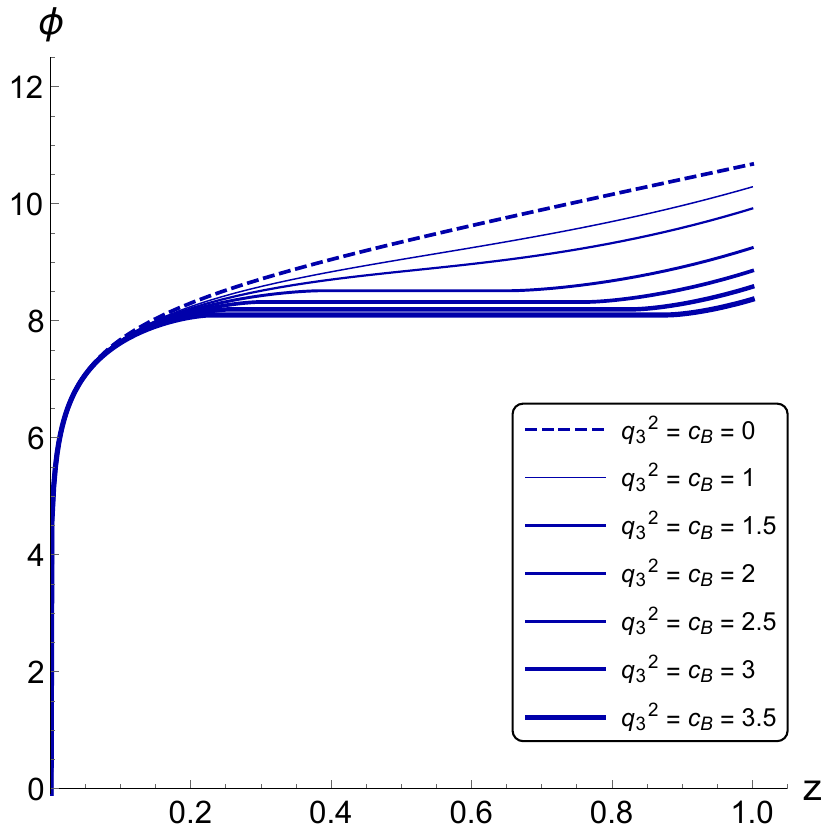} \\
  C \hspace{200pt} D \ \\
  \includegraphics[scale=0.24]{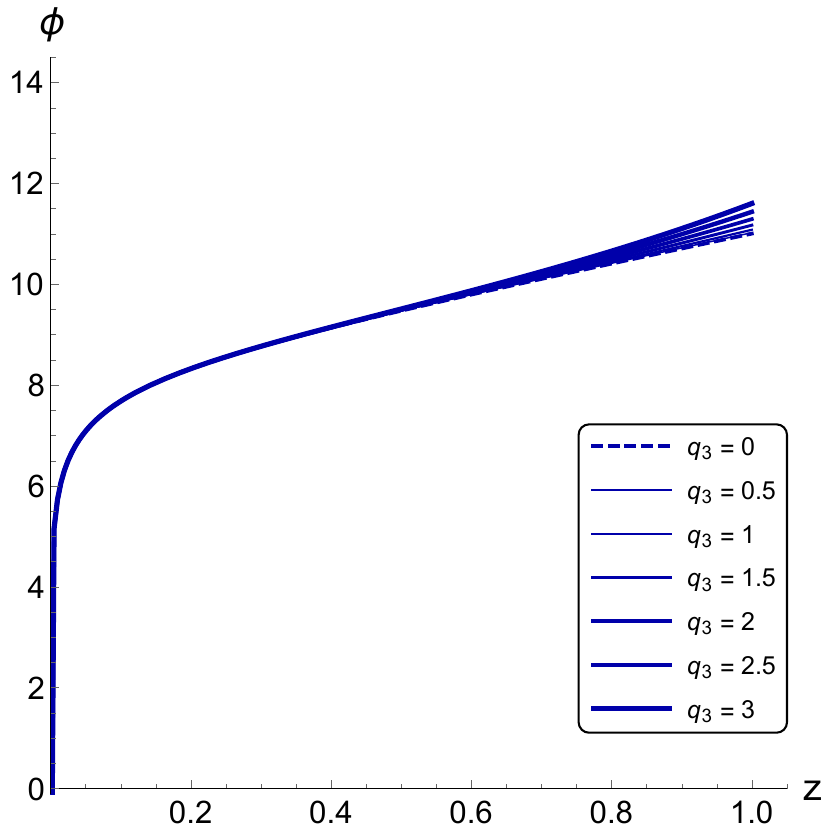} \
  \includegraphics[scale=0.24]{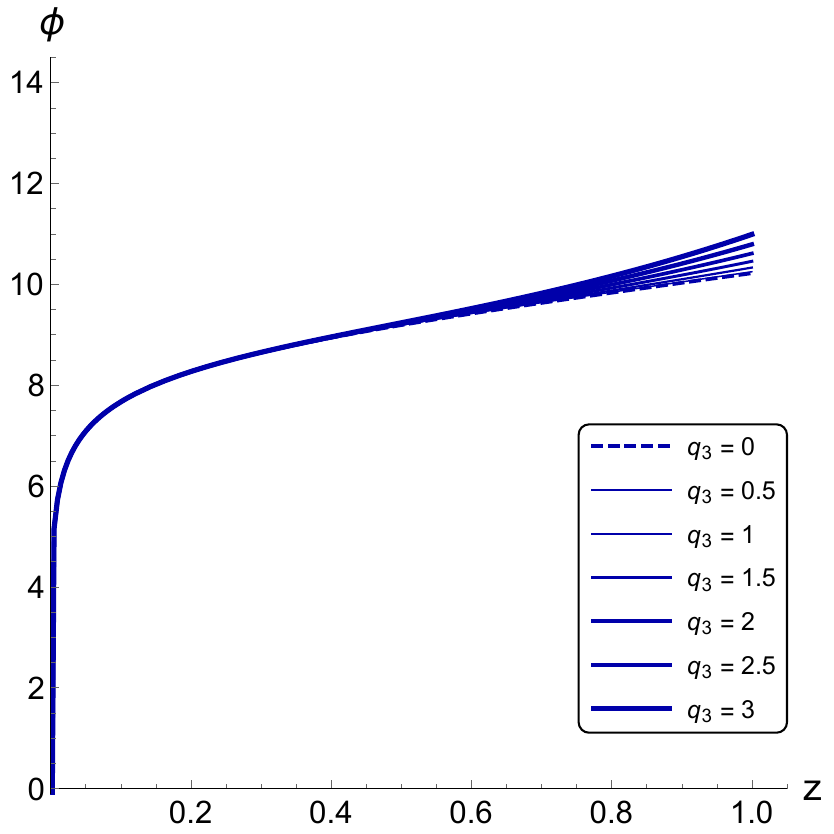} \quad
  \includegraphics[scale=0.24]{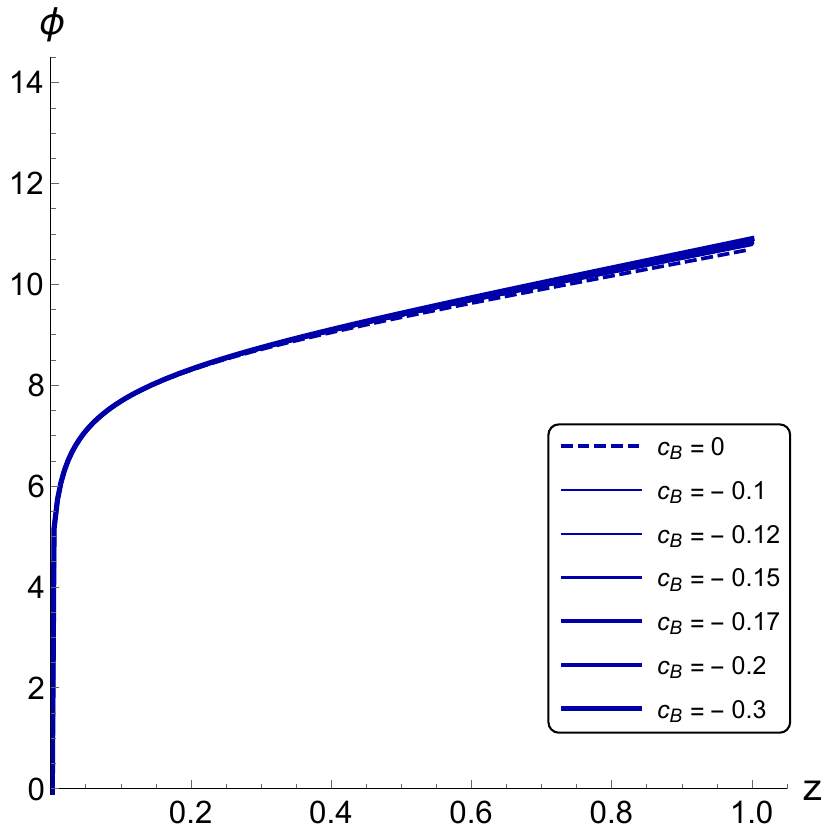} \
  \includegraphics[scale=0.24]{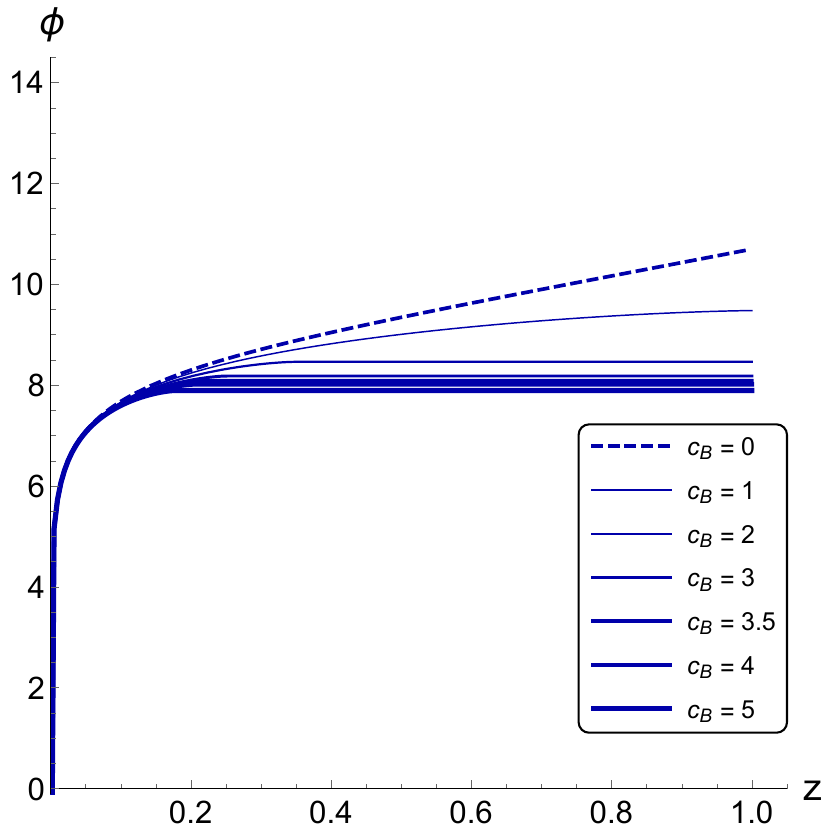} \\
  E \hspace{200pt} F \ \\
  \includegraphics[scale=0.24]{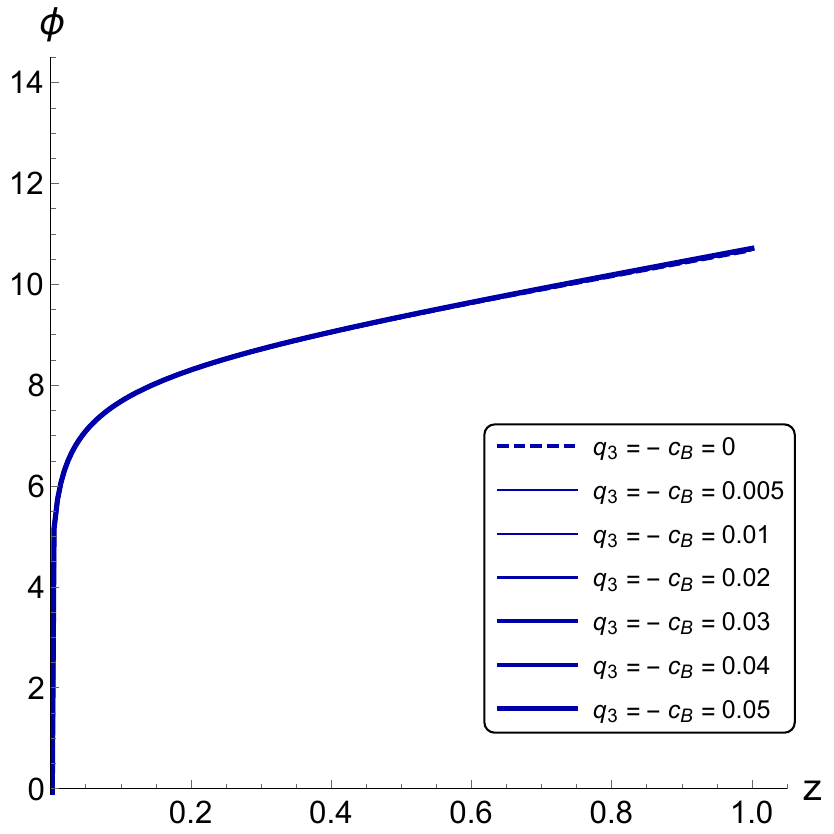} \
  \includegraphics[scale=0.24]{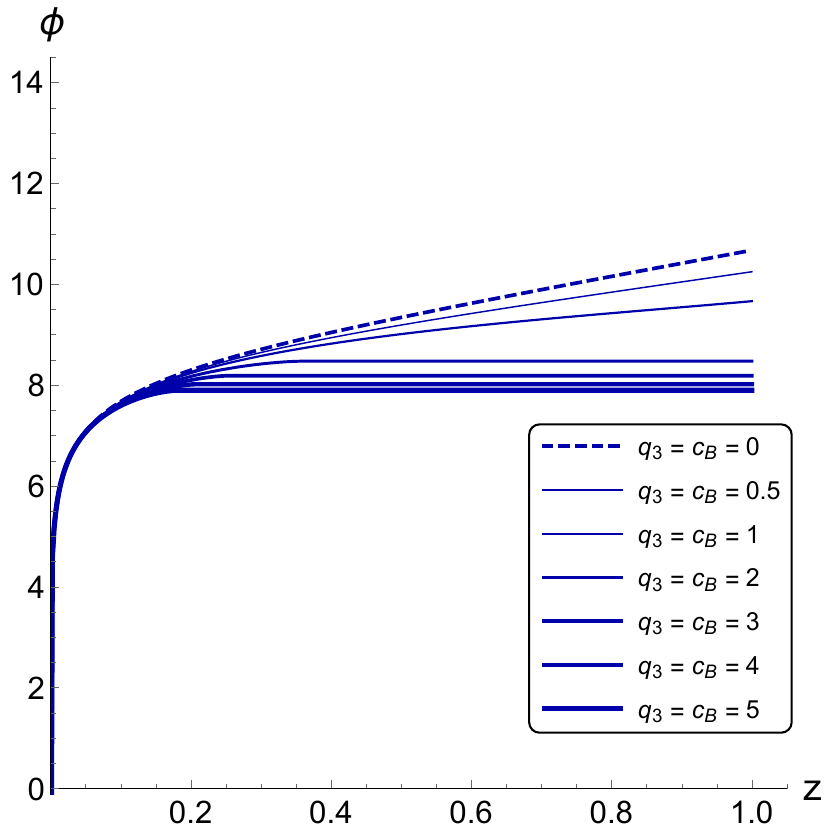} \quad
  \includegraphics[scale=0.24]{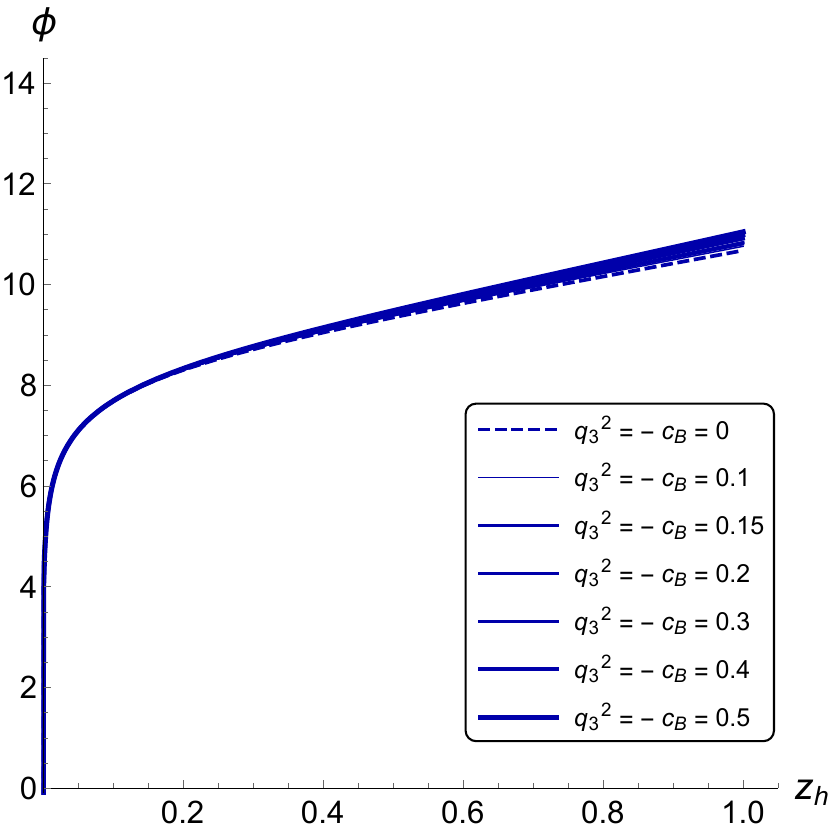} \
  \includegraphics[scale=0.24]{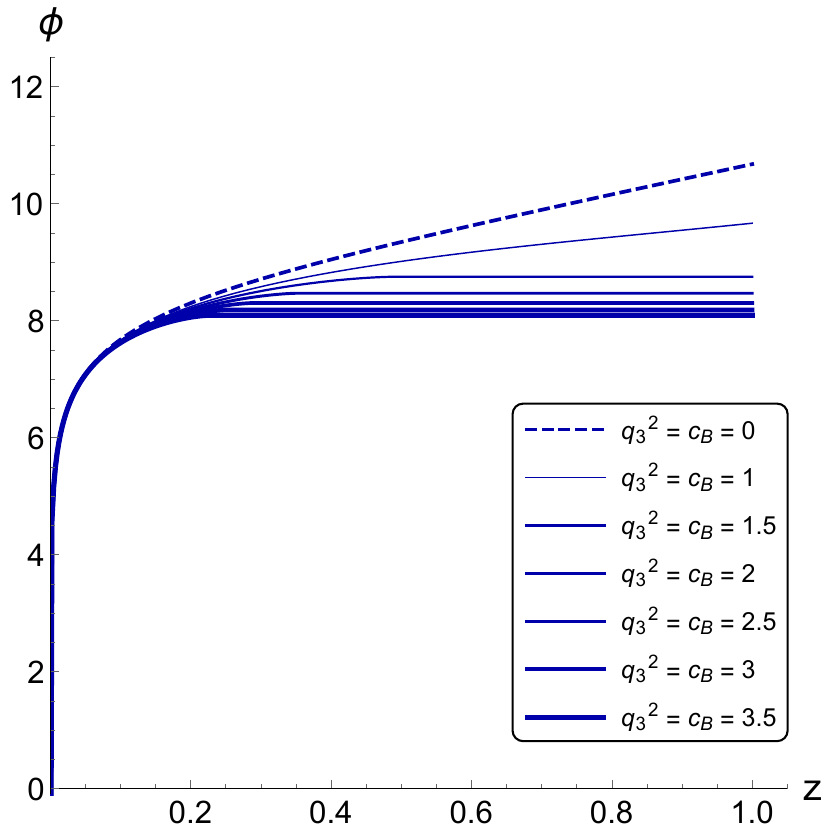} \\
  G \hspace{200pt} H
  \caption{Scalar field $\phi(z)$ in magnetic field with different
    $q_3$ (A,E) for $c_B = - \, 0.5$ (left) and $c_B = 0.5$ (right);
    with different $c_B$ for $q_3 = 0.5$ (B,F) for $c_B < 0$ (left)
    and $c_B > 0$ (right); for different $q_3 = \pm \, c_B$ (C,G); for
    different $q_3^2 = \pm \, c_B$ (D,H) for $d = 0.06 > 0.05$ (A-D)
    and $d = 0.01 < 0.05$ (E-H) in primary anisotropic case $\nu =
    4.5$, $a = 0.15$, $c = 1.16$.}
  \label{Fig:phiz-q3cB-nu45-mu0-z5}
\end{figure}

\begin{figure}[t!]
  \centering
  \includegraphics[scale=0.24]{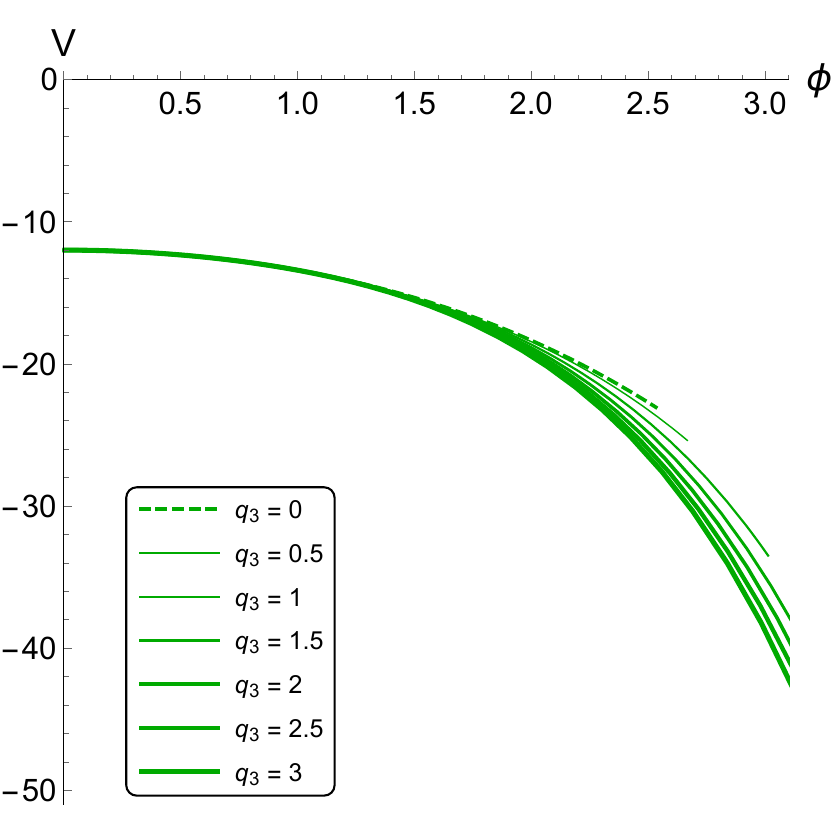} \
  \includegraphics[scale=0.24]{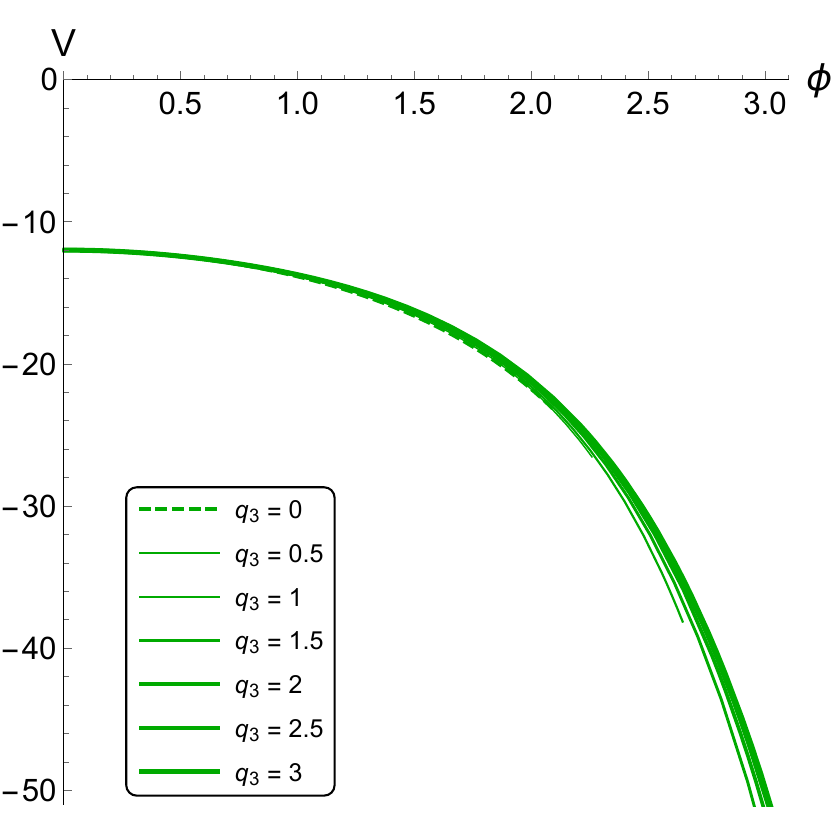} \quad
  \includegraphics[scale=0.24]{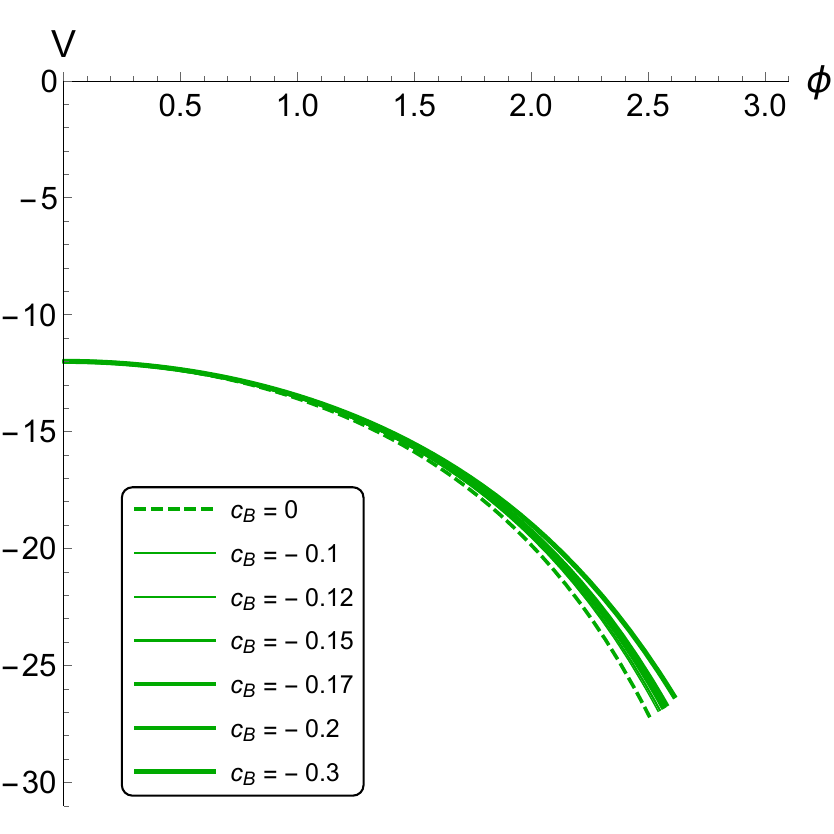} \
  \includegraphics[scale=0.24]{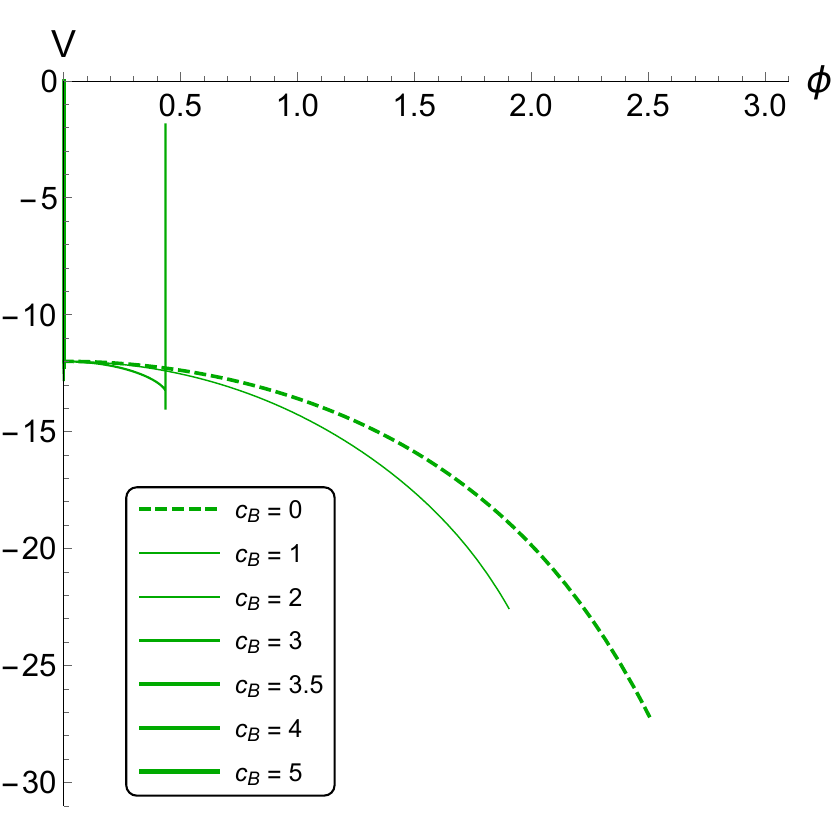} \\
  A \hspace{200pt} B \ \\
  \includegraphics[scale=0.24]{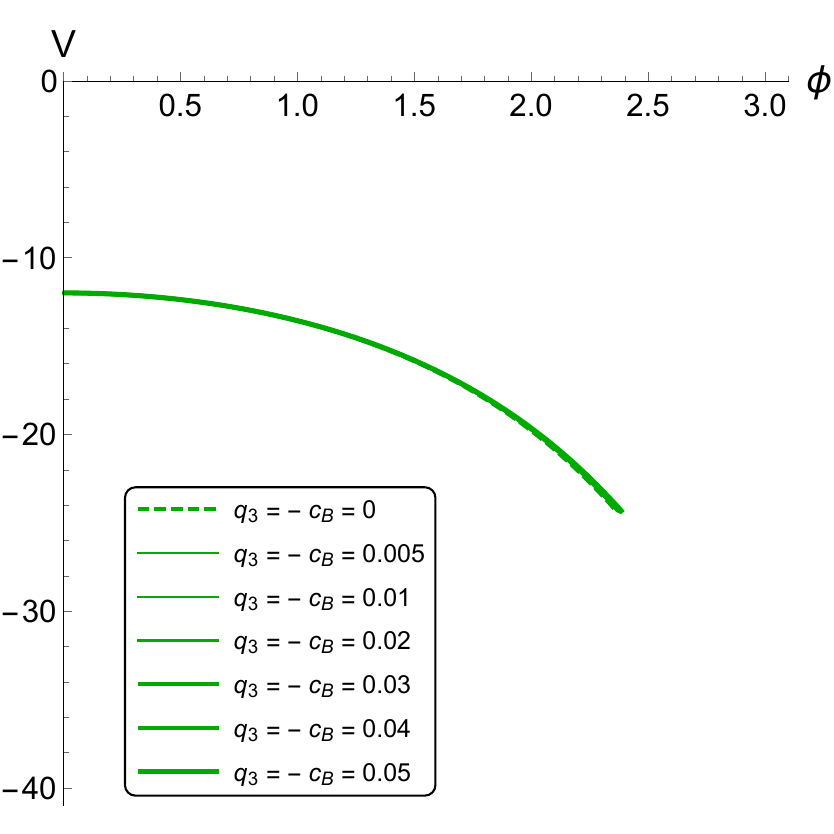} \
  \includegraphics[scale=0.24]{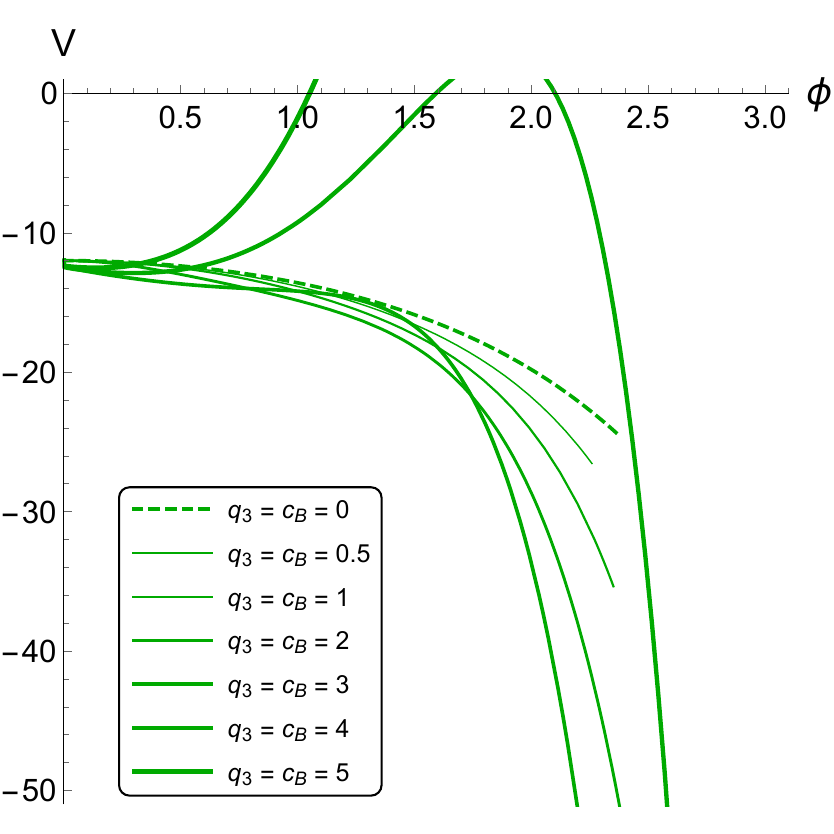} \quad
  \includegraphics[scale=0.24]{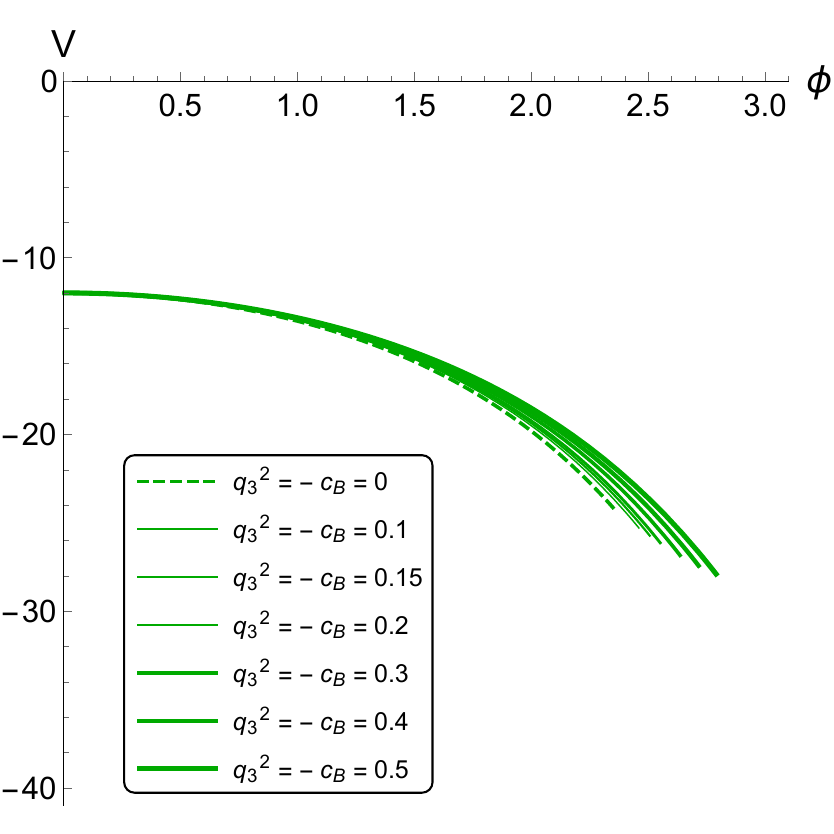} \
  \includegraphics[scale=0.24]{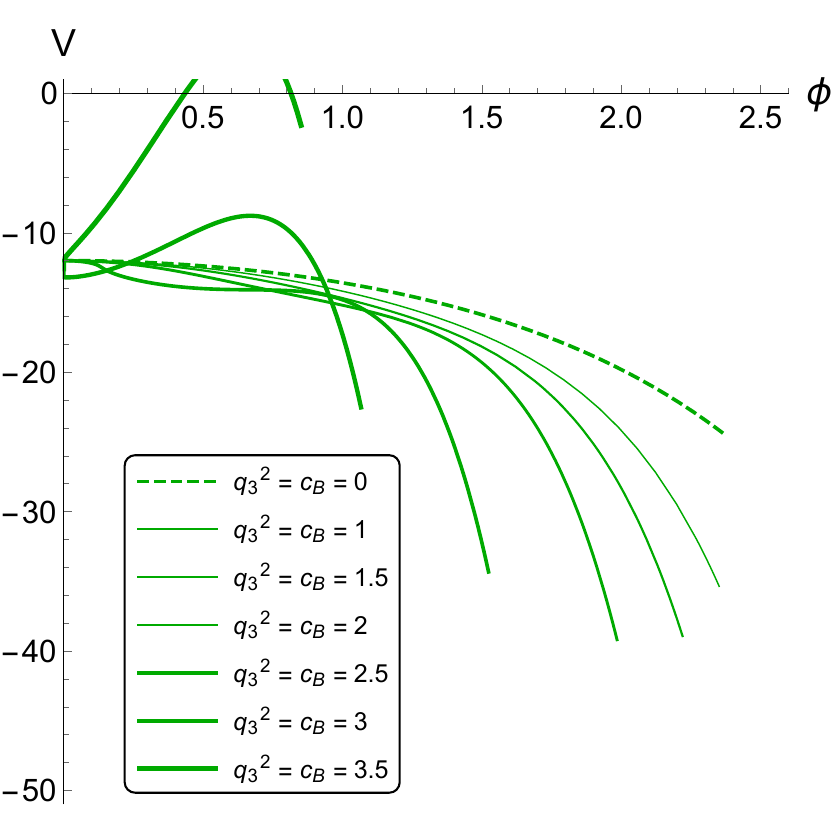} \\
  C \hspace{200pt} D \ \\
  \includegraphics[scale=0.24]{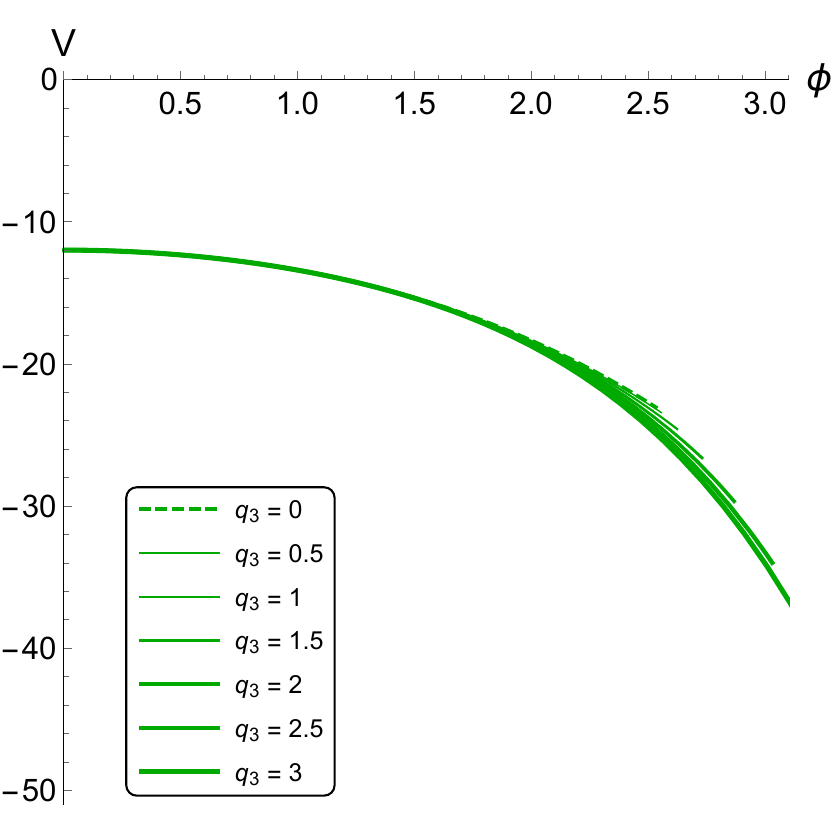} \
  \includegraphics[scale=0.24]{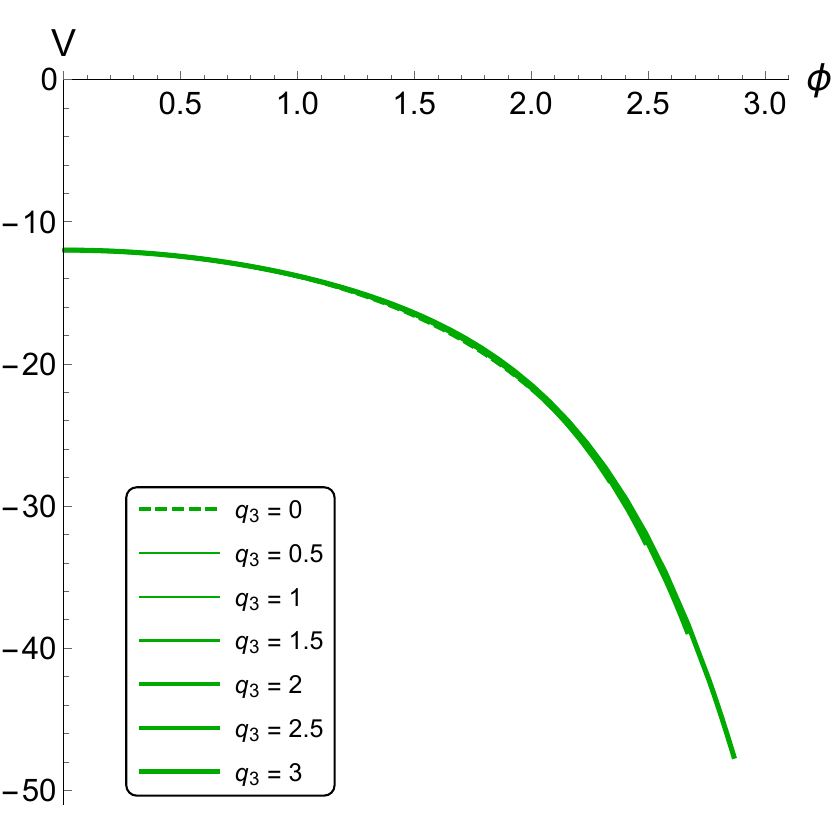} \quad
  \includegraphics[scale=0.24]{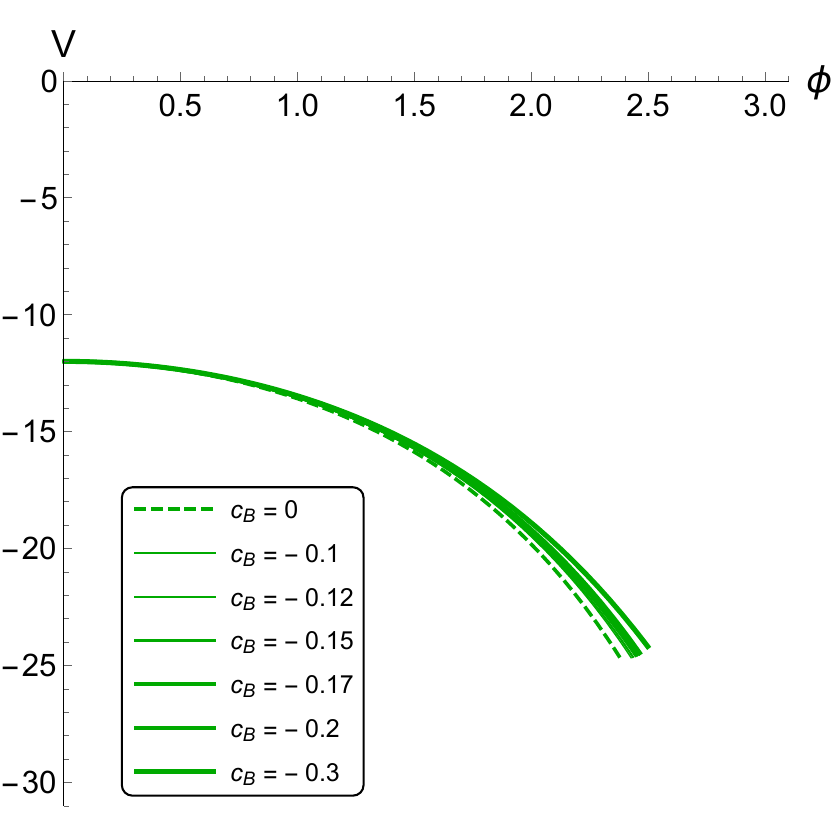} \
  \includegraphics[scale=0.24]{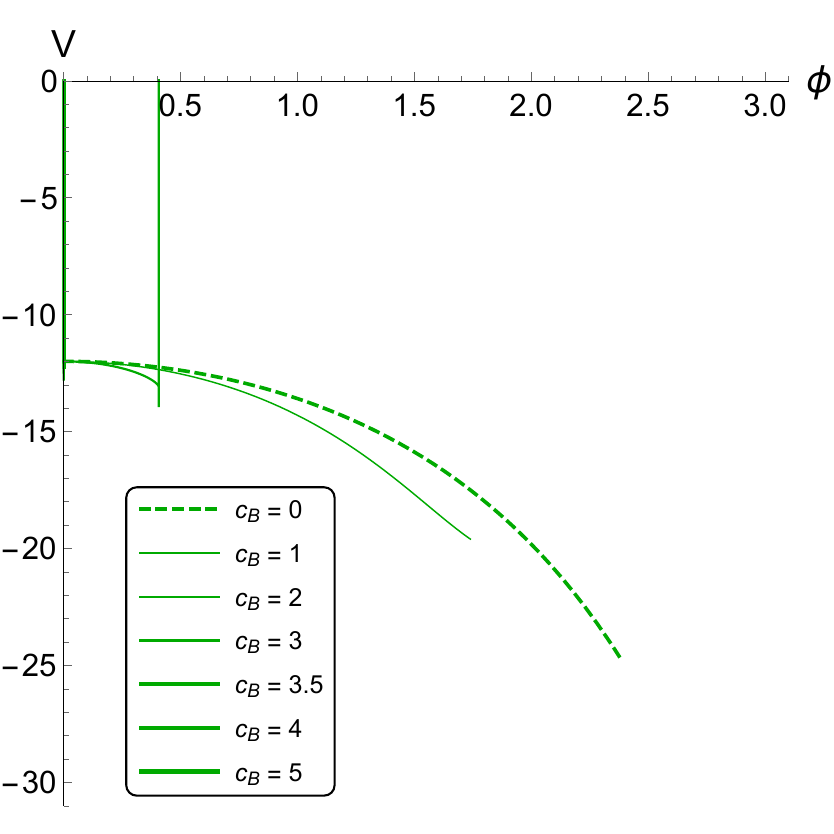} \\
  E \hspace{200pt} F \ \\
  \includegraphics[scale=0.24]{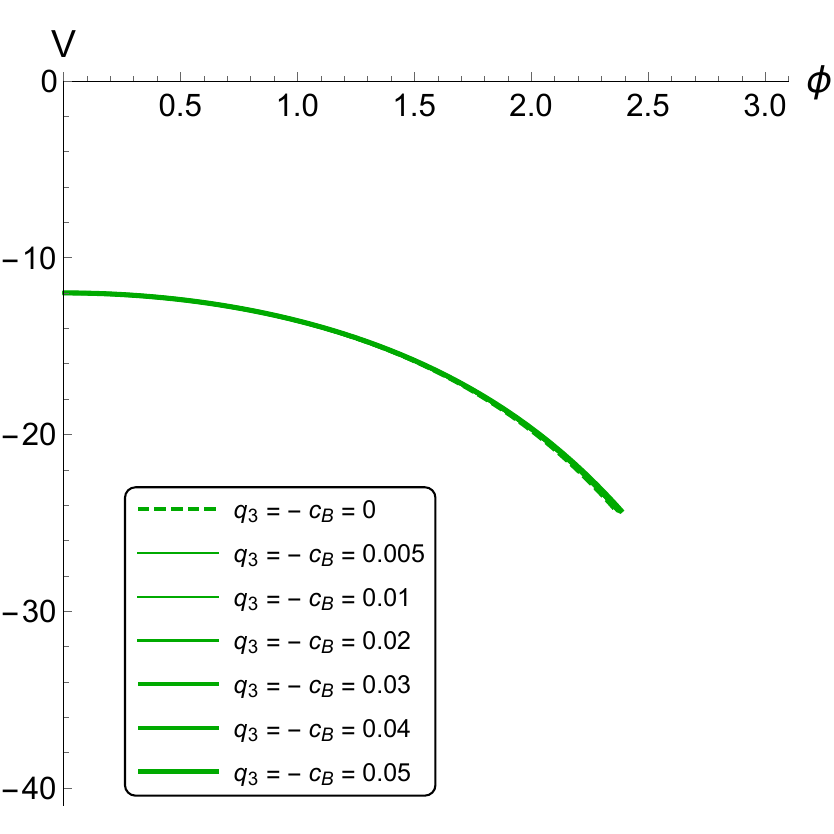} \
  \includegraphics[scale=0.24]{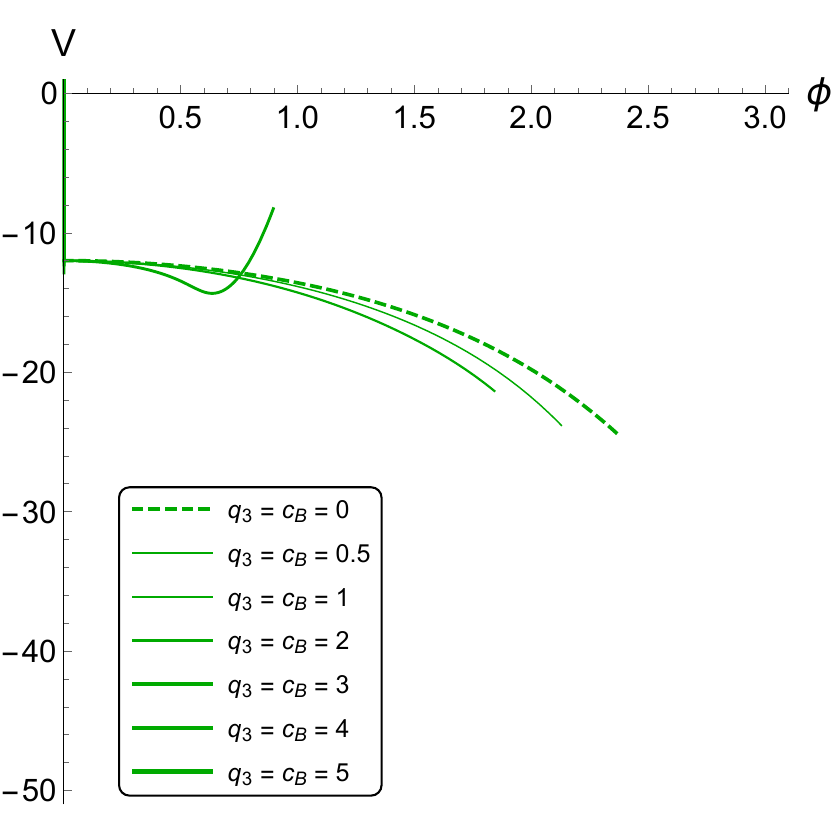} \quad
  \includegraphics[scale=0.24]{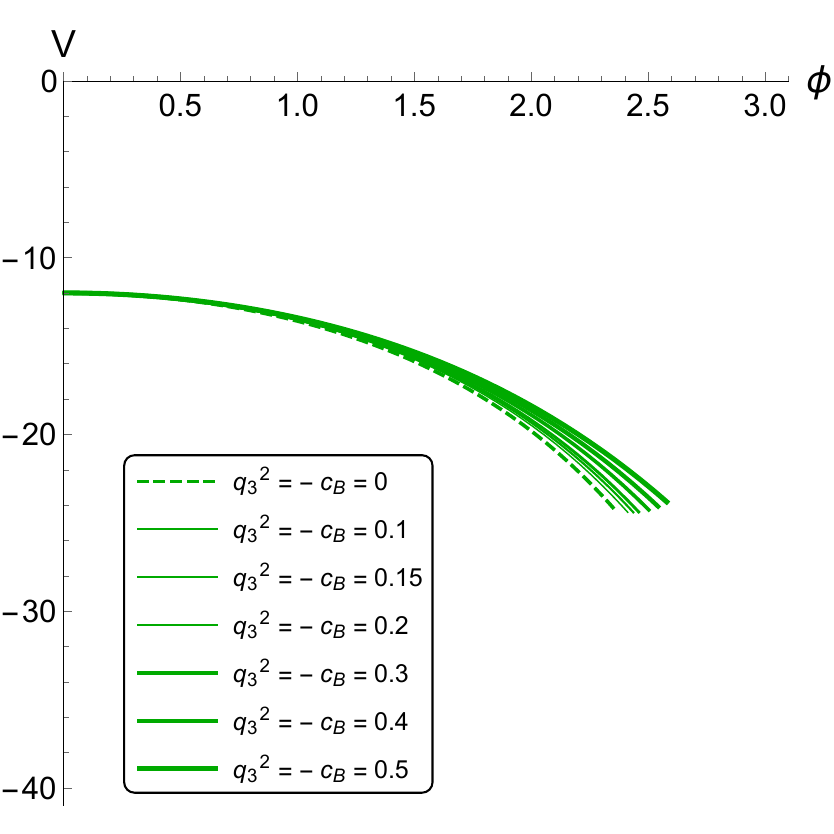} \
  \includegraphics[scale=0.24]{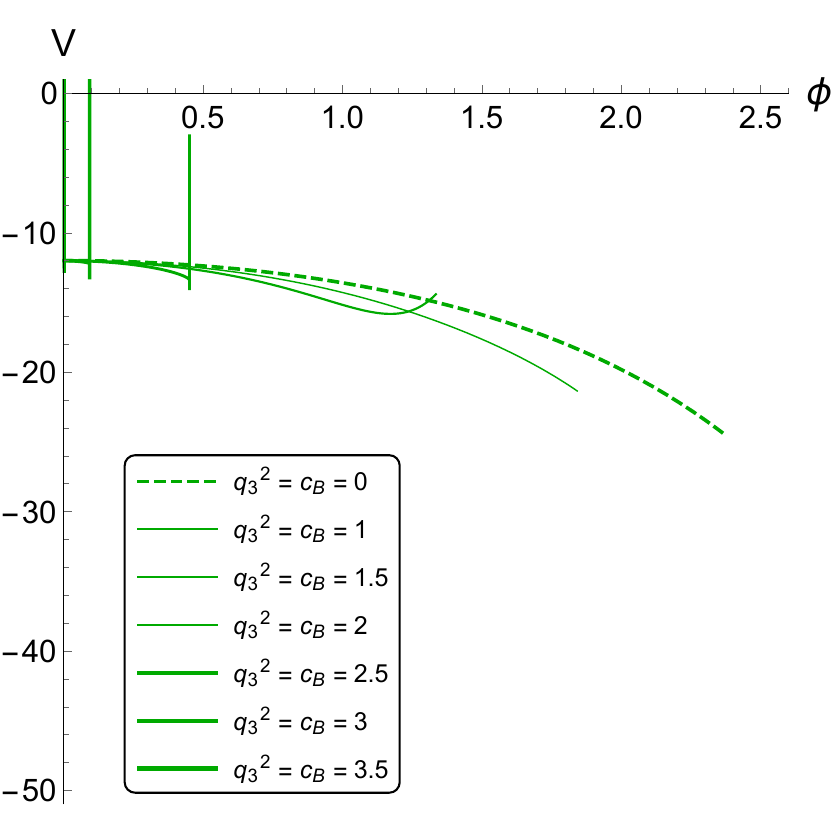} \\
  G \hspace{200pt} H
  \caption{Scalar potential $V(\phi)$ in magnetic field with different
    $q_3$ (A,E) for $c_B = - \, 0.5$ (left) and $c_B = 0.5$ (right);
    with different $c_B$ for $q_3 = 0.5$ (B,F) for $c_B < 0$ (left)
    and $c_B > 0$ (right); for different $q_3 = \pm \, c_B$ (C,G); for
    different $q_3^2 = \pm \, c_B$ (D,H) for $d = 0.06 > 0.05$ (A-D)
    and $d = 0.01 < 0.05$ (E-H) in primary isotropic case $\nu = 1$,
    $a = 0.15$, $c = 1.16$, $\mu = 0$.}
  \label{Fig:Vphi-q3cB-nu1-mu0-z5}
\end{figure}

\begin{figure}[t!]
  \centering
  \includegraphics[scale=0.24]{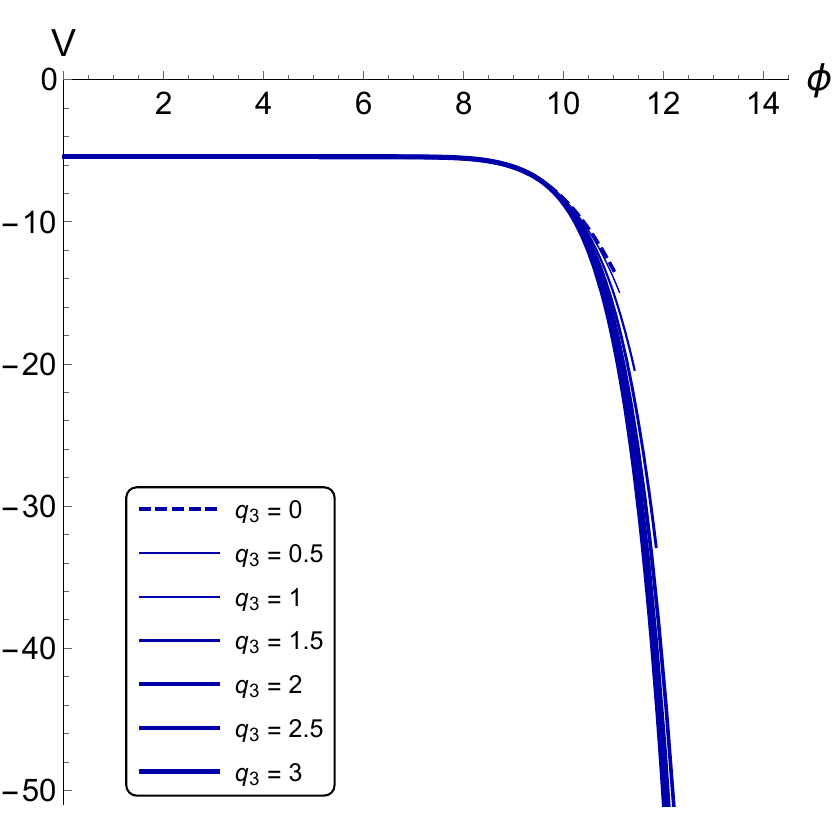} \
  \includegraphics[scale=0.24]{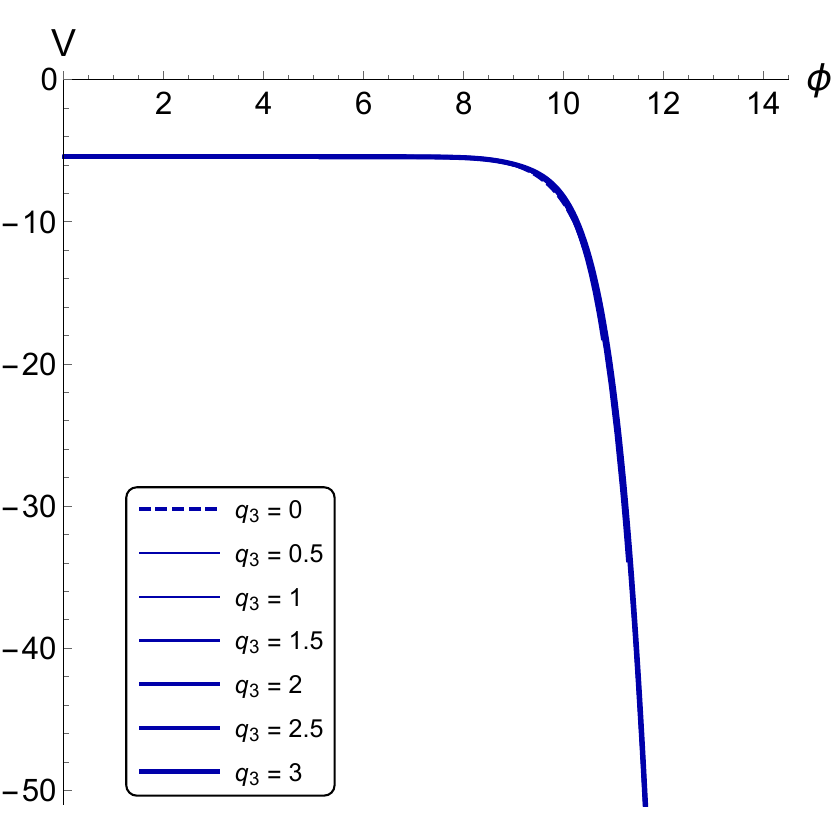} \quad
  \includegraphics[scale=0.24]{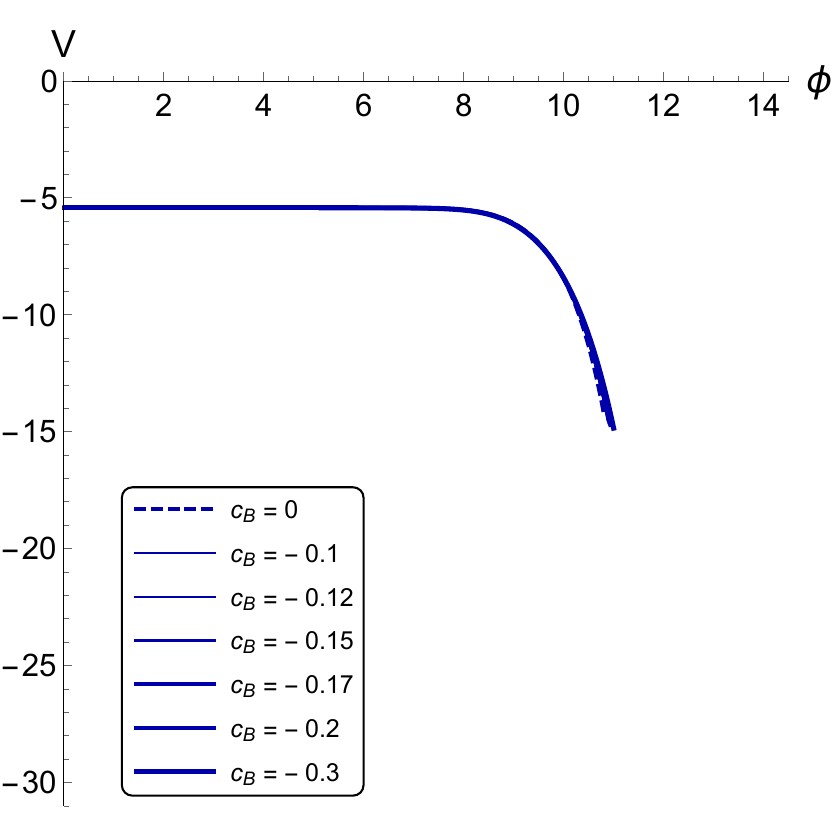} \
  \includegraphics[scale=0.24]{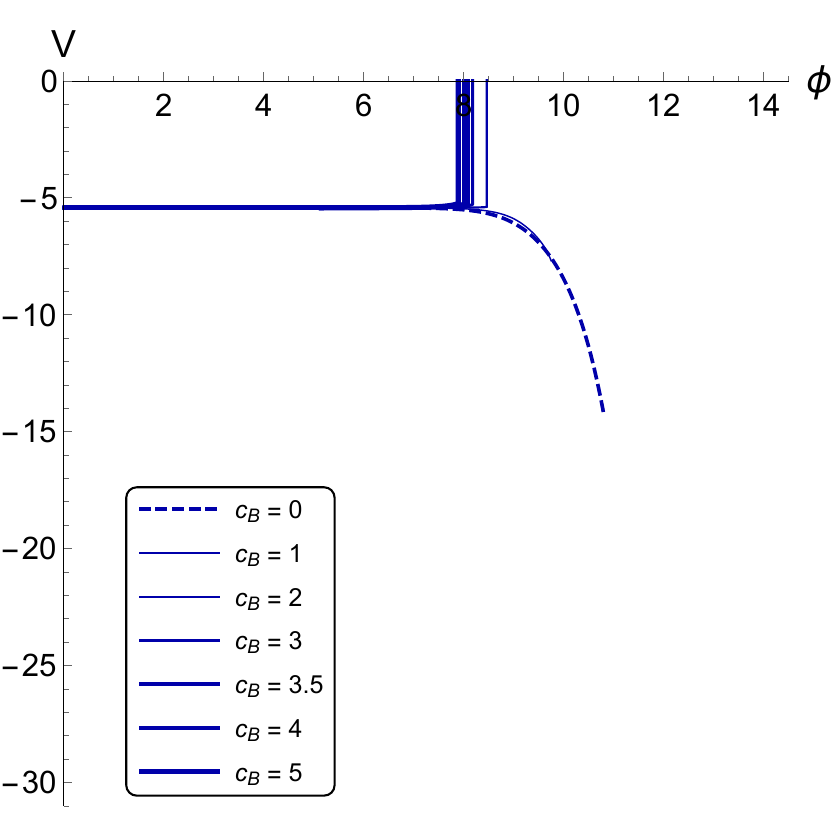} \\
  A \hspace{200pt} B \ \\
  \includegraphics[scale=0.24]{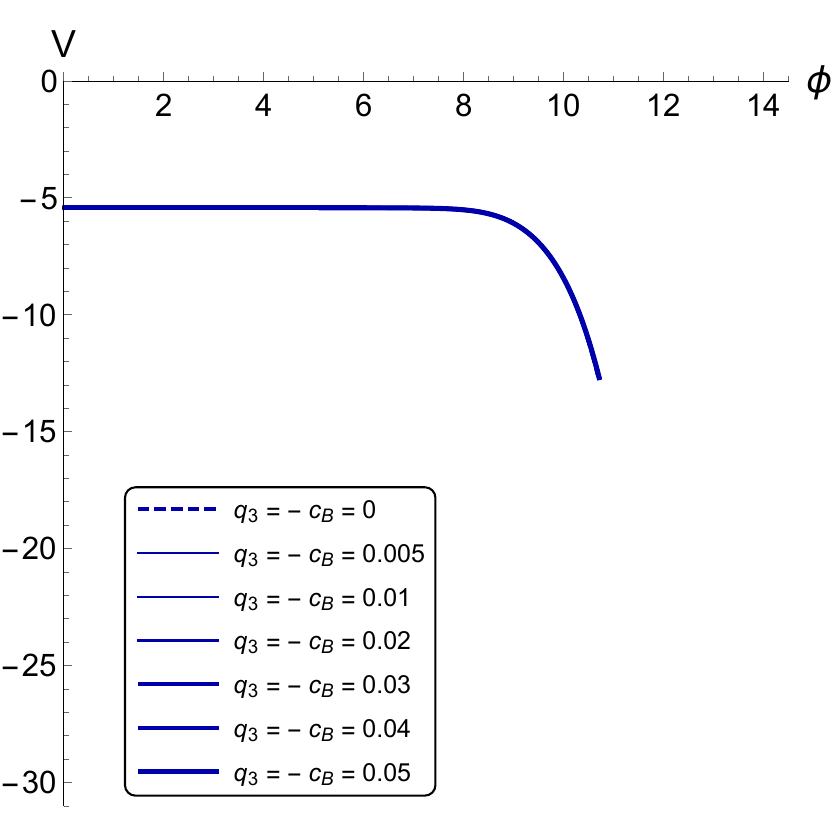} \
  \includegraphics[scale=0.24]{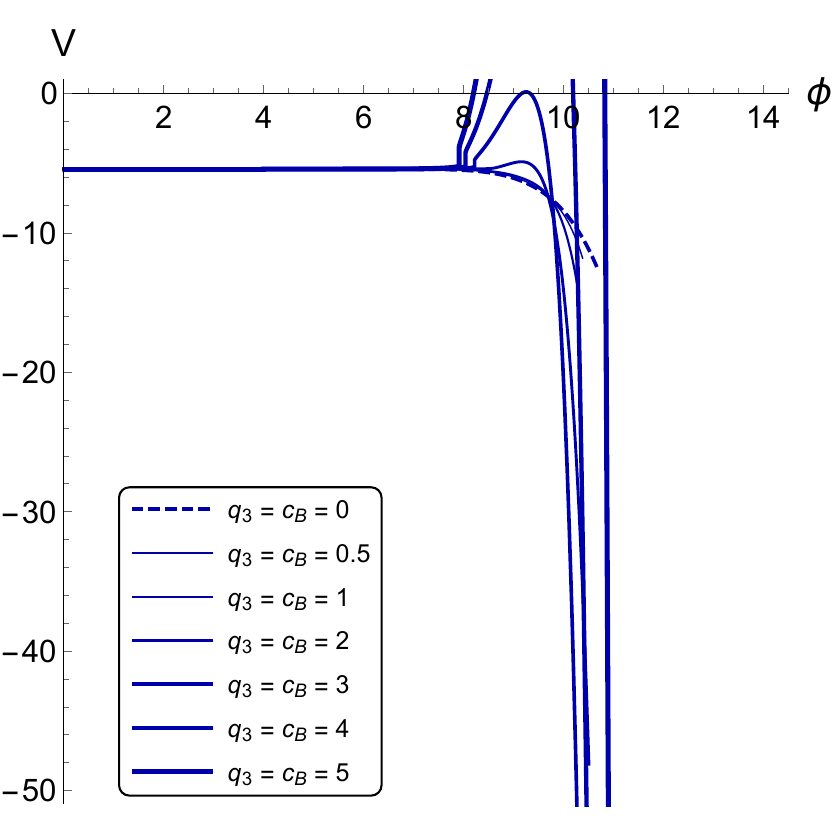} \quad
  \includegraphics[scale=0.24]{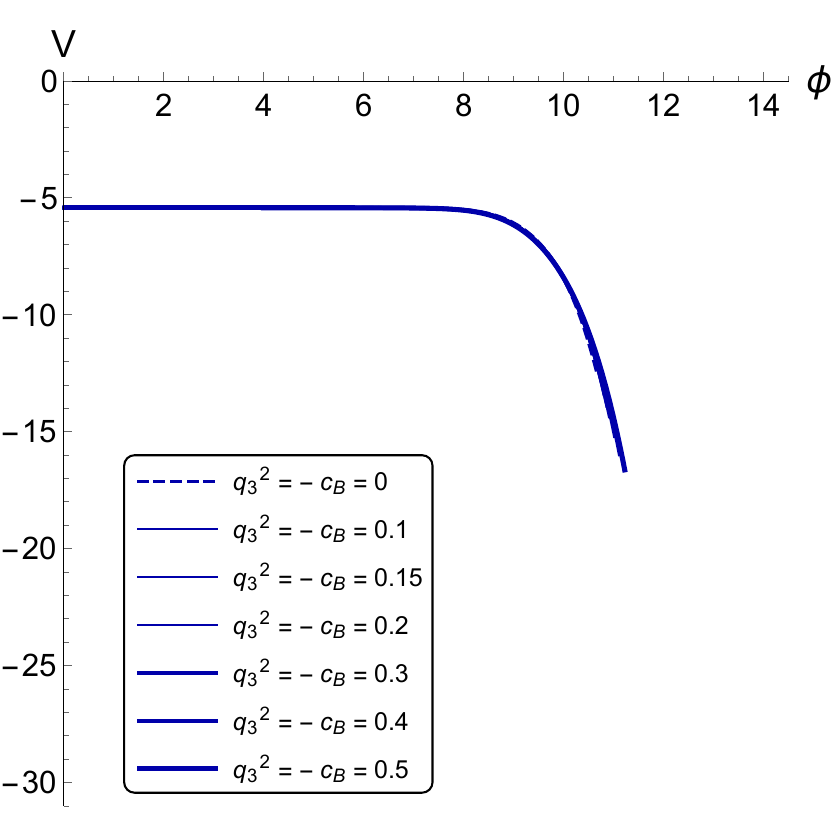} \
  \includegraphics[scale=0.24]{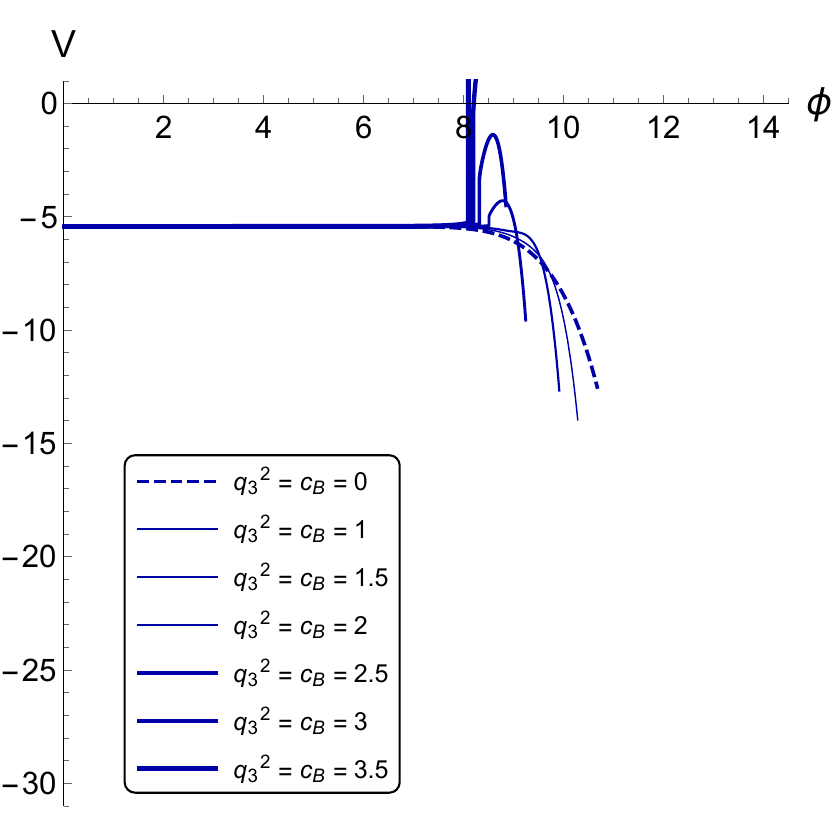} \\
  C \hspace{200pt} D \ \\
  \includegraphics[scale=0.24]{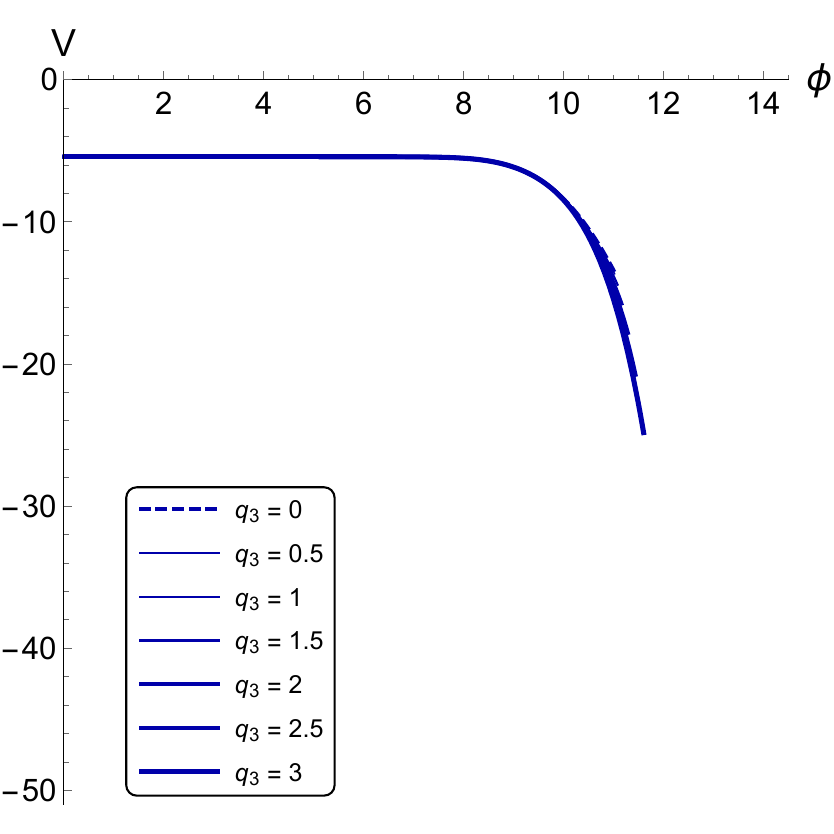} \
  \includegraphics[scale=0.24]{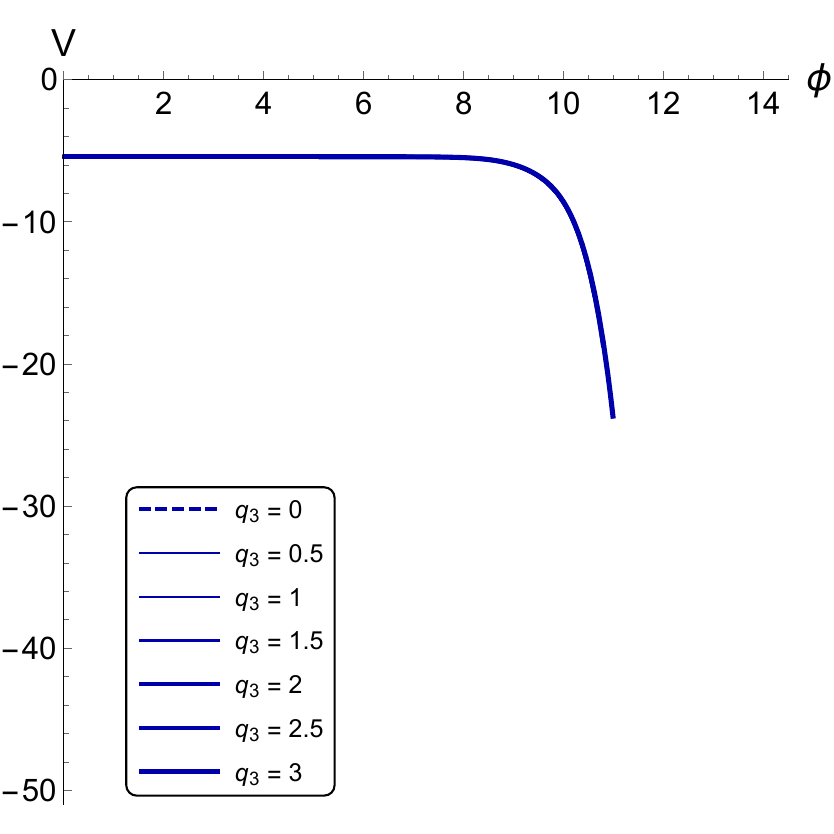} \quad
  \includegraphics[scale=0.24]{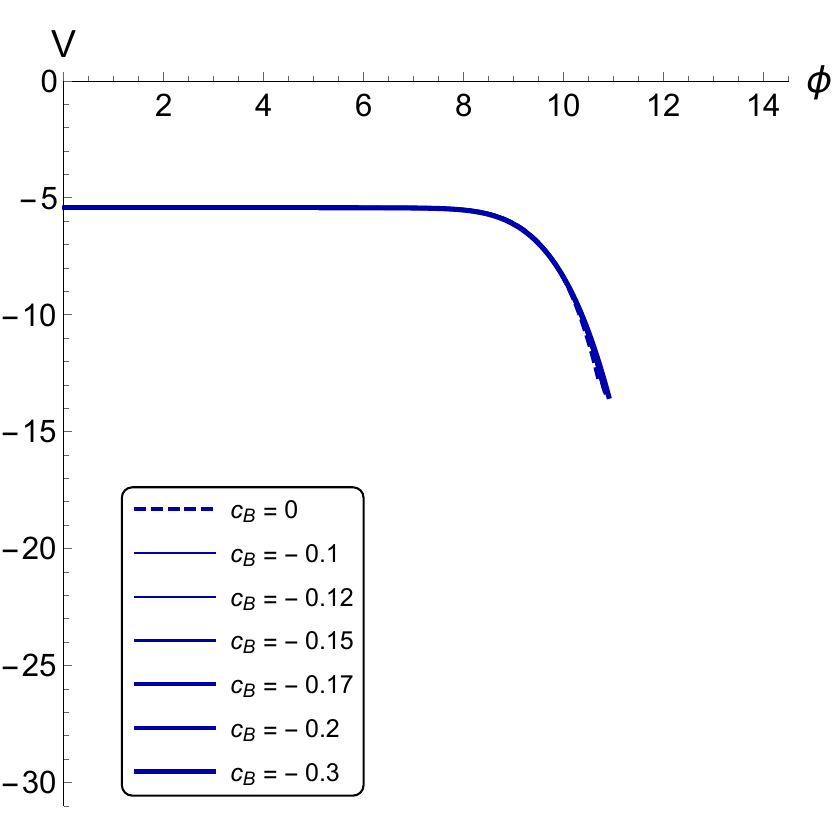} \
  \includegraphics[scale=0.24]{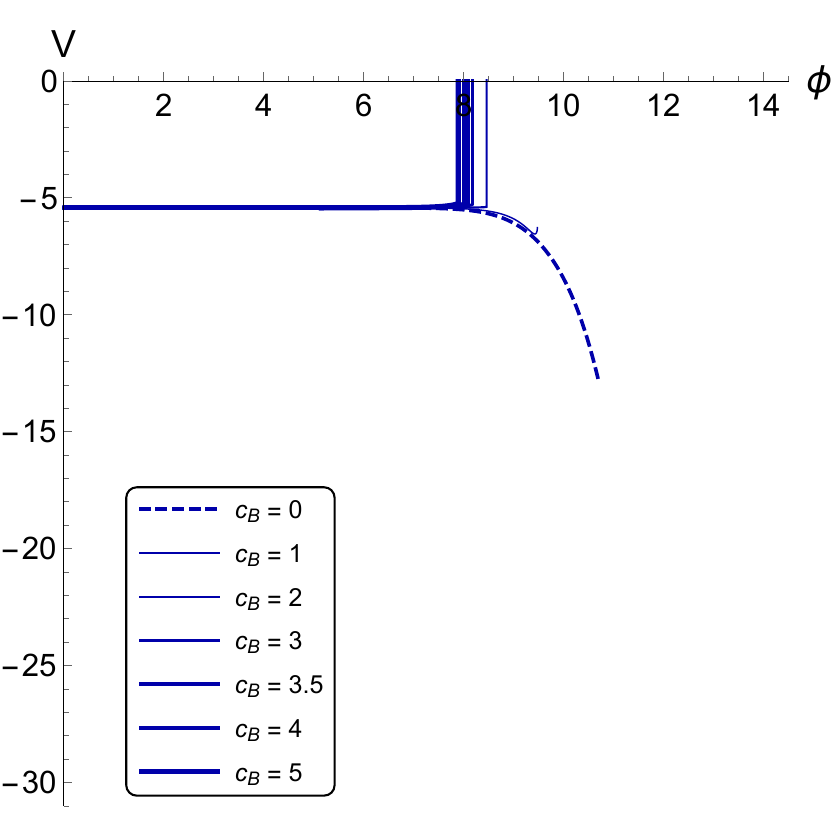} \\
  E \hspace{200pt} F \ \\
  \includegraphics[scale=0.24]{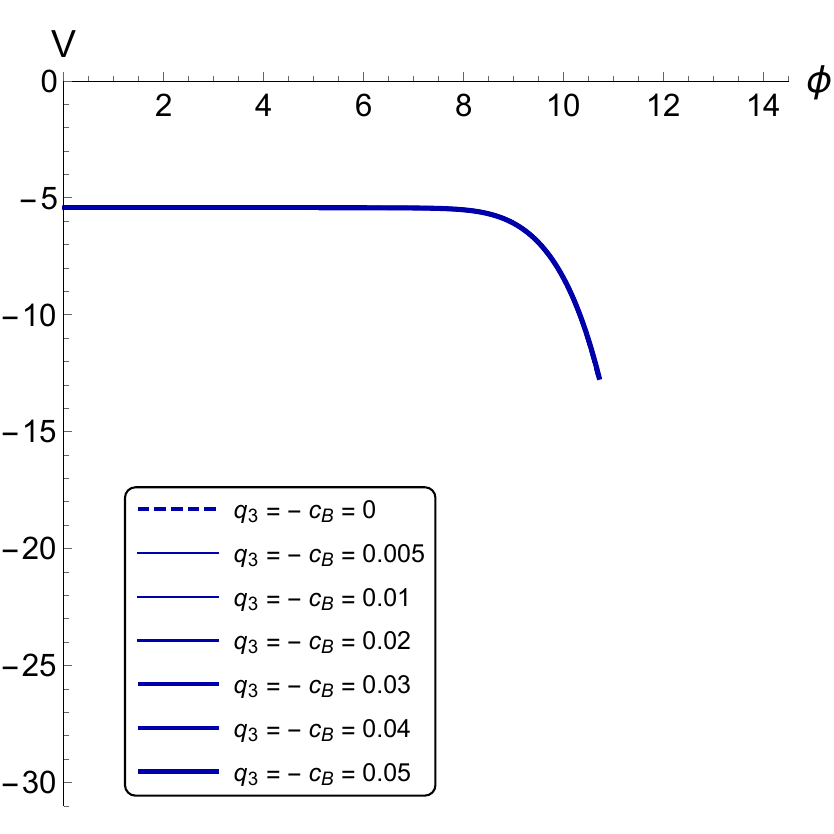} \
  \includegraphics[scale=0.24]{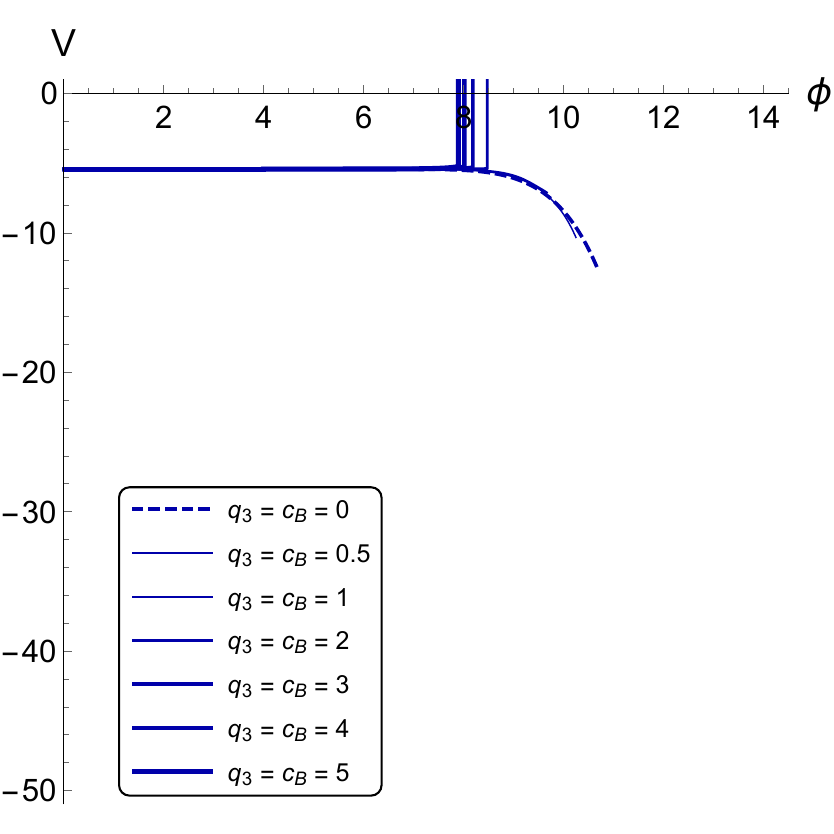} \quad
  \includegraphics[scale=0.24]{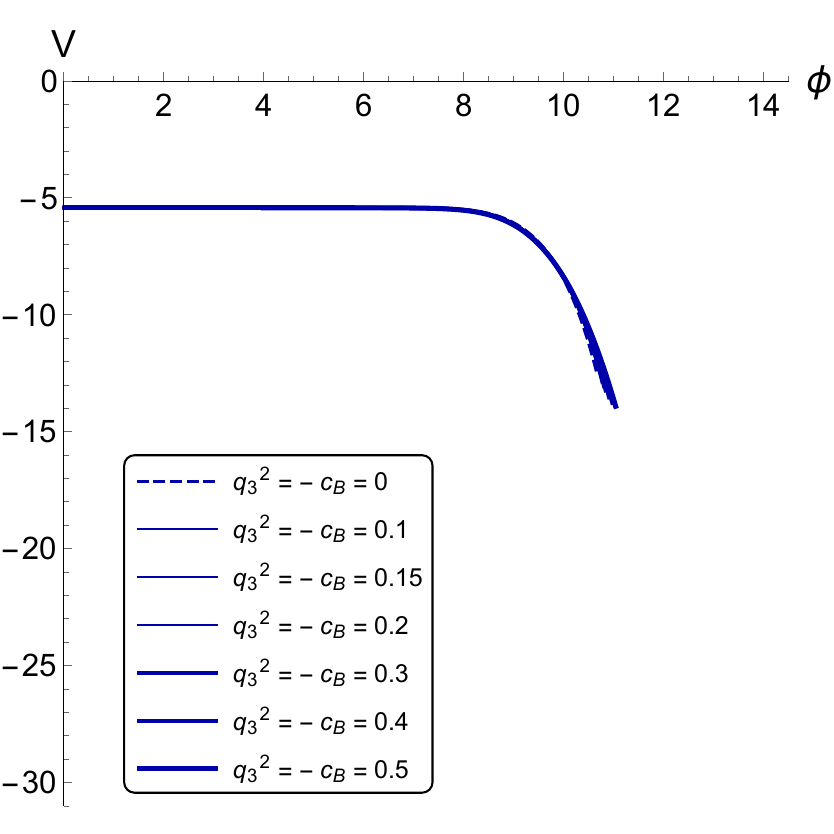} \
  \includegraphics[scale=0.24]{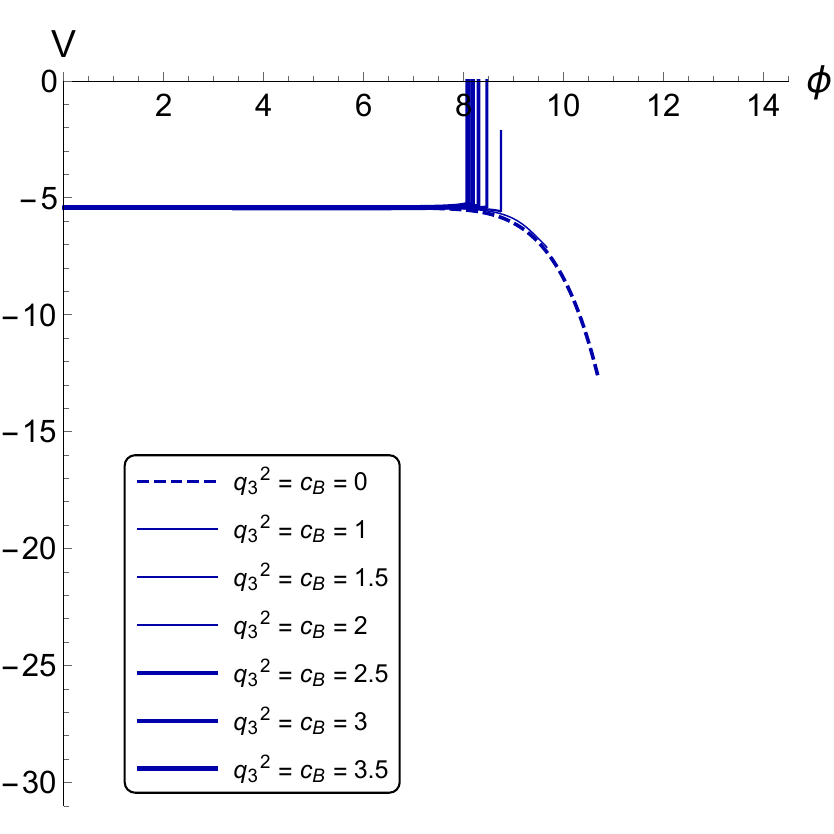} \\
  G \hspace{200pt} H
  \caption{Scalar potential $V(\phi)$ in magnetic field with different
    $q_3$ (A,E) for $c_B = - \, 0.5$ (left) and $c_B = 0.5$ (right);
    with different $c_B$ for $q_3 = 0.5$ (B,F) for $c_B < 0$ (left)
    and $c_B > 0$ (right); for different $q_3 = \pm \, c_B$ (C,G); for
    different $q_3^2 = \pm \, c_B$ (D,H) for $d = 0.06 > 0.05$ (A-D)
    and $d = 0.01 < 0.05$ (E-H) in primary anisotropic case $\nu =
    4.5$, $a = 0.15$, $c = 1.16$, $\mu = 0$.}
  \label{Fig:Vphi-q3cB-nu45-mu0-z5}
\end{figure}

\begin{figure}[t!]
  \centering
  \includegraphics[scale=0.24]{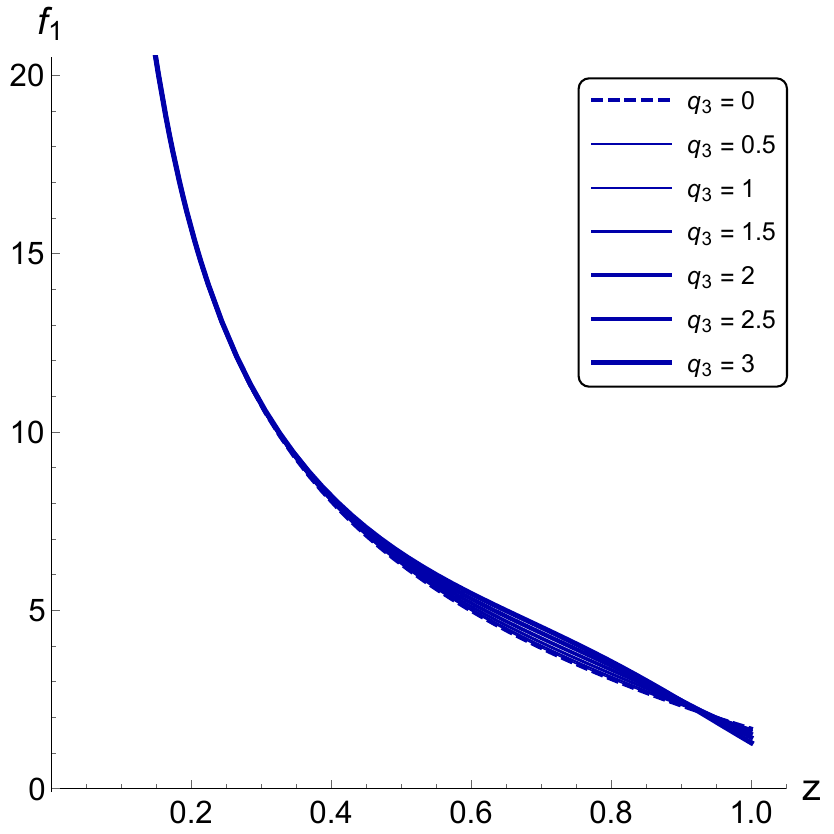} \
  \includegraphics[scale=0.24]{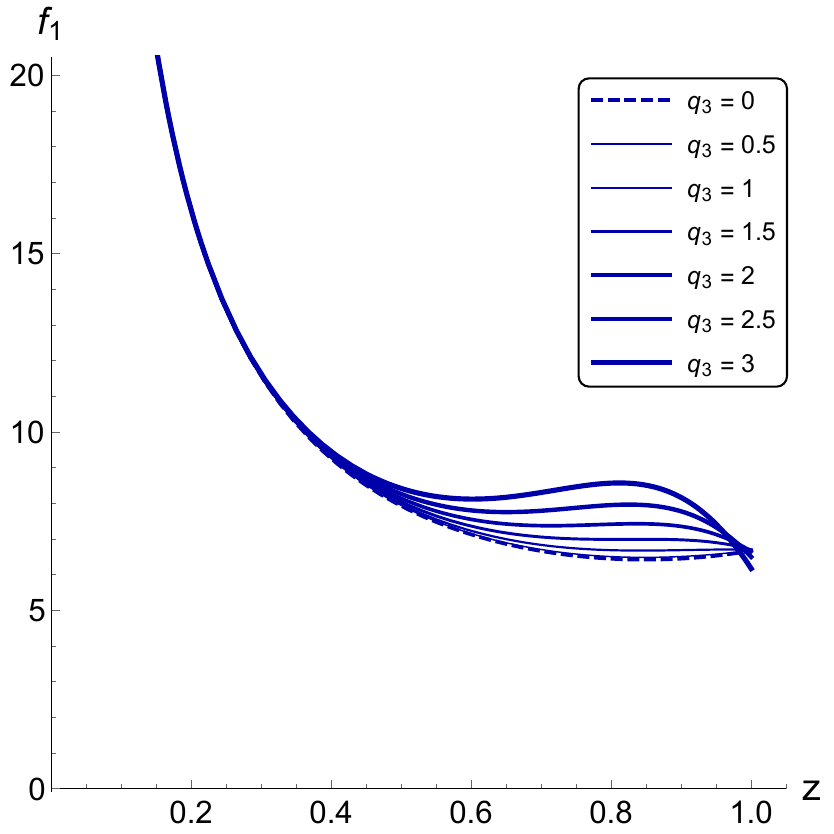} \quad
  \includegraphics[scale=0.24]{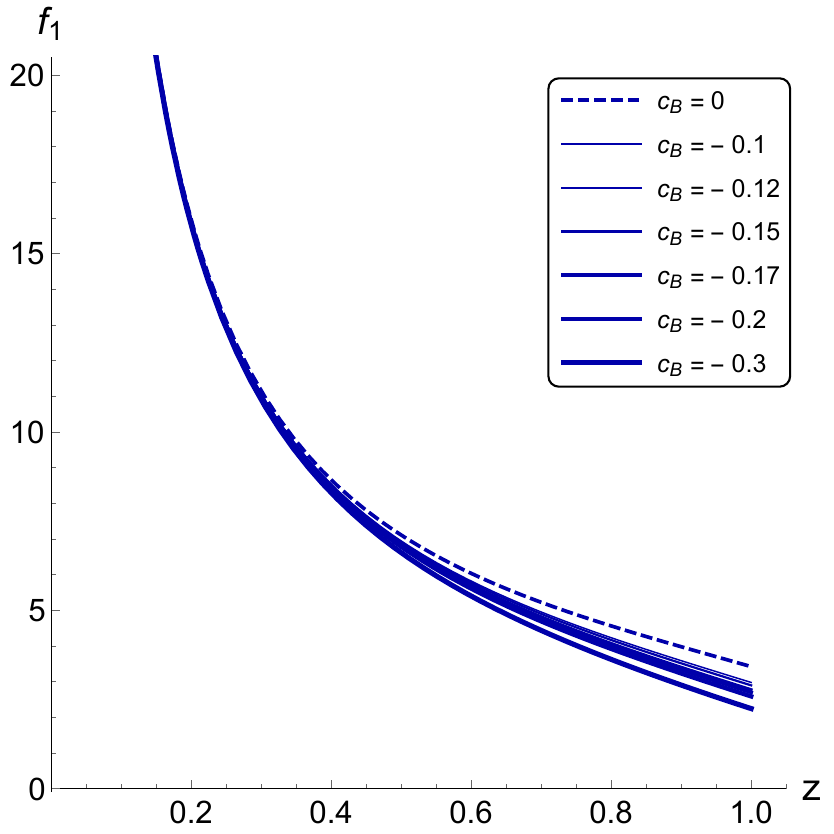} \
  \includegraphics[scale=0.24]{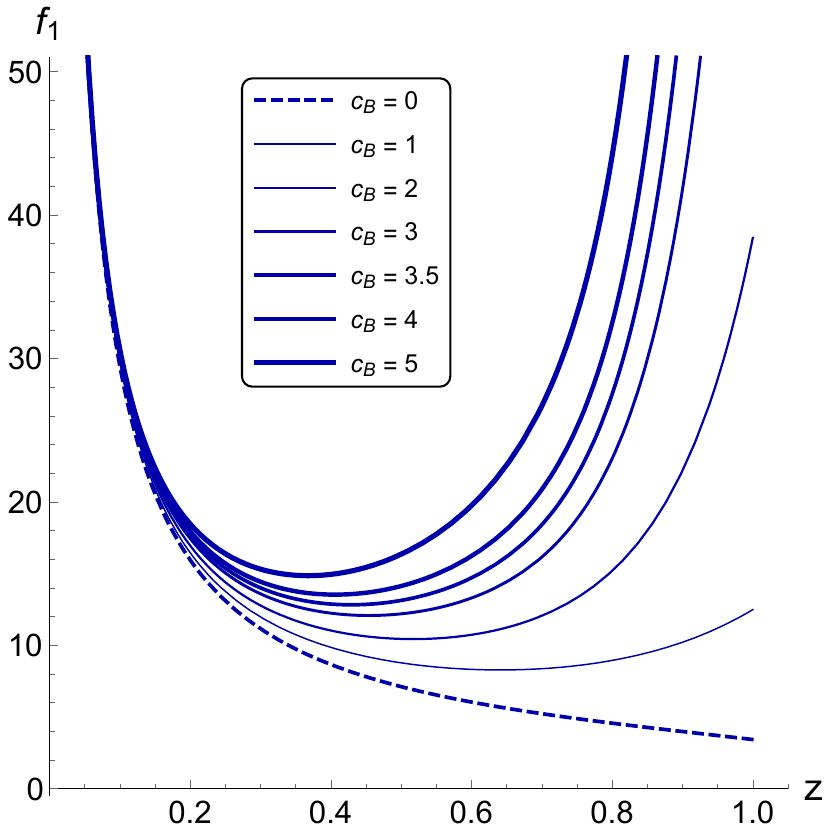} \\
  A \hspace{200pt} B \ \\
  \includegraphics[scale=0.24]{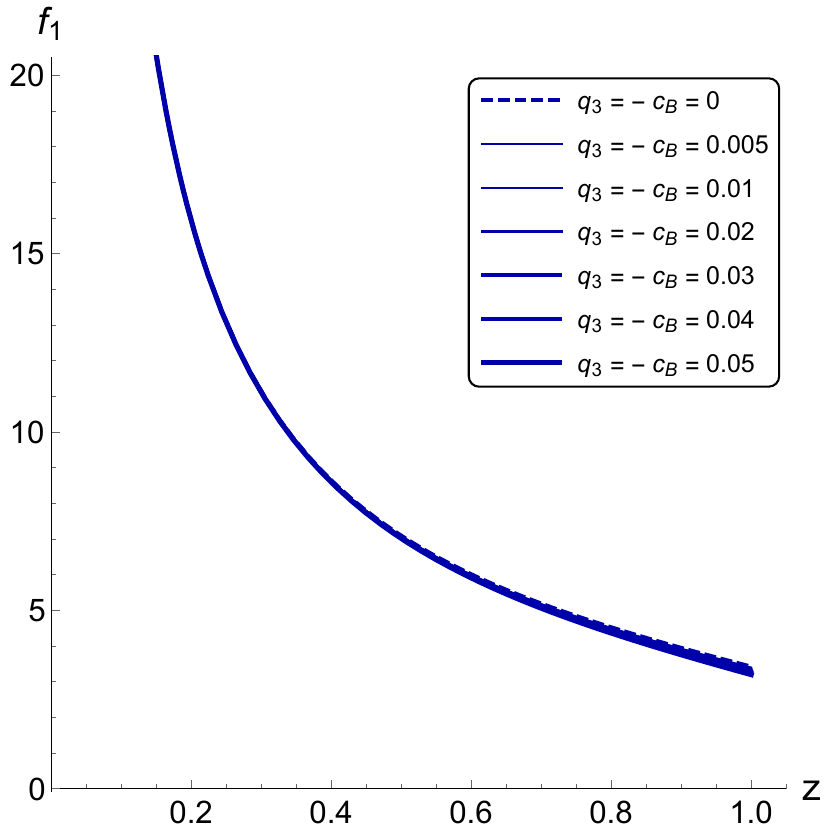} \
  \includegraphics[scale=0.24]{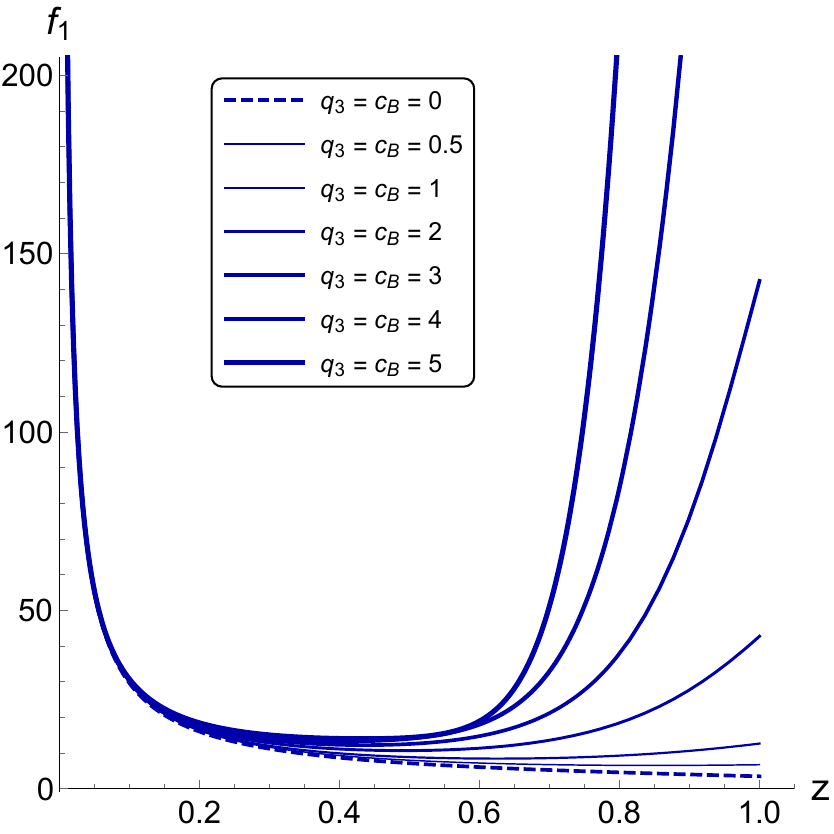} \quad
  \includegraphics[scale=0.24]{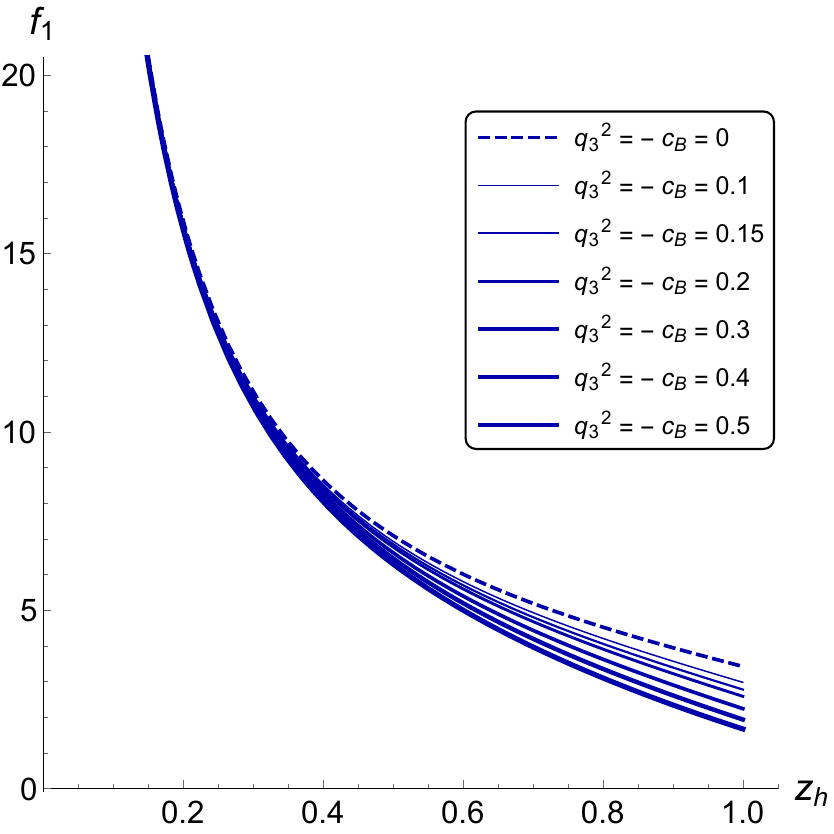} \
  \includegraphics[scale=0.24]{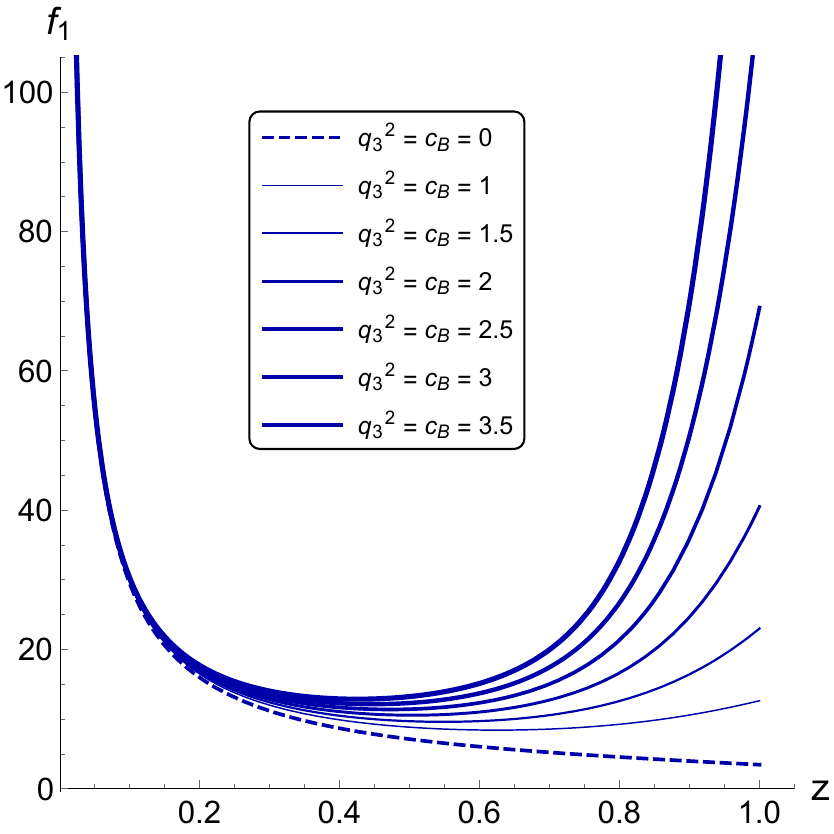} \\
  C \hspace{200pt} D \ \\
  \includegraphics[scale=0.24]{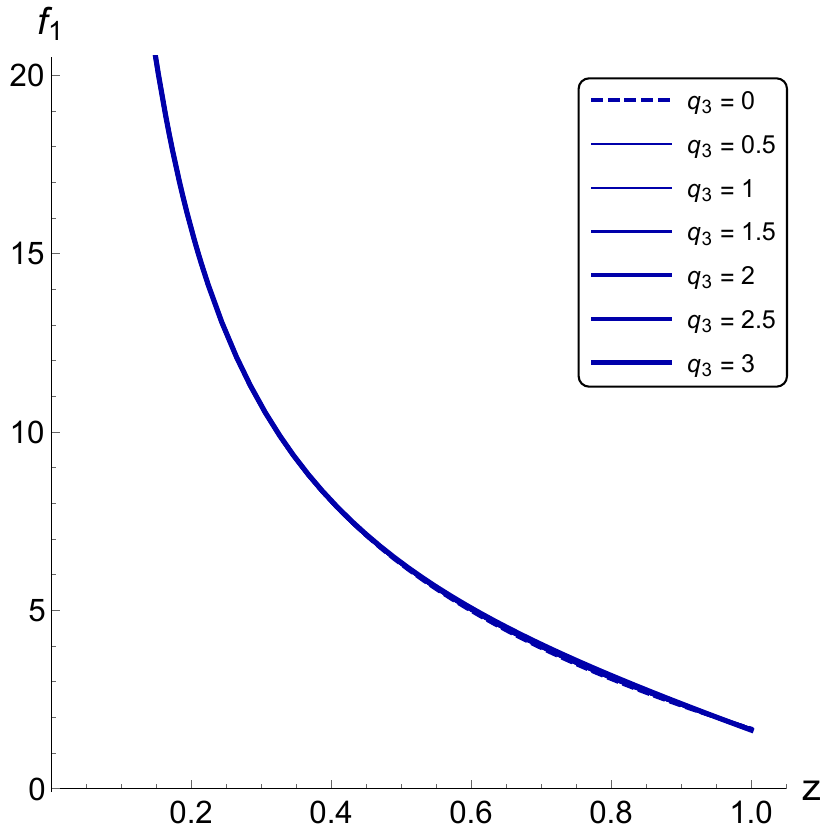} \
  \includegraphics[scale=0.24]{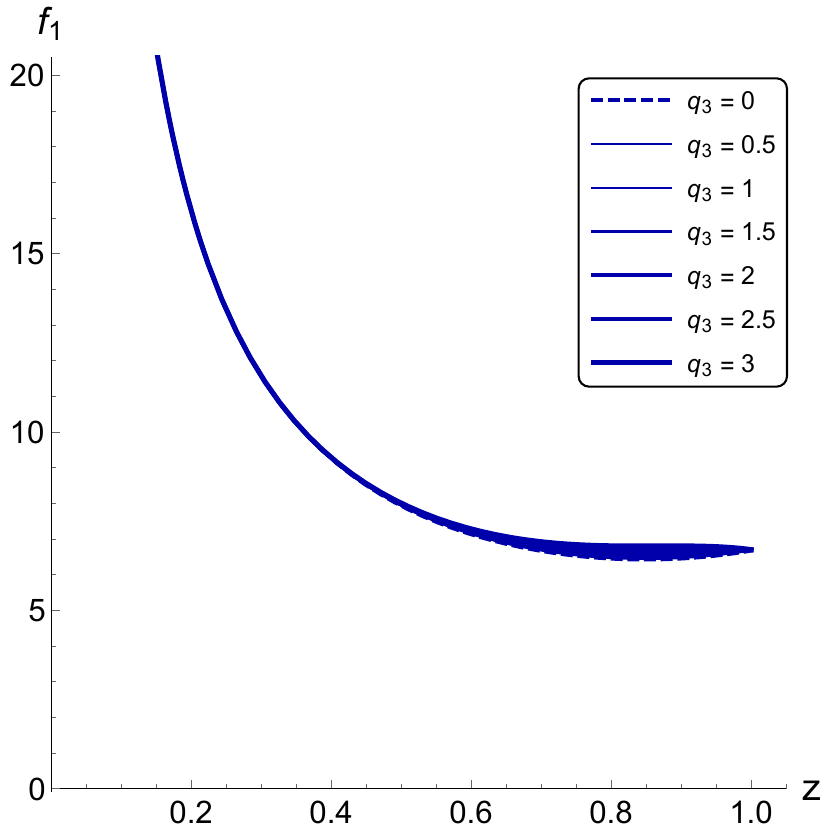} \quad
  \includegraphics[scale=0.24]{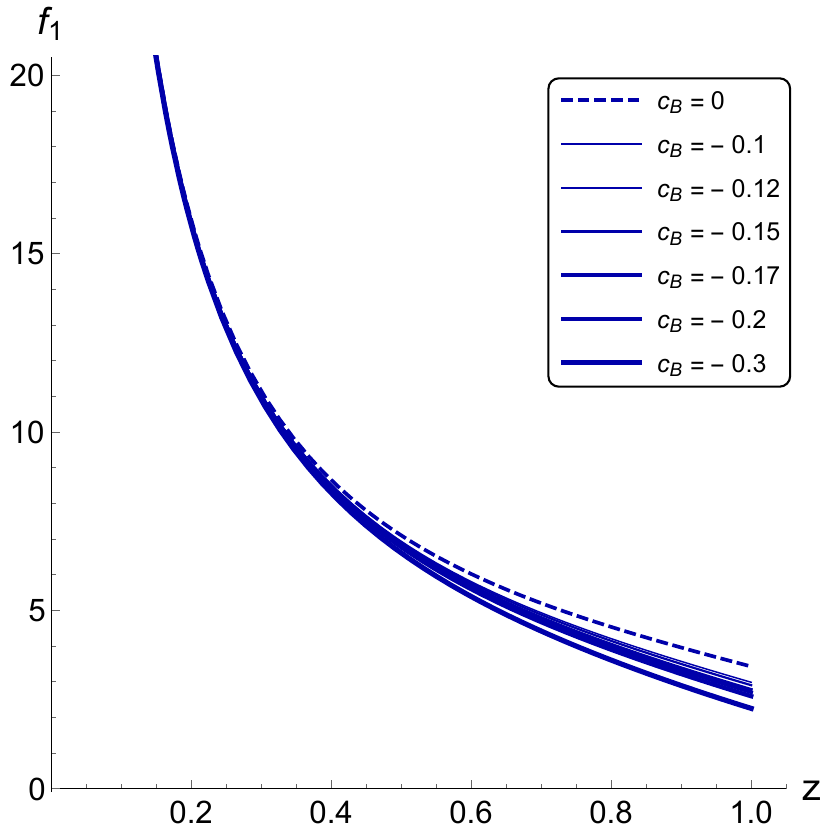} \
  \includegraphics[scale=0.24]{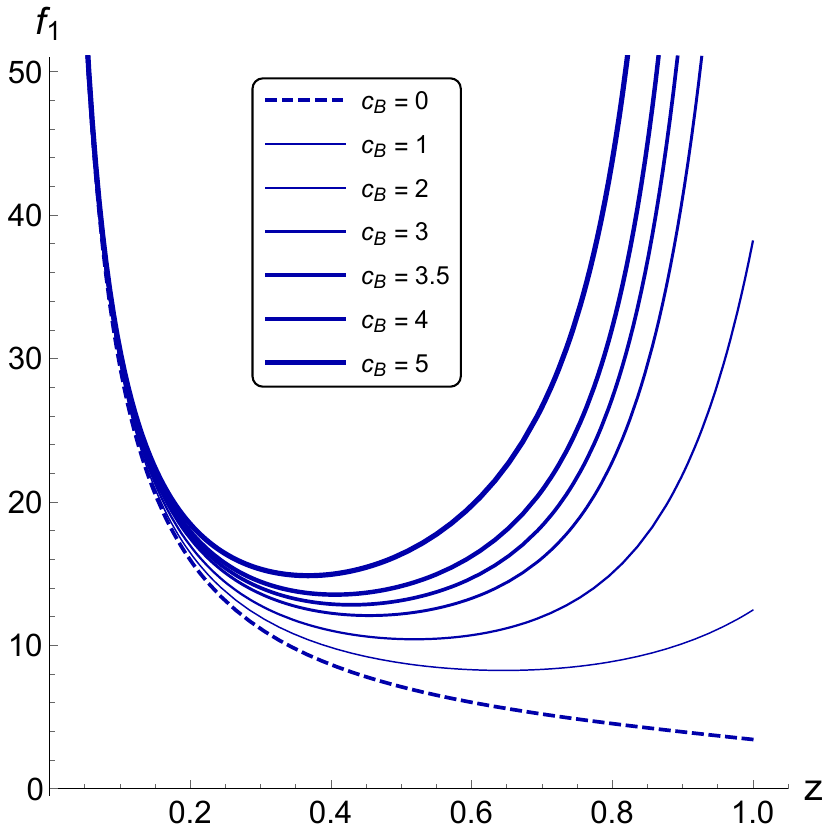} \\
  E \hspace{200pt} F \ \\
  \includegraphics[scale=0.24]{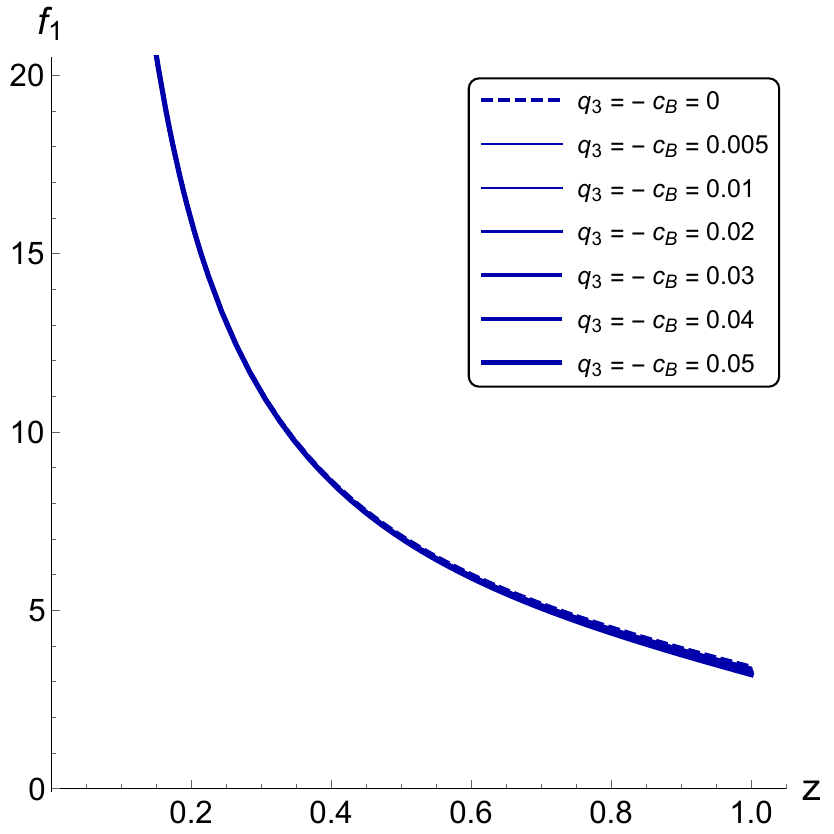} \
  \includegraphics[scale=0.24]{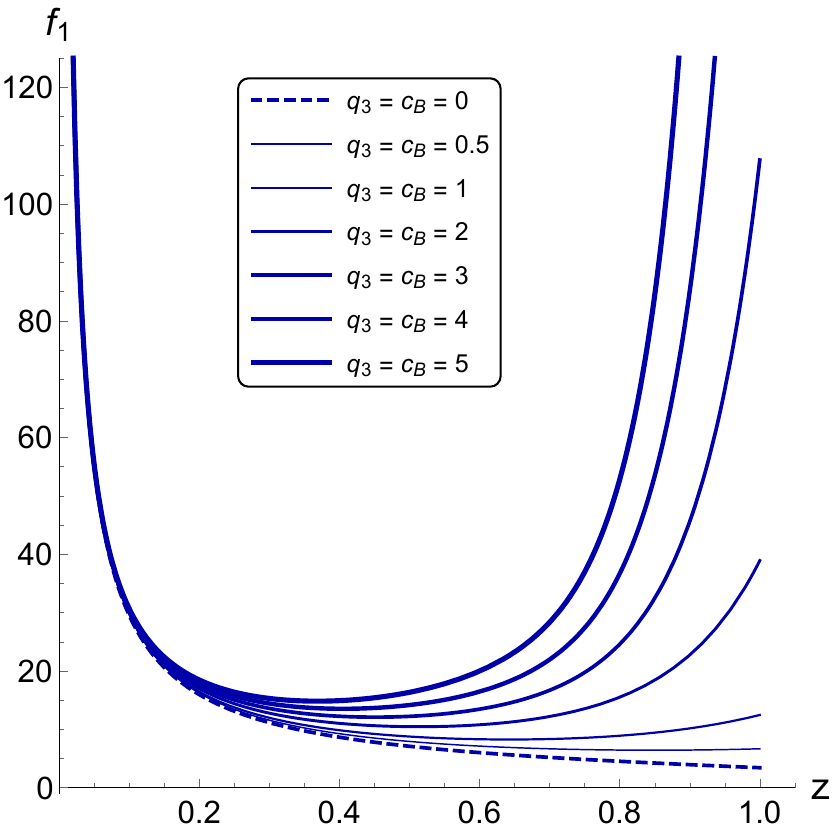} \quad
  \includegraphics[scale=0.24]{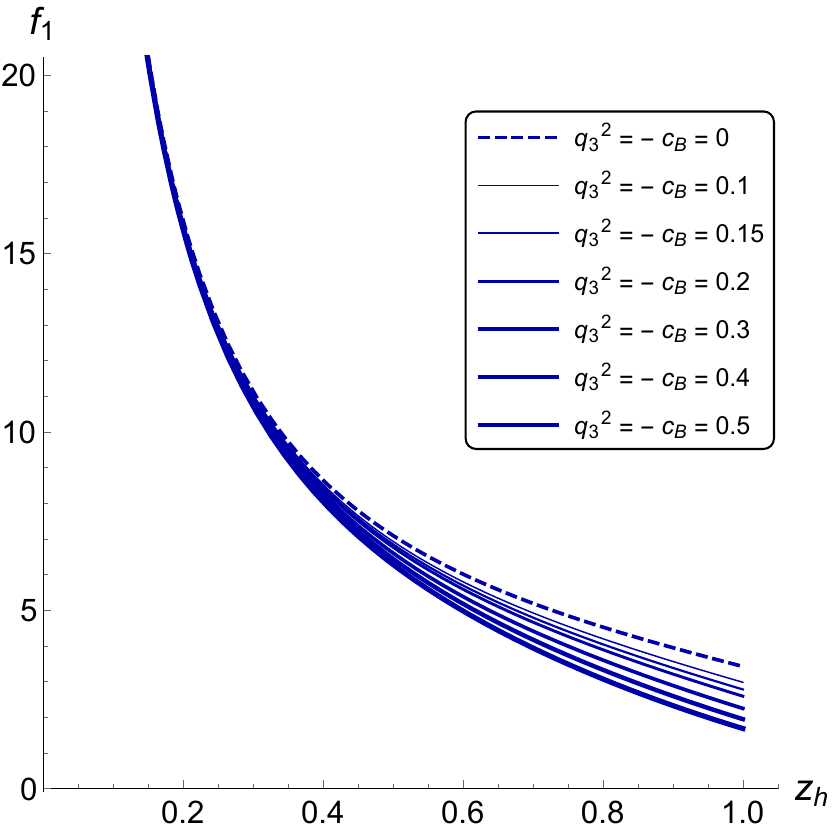} \
  \includegraphics[scale=0.24]{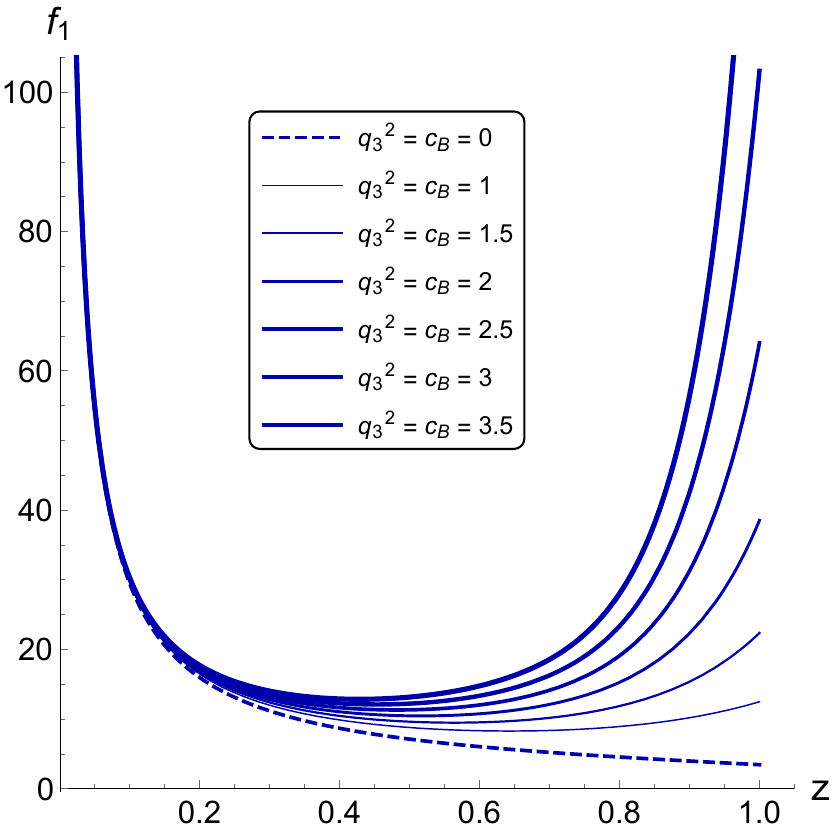} \\
  G \hspace{200pt} H
  \caption{Coupling function $f_1(z)$ in magnetic field with different
    $q_3$ (A,E) for $c_B = - \, 0.5$ (left) and $c_B = 0.5$ (right);
    with different $c_B$ for $q_3 = 0.5$ (B,F) for $c_B < 0$ (left)
    and $c_B > 0$ (right); for different $q_3 = \pm \, c_B$ (C,G); for
    different $q_3^2 = \pm \, c_B$ (D,H) for $d = 0.06 > 0.05$ (A-D)
    and $d = 0.01 < 0.05$ (E-H) in primary anisotropic case $\nu =
    4.5$, $a = 0.15$, $c = 1.16$, $\mu = 0$.}
  \label{Fig:f1z-q3cB-nu45-mu0-z5}
\end{figure}


\newpage

\

\newpage

\

\newpage

\

\newpage

\

\newpage

\

\newpage

\

\newpage

\

\newpage

\

\newpage

\section{Temperature and free energy. Plots}\label{appendixB}

\begin{figure}[h!]
  \centering 
  \includegraphics[scale=0.27]{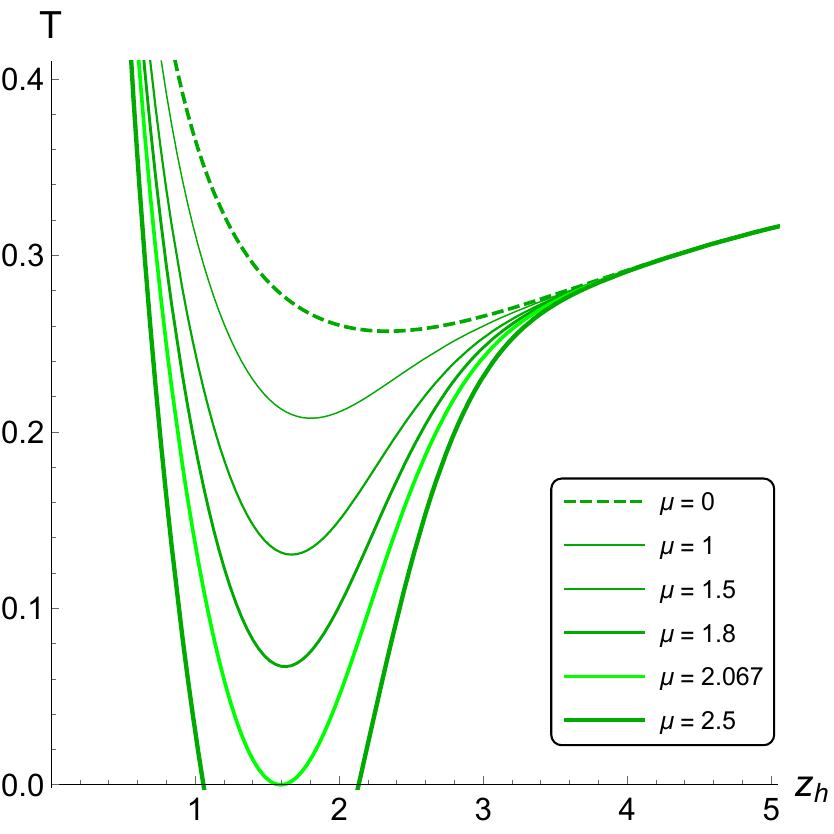} \quad
  \includegraphics[scale=0.27]{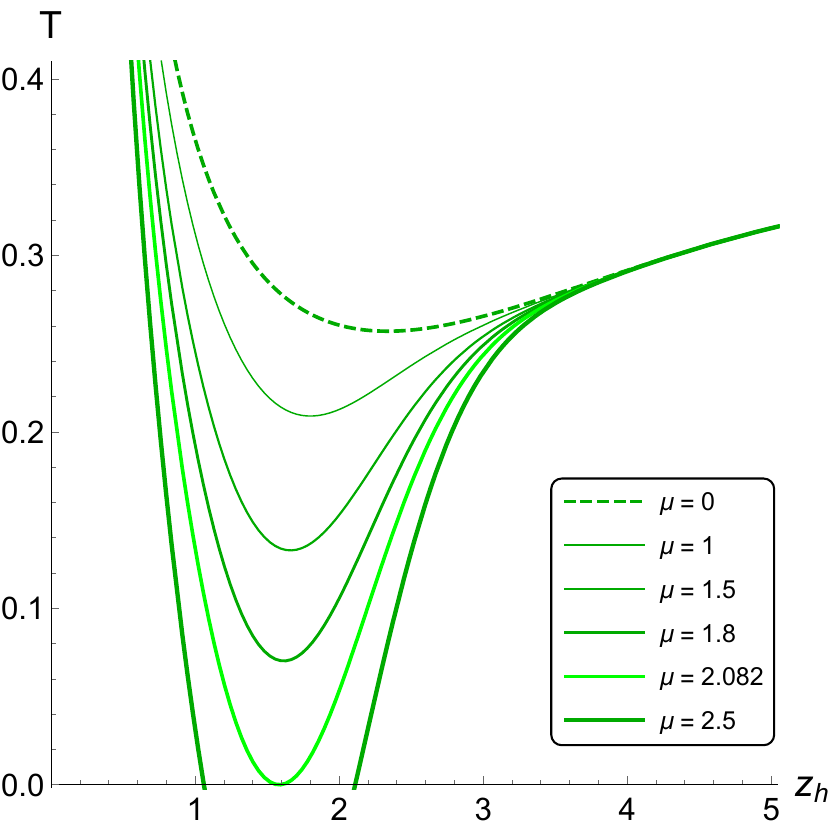} \quad
  \includegraphics[scale=0.27]{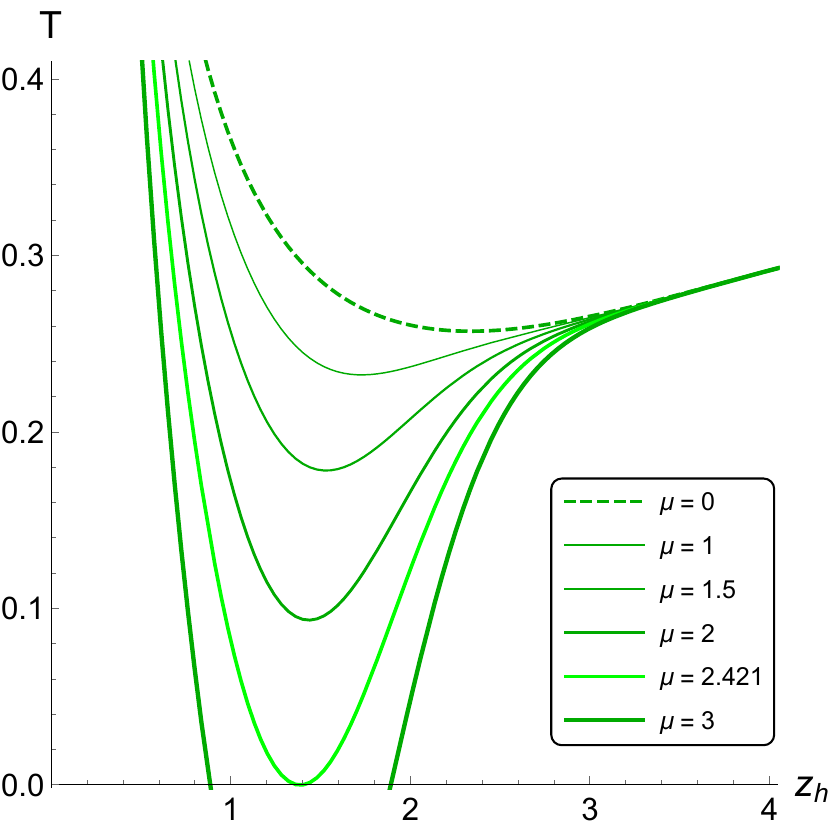} \\
  A \hspace{100pt} B \hspace{100pt} C \\
  \includegraphics[scale=0.27]{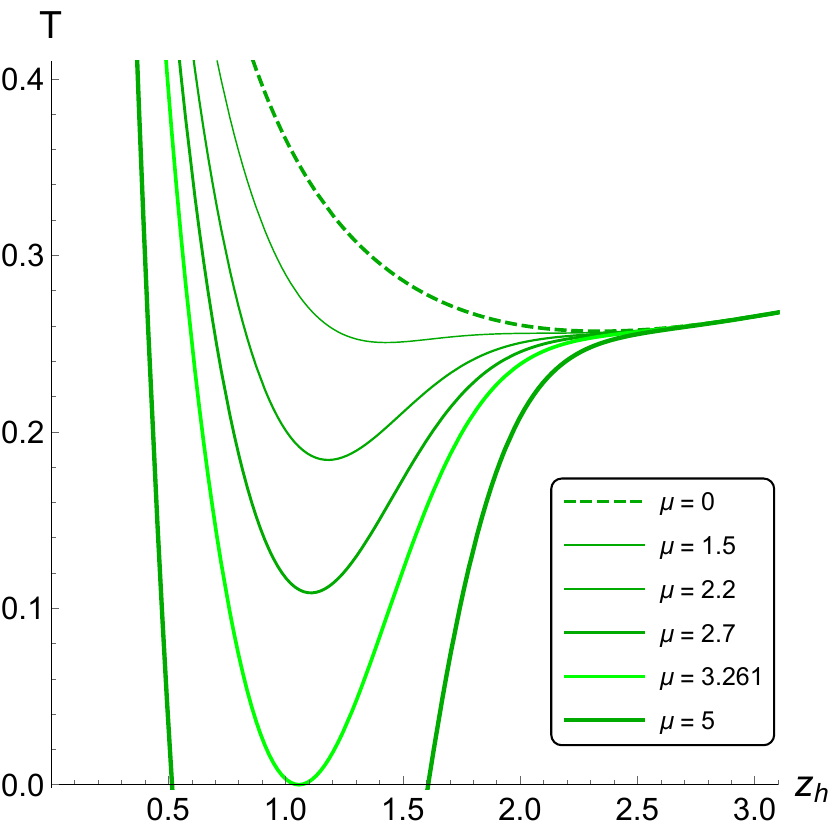} \quad
  \includegraphics[scale=0.27]{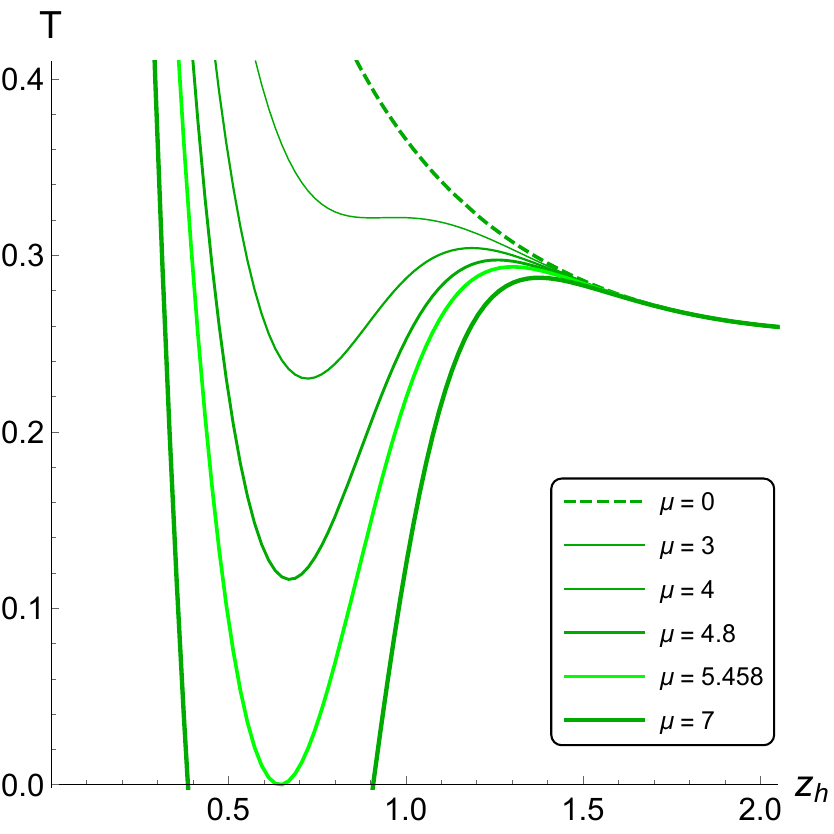} \quad
  \includegraphics[scale=0.27]{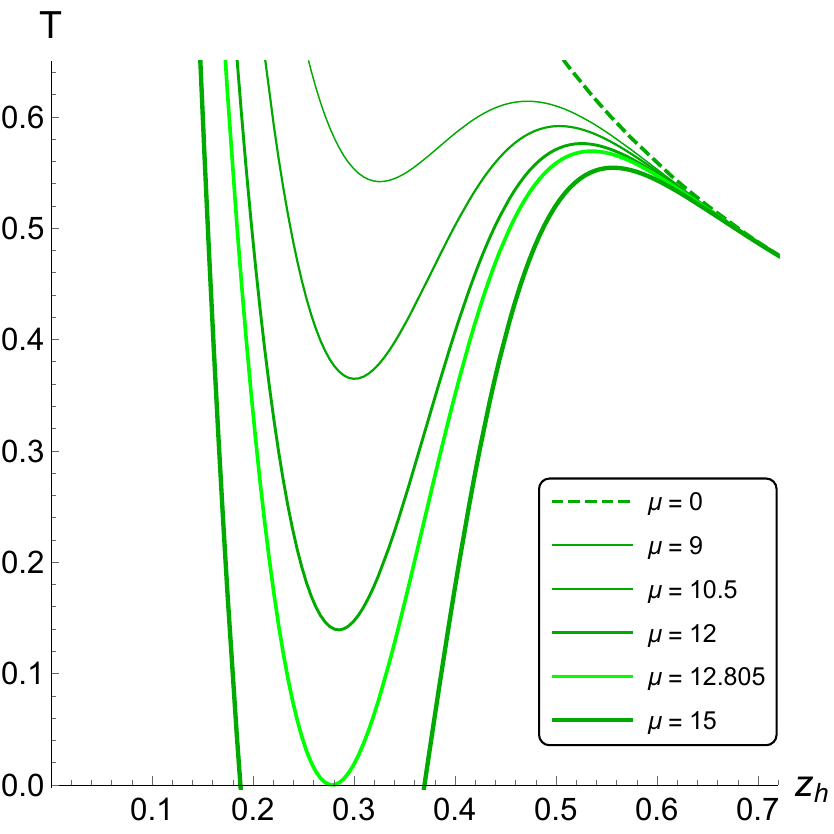} \\
  D \hspace{100pt} E \hspace{100pt} F
  \caption{Temperature $T(z_h,\mu)$ in magnetic field $q_3 = 0$ (A),
    $q_3 = 0.1$ (B), $q_3 = 0.5$ (C), $q_3 = 1$ (D), $q_3 = 2$ (E),
    $q_3 = 5$ (F); $\nu = 1$, $a = 0.15$, $c = 1.16$, $c_B = - \,
    0.01$, $d = 0$.}
  \label{Fig:Tzhmu-q3-nu1-d0-z5}
  \ \\
  \includegraphics[scale=0.27]{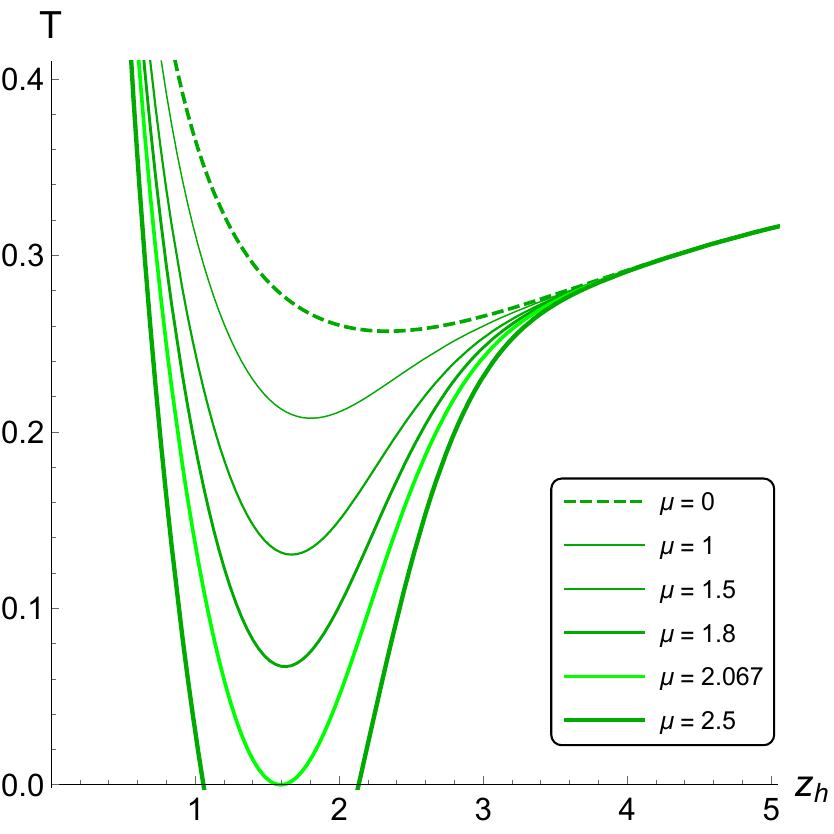} \quad
  \includegraphics[scale=0.27]{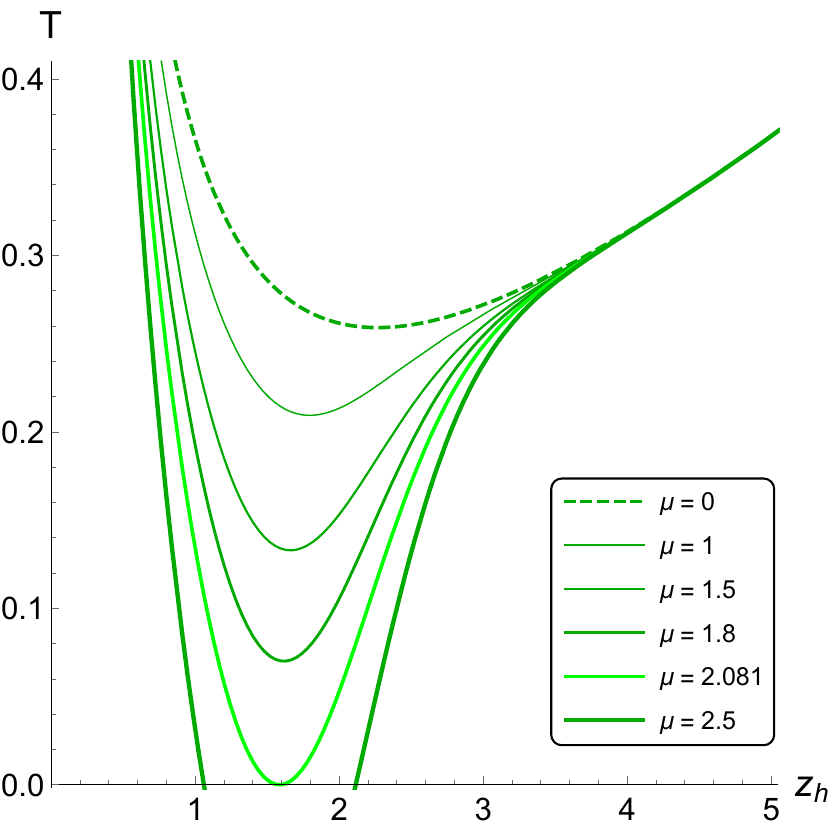} \quad
  \includegraphics[scale=0.27]{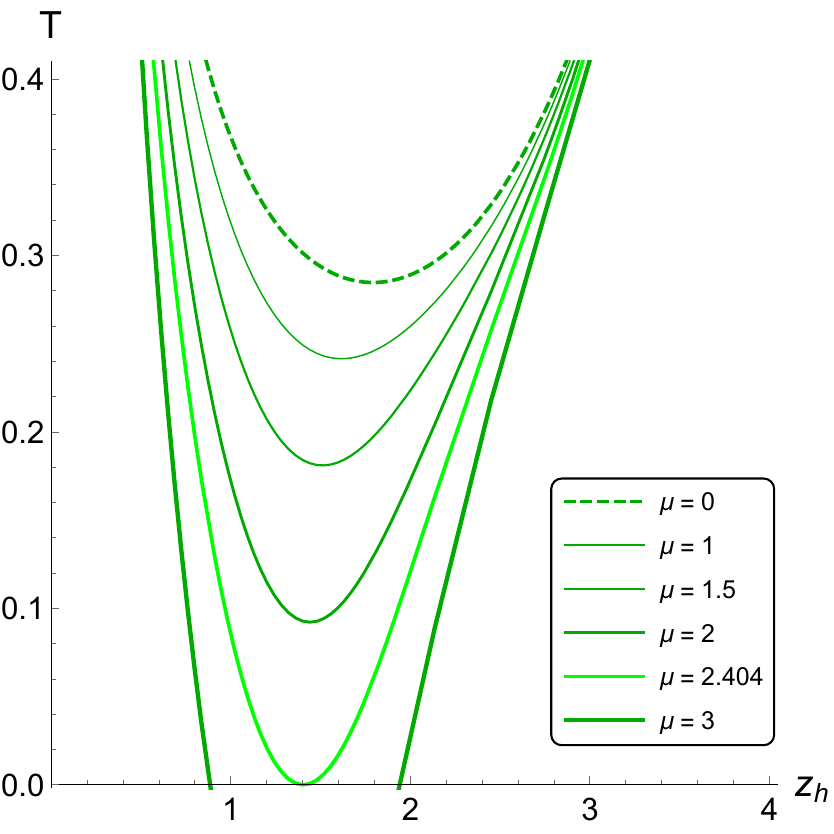} \\
  A \hspace{100pt} B \hspace{100pt} C \\
  \includegraphics[scale=0.27]{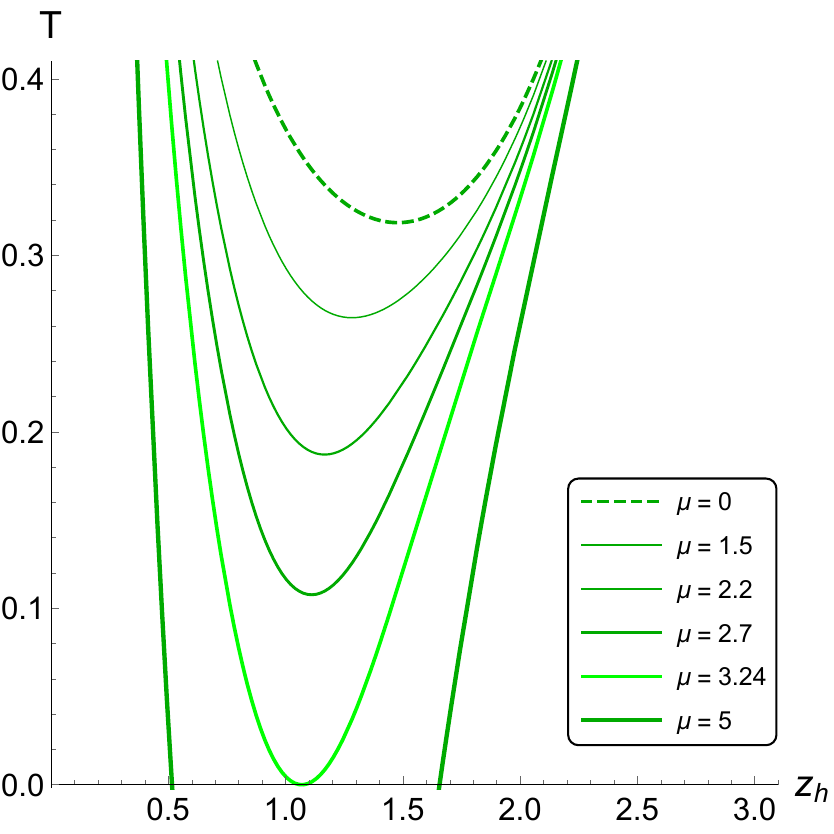} \quad
  \includegraphics[scale=0.27]{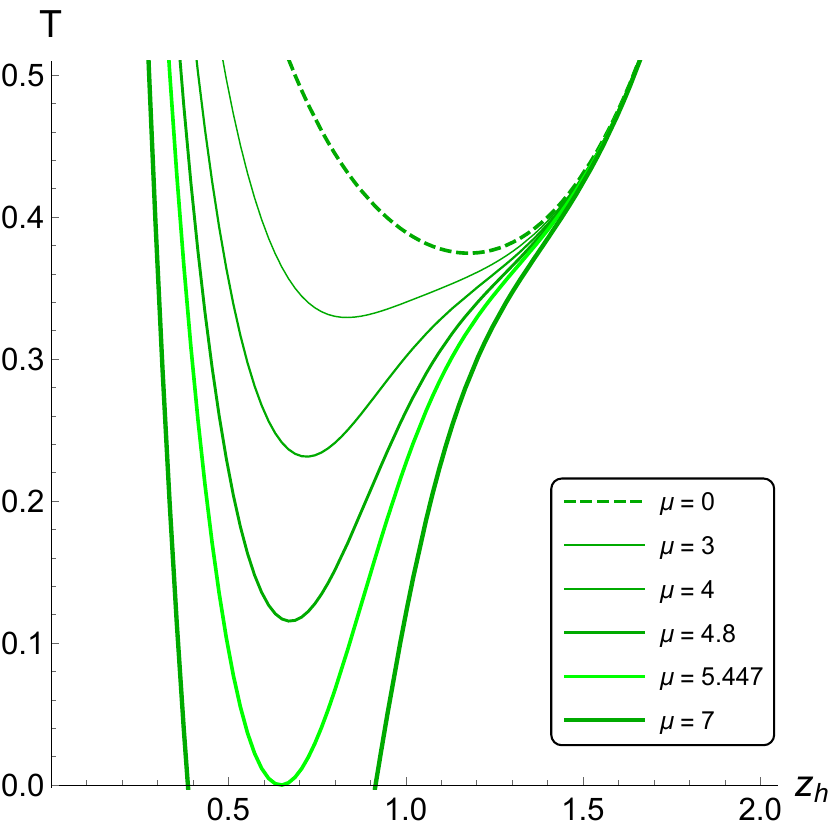} \quad
  \includegraphics[scale=0.27]{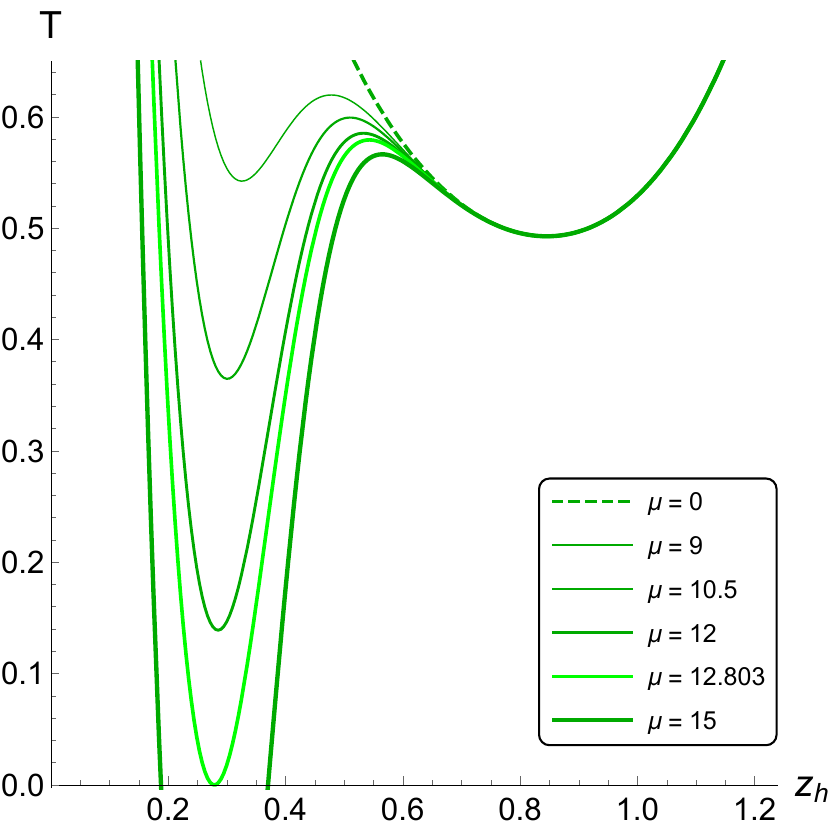} \\
  D \hspace{100pt} E \hspace{100pt} F
  \caption{Temperature $T(z_h,\mu)$ in magnetic field $q_3 = 0$ (A),
    $q_3 = 0.1$ (B), $q_3 = 0.5$ (C), $q_3 = 1$ (D), $q_3 = 2$ (E),
    $q_3 = 5$ (F); $\nu = 1$, $a = 0.15$, $c = 1.16$, $c_B = - \,
    0.01$, $d = 0.01$.}
  \label{Fig:Tzhmu-q3-nu1-d001-z5}
\end{figure}

\begin{figure}[t!]
  \centering 
  \includegraphics[scale=0.28]{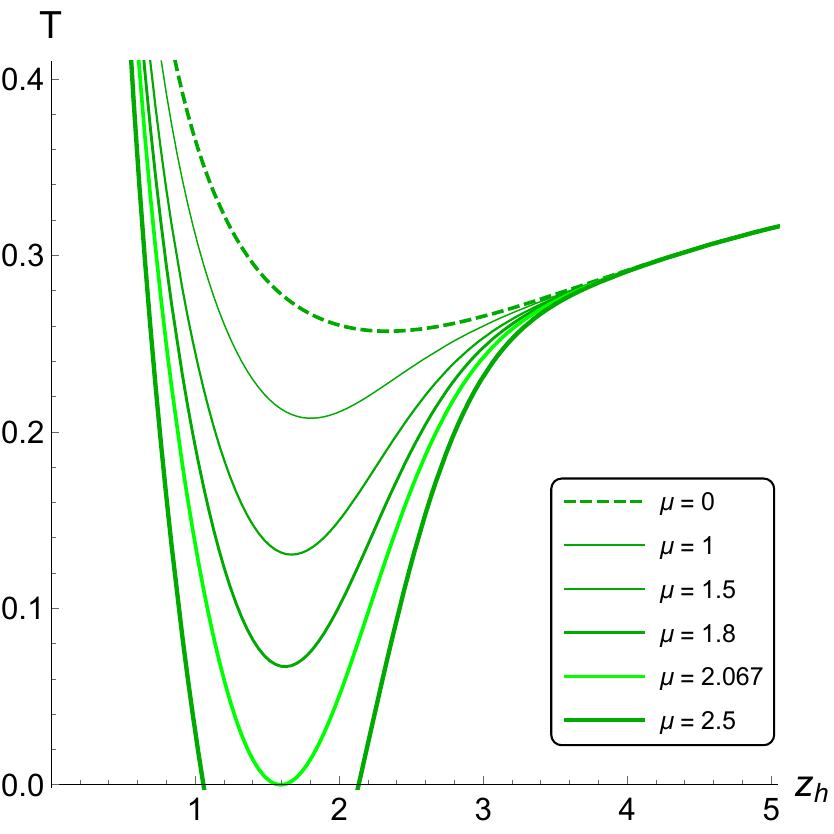} \quad
  \includegraphics[scale=0.28]{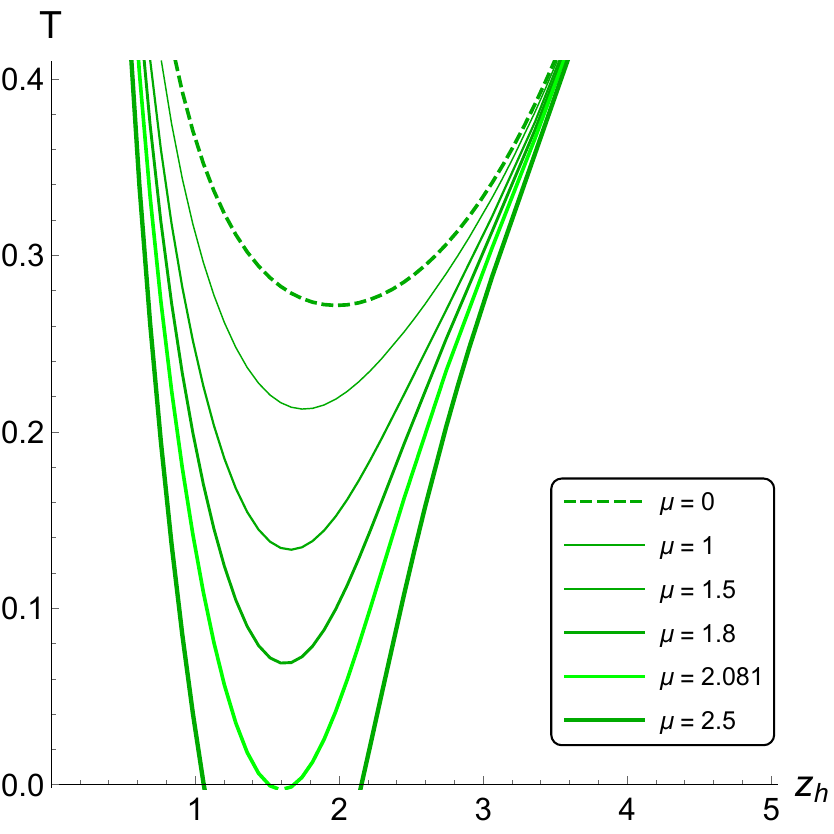} \quad
  \includegraphics[scale=0.28]{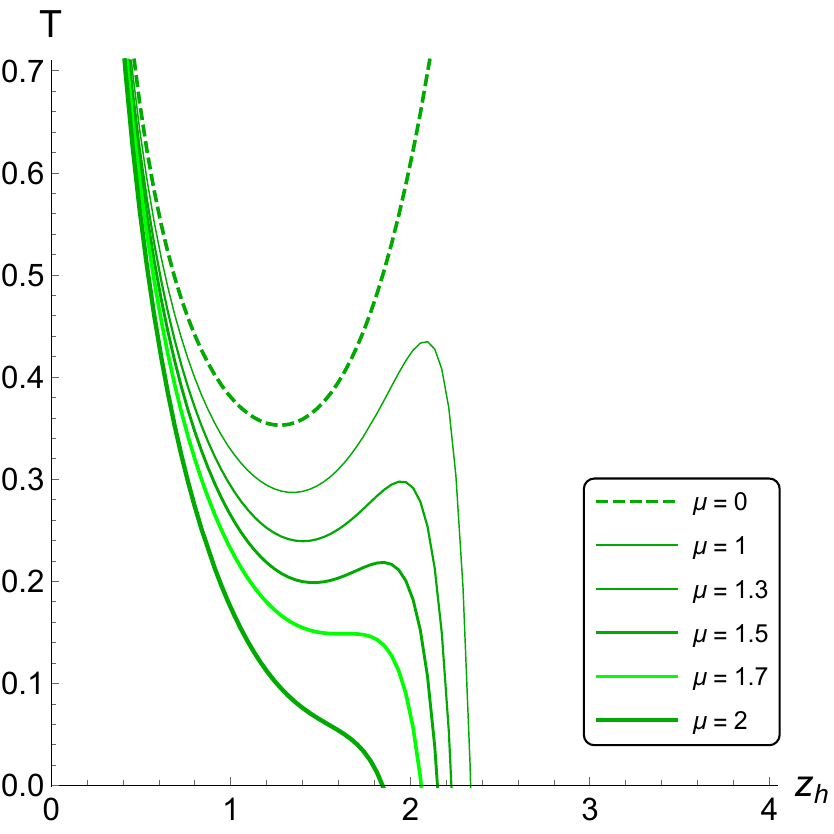} \\
  A \hspace{100pt} B \hspace{100pt} C \\
  \includegraphics[scale=0.28]{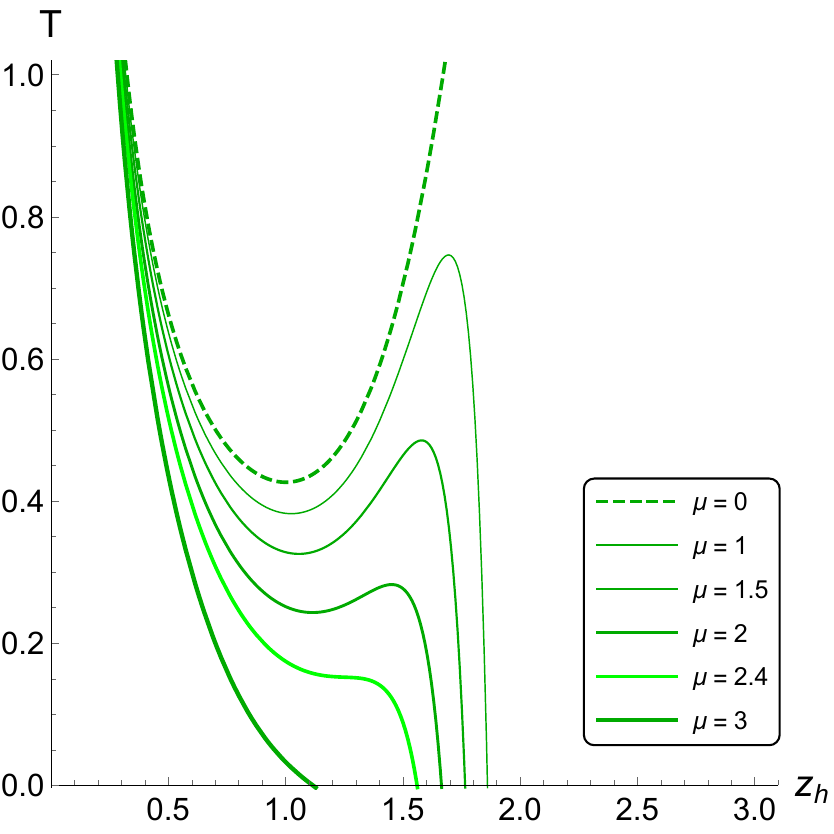} \quad
  \includegraphics[scale=0.28]{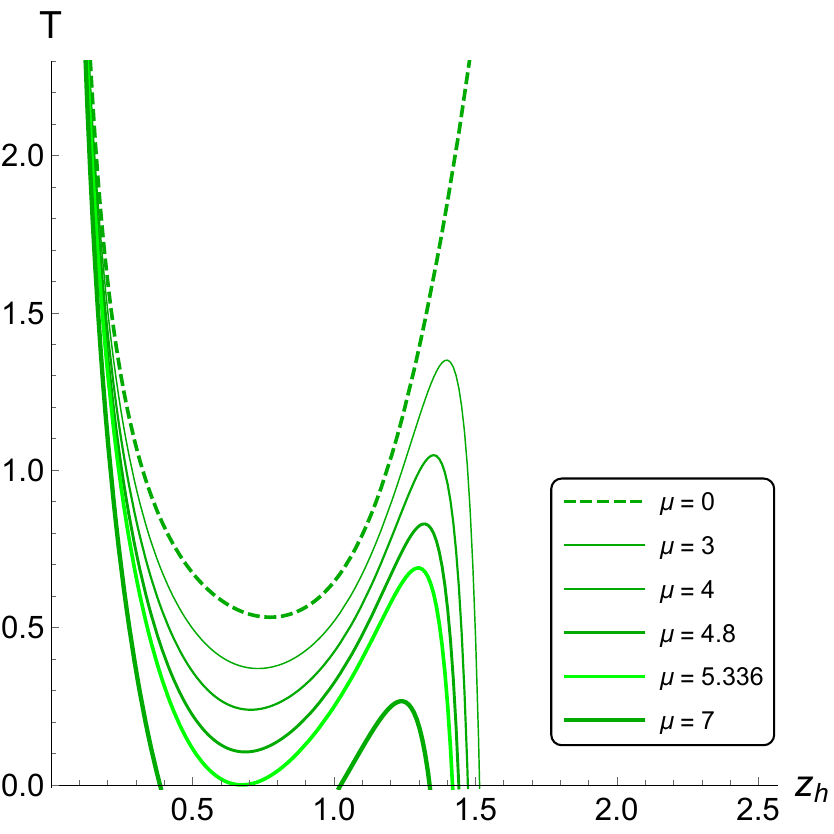} \quad
  \includegraphics[scale=0.28]{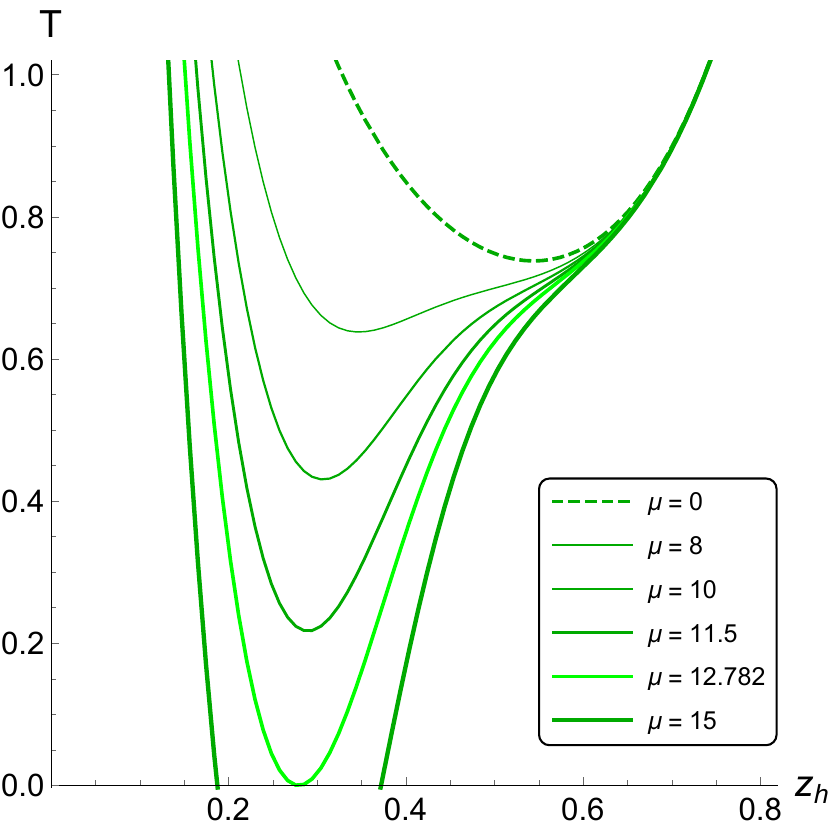} \\
  D \hspace{100pt} E \hspace{100pt} F
  \caption{Temperature $T(z_h,\mu)$ in magnetic field $q_3 = 0$ (A),
    $q_3 = 0.1$ (B), $q_3 = 0.5$ (C), $q_3 = 1$ (D), $q_3 = 2$ (E),
    $q_3 = 5$ (F); $\nu = 1$, $a = 0.15$, $c = 1.16$, $c_B = - \,
    0.01$, $d = 0.1$.}
  \label{Fig:Tzhmu-q3-nu1-d01-z5}
\end{figure}
\begin{figure}[h!]
  \centering 
  \includegraphics[scale=0.28]{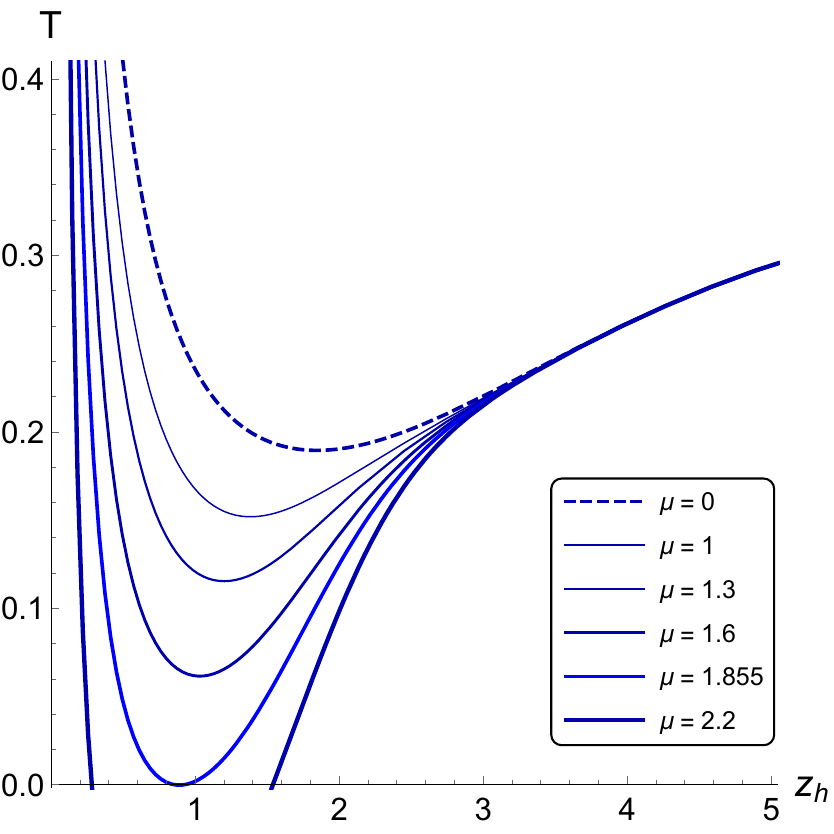} \quad
  \includegraphics[scale=0.28]{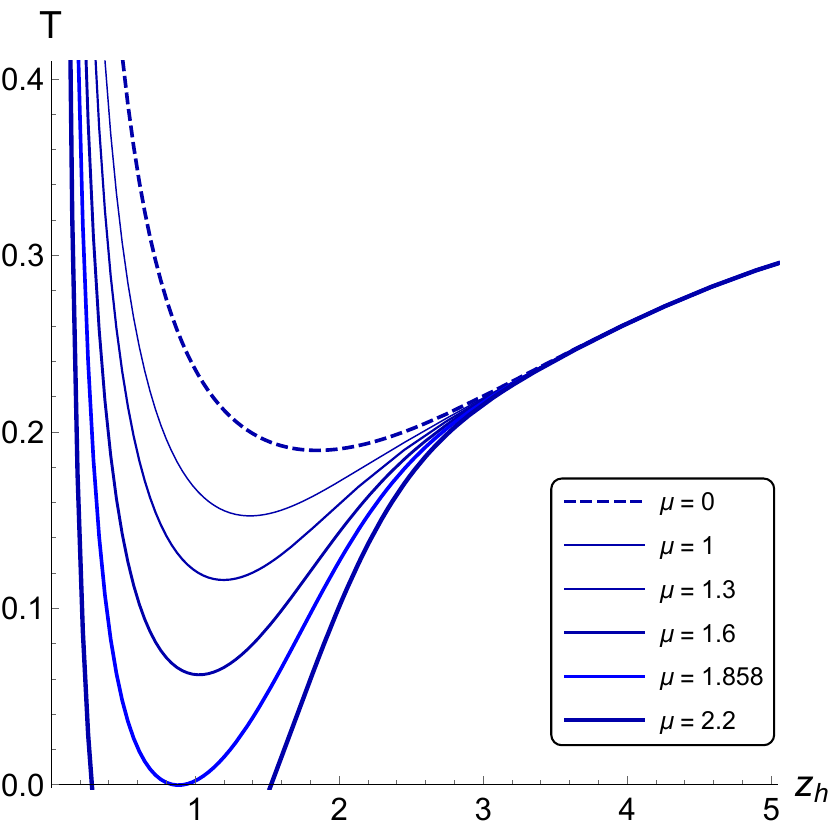} \quad
  \includegraphics[scale=0.28]{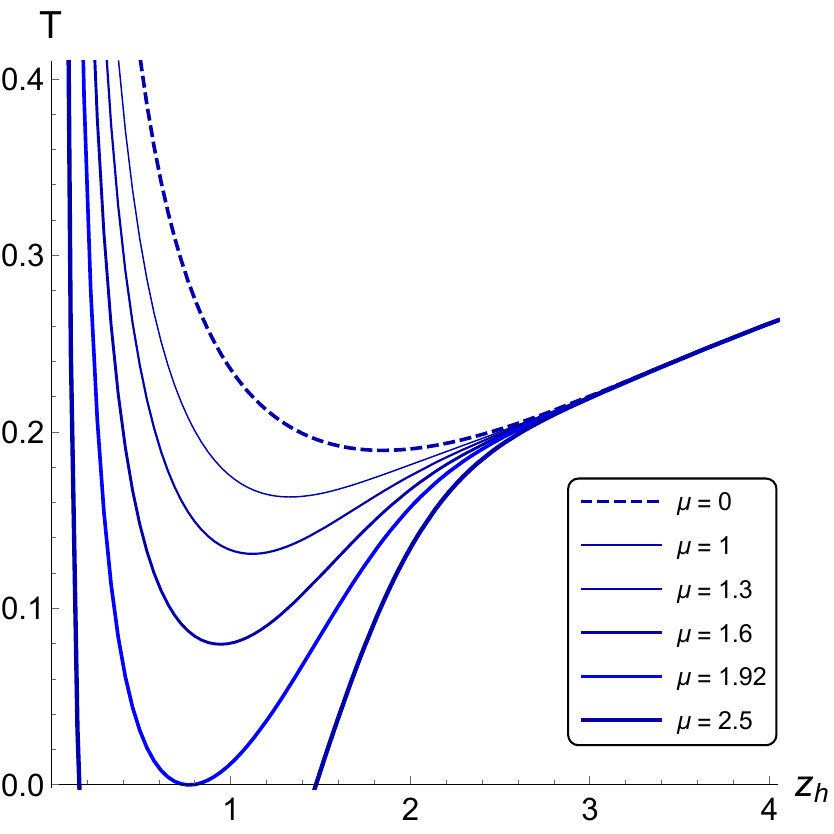} \\
  A \hspace{100pt} B \hspace{100pt} C \\
  \includegraphics[scale=0.28]{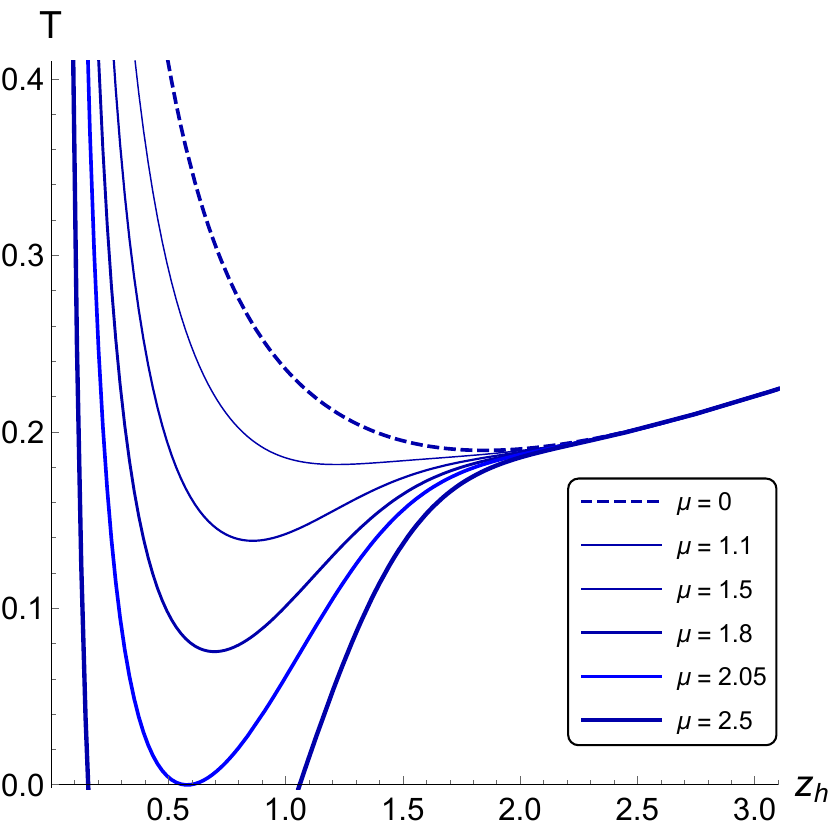} \quad
  \includegraphics[scale=0.28]{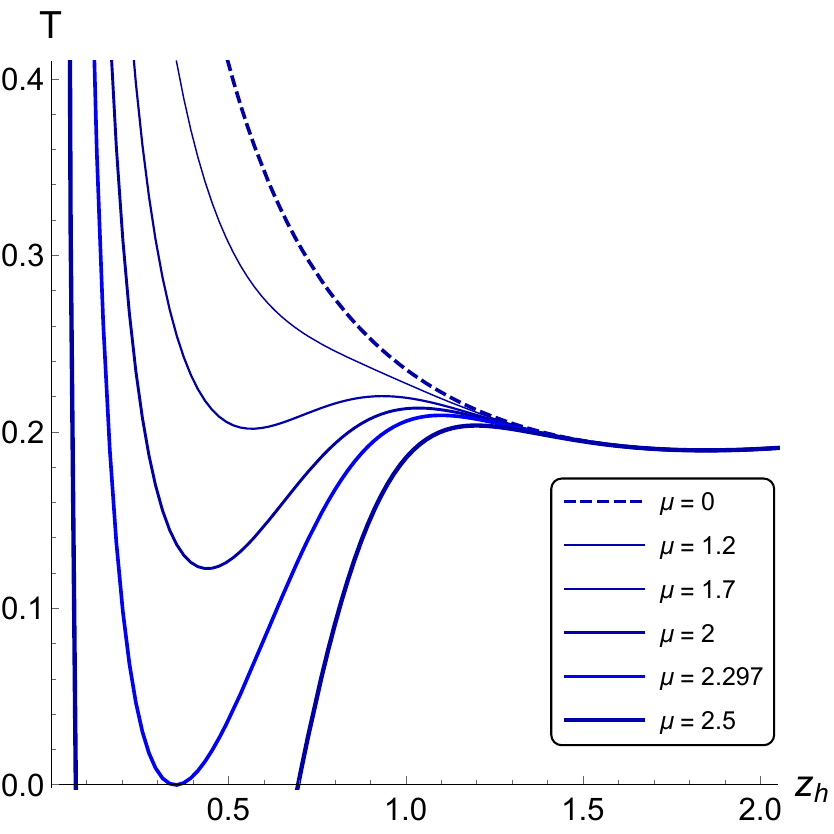} \quad
  \includegraphics[scale=0.28]{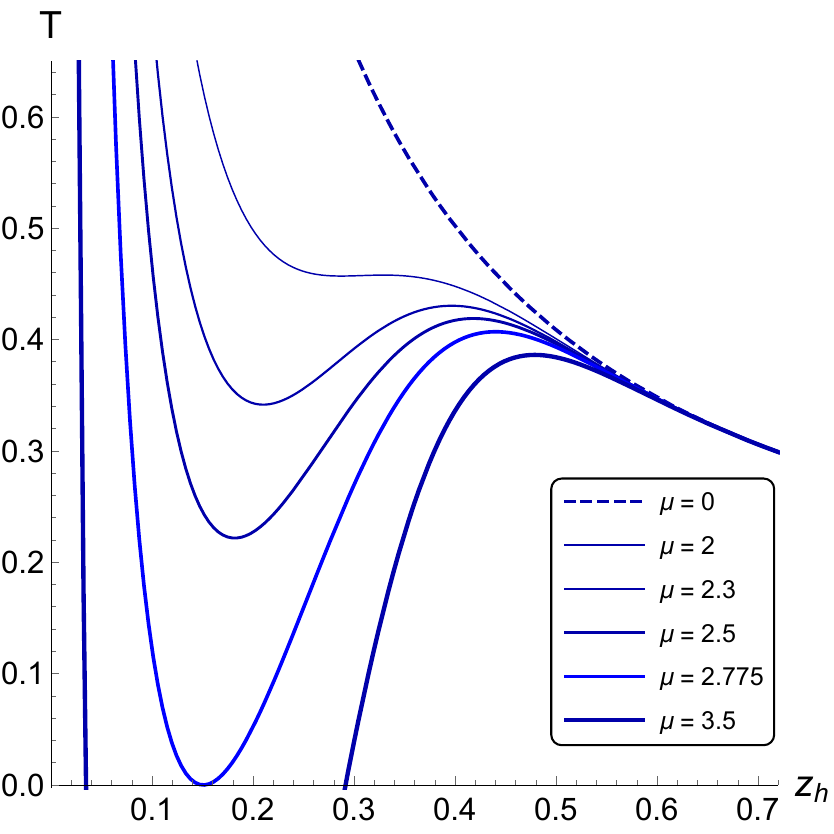} \\
  D \hspace{100pt} E \hspace{100pt} F
  \caption{Temperature $T(z_h,\mu)$ in magnetic field $q_3 = 0$ (A),
    $q_3 = 0.1$ (B), $q_3 = 0.5$ (C), $q_3 = 1$ (D), $q_3 = 2$ (E),
    $q_3 = 5$ (F); $\nu = 4.5$, $a = 0.15$, $c = 1.16$, $c_B = - \,
    0.01$, $d = 0$.}
  \label{Fig:Tzhmu-q3-nu45-d0-z5}
\end{figure}

\begin{figure}[t!]
  \centering 
  \includegraphics[scale=0.28]{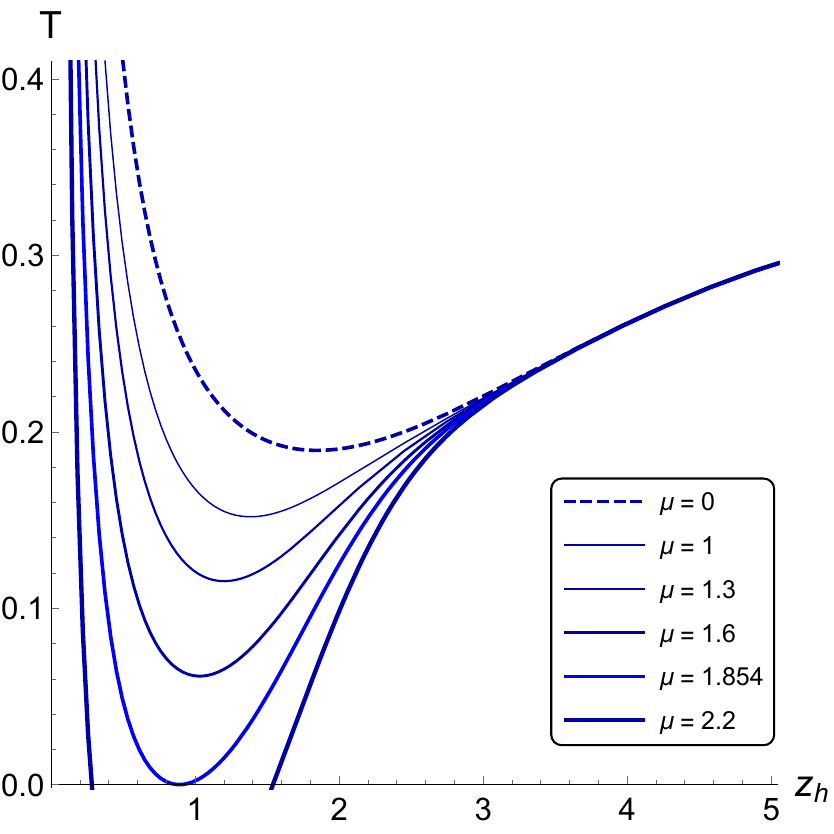} \quad
  \includegraphics[scale=0.28]{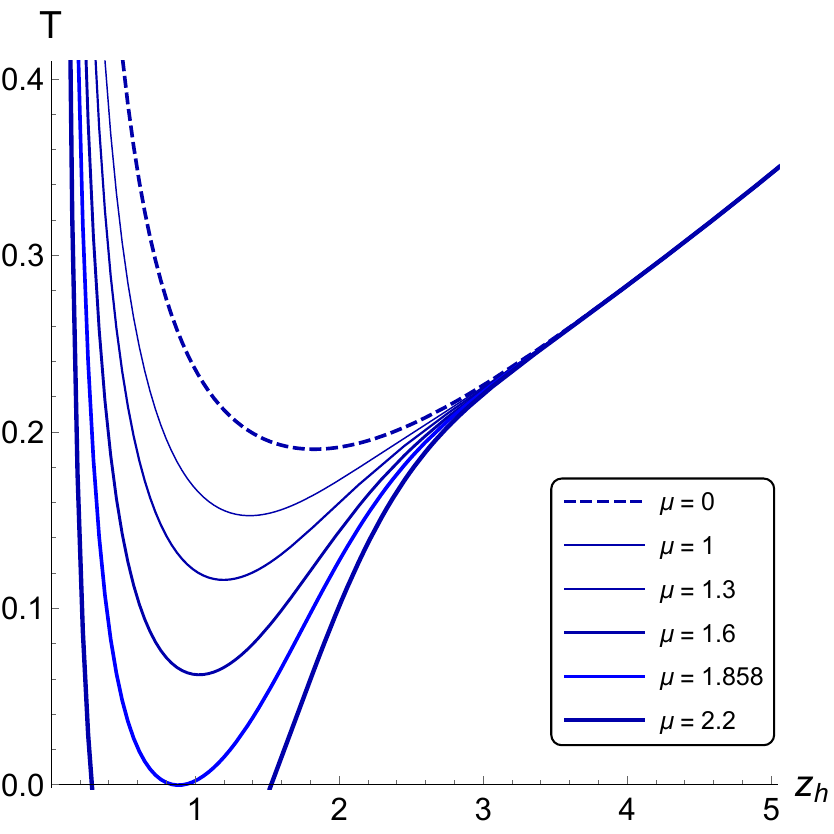} \quad
  \includegraphics[scale=0.28]{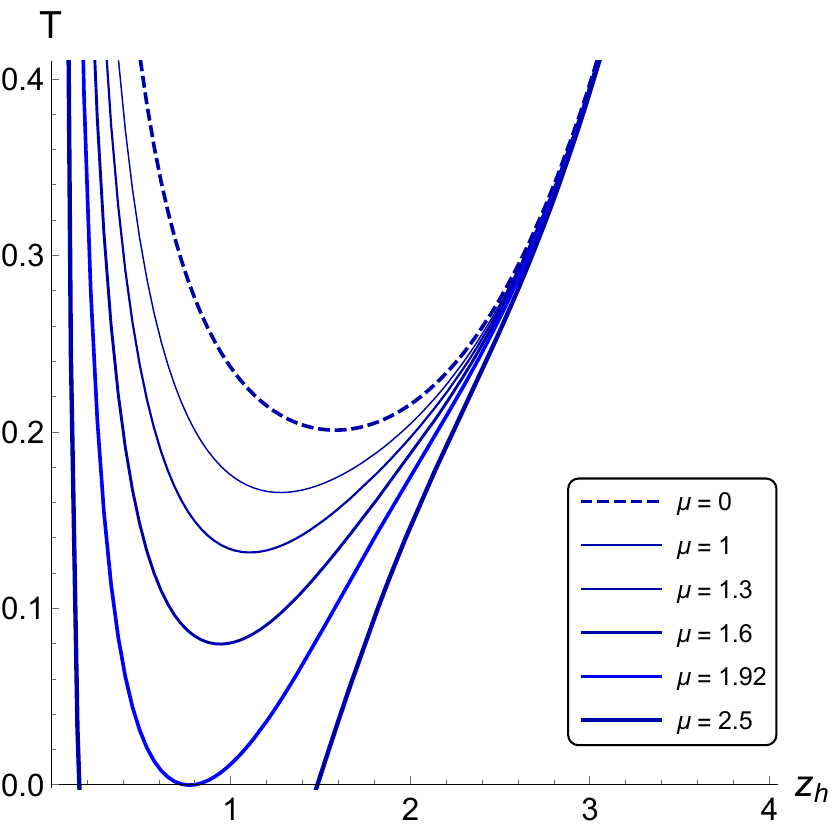} \\
  A \hspace{100pt} B \hspace{100pt} C \\
  \includegraphics[scale=0.28]{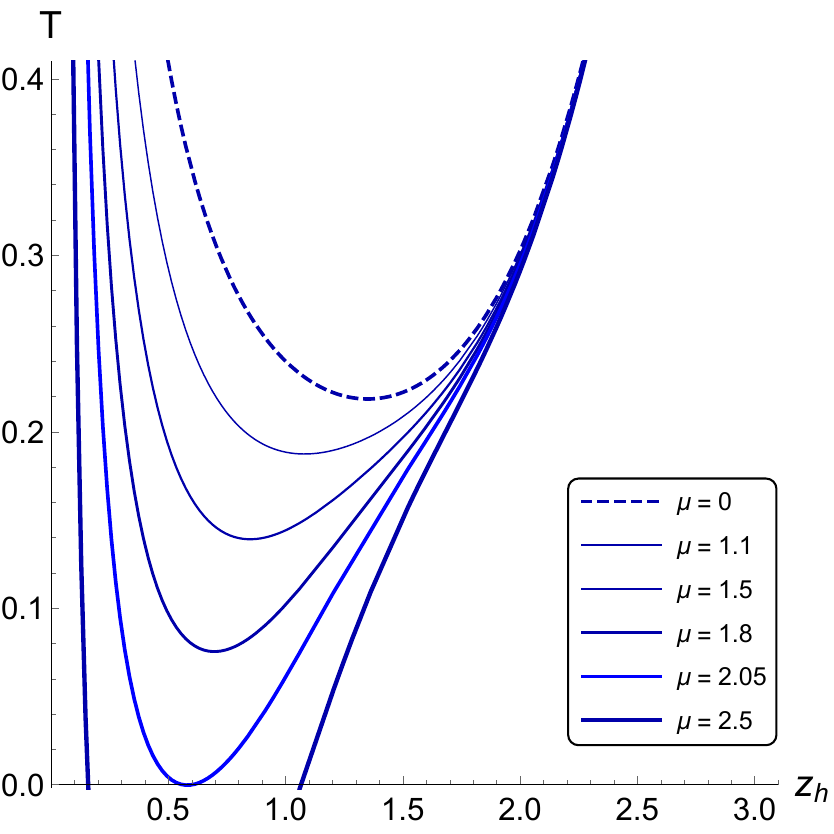} \quad
  \includegraphics[scale=0.28]{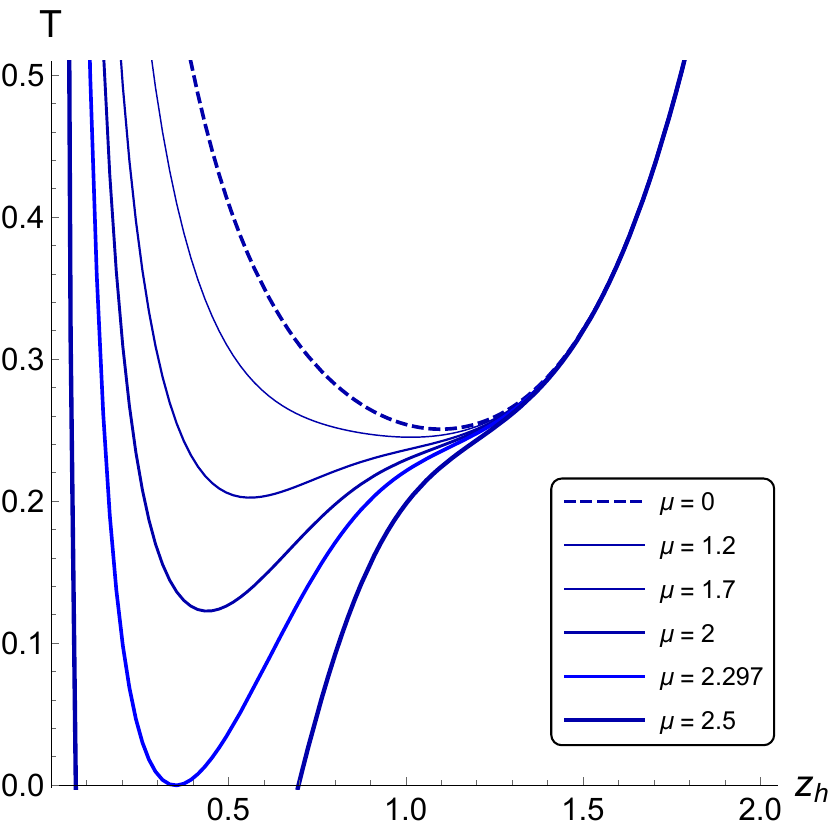} \quad
  \includegraphics[scale=0.28]{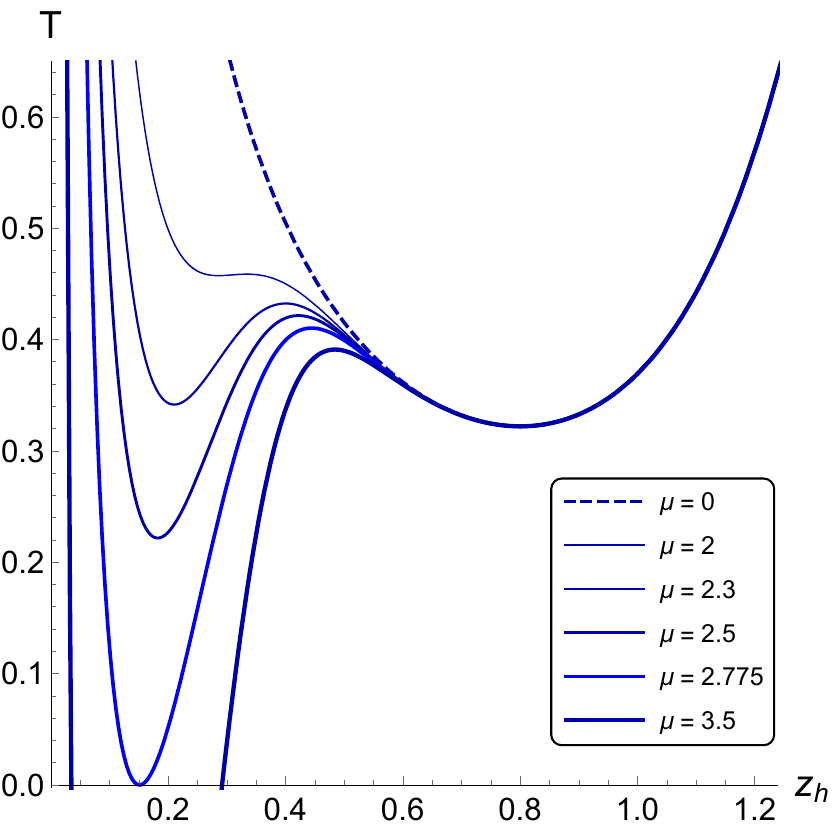} \\
  D \hspace{100pt} E \hspace{100pt} F
  \caption{Temperature $T(z_h,\mu)$ in magnetic field $q_3 = 0$ (A),
    $q_3 = 0.1$ (B), $q_3 = 0.5$ (C), $q_3 = 1$ (D), $q_3 = 2$ (E),
    $q_3 = 5$ (F); $\nu = 4.5$, $a = 0.15$, $c = 1.16$, $c_B = - \,
    0.01$, $d = 0.01$.}
  \label{Fig:Tzhmu-q3-nu45-d001-z5}
\end{figure}
\begin{figure}[h!]
  \centering 
  \includegraphics[scale=0.28]{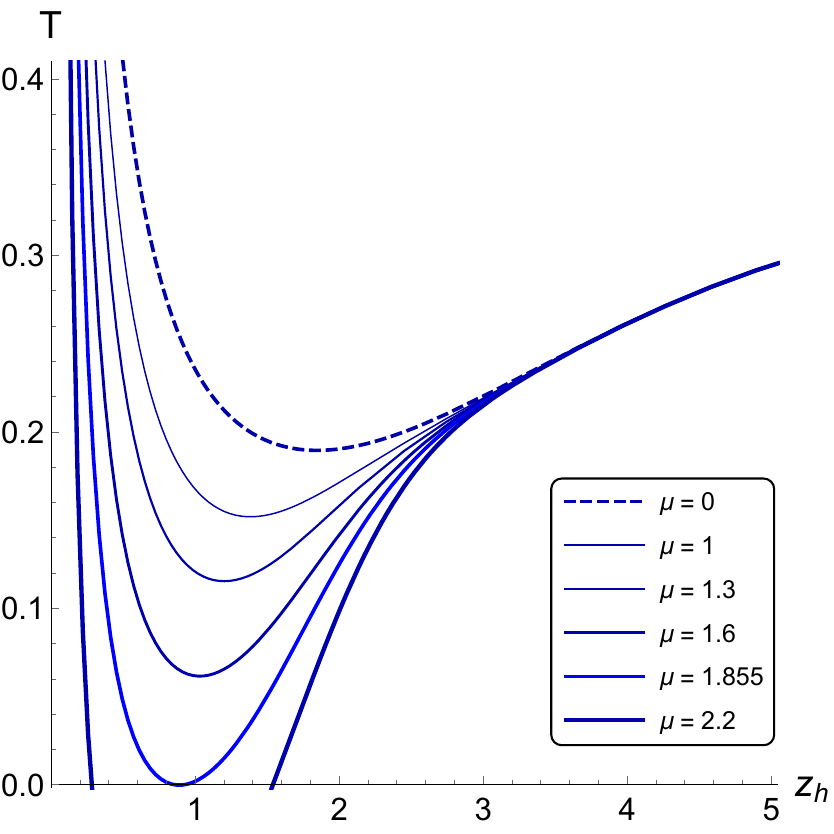} \quad
  \includegraphics[scale=0.28]{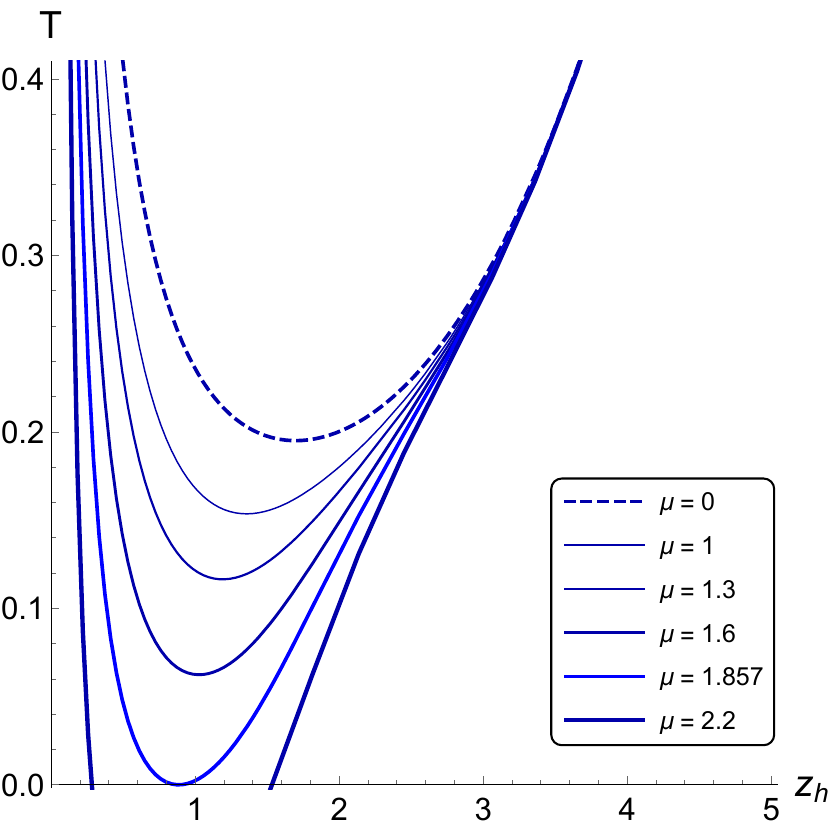} \quad
  \includegraphics[scale=0.28]{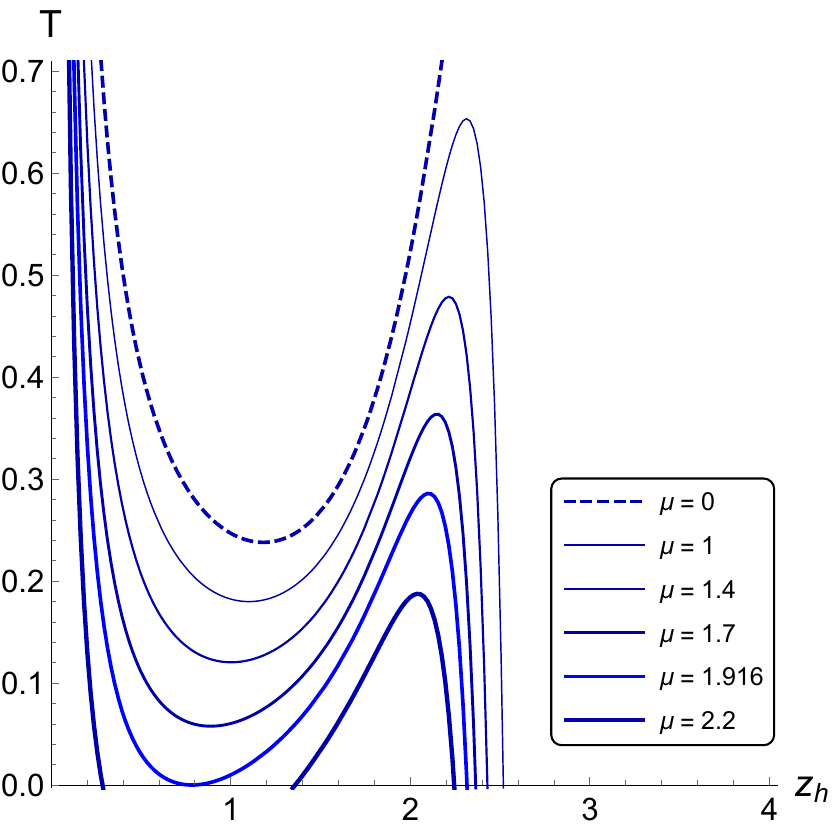} \\
  A \hspace{100pt} B \hspace{100pt} C \\
  \includegraphics[scale=0.28]{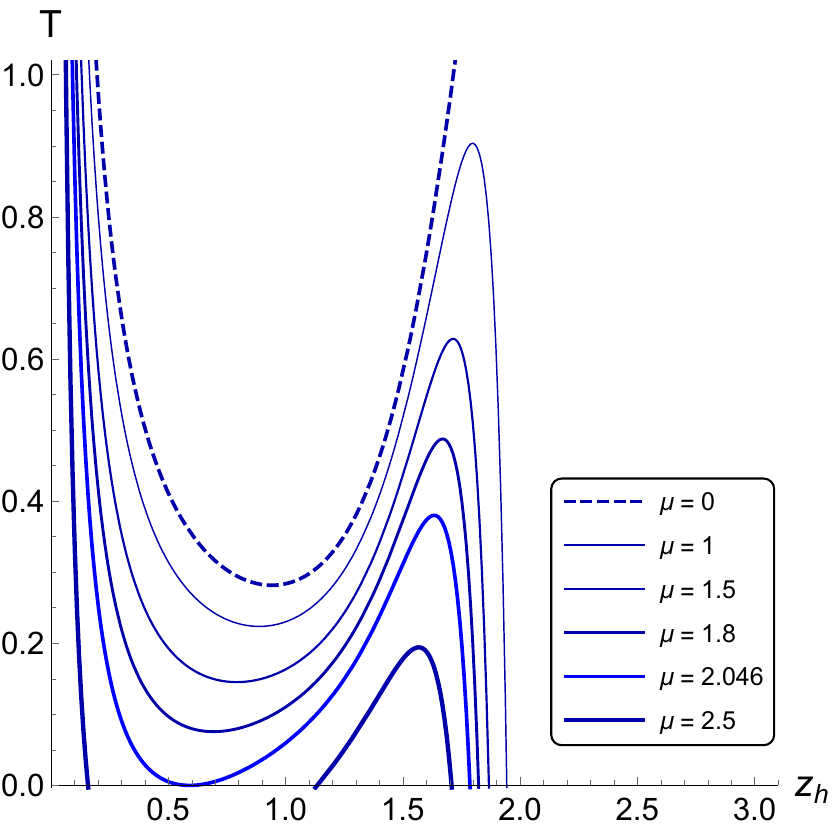} \quad
  \includegraphics[scale=0.28]{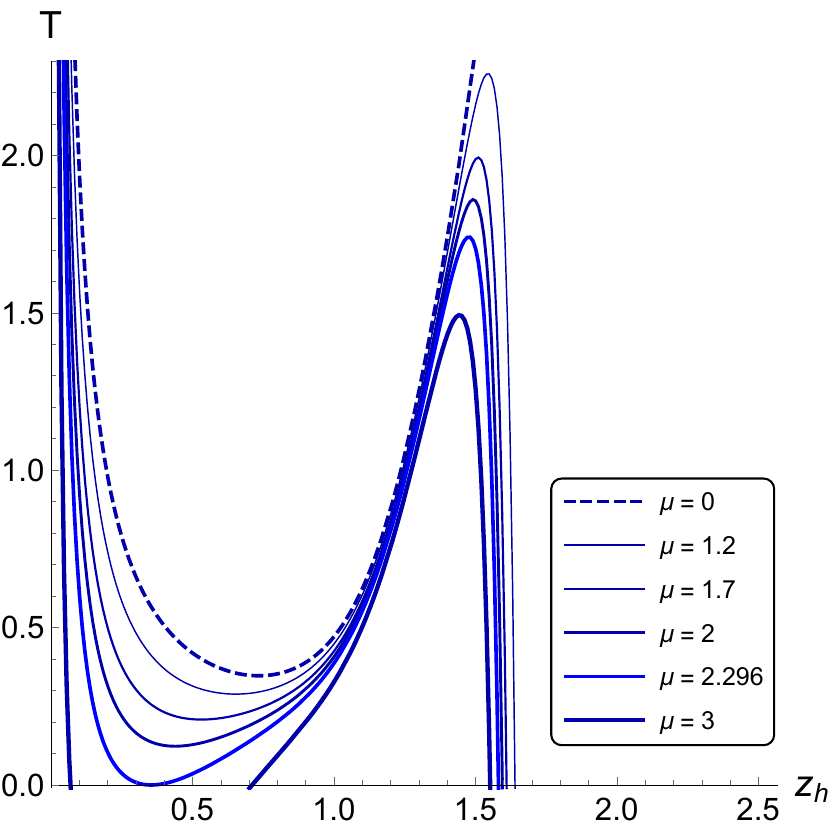} \quad
  \includegraphics[scale=0.28]{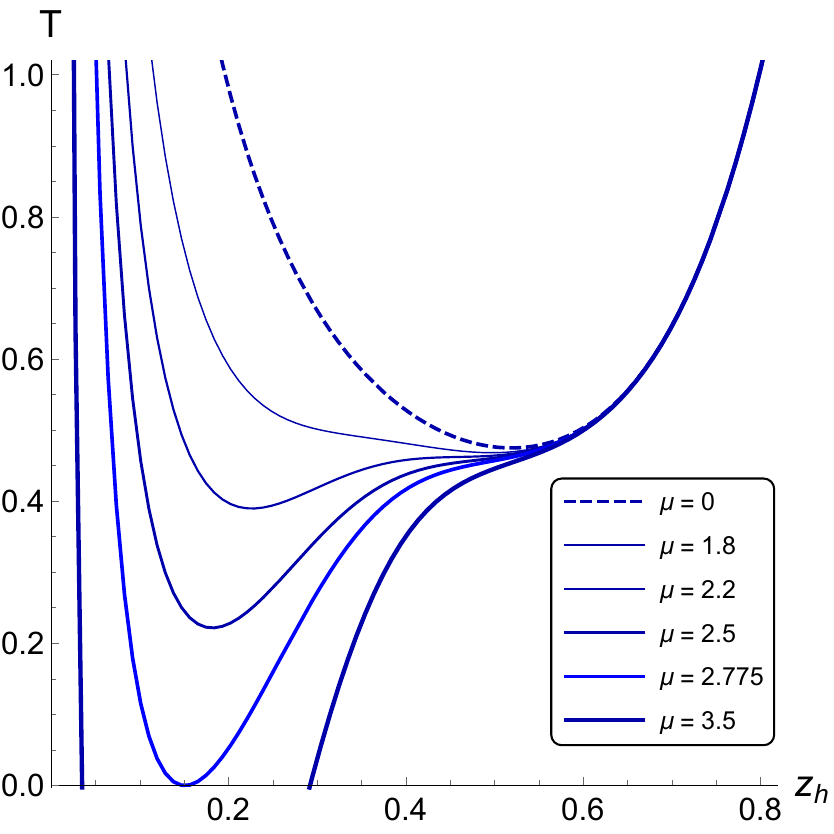} \\
  D \hspace{100pt} E \hspace{100pt} F
  \caption{Temperature $T(z_h,\mu)$ in magnetic field $q_3 = 0$ (A),
    $q_3 = 0.1$ (B), $q_3 = 0.5$ (C), $q_3 = 1$ (D), $q_3 = 2$ (E),
    $q_3 = 5$ (F); $\nu = 4.5$, $a = 0.15$, $c = 1.16$, $c_B = - \,
    0.01$, $d = 0.1$.}
  \label{Fig:Tzhmu-q3-nu45-d01-z5}
\end{figure}

\begin{figure}[t!]
  \centering 
  \includegraphics[scale=0.28]{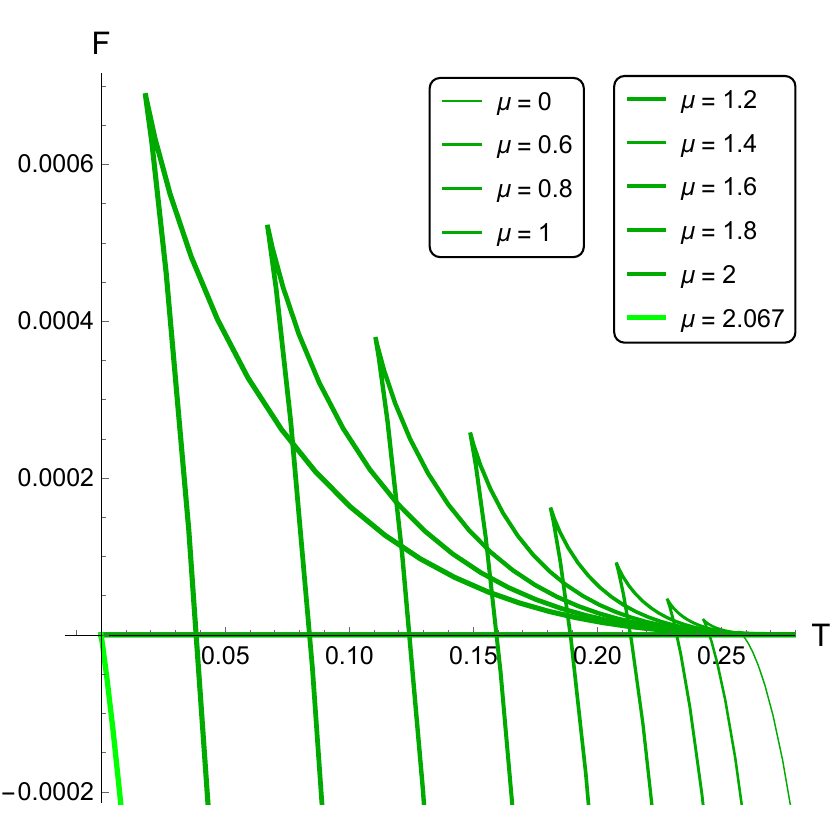} \quad
  \includegraphics[scale=0.28]{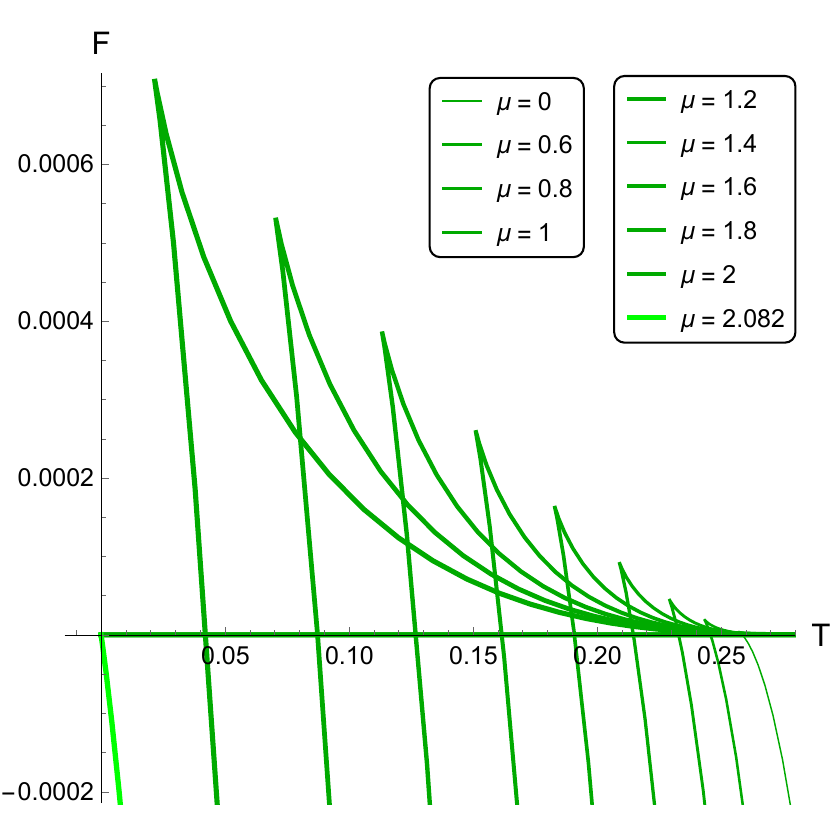} \quad
  \includegraphics[scale=0.28]{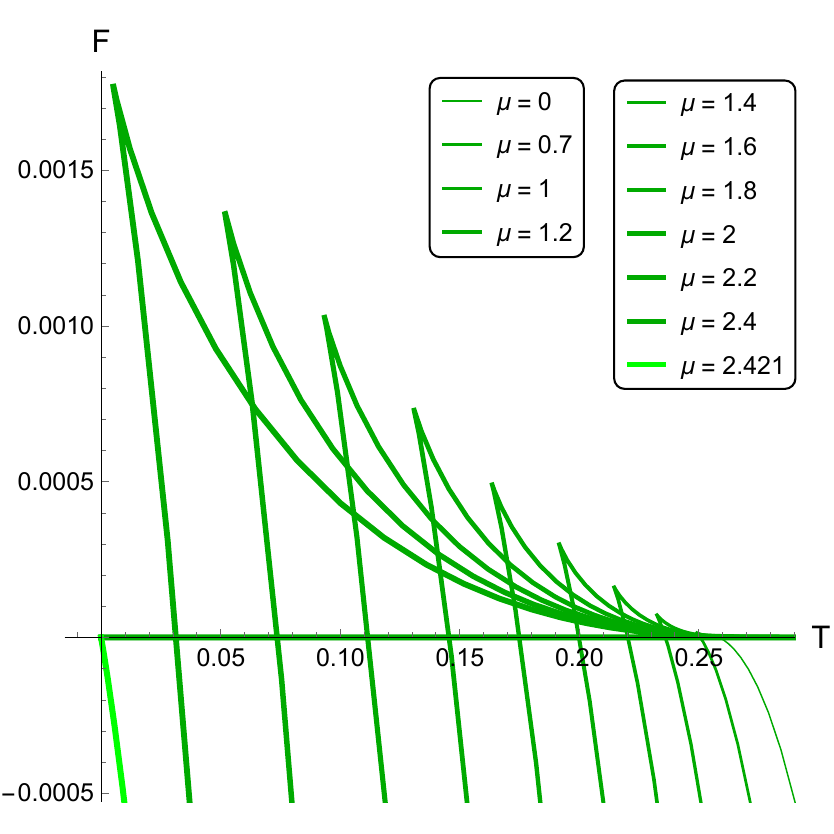} \\
  A \hspace{100pt} B \hspace{100pt} C \\
  \includegraphics[scale=0.28]{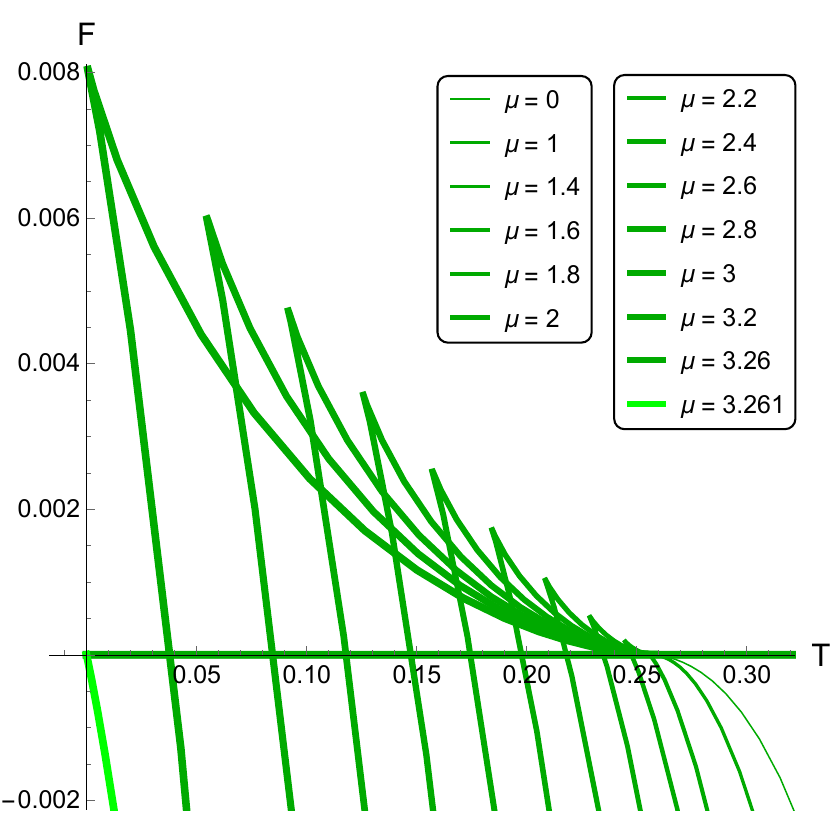} \quad
  \includegraphics[scale=0.28]{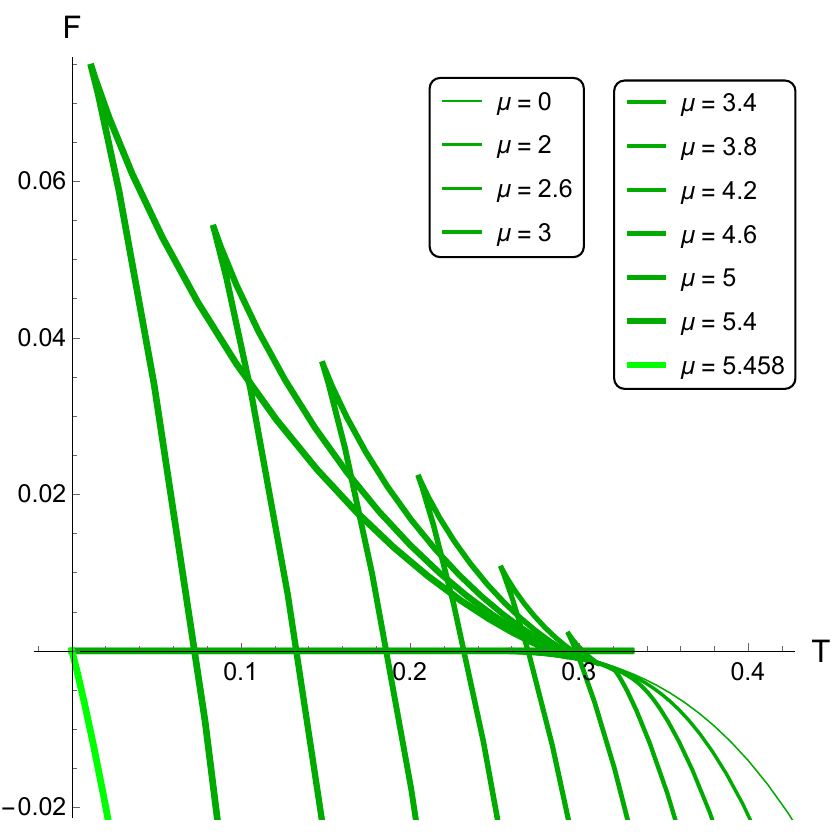} \quad
  \includegraphics[scale=0.28]{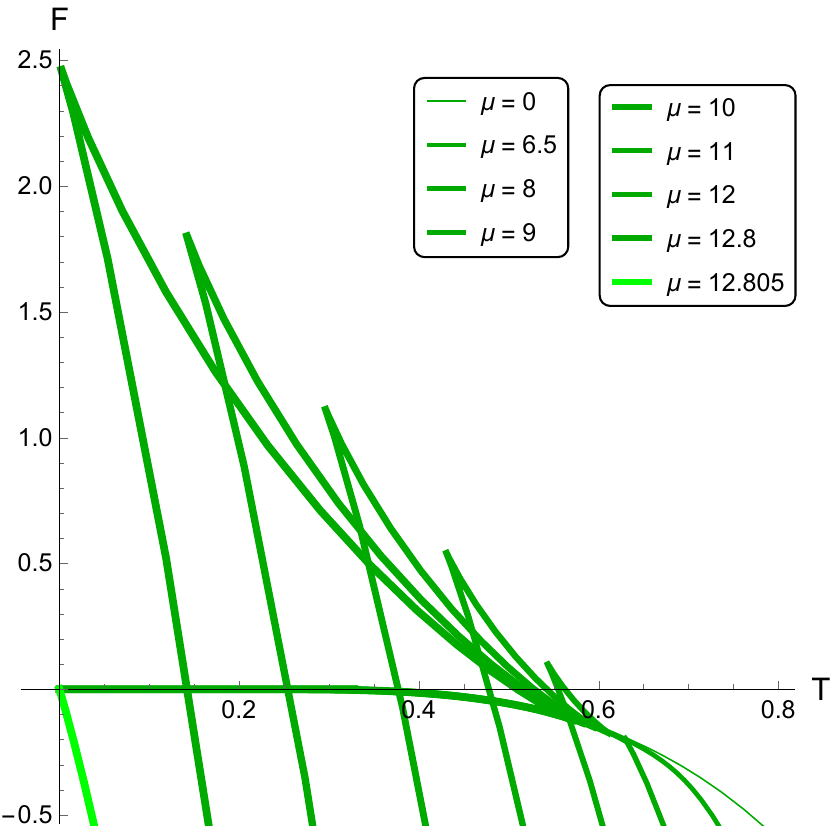} \\
  D \hspace{100pt} E \hspace{100pt} F
  \caption{Free energy $F(T,\mu)$ in magnetic field $q_3 = 0$ (A),
    $q_3 = 0.1$ (B), $q_3 = 0.5$ (C), $q_3 = 1$ (D), $q_3 = 2$ (E),
    $q_3 = 5$ (F); $\nu = 1$, $a = 0.15$, $c = 1.16$, $c_B = - \,
    0.01$, $d = 0$.}
  \label{Fig:FTmu-q3-nu1-d0-z5}
\end{figure}
\begin{figure}[h!]
  \centering 
  \includegraphics[scale=0.28]{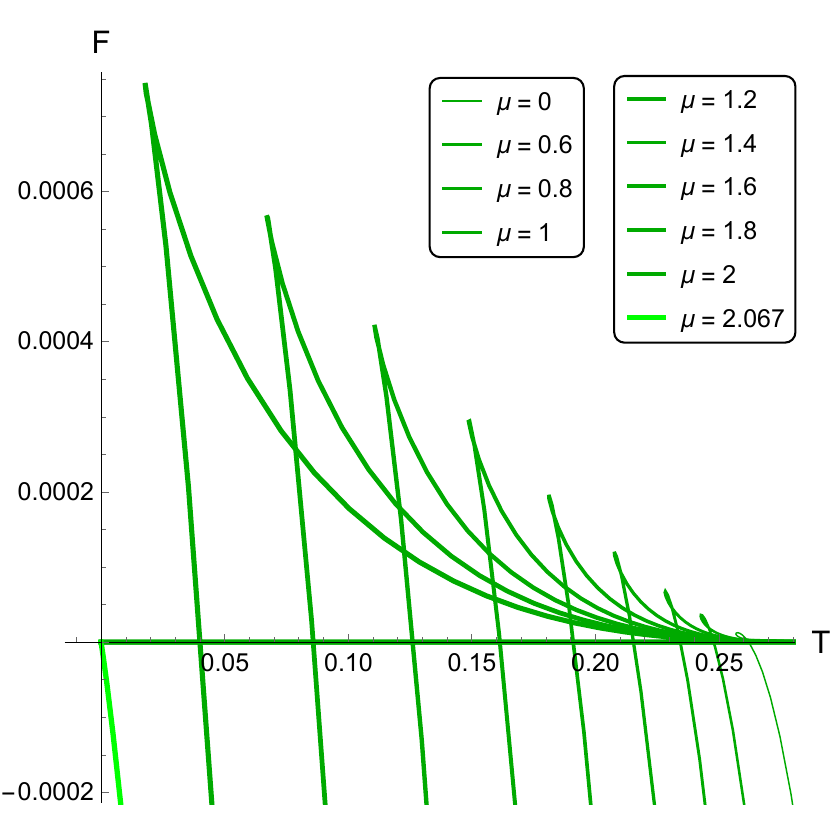} \quad
  \includegraphics[scale=0.28]{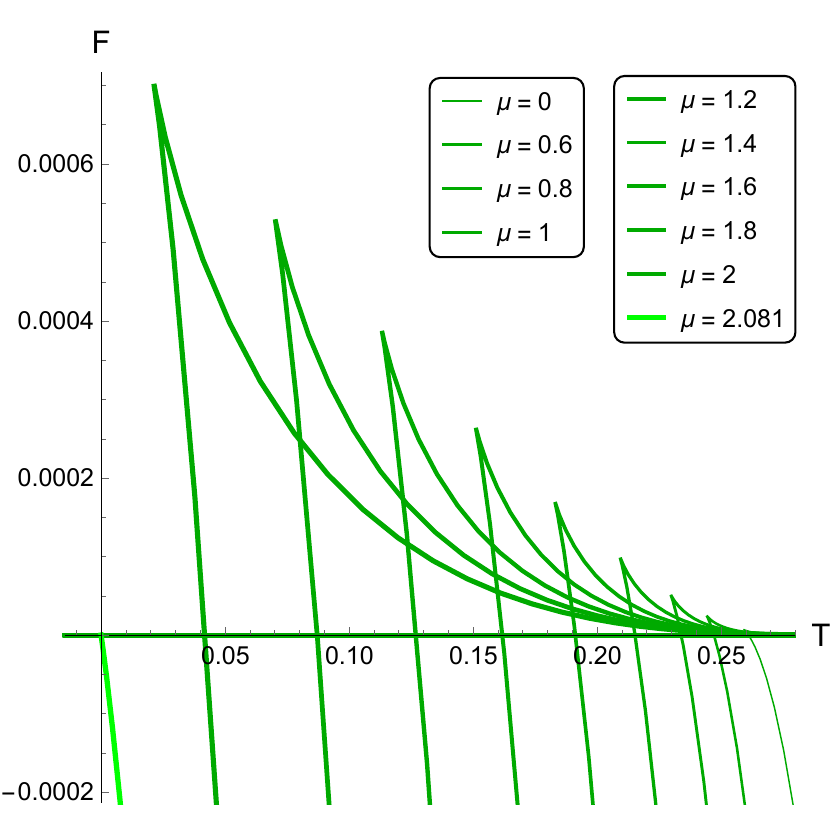} \quad
  \includegraphics[scale=0.28]{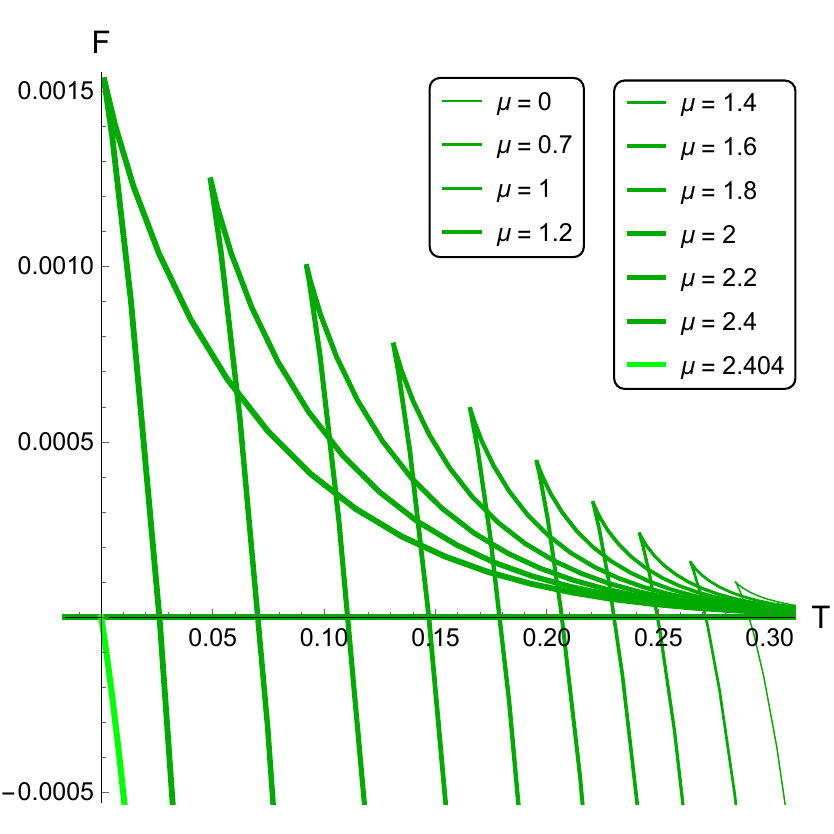} \\
  A \hspace{100pt} B \hspace{100pt} C \\
  \includegraphics[scale=0.28]{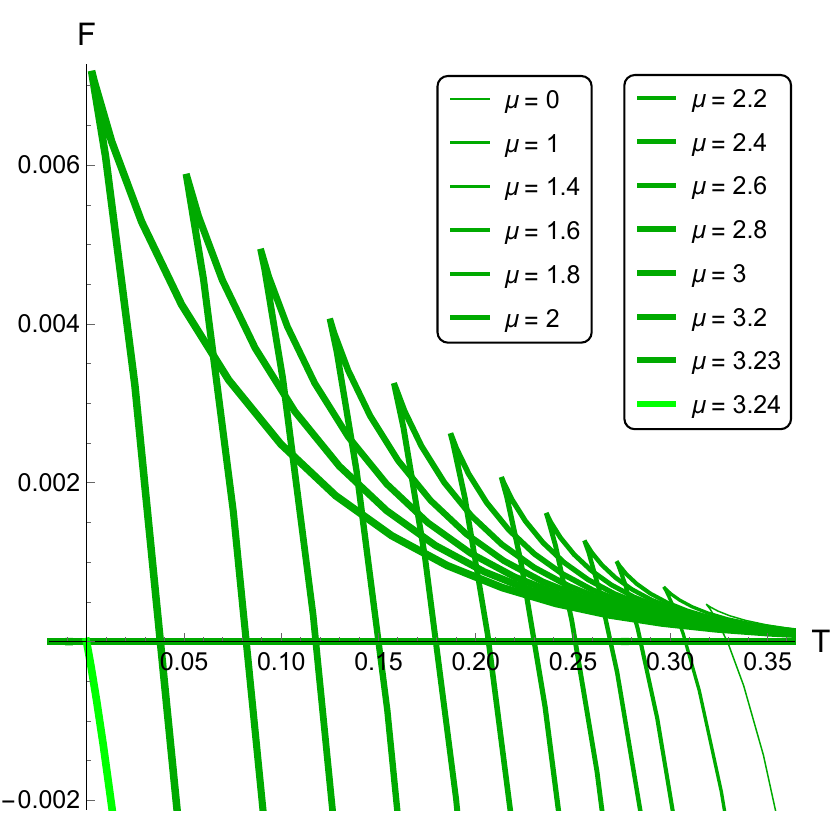} \quad
  \includegraphics[scale=0.28]{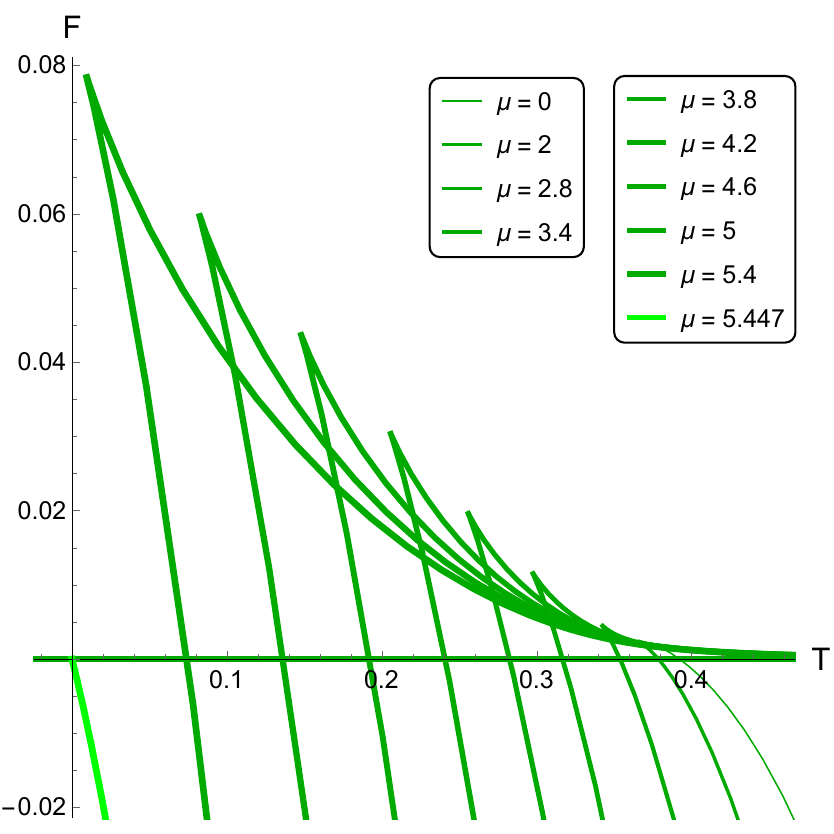} \quad
  \includegraphics[scale=0.28]{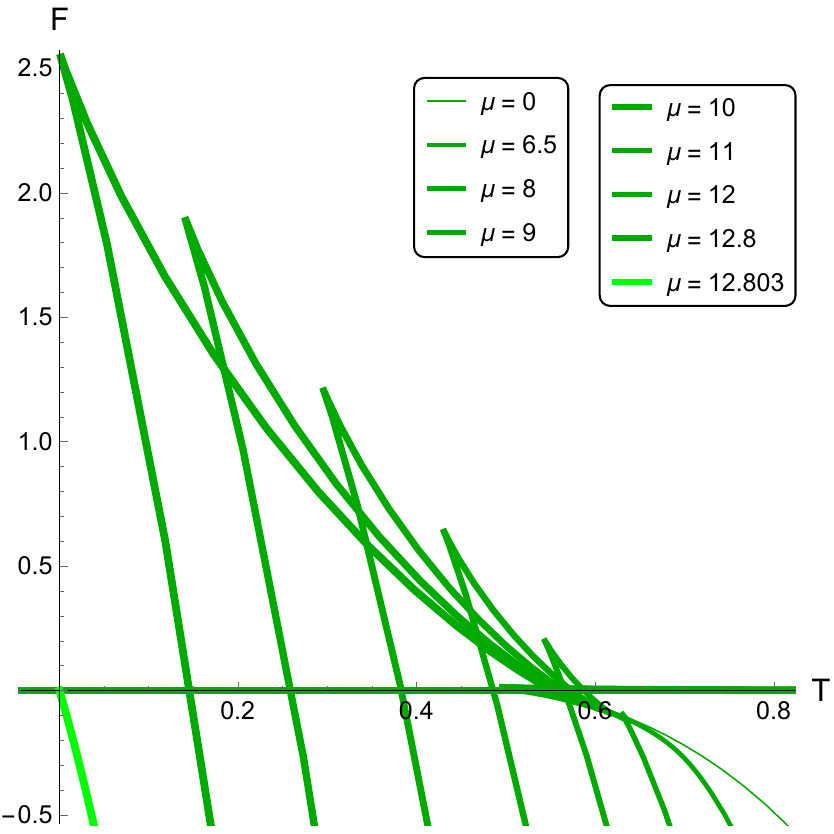} \\
  D \hspace{100pt} E \hspace{100pt} F
  \caption{Free energy $F(T,\mu)$ in magnetic field $q_3 = 0$ (A),
    $q_3 = 0.1$ (B), $q_3 = 0.5$ (C), $q_3 = 1$ (D), $q_3 = 2$ (E),
    $q_3 = 5$ (F); $\nu = 1$, $a = 0.15$, $c = 1.16$, $c_B = - \,
    0.01$, $d = 0.01$.}
  \label{Fig:FTmu-q3-nu1-d001-z5}
\end{figure}

\begin{figure}[t!]
  \centering 
  \includegraphics[scale=0.28]{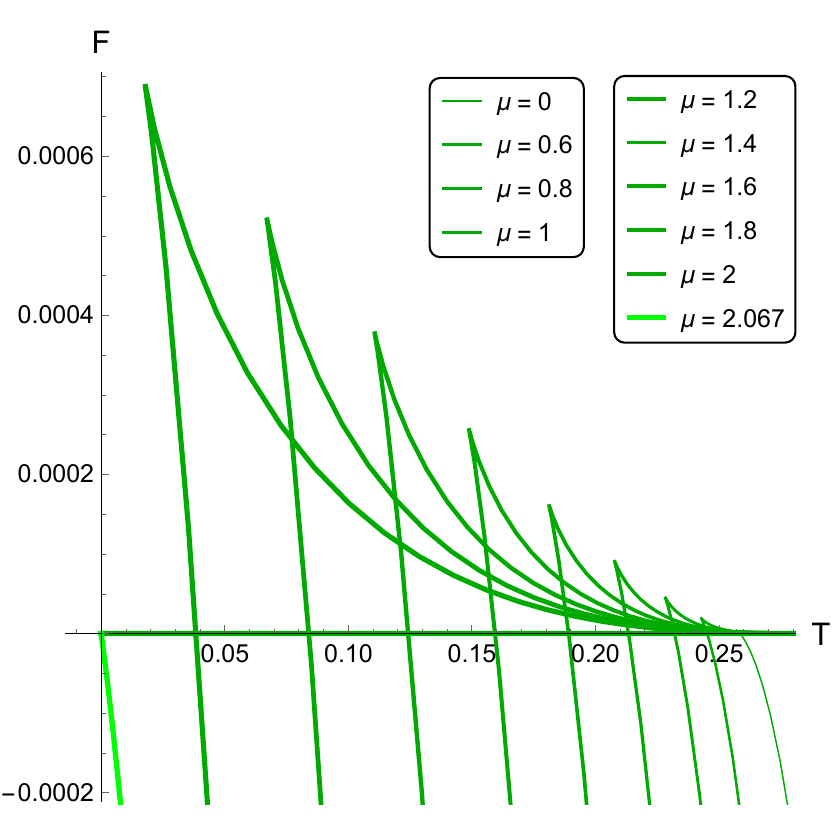} \quad
  \includegraphics[scale=0.28]{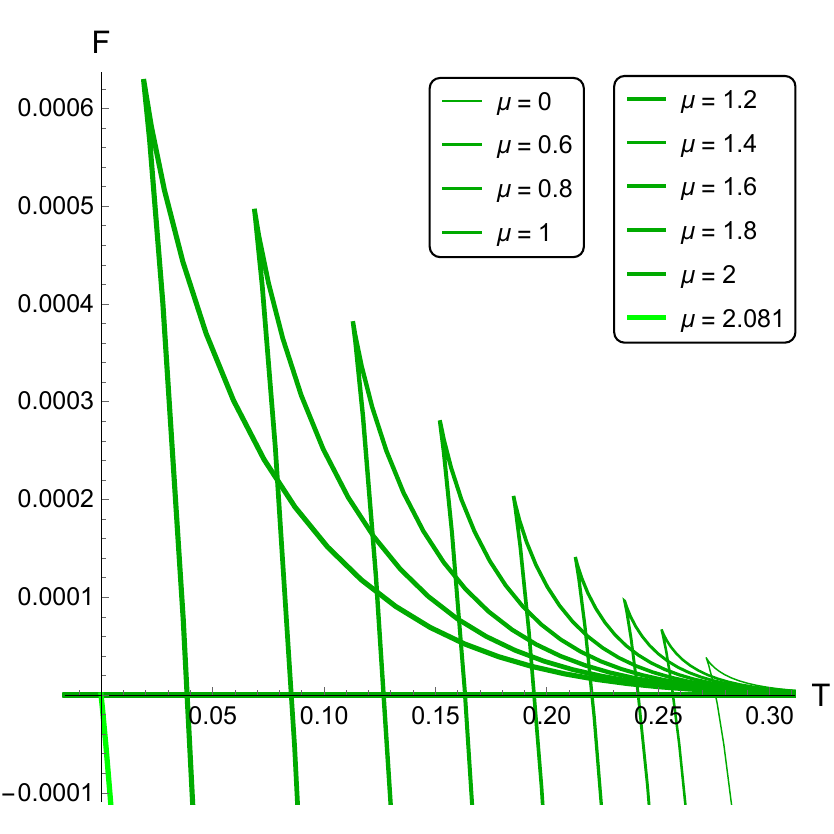} \quad
  \includegraphics[scale=0.28]{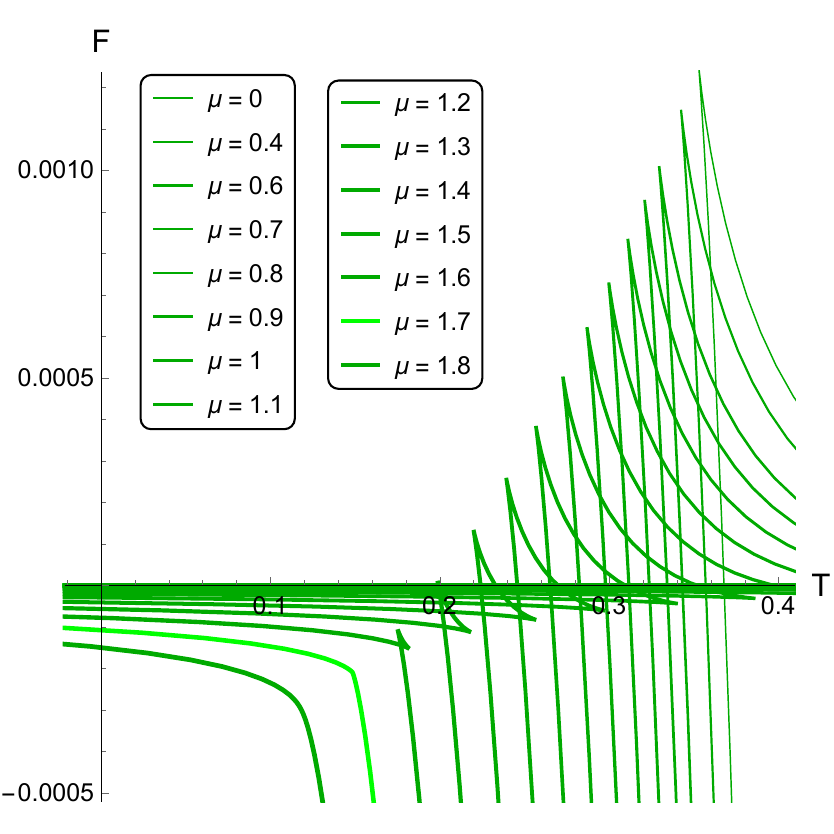} \\
  A \hspace{100pt} B \hspace{100pt} C \\
  \includegraphics[scale=0.28]{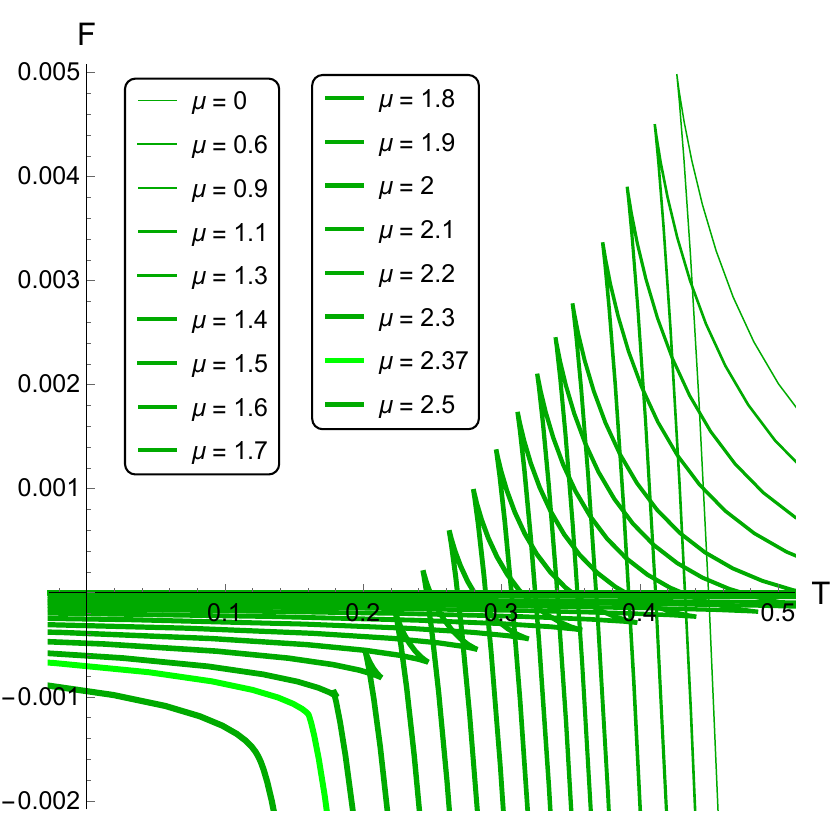} \quad
  \includegraphics[scale=0.28]{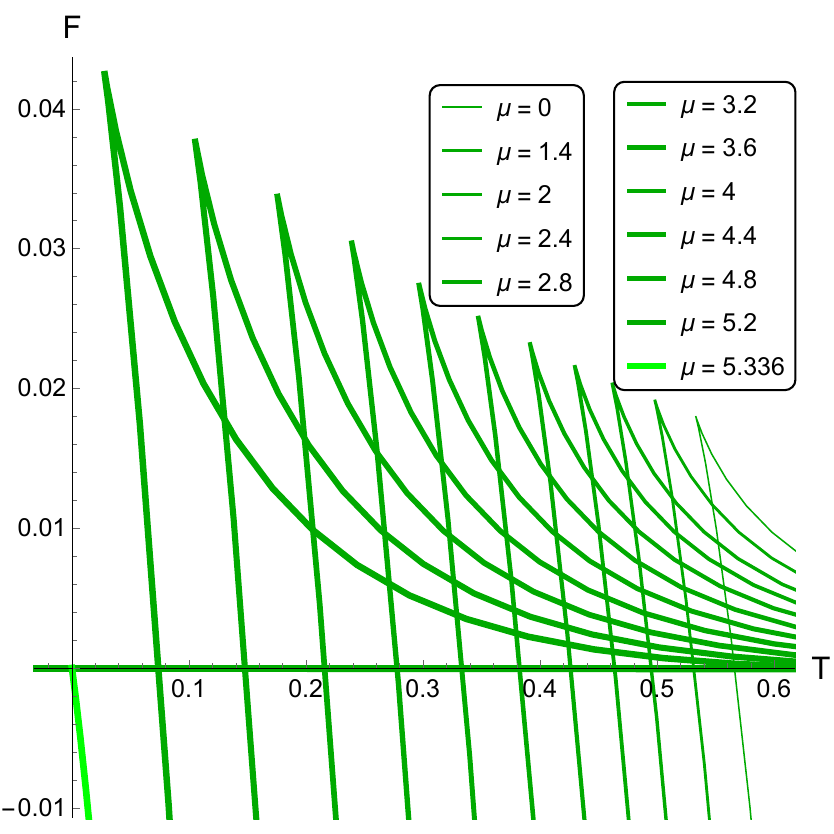} \quad
  \includegraphics[scale=0.28]{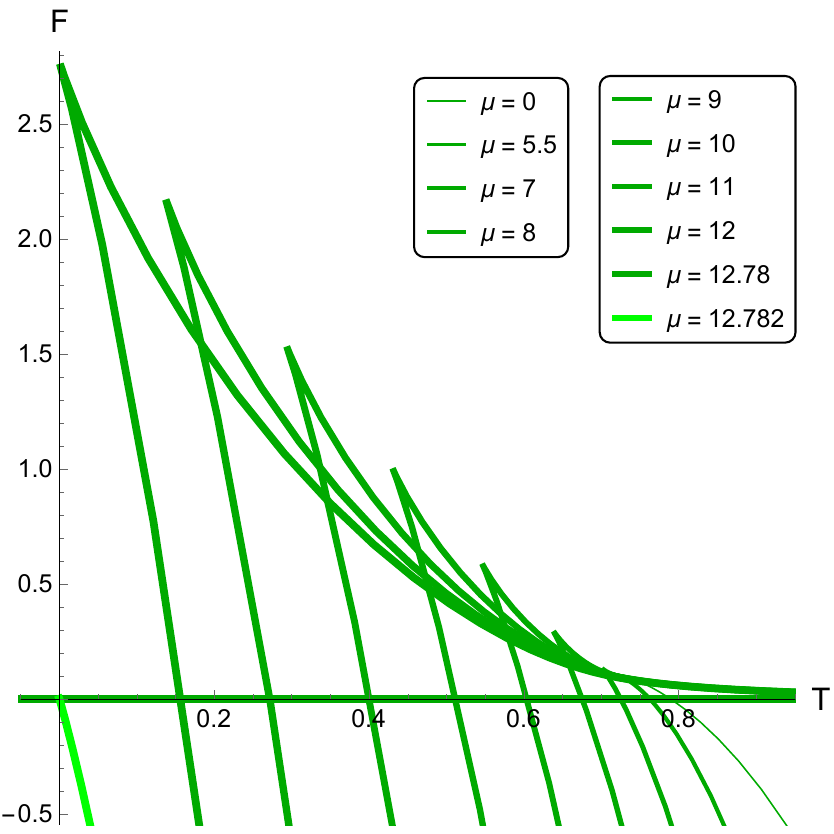} \\
  D \hspace{100pt} E \hspace{100pt} F
  \caption{Free energy $F(T,\mu)$ in magnetic field $q_3 = 0$ (A),
    $q_3 = 0.1$ (B), $q_3 = 0.5$ (C), $q_3 = 1$ (D), $q_3 = 2$ (E),
    $q_3 = 5$ (F); $\nu = 1$, $a = 0.15$, $c = 1.16$, $c_B = - \,
    0.01$, $d = 0.1$.}
  \label{Fig:FTmu-q3-nu1-d01-z5}
\end{figure}
\begin{figure}[h!]
  \centering 
  \includegraphics[scale=0.28]{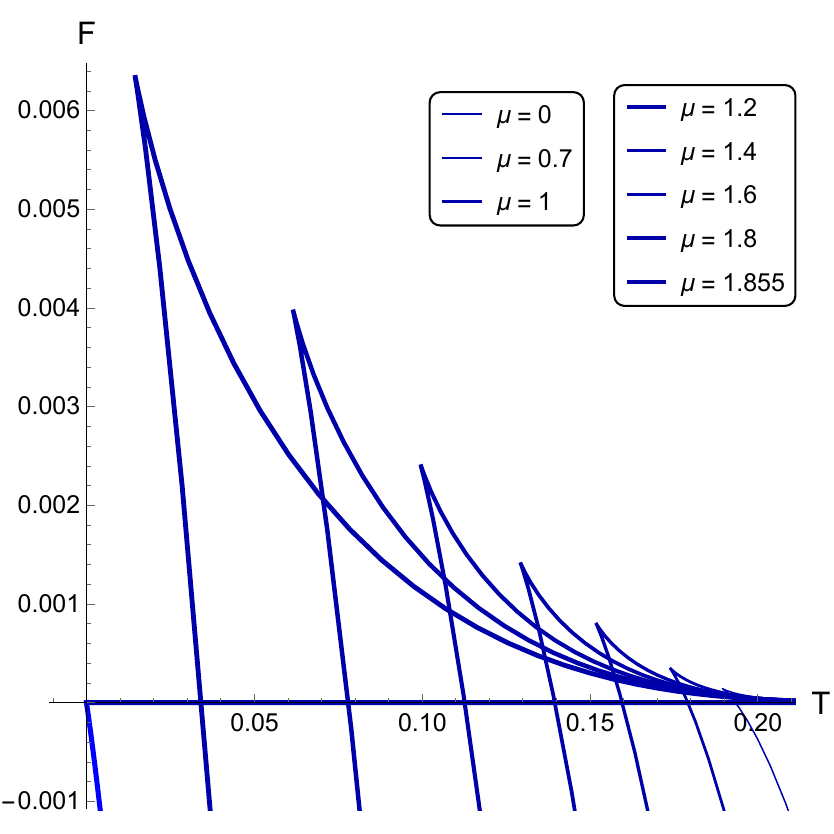} \quad
  \includegraphics[scale=0.28]{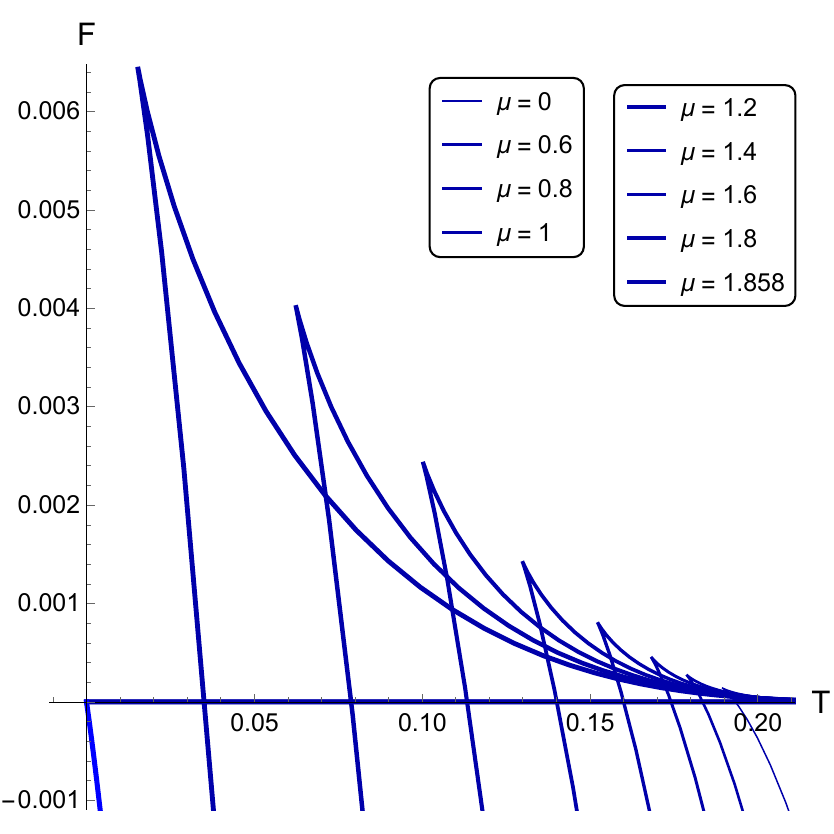} \quad
  \includegraphics[scale=0.28]{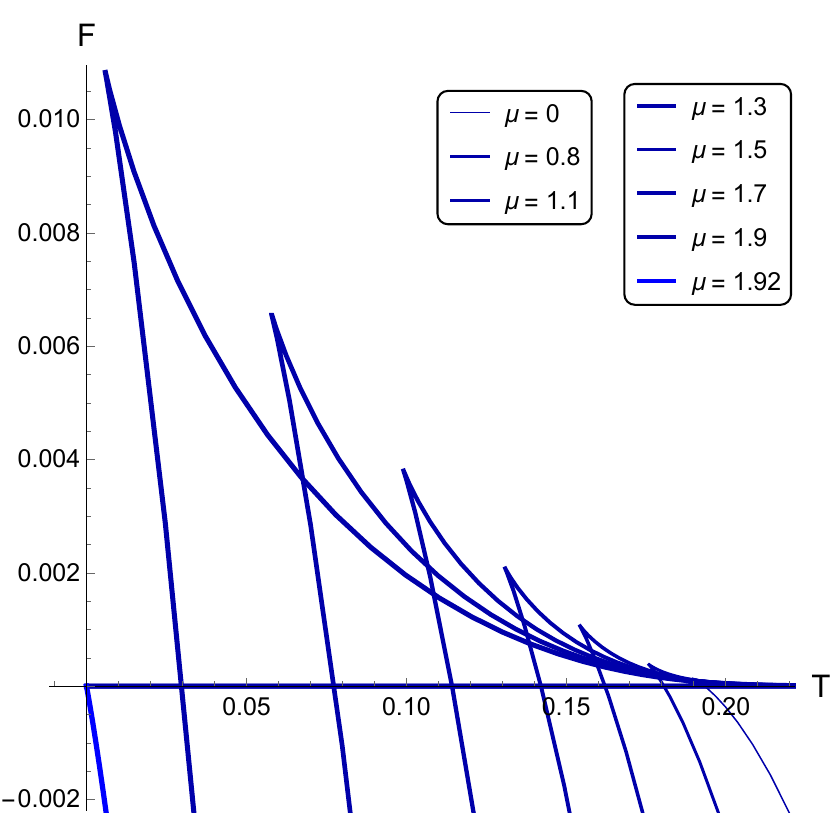} \\
  A \hspace{100pt} B \hspace{100pt} C \\
  \includegraphics[scale=0.28]{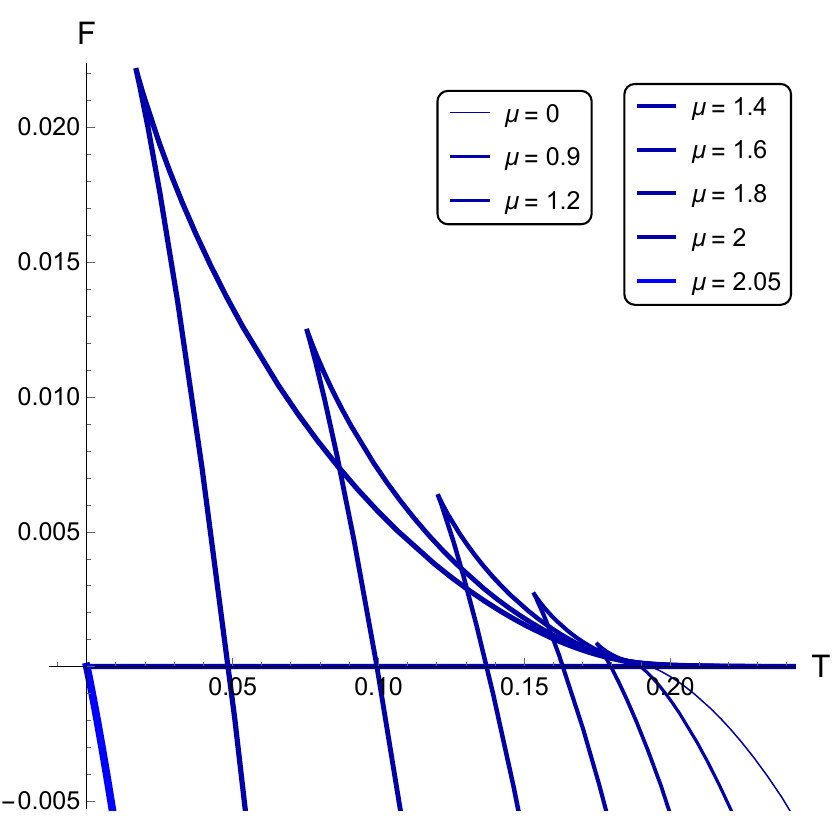} \quad
  \includegraphics[scale=0.28]{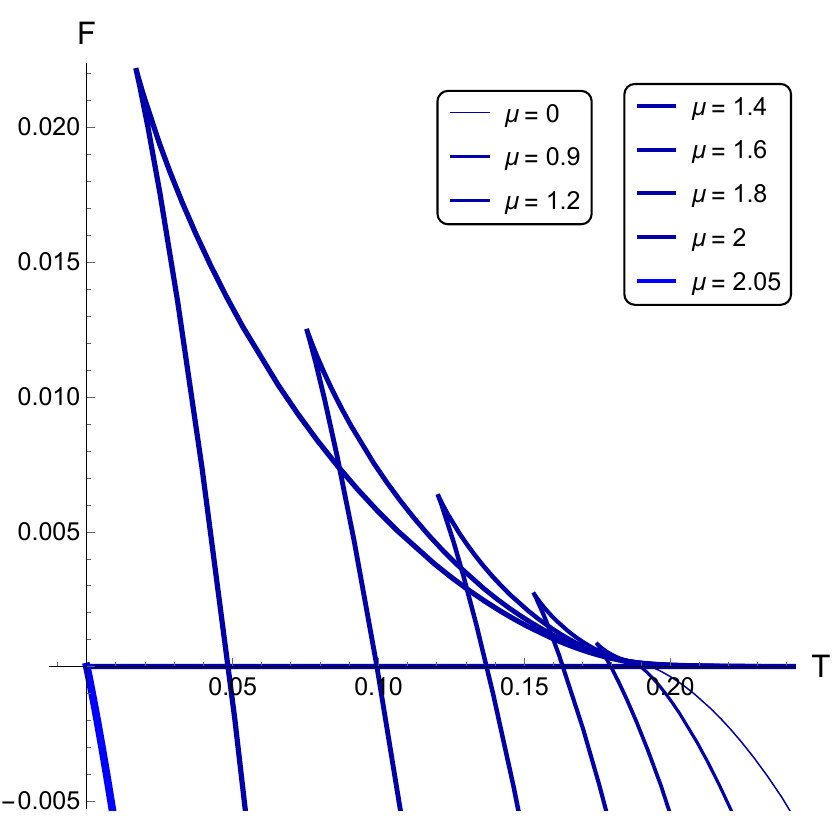} \quad
  \includegraphics[scale=0.28]{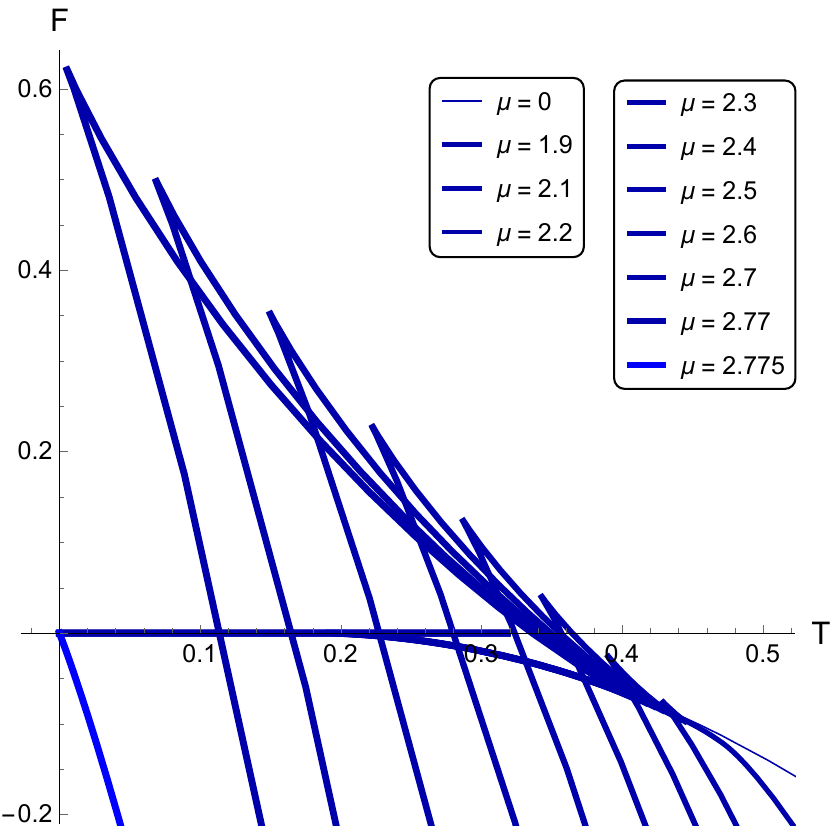} \\
  D \hspace{100pt} E \hspace{100pt} F
  \caption{Free energy $F(T,\mu)$ in magnetic field $q_3 = 0$ (A),
    $q_3 = 0.1$ (B), $q_3 = 0.5$ (C), $q_3 = 1$ (D), $q_3 = 2$ (E),
    $q_3 = 5$ (F); $\nu = 4.5$, $a = 0.15$, $c = 1.16$, $c_B = - \,
    0.01$, $d = 0$.}
  \label{Fig:FTmu-q3-nu45-d0-z5}
\end{figure}

\begin{figure}[t!]
  \centering 
  \includegraphics[scale=0.28]{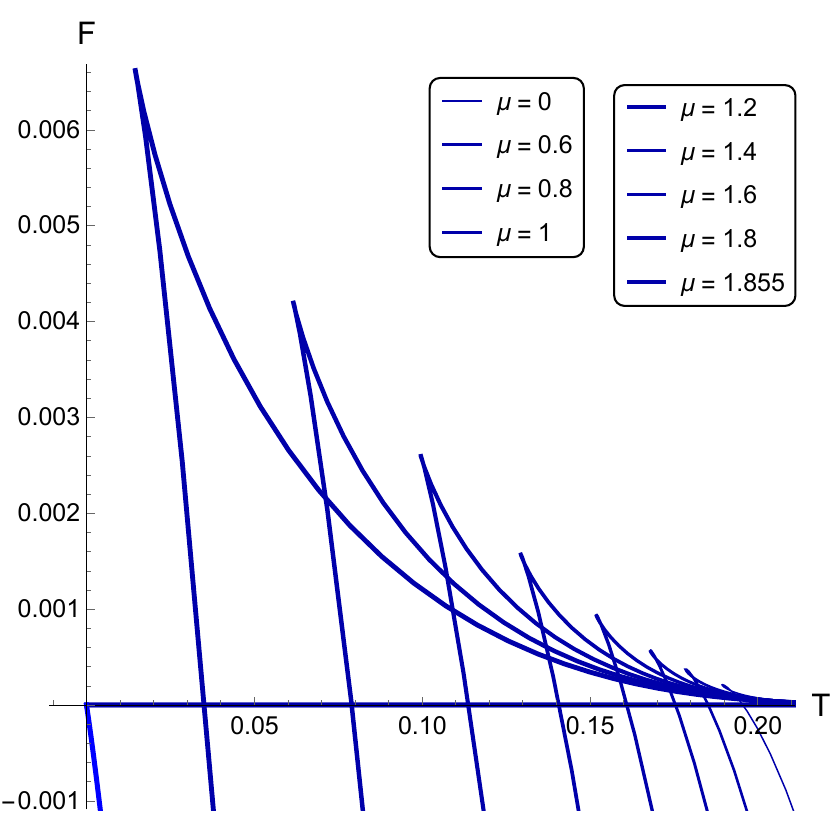} \quad
  \includegraphics[scale=0.28]{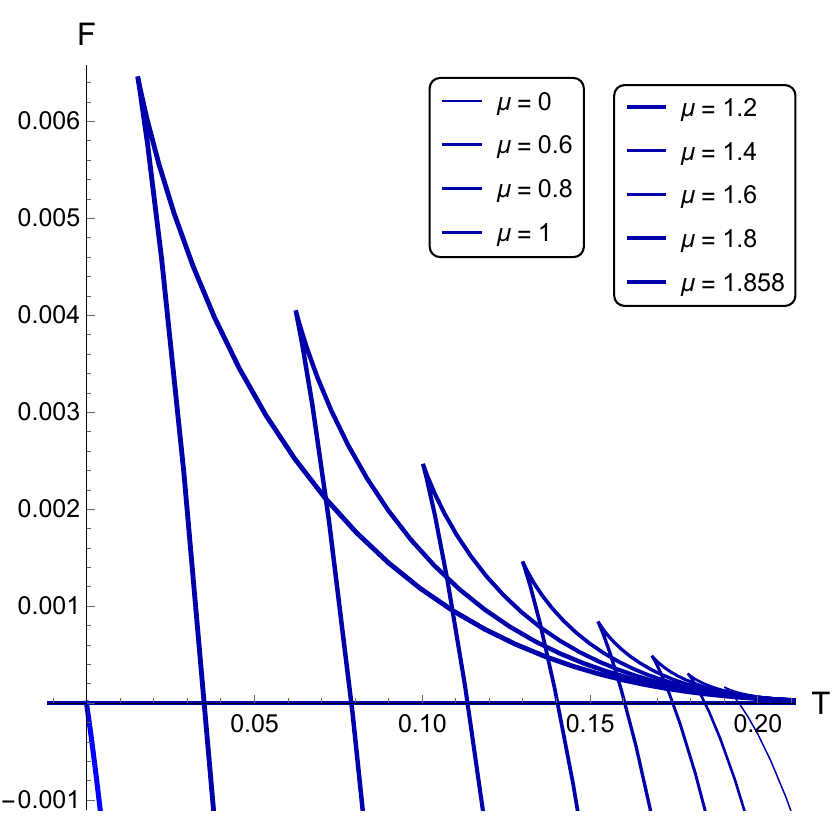} \quad
  \includegraphics[scale=0.28]{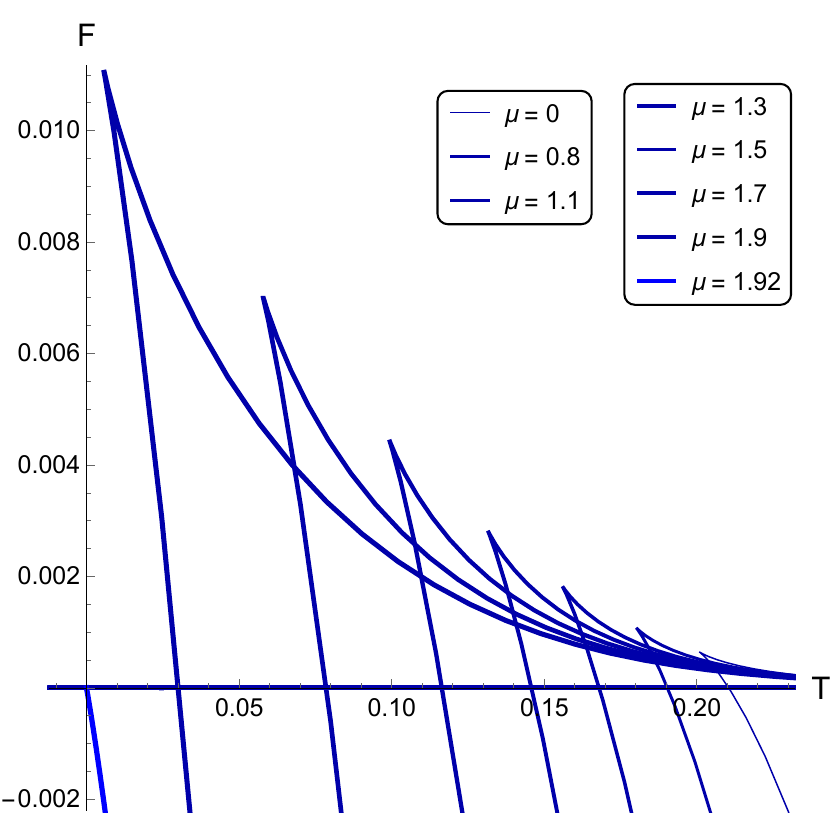} \\
  A \hspace{100pt} B \hspace{100pt} C \\
  \includegraphics[scale=0.28]{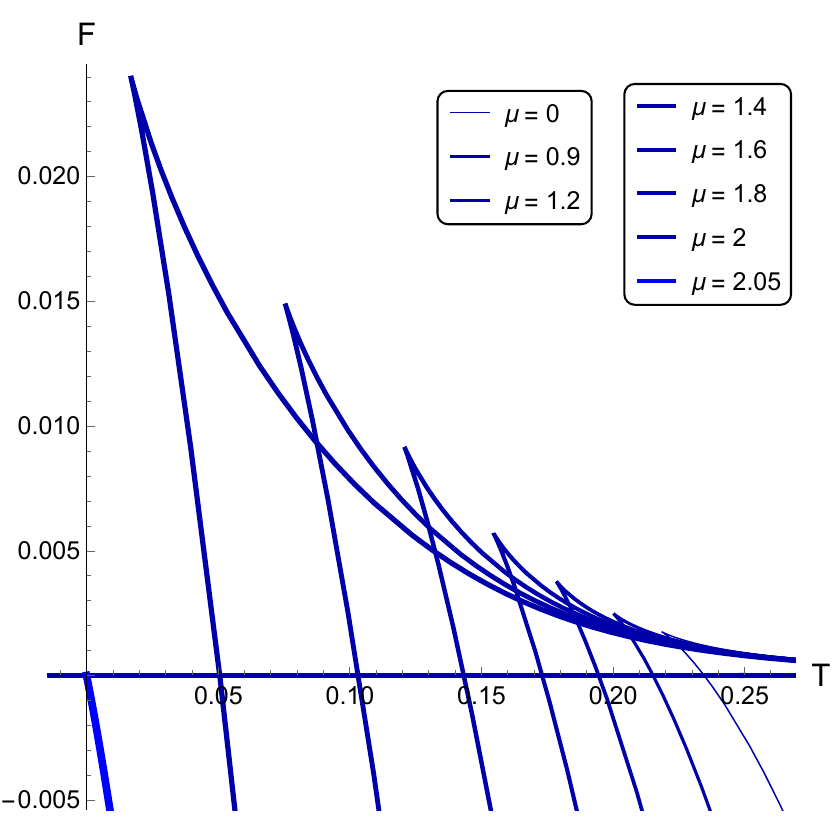} \quad
  \includegraphics[scale=0.28]{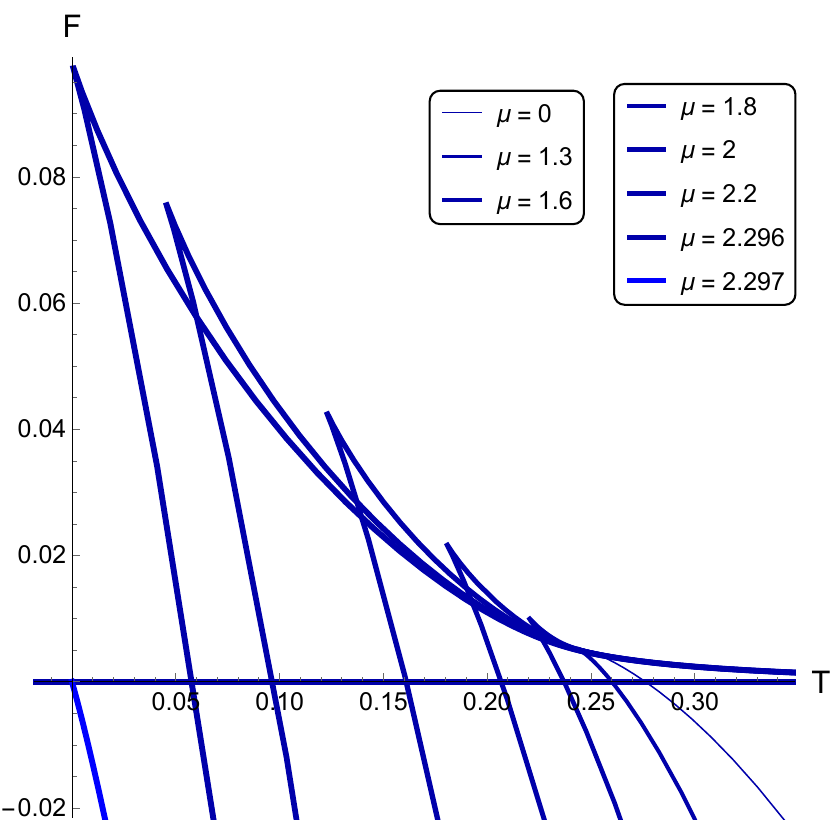} \quad
  \includegraphics[scale=0.28]{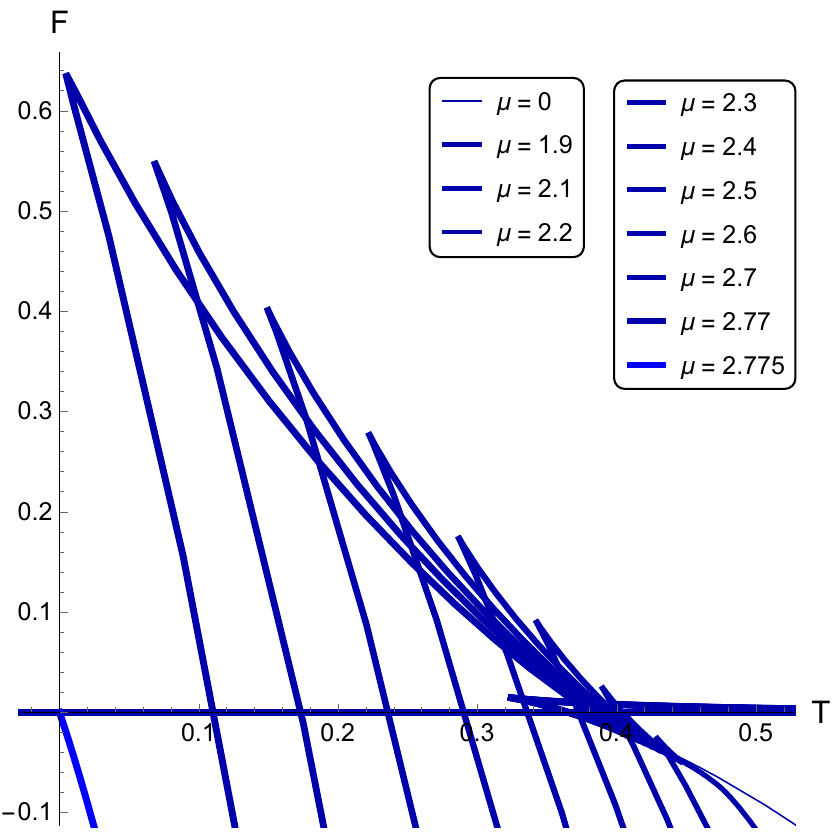} \\
  D \hspace{100pt} E \hspace{100pt} F
  \caption{Free energy $F(T,\mu)$ in magnetic field $q_3 = 0$ (A),
    $q_3 = 0.1$ (B), $q_3 = 0.5$ (C), $q_3 = 1$ (D), $q_3 = 2$ (E),
    $q_3 = 5$ (F); $\nu = 4.5$, $a = 0.15$, $c = 1.16$, $c_B = - \,
    0.01$, $d = 0.01$.}
  \label{Fig:FTmu-q3-nu45-d001-z5}
\end{figure}
\begin{figure}[h!]
  \centering 
  \includegraphics[scale=0.28]{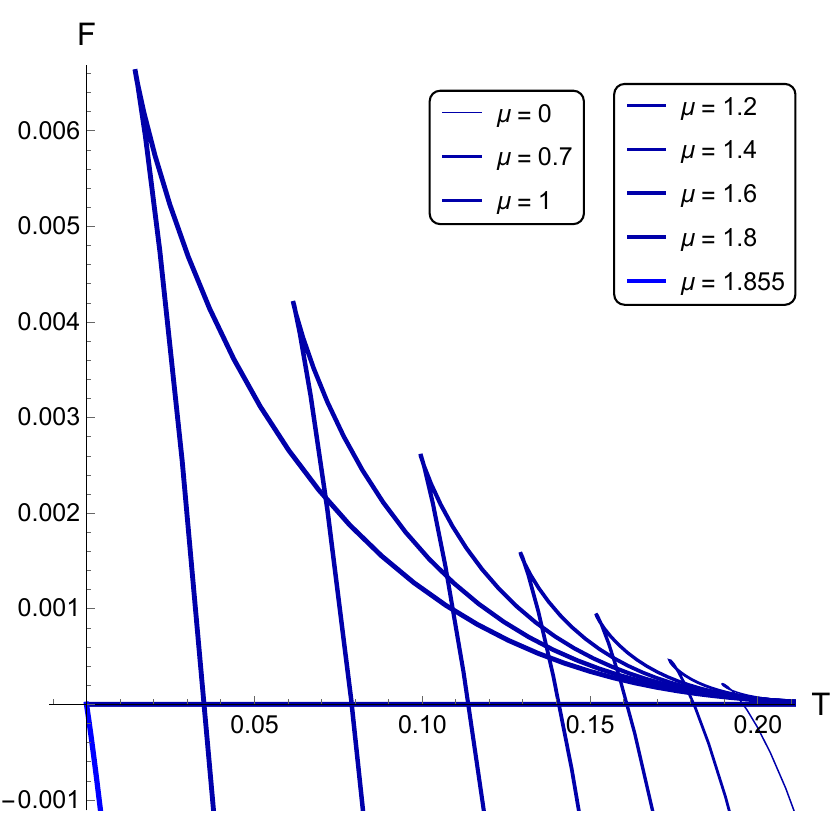} \quad
  \includegraphics[scale=0.28]{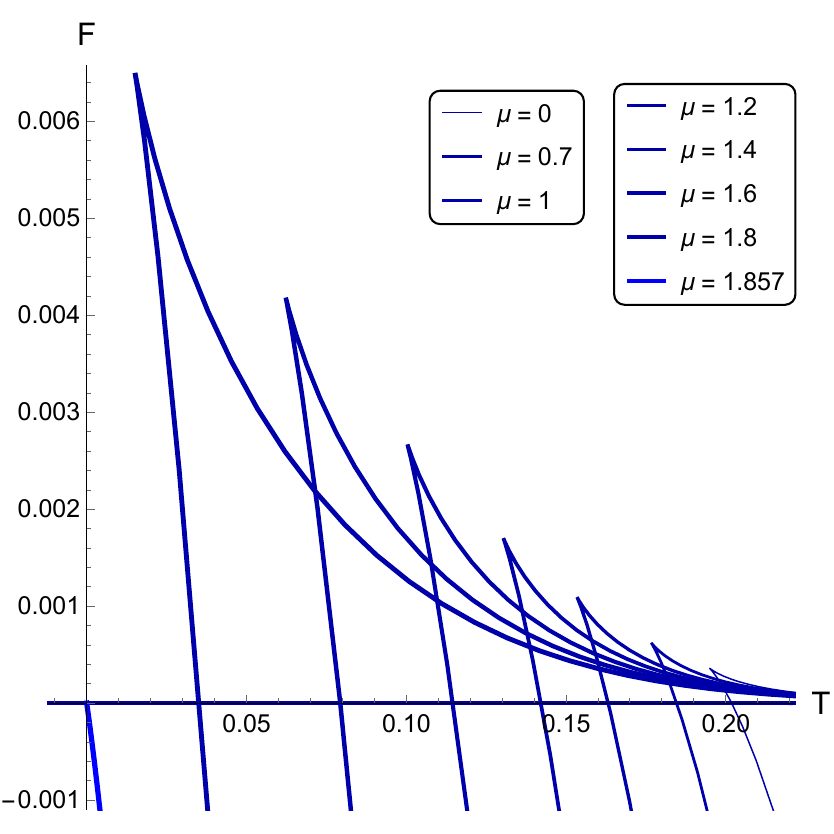} \quad
  \includegraphics[scale=0.28]{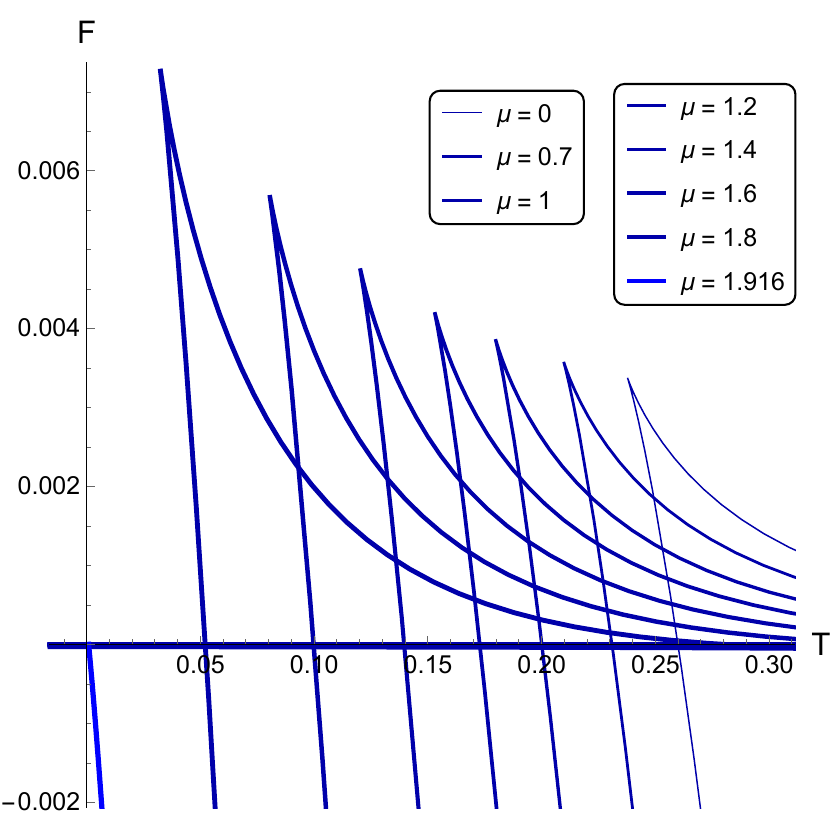} \\
  A \hspace{100pt} B \hspace{100pt} C \\
  \includegraphics[scale=0.28]{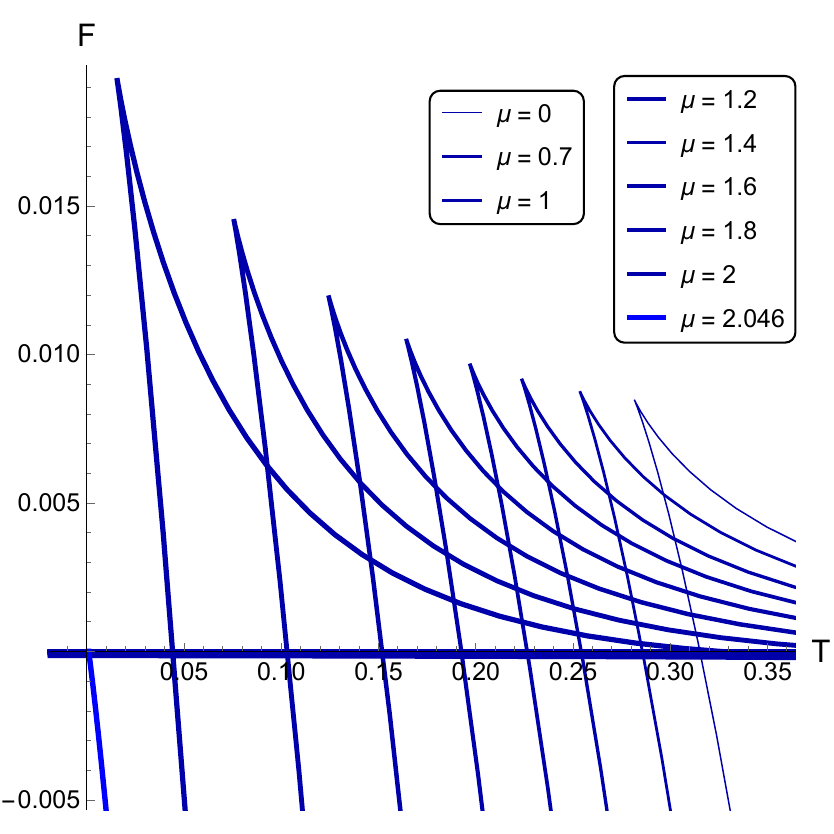} \quad
  \includegraphics[scale=0.28]{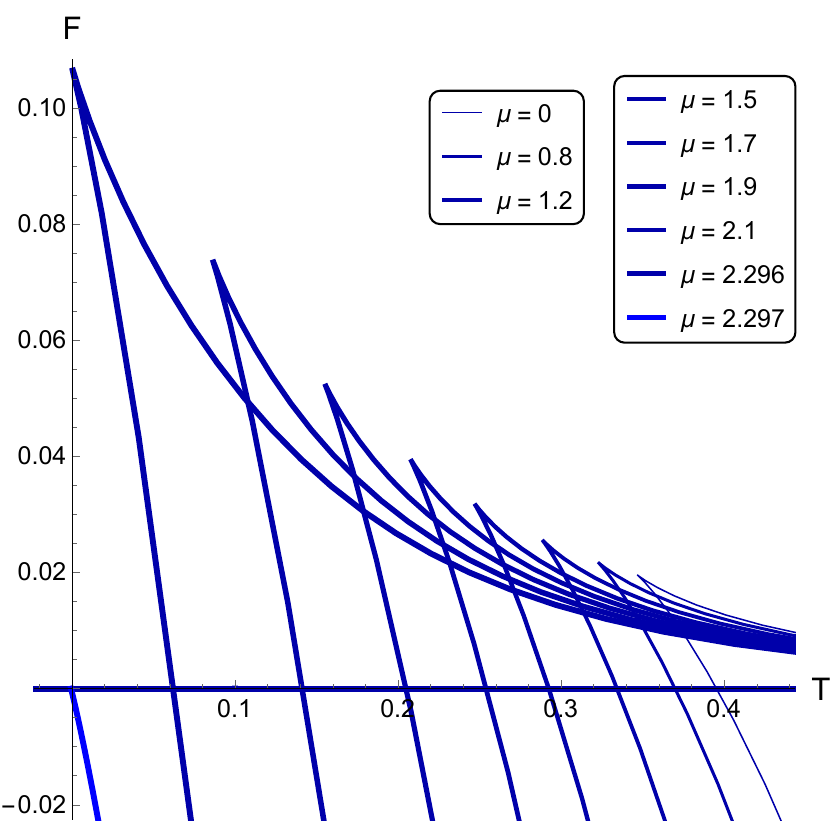} \quad
  \includegraphics[scale=0.28]{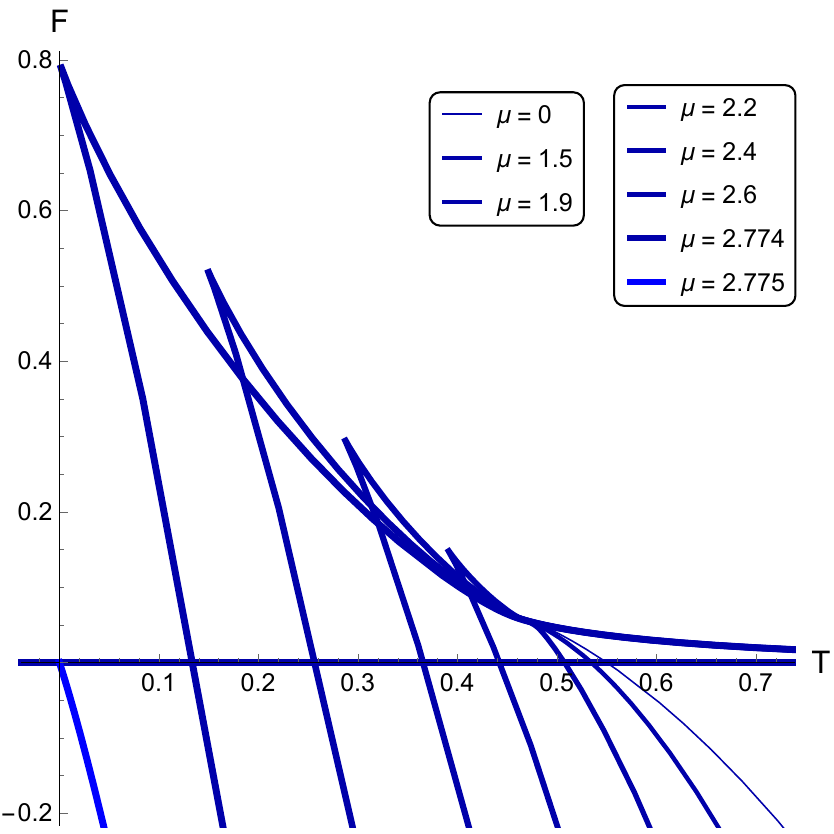} \\
  D \hspace{100pt} E \hspace{100pt} F
  \caption{Free energy $F(T,\mu)$ in magnetic field $q_3 = 0$ (A),
    $q_3 = 0.1$ (B), $q_3 = 0.5$ (C), $q_3 = 1$ (D), $q_3 = 2$ (E),
    $q_3 = 5$ (F); $\nu = 4.5$, $a = 0.15$, $c = 1.16$, $c_B = - \,
    0.01$, $d = 0.1$.}
  \label{Fig:FTmu-q3-nu45-d01-z5}
\end{figure}


\end{document}